\documentclass[iop]{emulateapj}
\usepackage{apjfonts}

\input epsf.tex

\begin{document}
\title{Stellar populations and evolution of early-type cluster galaxies: 
Constraints from optical imaging and spectroscopy of \lowercase{z}=0.5-0.9 galaxy clusters}
\author{Inger J{\o}rgensen, Kristin Chiboucas}
\affil{Gemini Observatory, 670 N.\ A`ohoku Pl., Hilo, HI 96720, USA}
\email{ijorgensen@gemini.edu, kchiboucas@gemini.edu}

\submitted{Accepted for publication in Astronomical Journal, January 9, 2013}

\begin{abstract}
We present an analysis of stellar populations and evolutionary history of galaxies in 
three similarly rich galaxy clusters MS0451.6-0305 ($z=0.54$), RXJ0152.7--1357 ($z=0.83$), and 
RXJ1226.9+3332 ($z=0.89$).
Our analysis is based on high signal-to-noise ground-based optical spectroscopy and 
Hubble Space Telescope imaging for a total of 17-34 members in each cluster. 
Using the dynamical masses together with the effective radii and the velocity dispersions we find
no indication of evolution of sizes or velocity dispersions with redshift at a given galaxy mass.
We establish the Fundamental Plane (FP) and scaling relations between absorption line indices
and velocity dispersions. 
We confirm that the FP is steeper at $z \approx 0.86$ compared to the low redshift FP,
indicating that under the assumption of passive evolution the formation redshift, $z_{\rm form}$, depends
on the galaxy velocity dispersion (or alternatively mass). 
At a velocity dispersion of $\sigma = \rm 125\,km\,s^{-1}$ (Mass = $10^{10.55} {\rm M}_\sun$)
we find $z_{\rm form} = 1.24 \pm 0.05$, while at $\sigma = \rm 225\,km\,s^{-1}$ (Mass = $10^{11.36} {\rm M}_\sun$) 
the formation redshift is $z_{\rm form} = 1.95_{-0.2}^{+0.3}$, for a Salpeter initial mass function.
The three clusters follow similar scaling relations between absorption line indices
and velocity dispersions as those found for low redshift galaxies. The zero point offsets
for the Balmer lines depend on cluster redshifts. However, the offsets indicate a 
slower evolution, and therefore higher formation redshift, than the zero point differences found from
the FP, if interpreting the data using a passive evolution model.
Specifically, the strength of the higher order Balmer lines H$\delta$ and H$\gamma$ implies
$z_{\rm form}>2.8$.
The scaling relations for the metal indices in general show small and in some cases 
insignificant zero point offsets, favoring high formation redshifts for a passive evolution model.
Based on the absorption line indices and recent stellar population models from Thomas et al.\
we find that MS0451.6-0305 has a mean metallicity [M/H] approximately 0.2 dex below 
that of the other clusters and our low redshift sample. We confirm our previous 
result that  RXJ0152.7-1357 has a mean abundance ratio $\rm [\alpha /Fe]$ approximately 
0.3 dex higher than that of the other clusters. 
The differences in [M/H] and $\rm [\alpha /Fe]$ between the high-redshift clusters and the low
redshift sample are inconsistent with a passive evolution scenario for early-type cluster 
galaxies over the redshift interval studied. 
Low-level star formation may be able to bring the metallicity of MS0451.6-0305 in agreement 
with the low redshift sample, while we speculate whether galaxy mergers can lead to 
sufficiently large changes in the abundance ratios for the RXJ0152.7-1357 galaxies to allow them
to reach the low redshift sample values in the time available.
\end{abstract}


\section{Introduction}

One of the main goals of studies of cluster galaxies is to establish their evolution 
in terms of the star formation history, the effect of mergers, and the changes in the morphologies
as a function of redshift and galaxy mass. 
Through these studies the aim is to understand how the galaxies evolve from the 
high redshift galaxies in proto clusters to the low redshift nearby galaxies as we 
observe them in rich clusters like the Coma and the Perseus cluster.

At the highest redshift the challenge is to find the proto clusters, and then study their properties 
and galaxy populations (see Hatch et al.\ (2011) for a 
recent survey aimed at identifying proto clusters). 
An important milestone in the evolution is when the red sequence of passively evolving galaxies 
is established and when the most massive galaxies are in place in the clusters, a process that
seems to take place between redshifts of $z=3$ and $z=2$ (e.g., Kodama et al.\ 2007; Zirm et al.\ 2008).
To follow the evolution over the last half of the age of the Universe,
the galaxies in rich clusters from redshift $z\approx 1$ to the present are studied in detail,
see e.g.\ J\o rgensen et al.\ (2005, 2006, 2007), Barr et al.\ (2005, 2006), van Dokkum \& van der Marel (2007), 
S\'{a}nchez-Bl\'{a}zquez et al.\ (2009), and Saglia et al.\ (2010), and references in these papers.
Finally, very detailed spatially resolved studies of the stellar populations and kinematics, which are
only possible for low redshift galaxies, provide the fingerprint of the evolutionary processes 
as they are seen at the present, e.g.\ Kuntschner et al.\ (2010), Cappellari et al.\ (2011).

Previous work focused on the studies of galaxy clusters between $z=1$ and the present has 
established that the Fundamental Plane (FP) zero point changes with redshift and found that the 
change for massive galaxies is in agreement with passive evolution, with a formation redshift $z_{\rm form} \approx 2$ or higher
(J\o rgensen et al.\ 2006, 2006; van Dokkum \& van der Marel 2007; Saglia et al.\ 2010, and references therein).
The formation redshift $z_{\rm form}$ should be understood as the approximate epoch of the last major star formation episode.
The FP is a tight empirical relation between the effective radius $r_e$, the mean surface brightness within that
radius $\langle I \rangle _e$, and the central velocity dispersion $\sigma$, linear in log-space
\begin{equation}
\log r_e = \alpha \log \sigma + \beta \log \langle I \rangle _e + \gamma
\end{equation}
(Dressler et al.\ 1987; Djorgovski et al.\ 1987; J\o rgensen et al.\ 1996).
The relation can be interpreted as a relation between galaxy masses and mass-to-light (M/L) ratios,
e.g.\ Faber et al.\ (1987)
\begin{equation}
\log M/L = \xi \log {\rm Mass} + \gamma '
\end{equation}
and can as such be used to study the evolution of galaxies as a function of redshift.
The galaxy dynamical masses are usually determined using the 
approximation ${\rm Mass} = 5 r_{\rm e} \sigma ^2 {\rm G^{-1}}$ (Bender et al.\ 1992).
Van Dokkum \& van der Marel (2007) summarized studies of the FP for intermediate redshift clusters
($z=0.2$ to $z\approx 1$) done prior to 2007 and used all available data
together. 
They concluded that for galaxies more massive than $\rm 10^{11} M_{\sun}$
the data are consistent with passive evolution with a formation redshift $z_{\rm form}=2.0$. 
More recently, Holden et al.\ (2010) focused on one cluster at $z=0.8$, while
Saglia et al.\ (2010) studied the FP for the large sample of ESO Distant Cluster Survey 
data (EDisCS).
Saglia et al.\ find some evidence of the FP coefficients changing with redshift, in agreement 
with our previous result for two $z=0.8-0.9$ clusters (J\o rgensen et al.\ 2006, 2007), while
Holden et al.\ do not find any significant difference between the FP slopes for their cluster sample
and the low redshift comparison sample. However, the Holden et al.\ sample becomes very sparse
at low masses and may therefore not be in contradiction with the results from Saglia et al.\ and 
J\o rgensen et al. Both Holden et al.\ and Saglia et al.\ find a zero point dependence with
redshift in agreement with the result from van Dokkum \& van der Marel (2007).

While the FP provides an important constraint for models for galaxy evolution, more detailed information 
regarding the stellar content and the star formation histories can potentially be extracted from 
measurements of the absorption line strengths.
Only few studies are available on the absorption line strengths of individual intermediate redshift 
galaxies. Kelson et al.\ (2001) studied the higher order Balmer lines for clusters between 
$z=0.8$ and the present and found that these are in agreement with passive evolution and
a formation redshift $z_{\rm form} >2.5$. This is a higher formation redshift than found from
recent investigations based on the FP. In a later study Kelson et al.\ (2006) used line strengths for
a large sample of early-type galaxies in a $z=0.33$ cluster to address the question 
of whether metallicity or abundance ratios vary with velocity dispersion. They find that both
the total metallicity and the nitrogen abundance (measured relative to 
the $\alpha$-element abundance) vary with velocity dispersion, while they also confirm 
that the galaxies are old.
S\'{a}nchez-Bl\'{a}zquez et al.\ (2009) used the EDisCS data, both stacked spectra
and line index measurements for individual galaxies, to investigate how
relations between line indices and velocity dispersions change with redshift. 
For galaxies with velocity dispersion above 175 $\rm km\,s^{-1}$ their data support a
formation redshift $z_{\rm form} > 1.4$, while the apparent lack of evolution for 
lower velocity dispersion galaxies compared to low redshift galaxies is best
explained by such galaxies entering the red sequence
in fairly large numbers (40 per cent of galaxies) at redshifts 0.75 to 0.45.
Similar conclusions have been reached by Bell et al.\ (2004), Brown et al.\ (2007) and 
Faber et al.\ (2007) based on studies of the luminosity functions of galaxies from $z$=1 to
the present. These authors find that the number density (or the total mass) on the
red sequence for galaxies at or below $L^{\star}$ has increased with a factor two since
$z$=1.
This leads to the question of how to select samples in the low redshift clusters that may be the
end-points of the galaxies in the intermediate redshift samples, or conversely how to correct
for this ``progenitor bias'' as discussed by van Dokkum \& Franx (2001). We return to this
question in the discussion (Section \ref{SEC-PASSIVE}).

The importance of the cluster environment for the galaxy properties and evolution has been
known and quantified since Dressler's (1980) classical study of the relation between
morphological types and cluster density. In intermediate redshift clusters a few recent
studies address the question of the cluster environment in connection with the 
star formation history.  Moran et al.\ (2005) present results on the FP and scaling relations between line indices
and velocity dispersions for a $z=0.4$ cluster.  
They find that the cluster center galaxies are older than those at larger cluster center distance
and that galaxies with very strong Balmer lines (for their velocity dispersion) are absent
in the very core of the cluster.
The role of cluster environment was also investigated by Demarco et al.\ (2010) using 
stacked spectra of galaxies in RXJ0152.7--1357 ($z=0.83$) finding support for the idea
that the dense cluster environment halts the star formation in the low mass galaxies as
they enter the cluster. Their data support a downsizing scenario in which the less massive galaxies
formed stars more recently than the more massive galaxies.

While many diverse studies have attempted to piece together a coherent picture of galaxy evolution
over the redshift range that covers the last half of the age of the Universe, the majority of the 
data sets are limited to a few clusters, do not reach high enough signal-to-noise (S/N) for the 
spectroscopy to allow individual galaxies to be studied, and/or cover only the more massive galaxies in the clusters.
Our project, the ``The Gemini/{\it HST} Galaxy Cluster Project'' (GCP), was designed to
overcome these issues, and the current paper is one in a series of papers from this project.
A detailed project description can be found in J\o rgensen et al.\ (2005).
Our emphasis is to investigate galaxy properties as a function of mass and redshift
based on high S/N ground-based spectroscopy and combined with imaging obtained
with either the Advanced Camera for Surveys (ACS) or the Wide Field Planetary Camera 2
on board the Hubble Space Telescope ({\it HST}).
The sample consists of 15 X-ray selected rich galaxy clusters covering 
a redshift interval from $z=0.15$ to $z=1.0$ and was originally selected using a 
lower limit on the X-ray luminosity of \mbox{$L_X(0.1-2.4{\rm keV}) = 2 \times 10^{44} {\rm ergs\,s^{-1}}$} 
for a cosmology with $q_0=0.5$ and $\rm H_0 = 50\,km\,s^{-1}\,Mpc^{-1}$.
Using the consistently calibrated X-ray data from Piffaretti et al.\ (2011) this 
limit corresponds to an X-ray luminosity in the 0.1-2.4 keV band 
within $R_{500}$ of $L_{500} = 10^{44} {\rm ergs\,s^{-1}}$ for a $\Lambda$CDM cosmology with 
$\rm H_0 = 70\,km\,s^{-1}\,Mpc^{-1}$, $\Omega_{\rm M}=0.3$, and $\Omega_{\rm \Lambda}=0.7$.
The characteristic radius $R_{500}$ is the radius within which the cluster over-density is 500 times the 
critical density at the cluster redshift. It is typically 1 Mpc for the clusters included in our sample.

For each cluster we obtain spectroscopy of 30-50 potential cluster members, usually resulting
in samples of 20 or more confirmed cluster members.
Our samples span from the highest mass galaxies at $\rm Mass \approx 10^{12.6} M_{\sun}$ 
down to galaxies with $\rm Mass = 10^{10.3} M_{\sun}$.
This is a factor five lower mass than reached by most similar studies of cluster galaxies
at $z>0.5$ and our samples in each cluster contain two to three times more galaxies than previous
cluster samples with this quality data at $z>0.5$, 
e.g.\ van Dokkum \& van der Marel (2007) and references herein.
We use line index measurements for the galaxies together
with velocity dispersions and two-dimensional photometry to establish scaling relations. 
The reader is referred to J\o rgensen et al.\ (2005) for 
a more complete description of the observing strategy of the project.

In this paper we use the ground-based spectroscopy and the {\it HST}/ACS imaging for the 
three massive galaxy clusters MS0451.6-0305 at $z=0.54$,
RXJ0152.7--1357 at $z=0.83$, and RXJ1226.9+3332 at $z=0.89$
to establish the FP and as well as scaling relations between absorption line strengths 
and velocity dispersion.
We then address the question of whether a passive evolution model with a formation redshift dependent
on the galaxy velocity dispersion (and the mass) can explain the data.
We establish the changes in ages, metallicity and abundance ratios as a function of redshifts
and compare the results with the passive evolution model.
Detailed discussion of models with ongoing star formation or merging in the redshift interval
covered by our data is beyond the scope of the present paper. We plan to return to this
topic in a future paper.
The full data set for the GCP is still in the process of being processed and analyzed.
We will present results for the full data set in future papers.

The reader mostly interested in the results, discussion and conclusion is advised to focus on 
Sections \ref{SEC-SCALINGREL}-\ref{SEC-CONCLUSION}.
Background information for the clusters can be found in Section \ref{SEC-CLUSTERS}.
The observational data are described in Sections \ref{SEC-DATAGMOS}-\ref{SEC-COMPDATA}, 
as well as in the Appendices. 
The adopted single stellar population (SSP) models and the passive evolution
model are covered in Section \ref{SEC-MODELS}.
In Section \ref{SEC-CLUSTERZ} we establish the cluster redshifts, the cluster membership, and the cluster velocity
dispersions. 
Section \ref{SEC-METHODSAMPLE} gives an overview over the methods used for the analysis and 
defines the final sample of galaxies used in the analysis.
The scaling relations are established in Section \ref{SEC-SCALINGREL}. 
In Section \ref{SEC-INDICES} we evaluate to what extent the SSP models reproduce the 
line index data and we derive distributions of the ages, metallicities [M/H] 
and abundance ratios $\rm [\alpha /Fe]$ for each of the clusters.
The evolution as a function of redshift and galaxy masses is established in Section \ref{SEC-EVOLUTION}.
In Section \ref{SEC-DISCUSSION} we discuss the results and evaluate different effects that
may lead to the measured changes with redshifts and galaxy masses.
The conclusions are summarized in Section \ref{SEC-CONCLUSION}.

Throughout this paper we adopt a $\Lambda$CDM cosmology with 
$\rm H_0 = 70\,km\,s^{-1}\,Mpc^{-1}$, $\Omega_{\rm M}=0.3$, and $\Omega_{\rm \Lambda}=0.7$.

\begin{deluxetable*}{llrrrrr}
\tablecaption{Cluster properties \label{tab-redshifts} }
\tablewidth{0pt}
\tablehead{
\colhead{Cluster} & \colhead{Redshift} & \colhead{$\sigma _{\rm cluster}$} & 
\colhead{$L_{500}$} & \colhead{$M_{500}$} & \colhead{$R_{500}$} & \colhead{N$_{\rm member}$} \\
 & & $\rm km~s^{-1}$ & $10^{44} \rm{erg\,s^{-1}}$ & $10^{14}M_{\sun}$ & Mpc \\
\colhead{(1)} & \colhead{(2)} & \colhead{(3)} & \colhead{(4)} & \colhead{(5)} & \colhead{(6)} & \colhead{(7)}
}
\startdata
Perseus = Abell 426\tablenotemark{a} & 0.0179 & $1277_{-78}^{+95}$ &   6.217 & 6.151 & 1.286 &  63 \\
Abell 194\tablenotemark{a,b}         & 0.0180 &  $480_{-38}^{+48}$ &   0.070 & 0.398 & 0.516 &  17 \\
Coma = Abell 1656\tablenotemark{a}   & 0.0231 & $1010_{-44}^{+51}$ &   3.456 & 4.285 & 1.138 & 116 \\
MS0451.6--0305\tablenotemark{c}    & $0.5398\pm 0.0010$ & $1450_{-159}^{+105}$ & 15.352 & 7.134 & 1.118 &  47 \\
RXJ0157.2--1357\tablenotemark{d}   & $0.8350\pm 0.0012$ & $1110_{-174}^{+147}$ &  6.291 & 3.222 & 0.763 &  29 \\
RXJ1226.9+3332\tablenotemark{c}    & $0.8908\pm 0.0011$ & $1298_{-137}^{+122}$ & 11.253 & 4.386 & 0.827 &  55 \\
\enddata
\tablecomments{Col.\ (1) Galaxy cluster; col.\ (2) cluster redshift; col.\ (3) cluster velocity dispersion;
col.\ (4) X-ray luminosity in the 0.1--2.4 keV band within the radius $R_{500}$, from Piffaretti et al.\ (2011); 
col.\ (5) cluster mass derived from X-ray data within the radius $R_{500}$, from Piffaretti et al.;
col.\ (6) radius within which the mean over-density of the cluster is 500 times the critical density at the 
cluster redshift, from Piffaretti et al.;
col.\ (7) Number of member galaxies for which spectroscopy is used in this paper.}
\tablenotetext{a}{Redshift and velocity dispersion from Zabludoff et al.\ (1990).}
\tablenotetext{b}{Abell 194 does not meet the X-ray luminosity selection criteria of the main cluster sample.}
\tablenotetext{c}{Redshifts and velocity dispersions from this paper.}
\tablenotetext{d}{Redshift and velocity dispersion from J\o rgensen.\ (2005). The velocity dispersions for the Northern and Southern sub-clusters are $(681 \pm 232)\, \rm {km\,s^{-1}}$ and 
$(866 \pm 266)\, \rm {km\,s^{-1}}$, respectively.}
\end{deluxetable*}

\begin{deluxetable}{lr}
\tablecaption{Gemini North Instrumentation \label{tab-inst} }
\tablewidth{230pt}
\tablehead{}
\startdata
Instrument      & GMOS-N       \\
CCDs            & 3 $\times$ E2V 2048$\times$4608 \\
r.o.n.\tablenotemark{a}          & (3.5,3.3,3.0) e$^-$       \\
gain\tablenotemark{a}            & (2.10,2.337,2.30) e$^-$/ADU  \\
Pixel scale     & 0.0727arcsec/pixel \\
Field of view   & $5\farcm5\times5\farcm5$ \\
Imaging filters & $g'$$r'$$i'$$z'$ \\
Grating         & R400\_G5305 \\
Spectroscopic filter & OG515\_G0306 \\
Wavelength range\tablenotemark{b} & 5000-10000\AA  \\
\enddata
\tablenotetext{a}{Values for the three detectors in the array.}
\tablenotetext{b}{The exact wavelength range varies from slitlet to slitlet.}
\end{deluxetable}

\begin{figure*}
\epsfxsize 17.5cm
\epsfbox{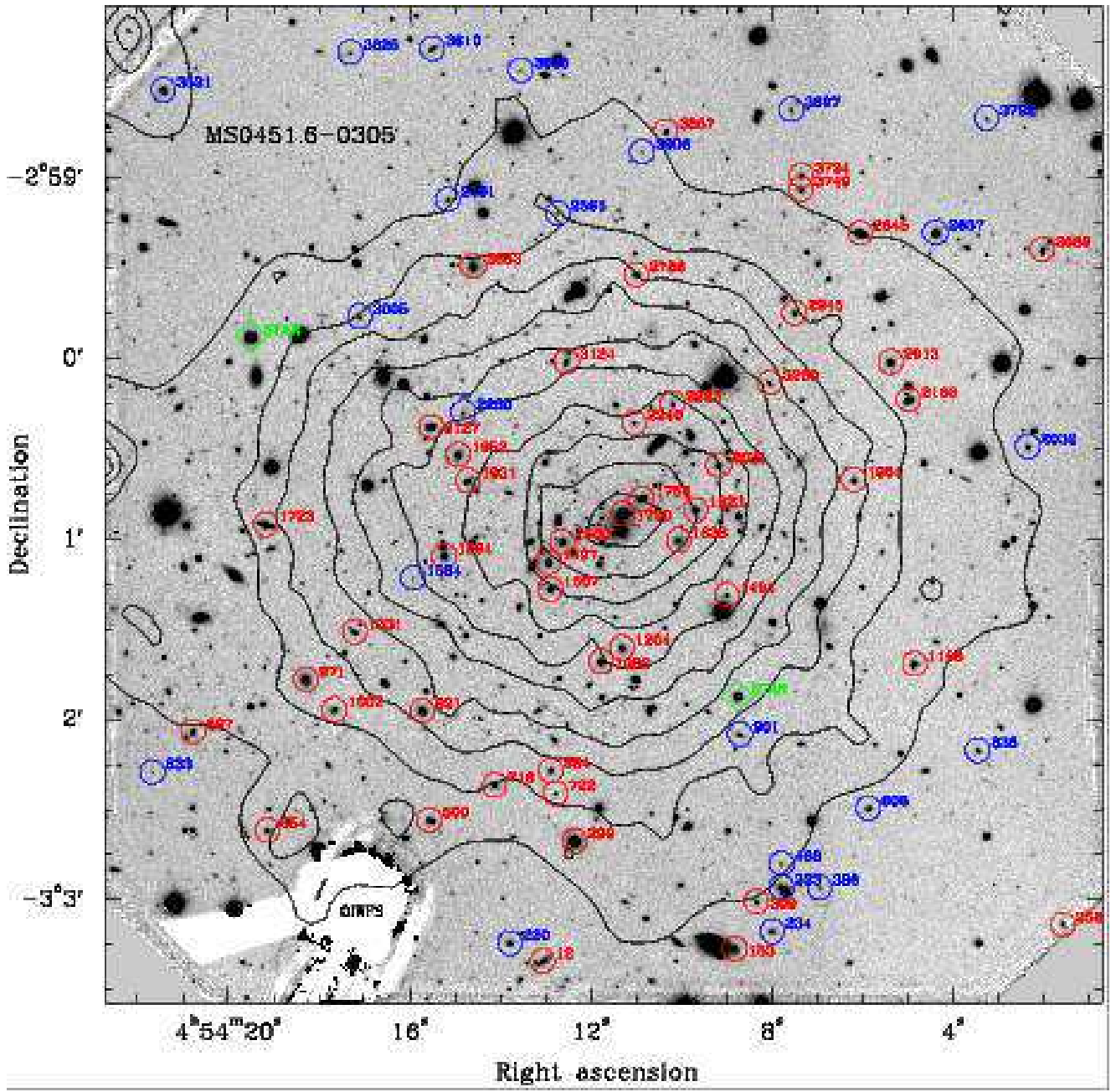}
\caption[]{
The GMOS-N $i'$-band image of MS0451.6--0305 with contours of the {\it XMM-Newton} data overlaid. 
The field covers approximately 5.5 arcmin $\times$ 5.5 arcmin.
Red circles -- confirmed cluster members labeled with their ID number;
blue circles -- non-members with spectroscopy labeled with their ID number;
green diamonds -- the two blue stars included in each mask to facilitate correction of the spectra 
for telluric absorption lines.
The X-ray image is the sum of the images from the two {\it XMM-Newton} EPIC-MOS cameras.
The X-ray image has been smoothed such that the structure seen is significant
at the 3 $\sigma$ level or higher. 
The spacing between the contours is logarithmic with a factor 1.5 between each contour.
The GMOS-N on-instrument wavefront sensor used for guiding vignets the field in the lower left, 
marked with ``OIWFS''.
\label{fig-MS0451grey} }
\end{figure*}

\begin{figure*}
\epsfxsize 17.5cm
\epsfbox{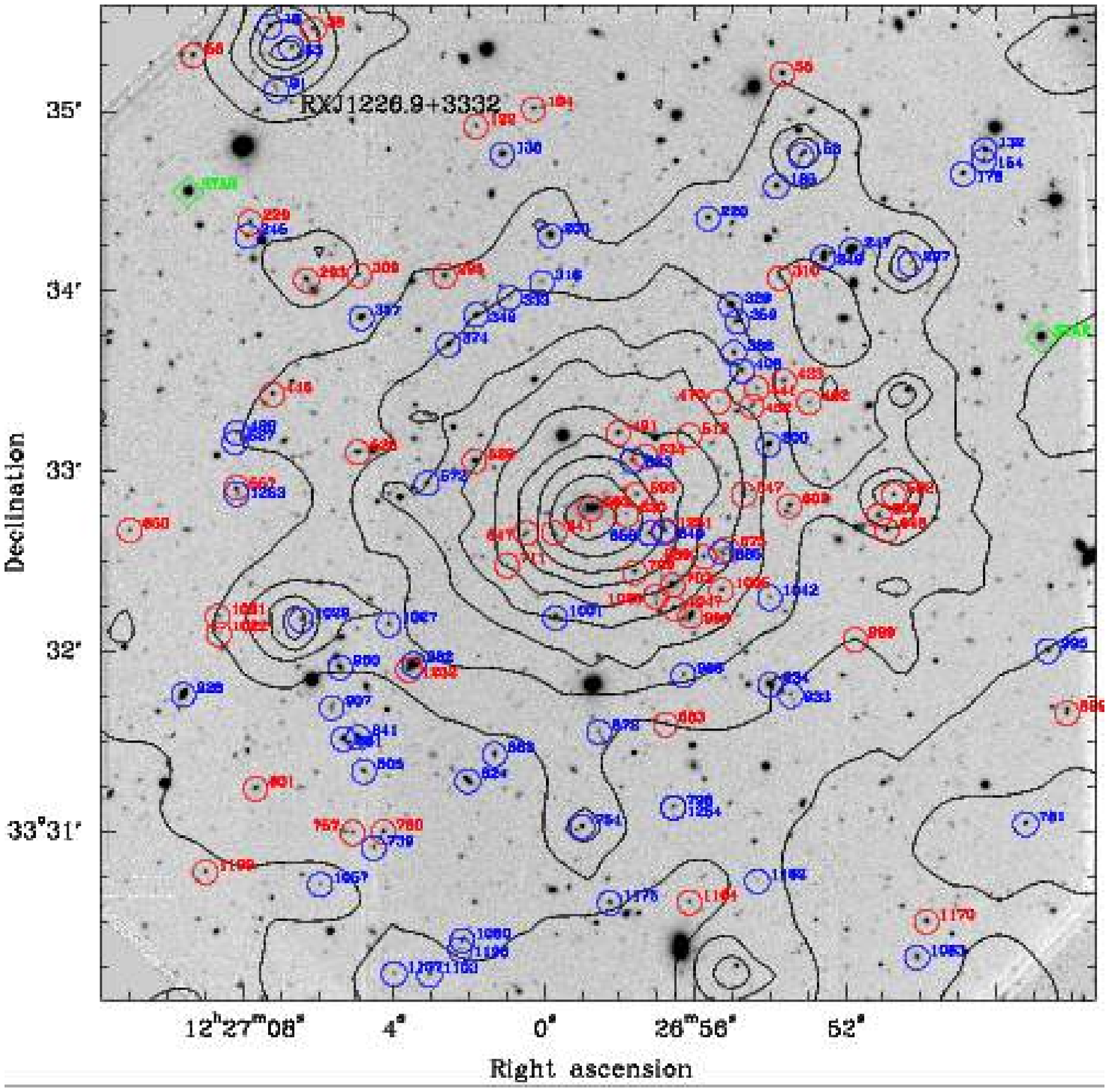}
\caption[]{
The GMOS-N $i'$-band image of RXJ1226.9+3332 with contours of the {\it XMM-Newton} data overlaid. 
The field covers approximately 5.5 arcmin $\times$ 5.5 arcmin.
Symbols and X-ray contours as on Fig. \ref{fig-MS0451grey}.
The structure in the X-ray emission seen in the outer part of the cluster can be associated with AGNs, 
either background (ID 53, 153, 754) or in the cluster (ID 592). 
The object ID 1029 is a blue point source, possible also 
a background AGN, though the spectrum has too low signal-to-noise to allow a determination of the redshift.
\label{fig-RXJ1226grey} }
\end{figure*}

\section{The three clusters: Background information \label{SEC-CLUSTERS}}

The three clusters MS0451.6-0305 ($z=0.54$), RXJ0152.7--1357 ($z=0.83$), and 
RXJ1226.9+3332 ($z=0.89$) are among the highest X-ray luminosity clusters in our sample. 
Table \ref{tab-redshifts} summarizes the cluster properties for the three clusters, as well
as the three low redshift comparison clusters, Perseus, Coma and Abell 194.

MS0451.6-0305 is the most X-ray luminous galaxy cluster included in
the Einstein Extended Medium Sensitivity Survey (Gioia \& Luppino 1994). 
In the MAssive Cluster Survey (MACS) that consists of the most X-ray luminous galaxy clusters 
from the {\it ROSAT} Bright Source Catalog (Ebeling et al.\ 2007) the cluster is among the 
most X-ray luminous at $z>0.5$.
Over the past 20 years this very rich cluster has been the topic of much research and
has been observed with several space borne observatories 
(e.g.\ {\it HST}/WFPC2, {\it HST}/ACS, {\it GALEX}, {\it Chandra}, {\it XMM}-Newton).
The cluster was included in the Canadian Network for Observation Cosmology (CNOC) survey,
which contains redshifts for 46 cluster members. The velocity dispersion of $1330\,{\rm km\,s^{-1}}$ was
the largest found for clusters in the survey (Ellingson et al.\ 1998; Borgani et al.\ 1999).
The cluster has been the target for a wide field {\it HST}/ACS imaging survey, which was used
by Moran et al.\ (2007ab) together with optical spectroscopy to study both the morphological evolution
and the star formation history of the member galaxies. Of most importance to our study of this cluster
is that Moran et al.\ (2007b) found that the early-type galaxies are best modeled with a truncated
star formation history where star formation has stopped 5 Gyr prior. With our adopted
cosmology that corresponds to a formation redshift of $z_{\rm form} \approx 2$.
Figure \ref{fig-MS0451grey} shows the {\it XMM-Newton} data together with our $i'$-band 
images of MS0451.6--0305. The X-ray surface brightness map supports that the
cluster is a relaxed structure with no significant substructure. However, gravitational
lensing studies (Comerford et al.\ 2010) and higher spatial resolution X-ray data from 
{\it Chandra} that shows the brightest cluster galaxy slightly offset from the peak
X-ray emission (Borys et al.\ 2004) both indicate that the cluster ought to be regarded as unrelaxed.

The massive cluster of galaxies RXJ0152.7--1357 was discovered from {\it ROSAT}
data by three different surveys: The {\it ROSAT} Deep Cluster Survey (RDCS) and the Wide Angle
{\it ROSAT} Pointed Survey (WARPS) (see Ebeling et al.\ 2000),
as well as the Bright Serendipitous High-Redshift Archival Cluster (SHARC) survey 
(Nichol et al.\ 1999).
X-ray observations from {\it XMM-Newton} and {\it Chandra} (Jones et al.\ 2004; 
Maughan et al.\ 2003) show that the cluster consists of two sub-clumps and support the view that 
RXJ0152.7--1357 is in the process of merging from two clumps of roughly equal mass. 
See J\o rgensen et al.\ (2005) for a figure of the {\it XMM-Newton} data overlayed 
our $i'$-band imaging data.
Demarco et al.\ (2010) used spectroscopic and photometric data to study the star formation
history in RXJ0152.7--1357 and found that the data support a downsizing scenario in which
there is a $\approx 1.5$ Gyr age difference between the older high mass (passive) galaxies 
(Mass $> 10^{10.9} M_{\sun}$)
and the younger low mass galaxies (Mass $\rm < 10^{10.4} M_{\sun}$), and the low mass
galaxies just in the last 1 Gyr stopped forming stars. For our adopted cosmology, this is equivalent
to formation redshifts of $z_{\rm form}=1.6$ and 1.1 for the high mass and low mass galaxies,
respectively.

RXJ1226.9+3332 is also one of the most X-ray luminous clusters at $z>0.5$.
The cluster as discovered in the WARPS survey (Ebeling et al.\ 2001).
Figure \ref{fig-RXJ1226grey} shows the {\it XMM-Newton} data together with our $i'$-band imaging.
The sub-structures in X-ray surface brightness for RXJ1226.9+3332 are associated with AGNs, 
see Section \ref{SEC-CLUSTERZ}, and the X-ray structure of the cluster is that of a relaxed structure.  
However, from the temperature map of the X-ray gas based on {\it XMM-Newton} data Maughan et al.\ (2007) find 
evidence for a recent merger event as an off-center sub-clump of X-ray gas that shows 
higher X-ray temperature and is associated with a local over-density in the galaxy distribution
about 45 arc seconds to the south west of the cluster center.

Compared to our chosen low redshift rich clusters of galaxies, Coma and Perseus, RXJ0152.7--1357 
is similar in X-ray luminosity to Perseus, while MS0451.6-0305 and RXJ1226.9+3332 have X-ray
luminosities 2-2.5 times that of Perseus. The X-ray luminosity of Coma is significantly
lower, being about 55 percent of that of Perseus (see Table \ref{tab-redshifts}).
As summarized above, substructure is also present to varying degree in the intermediate redshift clusters.
Thus, while all five clusters are very massive, environmental differences may still play a role
in the evolution of their galaxy population.

\section{Observational data - ground-based \label{SEC-DATAGMOS}}

Imaging and spectroscopy of the three clusters were obtained for the GCP with the 
Gemini Multi-Object Spectrograph on Gemini North (GMOS-N).
See Hook et al.\ (2004) for a detailed description of GMOS-N.
The instrument information is listed in Table \ref{tab-inst}, while
the Gemini program IDs and observing dates are given in Table \ref{tab-GMOSobs}.
The observations are summarized in Tables \ref{tab-imdata} and \ref{tab-spdata}.
The details of the observations of RXJ0152.7--1357 are given in J\o rgensen et al.\ (2005). However,
for completeness this cluster is also included in the summary tables in the current paper, 
with the data reproduced from J\o rgensen et al.
All programs executed in Director's Discretionary time or in the usual queue were obtained specifically for GCP.
These data and the data obtained in engineering time are in the following referred to as the ``GCP data''.

In addition we use public spectroscopic data for 
RXJ1226.9+3332 from programs GN-2003A-C-1 and GN-2004A-C-8 available from the 
Gemini Science Archive (GSA), see Table \ref{tab-GMOSobs}.
These data are referred to as the ``GSA data''. The imaging data from these programs 
were not used in the present paper.

\begin{deluxetable*}{lllll}
\tablecaption{GMOS-N Observations\label{tab-GMOSobs} }
\tablewidth{0pt}
\tablehead{
\colhead{Cluster} & \colhead{Program ID} & \colhead{Dates [UT]} & \colhead{Data type} & \colhead{Program type} } \\
\startdata
MS0451.6--0305  & GN-2001B-DD-3  & 2001 Dec 25 & imaging & DD\tablenotemark{a} \\
                & GN-2002B-Q-29  & 2002 Sep 12 to 2002 Sept 16 & imaging & queue \\
                & GN-2003B-Q-21  & 2003 Dec 24 & imaging & queue \\
                & GN-2002B-DD-4  & 2002 Dec 31 to 2003 Jan 2 & spectroscopy & DD in queue \\
                & GN-2003B-DD-3  & 2003 Dec 19 to 2003 Dec 23 & spectroscopy & DD in queue \\
RXJ1226.9+3332  & GN-2003A-DD-4  & 2003 January 31 to 2003 May 6 & imaging & DD in queue \\
                & GN-2003A-SV-80 & 2003 March 13 & imaging & engineering \\
                & GN-2004A-Q-45  & 2004 February 17 to UT 2004 July 20 & spectroscopy & queue \\
                & GN-2003A-C-1   & 2003 April 29 to 2003 May 1 & spectroscopy & classical \\
                & GN-2004A-C-8   & 2004 March 17 to 2004 March 18 & spectroscopy & classical \\
\enddata
\tablenotetext{a}{Director's Discretionary time.}
\end{deluxetable*}

\begin{deluxetable*}{lllrccc}
\tablecaption{GMOS-N Imaging Data \label{tab-imdata} }
\tablewidth{0pc}
\tablehead{
\colhead{Cluster} & \colhead{Program ID} & \colhead{Filter} & \colhead{Exposure time} & \colhead{FWHM\tablenotemark{a}} & \colhead{Sky brightness} & \colhead{Galactic extinction} \\
\colhead{} & \colhead{} & \colhead{} & \colhead{} & \colhead{(arcsec)} & \colhead{(mag arcsec$^{-2}$)} & \colhead{(mag)} }
\startdata
MS0451.6--0305  
& GN-2003B-Q-21 & $g'$     & 6 $\times$ 600sec   & 0.80 & 22.28 & 0.127 \\
& GN-2002B-Q-29 & $r'$     & 15 $\times$ 600sec (2 frames in twilight)  & 0.57 & 20.87 & 0.099 \\
& GN-2001B-DD-3,2B-Q-29 & $i'$     & 6 $\times$ 600sec (dark sky) & \\
&&                                 & + 2 $\times$ 300sec (grey sky) & 0.71 & 18.43 & 0.078 \\
& GN-2001B-DD-3 & $z'$     & 19 $\times$ 600sec (bright sky)         & 0.72 & 18.52 & 0.064 \\
RXJ0152.7--1357 & GN-2002B-Q-29,SV-90
 & $r'$     & 12 $\times$ 600sec           & 0.68 & 20.65 & 0.042\\
&& $i'$     & 7 $\times$ 450sec (dark sky) & 0.56 & 19.63 & 0.033 \\
&&          & + 100 $\times$ 120sec (bright sky) & \\
&& $z'$     & 13 $\times$ 450sec (dark sky) & 0.59 & 19.16 & 0.027 \\
&&          & + 14 $\times$ 450sec (bright sky) & \\ 
RXJ1226.9+3332 & GN-2003A-DD-4,SV-80
 & $r'$     & 9 $\times$ 600sec                     & 0.75 & 21.30 & 0.056 \\
&& $i'$     & 7 $\times$ 300sec + 3 $\times$ 360sec & 0.78 & 20.58 & 0.044 \\
&& $z'$     & 29 $\times$ 120sec (grey sky)         & 0.68 & 19.90 & 0.036 \\
\enddata
\tablenotetext{a}{Image quality measured as the average FWHM of 7-10 stars in the field from the final stacked images.}
\end{deluxetable*}

\begin{deluxetable*}{llrrrrrrr}
\tablecaption{GMOS-N Spectroscopic Data \label{tab-spdata} }
\tablewidth{0pc}
\tablehead{
\colhead{Cluster} & \colhead{Program ID} & \colhead{Exposure time} & \colhead{$N_{\rm exp}$\tablenotemark{a}} & \colhead{FWHM\tablenotemark{b}} & \colhead{$\sigma _{\rm inst}$\tablenotemark{c}} & \colhead{Aperture\tablenotemark{d}} & \colhead{Slit lengths} & \colhead{S/N\tablenotemark{e}} \\ 
\colhead{}        & \colhead{} & \colhead{} & \colhead{} & \colhead{(arcsec)} & \colhead{} & \colhead{(arcsec)} & \colhead{(arcsec)} }
\startdata
MS0451.6--0305 & GN-2002B-DD-4, \\
&  GN-2003B-DD-3 & 40,500 sec +32,400 sec & 27 & 0.78 & 3.172\AA, 143 $\rm km\,s^{-1}$ & $1\times 1.40$, 0.68 & 4.1--13.5 & 77\\
RXJ0152.7--1357 & GN-2002B-Q-29 & 77,960 sec & 25 & 0.65 & 3.065\AA, 116 $\rm km\,s^{-1}$ & $1\times 1.15$, 0.62 & 5--14 & 31\\
RXJ1226.9+3332  & GN-2004A-Q-45 & 64,800 sec + 64,800 sec & 72  & 0.68 &  3.076\AA, 113 $\rm km\,s^{-1}$ & $1\times 0.85$, 0.53 & 2.75 & 48\\
RXJ1226.9+3332  & GN-2003A-C-1 & 12,000 sec + 12,000 sec & 10  & 0.66 &  3.856\AA, 142 $\rm km\,s^{-1}$ & $1.25\times 0.85$, 0.60 & 2.5--14 & 29 \\
RXJ1226.9+3332  & GN-2004A-C-8 & 7,200 sec & 4\,\tablenotemark{f} & 0.65  & 3.856\AA, 142 $\rm km\,s^{-1}$ & $1.25\times 0.85$, 0.60 & 2.5--14 & 15 \\
\enddata
\tablenotetext{a}{Number of individual exposures.}
\tablenotetext{b}{Image quality measured as the average FWHM at 8000{\AA} of either the blue stars included in the masks (GN-2002B-DD-4, GN-2002B-Q-29, GN-2003B-DD-3, GN-2003B-Q-21 and GN-2004A-Q-45) or of the QSOs/AGNs included in the masks (GN-2004A-C-8). For GN-2003A-C-1 the FWHM is measured at 7930{\AA} from the one QSO included in the mask.}
\tablenotetext{c}{Median instrumental resolution derived as sigma in Gaussian fits to the sky lines of the stacked 
spectra. The second entry is the equivalent resolution in $\rm km\,s^{-1}$ at 4300{\AA} in the rest frame of the clusters.}  
\tablenotetext{d}{The first entry is the rectangular extraction aperture 
(slit width $\times$ extraction length). The second entry is the radius in an equivalent 
circular aperture, $r_{\rm ap}= 1.025 (\rm {length} \times \rm{width} / \pi)^{1/2}$, cf.\ J\o rgensen et al.\ (1995b).}
\tablenotetext{e}{Median S/N per {\AA}ngstrom in the rest frame of the cluster. }
\tablenotetext{f}{A fifth exposure was taken but contains no significant signal from the science targets.}
\end{deluxetable*}

The imaging for each cluster covers one GMOS-N field, which is approximately 5.5\,arcmin $\times$ 5.5\,arcmin.
All spectroscopic observations were obtained using GMOS-N in the multi-object spectroscopic mode.
One GMOS mask was used for RXJ0152.7--1357 (J\o rgensen et al.\ 2005), while for
MS0451.6-0305 and RXJ1226.9+3332 observations were obtained using two GMOS masks in 
order to cover more objects, while including the faintest in both masks.
The GSA data was obtained with three masks for RXJ1226.9+3332.
There is some overlap in spectroscopic targets between the GPC data and the GSA data. 
We use these targets to ensure consistent calibration of the velocity dispersions and 
line indices as described in the Appendix \ref{SEC-SPECPARAM}.
However, in the analysis we give our higher S/N GCP data preference over the GSA data for 
targets in common.

All spectroscopic observations for the GCP used the R400 grating and a slit width of 1 arcsec,
while the GSA data were obtained with R400 and a slit width of 1.25 arcsec. 
The resulting instrumental resolution in the rest frames of the clusters
are listed in Table \ref{tab-spdata}. For program GN-2004A-Q-45 we used
the nod-and-shuffle mode of GMOS-N. All other spectroscopic data were taken in the conventional mode.

For the cluster members the S/N per {\AA}ngstrom in the rest frame of the galaxies was derived from the
rest frame wavelength interval 4100-4600 {\AA} for RXJ0152.7--1357 and RXJ1226.9+3332 
and from 4100-5500 {\AA} for  MS0451.6--0305. The median S/N for spectra in each of the clusters is 
listed in Table \ref{tab-spdata}.
The S/N for the individual galaxies in MS0451.6--0305 and RXJ1226.9+3332 are listed in 
Tables \ref{tab-MS0451kin} and \ref{tab-RXJ1226kin} in the Appendix \ref{SEC-SPECTROSCOPY}. 
The information for RXJ0152.7--1357 can be found in J\o rgensen et al.\ (2005).

\subsection{Imaging}

The GMOS-N imaging data for MS0451.6--0305 and RXJ1226.9+3332 were reduced and co-added using the same methods as for the 
RXJ0152.7--1357 imaging data (J\o rgensen et al.\ 2005). One co-added image was produced
for each filter and field. The co-added images were then processed with SExtractor v.2.1.6
(Bertin \& Arnouts 1996) as described in J\o rgensen et al.

The photometry was standard calibrated using the magnitude zero points and color terms
derived in J\o rgensen (2009). The absolute accuracy of the standard calibration is
expected to be 0.035 to 0.05 mag, as described in that paper.
Tables \ref{tab-photdataMS0451} and \ref{tab-photdataRXJ1226} list the photometry for 
MS0451.6--0305 and RXJ1226.9+3332 calibrated to the Sloan Digital Sky Survey (SDSS) system. 
Similar photometry for RXJ0152.7--1357 can be found in J\o rgensen et al.\ (2005). 
The $r'$-band imaging for MS0451.6--0305 was obtained in significantly better seeing than
the other bands, and therefore is significantly deeper. Object detection was done in
the $r'$-band. Due to the difference in seeing the colors involving the $r'$ band are expected
to be affected by a systematic error of $\approx 0.05 $ mag, with the galaxies being too blue.
As this systematic error has no significant effect on our analysis,
we have chosen not correct the colors for the difference in seeing.

The Galactic extinction in the direction of the three clusters is $A_B=0.14$, 0.064, and 0.085
for MS0451.6--0305, RXJ0152.7--1357, and RXJ1226.9+3332, respectively (Schlegel et al.\ 1998).
Using the effective wavelength of the filters used for the photometry and the calibration
from Cardelli et al.\ (1989) we derive the extinction on those filters, see Table \ref{tab-imdata}.
The photometry in Tables \ref{tab-photdataMS0451} and \ref{tab-photdataRXJ1226} 
have not been corrected for the Galactic extinction.

We have compared the GMOS-N photometry with available literature data for the fields, details can be found
in Appendix \ref{SEC-IMAGING}. 
Based on these comparisons we conclude that the photometry is calibrated
to an external accuracy of 0.05 mag.

\begin{figure*}
\epsfxsize 16.5cm
\epsfbox{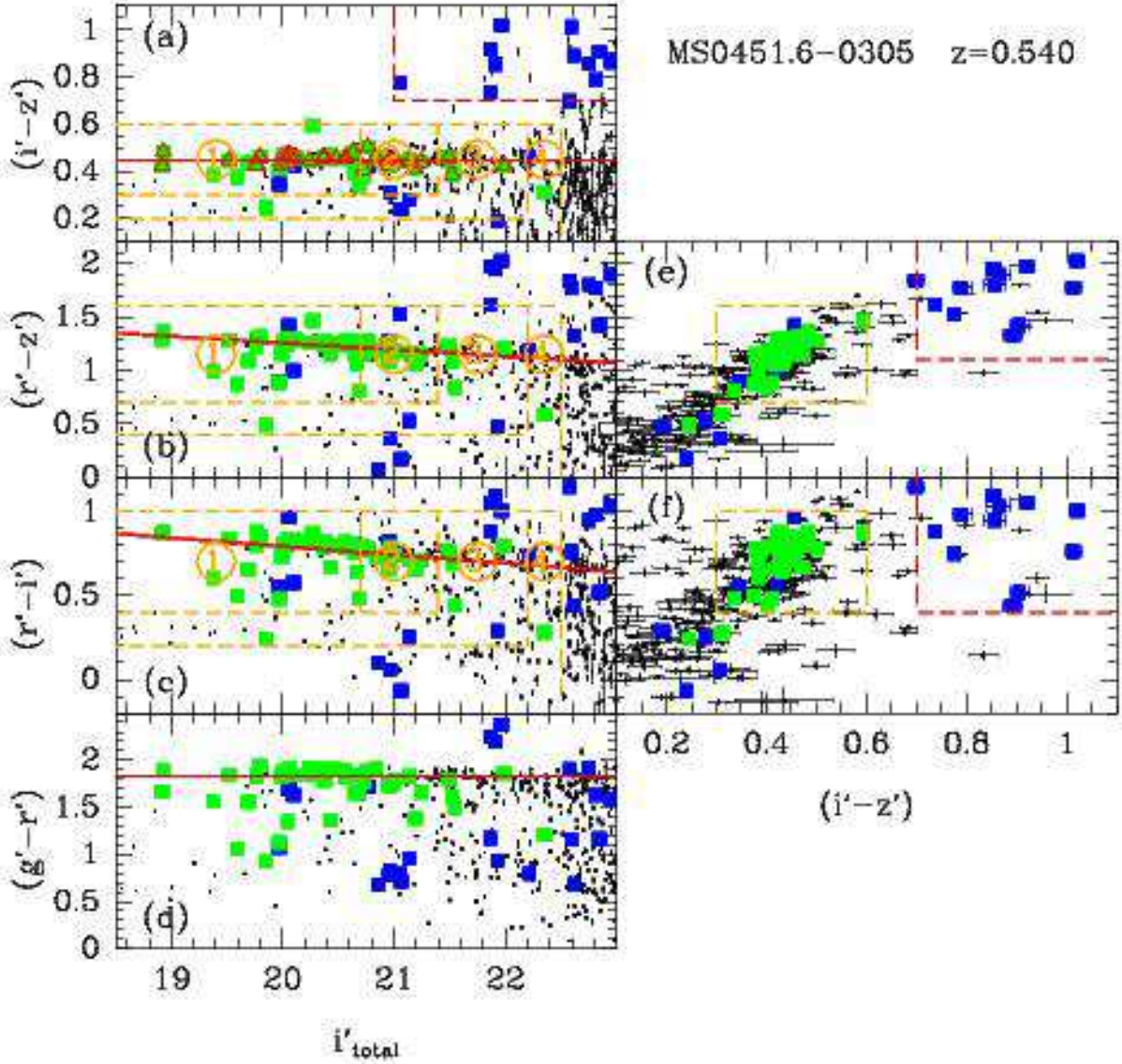}
\caption[]{MS0451.6-0305: Color-magnitude and color-color diagrams. Only galaxies
({\it class\_star} $<$ 0.80 in the $i'$-filter) with $i'\le 23$ mag are shown.
The magnitudes are the total magnitudes, while all colors are aperture colors.
The photometry has been corrected for the Galactic extinction, see Section 3.1.
Green filled boxes -- confirmed cluster members in our spectroscopic sample;
blue filled boxes -- non-members with spectroscopy;
small black points -- galaxies without spectroscopy.
Galaxies included in the analysis are on panel (a) shown with red open triangles over-plotted the green boxes.
Red lines -- the red sequence. The slope for the
color-magnitude relation in $(i'-z')$ is not significantly different from zero.
Thus, the line marks the median color.
The orange dashed lines and circled numbers show the object classes, see Section 3.2.
The orange dashed lines on panels (e) and (f) outline the sample limits in the colors for objects
in classes 1 and 2, see Section 3.2.
The red dashed lines on panels (a), (e) and (f) outline the sample limits for the background
galaxy sample described in Section 3.2.
\label{fig-CMMS0451} }
\end{figure*}

\begin{figure*}
\epsfxsize 16.5cm
\epsfbox{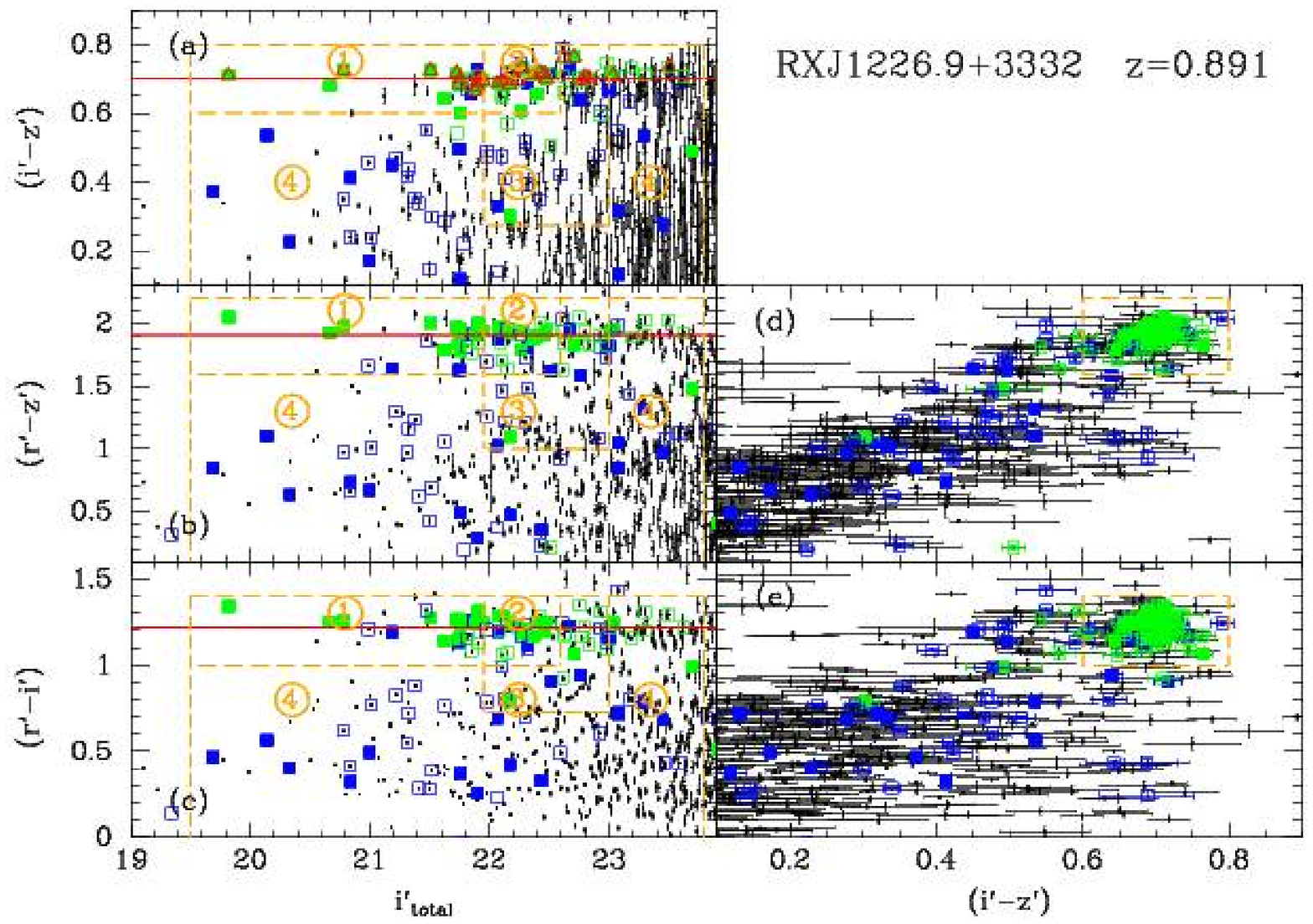}
\caption[]{RXJ1226.9+3332: Color-magnitude and color-color diagrams. Only galaxies
({\it class\_star} $<$ 0.80 in the $i'$-filter) with $i'\le 23.9$ mag are shown.
The magnitudes are the total magnitudes, while all colors are aperture colors.
The photometry has been corrected for the Galactic extinction, see Section 3.1.
Green filled boxes -- confirmed cluster members in our spectroscopic sample;
blue filled boxes -- non-members with spectroscopy;
green open boxes -- confirmed cluster members from the GSA data;
blue open boxes -- non-members with spectroscopy from the GSA data; 
small black points -- galaxies without spectroscopy.
Galaxies included in the analysis are on panel (a) shown with red open triangles over-plotted the green boxes.
Red lines -- the median colors for the red sequence. The slopes for the
color-magnitude relations are not significantly different from zero.
The orange dashed lines and circled numbers show the object classes, see Section 3.2.
The orange dashed lines on panels (d) and (e) outline the sample limits in the colors for objects
in classes 1 and 2, see Section 3.2.
\label{fig-CMRXJ1226} }
\end{figure*}

\subsection{Spectroscopy \label{SEC-SPECSEL}}

The principles of the spectroscopic sample selection for the project are 
described in J\o rgensen et al.\ (2005). In brief, sample selection is 
based on the GMOS photometry.  Stars and galaxies are separated using the 
SExtractor classification parameter {\it class\_star} derived from 
the image in the $i'$-filter.
For the purpose of selecting targets for the spectroscopic observations
we choose a threshold of 0.80, i.e., objects with {\it class\_star}$<$0.80
in the $i'$-image are considered galaxies.
In practice this excludes galaxies with effective radius $r_{\rm e} \le 0.15$ arcsec
($r_{\rm e} \le 1$ kpc) .
However, it was more important for the planning of our observations that the
selected spectroscopic targets had a high probability of being galaxies
rather than foreground stars.
Then we use the total magnitude in $i'$ and the available colors to define
four classes of objects in each field. The aim is to use a color selection that includes
likely cluster members.
Table \ref{tab-spsel} summarizes the definition of these four classes
for each of the three clusters.
The classes are shown on Figures \ref{fig-CMMS0451} and \ref{fig-CMRXJ1226}.
For RXJ1226.9+3332 the $(i'-z')$ colors are used as the primary color selection,
while for MS0451.6--0305 both $(r'-z')$ and $(i'-z')$ colors are used.
The classes are used in the selection of the spectroscopic sample as described in 
J\o rgensen et al.\ (2005). Objects in classes 1 and 2 are likely to be cluster members.
For each field roughly equal number of objects from each of these
two classes are included. 
Empty spaces in the mask designs are filled with class 3 objects
if possible. This class may include blue cluster members or cluster members too faint
to derive all spectroscopic parameters due to the resulting S/N of the spectroscopic
observations. Any remaining space in the mask designs is filled with class 4
objects. The brighter and bluer objects in this class are in general not expected 
to be members of the cluster, while the fainter and redder objects in this class
may be faint cluster members.
The selection criteria for galaxies included in the analysis is described in Section \ref{SEC-METHODSAMPLE}.
Here we note that any observed galaxy for which the properties meet the final selection criteria
is included in the analysis, independent of its class in this initial selection used
for the spectroscopic observations. We have visualized this by marking galaxies included
in the analysis with red triangles on Figures \ref{fig-CMMS0451}a and \ref{fig-CMRXJ1226}.

Redshift data for MS0451.6--0305 from the CNOC survey 
(Ellingson et al.\ 1998) were used to optimize inclusion of cluster members. 
In the mask design galaxies that are known cluster 
members based on these data were given preference over galaxies without redshift information.
Known non-members were assigned to class 4 and only included in the mask if no object from
the other classes could be used to fill the available space.

The masks for MS0451.6--0305 also include 12 objects selected to have $(i'-z') \ge 0.7$
and $(r'-z') \ge 1.1$
and therefore be candidate background galaxies with redshifts larger than about 0.75 and
passive stellar populations.
These are enclosed by the red box on Figure \ref{fig-CMMS0451}a, e and f, and not discussed
further in this paper though we do derive and list their redshifts.

The sample selection for the GSA data used for RXJ1226.9+3332 is not known.
We note that in addition to cluster members on the red sequence, the data 
contain a number of bluer member and non-member galaxies,
as well as some active-galactic nuclei (AGN).

\begin{deluxetable*}{lcl}
\tablecaption{Selection criteria for spectroscopic samples \label{tab-spsel} }
\tablewidth{0pc}
\tablehead{
\colhead{Cluster} & \colhead{Obj.Class} & \colhead{Selection criteria} }
\startdata
MS0451.6--0305\tablenotemark{a} & 
1 & $18 \le i' \le 20.7 ~ \wedge ~ 0.3 \le (i'-z') \le 0.6 ~ \wedge ~ 0.7 \le (r'-z') \le 1.6 $ \\
& 2 & $20.7 < i' \le 21.4 ~ \wedge ~ 0.3 \le (i'-z') \le 0.6 ~ \wedge ~ 0.7 \le (r'-z') \le 1.6 $  \\
& 3 & $(18.0 \le i' \le 21.4 ~ \wedge ~ 0.2 \le (i'-z') < 0.3 ~ \wedge ~ 0.4 \le (r'-z') < 0.7 ) ~ \vee ~$ \\
&   & $(21.4 \le i' \le 22.2 ~ \wedge ~ 0.2 \le (i'-z') \le 0.6 ~ \wedge ~ 0.4 \le (r'-z') \le 1.6)$ \\
 & 4 & $(18.0 \le i' \le 22.2 ~ \wedge ~ (i'-z') < 0.2) ~ \vee ~ (22.2 < i' \le 22.5 ~ \wedge ~ (i'-z') < 0.6)$ \\
RXJ0152.7--1357\tablenotemark{b} & 1 & $20.35 \le i' \le 21.35 ~ \wedge ~ (i'-z')\ge 0.7$ \\
 & 2 & $21.35 < i' \le 22.35 ~ \wedge ~ (i'-z')\ge 0.7$ \\
 & 3 & $20.35 \le i' \le 22.35 ~ \wedge ~ 0.45 \le (i'-z') < 0.7$ \\
 & 4 & $(i'<20.35 ~ \wedge ~ ~ 0.45 \le (i'-z') < 0.7) \vee ~ (i'>22.35 ~ \wedge ~ (i'-z')\ge 0.7)$ \\
RXJ1226.9+3332  & 1 &  $19.5 \le i' \le 21.9 ~ \wedge ~ 0.6 \le (i'-z')\le 0.8 ~ \wedge ~ 1.0 \le (r'-i')\le 1.4$  \\
 & 2 & $21.9 < i' \le 22.6 ~ \wedge ~ 0.6 \le (i'-z')\le 0.8 ~ \wedge ~ 1.0 \le (r'-i')\le 1.4$  \\
 & 3 & $(21.9 \le i' \le 22.6 ~ \wedge ~ 0.275 \le (i'-z') \le 0.6) ~ \vee ~$ \\
 &   & $(22.6 \le i' \le 23.0 ~ \wedge ~ 0.275 \le (i'-z') \le 0.8)$ \\
 & 4 & $(19.5 \le i' \le 21.9 ~ \wedge ~ (i'-z') < 0.6) ~ \vee ~ (21.9 < i' \le 23.0 ~ \wedge ~ (i'-z') < 0.275) ~ \vee ~$ \\
 &   & $(23.0 < i' \le 23.8 ~ \wedge ~ (i'-z') \le 0.8)$ \\
\enddata
\tablenotetext{a}{At the time of the sample selection for MS0451.6--0305 only $r'$, $i'$ and $z'$ 
imaging was available}
\tablenotetext{b}{Selection criteria for RXJ0152.7--1357 reproduced from J\o rgensen et al.\ (2005).}
\end{deluxetable*}

The spectroscopic samples are marked on  Figures \ref{fig-MS0451grey} and \ref{fig-RXJ1226grey}.
For each mask in the GCP observations, two blue stars were included in order to obtain a good correction
for the telluric absorption lines. The blue stars are also marked on figures.


The MS0451.6-0305 spectroscopic data were reduced in the same way as done for the 
RXJ0152.7--1357 data (J\o rgensen et al.\ 2005),
The only difference was that we applied our established correction for the charge 
diffusion in the red as described in Appendix \ref{SEC-SPECTROSCOPY}. 
The correction slightly improves the sky subtraction for these short slits.

The RXJ1226.9+3332 GSA data were reduced in a similar way, with the exception that for some 
of the very short slitlets with the objects very close to the end of the slitlet, 
sky subtraction was possible only after the wavelength calibration had been applied.
This in general leads to poorer sky subtraction as the (strong) sky-lines are being 
interpolated onto the rectified wavelength scale before the sky signal is subtracted out.

The RXJ1226.9+3332 GCP data were obtained using GMOS-N in the nod-and-shuffle mode 
(Glazebrook \& Bland-Hawthorn 2001).
The reduction of these data involved construction of special flat fields that take into
account the shuffling of the data on the detector array, as well as a correction for 
the effect of charge diffusion in the far red wavelength region. 

The details of the reductions of all the spectroscopic data are described in Appendix \ref{SEC-SPECTROSCOPY}. 
The final data products are cleaned and averaged spectra that
have been wavelength calibrated and calibrated to a relative flux scale.
Both extracted one-dimensional (1D) and the two-dimensional spectra are
kept after the basic reductions. However, in this paper we use only the 1D spectra.

The co-added 1D spectra were used for deriving the redshifts, velocity dispersions,
absorption line indices, and emission line equivalent widths of the galaxies. 
The details are described in J\o rgensen et al.\ (2005) and in Appendix \ref{SEC-SPECTROSCOPY}. 
In the following analysis we use the Lick/IDS absorption line indices for 
CN$_2$, Fe4383, C4668, Mg{\it b}, Fe5270 and Fe5335 (Worthey et al.\ 1994).
We use $\rm H\beta _G$ (see Gonz\'{a}lez 1993; J\o rgensen 1997) in place of the Lick/IDS
definition for $\rm H\beta$.
We also use the higher order Balmer line indices $\rm H\delta _A$ and $\rm H\gamma _A$
(Worthey \& Ottaviani 1997) and the CN3883 index defined by Davidge \& Clark (1994).
The uncertainties on the line indices were estimated both from the local S/N of the spectra
and from internal comparisions of indices derived from stacking subsets of the frames
available for each cluster. In the following we use the uncertainties from the latter 
method as these are larger than uncertainties based on the S/N. Table \ref{tab-compline}
in the appendix summarizes the uncertainties.

For galaxies with detectable emission from  [\ion{O}{2}] we determined the equivalent width 
of the [\ion{O}{2}]$\lambda\lambda$3726,3729 doublet,
in the following referred to as the ``[\ion{O}{2}] line''. 

Tables \ref{tab-MS0451line} and  \ref{tab-RXJ1226line} in Appendix \ref{SEC-SPECTROSCOPY} 
list the derived line indices and
the measurements of the  [\ion{O}{2}] equivalent widths for members of MS0451.6--0305 and 
RXJ1226.9+3336. The data for RXJ0152.7--1357 are available in J\o rgensen et al.\ (2005).

In the analysis involving the two higher order Balmer line indices $\rm H\delta _A$ and $\rm H\gamma _A$
we use the combined index as defined by Kuntschner (2000) 
$({\rm H\delta _A + H\gamma _A})' \equiv -2.5~\log \left ( 1.-({\rm H\delta _A + H\gamma _A})/(43.75+38.75) \right )$.
For the iron indices in the visible region we use the average iron index 
$\rm \langle Fe \rangle \equiv (Fe5270+Fe5335)/2$ rather than the individual indices.


\section{Observational data - HST \label{SEC-DATAHST}}

The three clusters have publicly available data obtained with {\it HST}/ACS. 
Table \ref{tab-hstdata} summarizes the data used in this paper.
A large mosaic of fields was observed for MS0451.6--0305. However, we only use 
the data that covers our spectroscopic sample.

For RXJ0152.7--1357 and RXJ1226.9+3332 we use the photometric parameters derived
as part of the project and published in Chiboucas et al.\ (2009). 
The data for MS0451.6--0305 were processed in the same way. Specifically, we stacked
the images using the drizzle technique (Fruchter \& Hook 2002) and derived effective radii, magnitudes and
surface brightnesses using the fitting program GALFIT (Peng et al.\ 2002).
Table \ref{tab-photMS0451HST} in Appendix  \ref{SEC-IMAGING} 
lists the derived parameters for the spectroscopic sample.
We have compared the derived parameters with data from S.\ Moran (private communication)
and find our photometry to be consistent with that of Moran to within 0.04 mag, 
see Appendix \ref{SEC-IMAGING} for details.

The HST photometry was transformed to the SDSS $i'$ as described in  Appendix \ref{SEC-IMAGING}.
Then the photometry was calibrated  to rest frame B-band, using the colors of the galaxies
determined from the GMOS-N photometry. The calibration was established using
stellar population models from Bruzual \& Charlot (2003) as described in J\o rgensen et al.\ (2005).
The calibrations at the cluster redshifts of all three clusters are listed in Appendix \ref{SEC-IMAGING},
Table \ref{tab-restB}.

\section{Low redshift comparison data \label{SEC-COMPDATA} }

As in J\o rgensen et al.\ (2005), we use our data for
galaxies in Coma, Perseus and Abell 194 as the low redshift comparison sample.
The line indices and velocity dispersions for the Coma cluster galaxies are
published in J\o rgensen (1999). We use the same B-band photometry for the cluster
as in J\o rgensen et al.\ (2005).
The spectroscopic data for Perseus and Abell 194 are the same as used in J\o rgensen et al.\ (2005).
Table \ref{tab-compdata} summarizes the number of galaxies in each of the cluster with
available velocity dispersions and line indices. The Coma cluster data do not include
line indices blue-wards of 4700 {\AA}.
The sample selection for the Coma cluster sample is detailed in J\o rgensen (1999). 
The Perseus and Abell 194 samples were selected in a similar fashion. Briefly, the initial
selection was based on morphological classifications, excluding spiral and irregular galaxies.
Apparent magnitude limits were applied, which with our adopted cosmology are equivalent
to an absolute magnitude limit of $M_{\rm B} = -18.6$ mag for all the low redshift clusters.
The Coma cluster sample is 93 percent complete to this limit. The Perseus and Abell 194
samples span the full magnitude range from the brightest cluster galaxy to $M_{\rm B} = -18.6$ mag
but are not intended to be complete.
Galaxies with emission lines or with $\rm Mass< 10^{10.3} M_{\sun}$ were excluded from the analysis.
Despite initially being selected morphologically, the resulting samples are equivalent to a
color selection, followed by exclusion of spirals, irregulars, emission line galaxies and low
mass galaxies. Thus, the selection criteria match those used for the selection of the
intermediate redshift galaxies included in the analysis, see Section \ref{SEC-METHODSAMPLE}.

The Coma cluster galaxies have $\rm Mg_2$ 
on the Lick/IDS system
measured for all galaxies with measured velocity dispersion, 
but Mg{\it b} only for those with original data from J\o rgensen (1999). The older literature
data, which were calibrated to consistency in that paper, did not include published Mg{\it b} measurements.
Due to the very broad continuum bands for $\rm Mg_2$ consistent calibration of 
this index to the Lick/IDS system relies on sufficient duplicate observations 
of galaxies or stars for which $\rm Mg_2$ is already calibrated. 
Further, we have no access to  the spectra on which the literature data for $\rm Mg_2$ are based. 
It is therefore not an option to remeasure $\rm Mg_2$ in a system that relies on flux calibrated
spectra rather than in the Lick/IDS system.
To take advantage of the old Lick/IDS $\rm Mg_2$ measurements, we therefore choose
to calibrate them to Mg{\it b}, which has narrower passbands and
can reliably be calibrated from spectra on a relative flux scale. 
The disadvantage is the slightly higher relative uncertainty on Mg{\it b} 
compared to that of $\rm Mg_2$.
For those galaxies in the Coma sample without Mg{\it b}, we transform $\rm Mg_2$ to Mg{\it b} using the relation
\begin{equation}
\log {\rm Mg{\it b}} = (1.72 \pm 0.17) {\rm Mg_2} + 0.165
\end{equation}
established from the 68 galaxies in the Coma sample (J\o rgensen 1999) for which both indices 
are available. The rms scatter of the relation is 0.033, all of which can be explained by
the measurement uncertainties on the two indices. Thus, the relation has no significant intrinsic scatter.

\begin{deluxetable*}{lclcc}
\tablecaption{HST/ACS imaging data\label{tab-hstdata} }
\tablewidth{0pc}
\tablehead{
\colhead{Cluster} & \colhead{No. of fields} & \colhead{Filter} & \colhead{Total $t_{exp}$(s)} & \colhead{Program ID} }
\startdata
MS0451.6--0305  & 6 & F814W & 4072 & 9836 \\
RXJ0152.7--1357 & 4 & F775W & 4800 & 9290 \\
RXJ1226.9+3332  & 4 & F814W & 4000 & 9033 \\
\enddata
\end{deluxetable*}

\begin{deluxetable}{lrrrrr}
\tablecaption{Low Redshift Comparison Data \label{tab-compdata} }
\tablewidth{0pc}
\tablehead{
\colhead{Cluster} & \colhead{Redshift} & \colhead{N($\log \sigma$)} &
\colhead{N(blue)} & \colhead{N(H$\beta _{\rm G}$)}  & \colhead{N(Mg{\it b},$\rm \langle Fe \rangle$)} 
}
\startdata
Perseus & 0.018 & 63 & 51 & 58 & 63 \\
A0194   & 0.018 & 17 & 14 & 10 & 10 \\
Coma    & 0.024 & 116 & \nodata & 90 & 68\tablenotemark{a} \\
\enddata
\tablenotetext{a}{115 galaxies in Coma have measurements of $\rm Mg_2$}
\tablecomments{ N($\log \sigma$) -- number of galaxies with velocity dispersion measurements; 
N(blue) -- number of galaxies with measurements of line indices blue-wards of 4700 {\AA}. 
N(H$\beta _{\rm G}$) -- number of galaxies with measurement of H$\beta _{\rm G}$.
N(Mg{\it b}, $\rm \langle Fe \rangle$) -- number of galaxies with measurement of Mg{\it b} and $\rm \langle Fe \rangle$.}
\end{deluxetable}


\begin{deluxetable*}{llrl}
\tablecaption{Predictions from Single Stellar Population Models \label{tab-models} }
\tablewidth{0pc}
\tablehead{
\multicolumn{2}{l}{Relation} & \colhead{rms} & \colhead{Reference} \\
\multicolumn{2}{l}{(1)} & \colhead{(2)} & \colhead{(3)} }
\startdata
$\rm \log M/L_B $ &$= 0.935 \log {\rm age} + 0.337 {\rm [M/H]} - 0.053 $  & 0.022 & Maraston 2005 \\
$\rm (H\delta _A + H\gamma _A)' $ &$= -0.126 \log {\rm age}  -0.106 {\rm [M/H]}  +0.091 {\rm [\alpha/Fe]} + 0.017 $\, \tablenotemark{a}  &0.011 &  Thomas et al. \\
$\log {\rm H\beta _G} $ &$= -0.230 \log {\rm age} - 0.108 {\rm [M/H]} + 0.053  {\rm [\alpha/Fe]} + 0.519$ & 0.014 & Thomas et al. \\
$\rm CN3883 $ &$= 0.076 \log {\rm age} + 0.150 {\rm [M/H]} +0.072 {\rm [\alpha/Fe]} + 0.132$  & 0.010 & Thomas et al. + Eq. \ref{eq-CN} \\
$\rm CN3883 $ &$= 0.084 \log {\rm age} + 0.175 {\rm [M/H]} +0.071 {\rm [\alpha/Fe]} +0.177 {\rm [C/\alpha ]} + 0.117 {\rm [N/\alpha ]}+ 0.123$  & 0.015 & Thomas et al. + Eq. \ref{eq-CN} \\
$\log {\rm Fe4383} $ &$= 0.272 \log {\rm age} + 0.331 {\rm [M/H]} - 0.356 {\rm [\alpha/Fe]} + 0.400$  & 0.030 &  Thomas et al. \\
$\log {\rm C4668}  $ &$= 0.121 \log {\rm age} + 0.484 {\rm [M/H]} + 0.116 {\rm [\alpha/Fe]} + 0.581$  & 0.030 &  Thomas et al. \\
$\log {\rm C4668}  $ &$= 0.108 \log {\rm age} + 0.475 {\rm [M/H]} + 0.100 {\rm [\alpha/Fe]} + 0.764 {\rm [C/\alpha ]} - 0.072 {\rm [N/\alpha ]} + 0.596$  & 0.031 &  Thomas et al. \\
$\log {\rm [C4668\,Fe4383]} $ &$= 0.212 \log {\rm age} + 0.594 {\rm [M/H]} + 0.715$\, \tablenotemark{b}  & 0.037 &  Thomas et al. \\
$\log {\rm Mg}b    $ &$= 0.242 \log {\rm age} + 0.322 {\rm [M/H]} + 0.242 {\rm [\alpha/Fe]} + 0.262$  & 0.021 &  Thomas et al. \\
$\log \langle {\rm Fe} \rangle $ &$= 0.122 \log {\rm age} + 0.260 {\rm [M/H]} - 0.243 {\rm [\alpha/Fe]} + 0.336$\, \tablenotemark{c}  & 0.009 &  Thomas et al. \\
$\log {\rm [MgFe]} $ &$= 0.182 \log {\rm age} + 0.292 {\rm [M/H]} + 0.299$\, \tablenotemark{d}  & 0.012 &  Thomas et al. \\
\enddata
\tablenotetext{a}{ $({\rm H\delta _A + H\gamma _A})' \equiv -2.5~\log \left ( 1.-({\rm H\delta _A + H\gamma _A})/(43.75+38.75) \right )$, Kuntschner (2000).
}
\tablenotetext{b}{${\rm [C4668\,Fe4383] \equiv C4668 \cdot (Fe4383)^{1/3}}$, see Section \ref{SEC-MODELS}.}
\tablenotetext{c}{$\langle {\rm Fe} \rangle \equiv ({\rm{Fe5270+Fe5335})/2}$.}
\tablenotetext{d}{${\rm [MgFe] \equiv (Mg{\it b} \cdot \langle Fe \rangle)^{1/2}}$, Gonz\'{a}lez (1993).}
\tablecomments{ (1) Relation established from the published model values. 
${\rm [M/H]}\equiv \log Z/Z_\sun$ is the total metallicity relative to solar.
$[\alpha /\rm{Fe}]$ is the abundance of the $\alpha$-elements relative to iron, and relative to the
solar abundance ratio. The age is in Gyr. The M/L ratios are stellar M/L ratios in solar units. 
(2) Scatter of the model values relative to the relation. (3) Reference for the model values. }
\end{deluxetable*}

\begin{figure}
\epsfxsize 8.5cm
\epsfbox{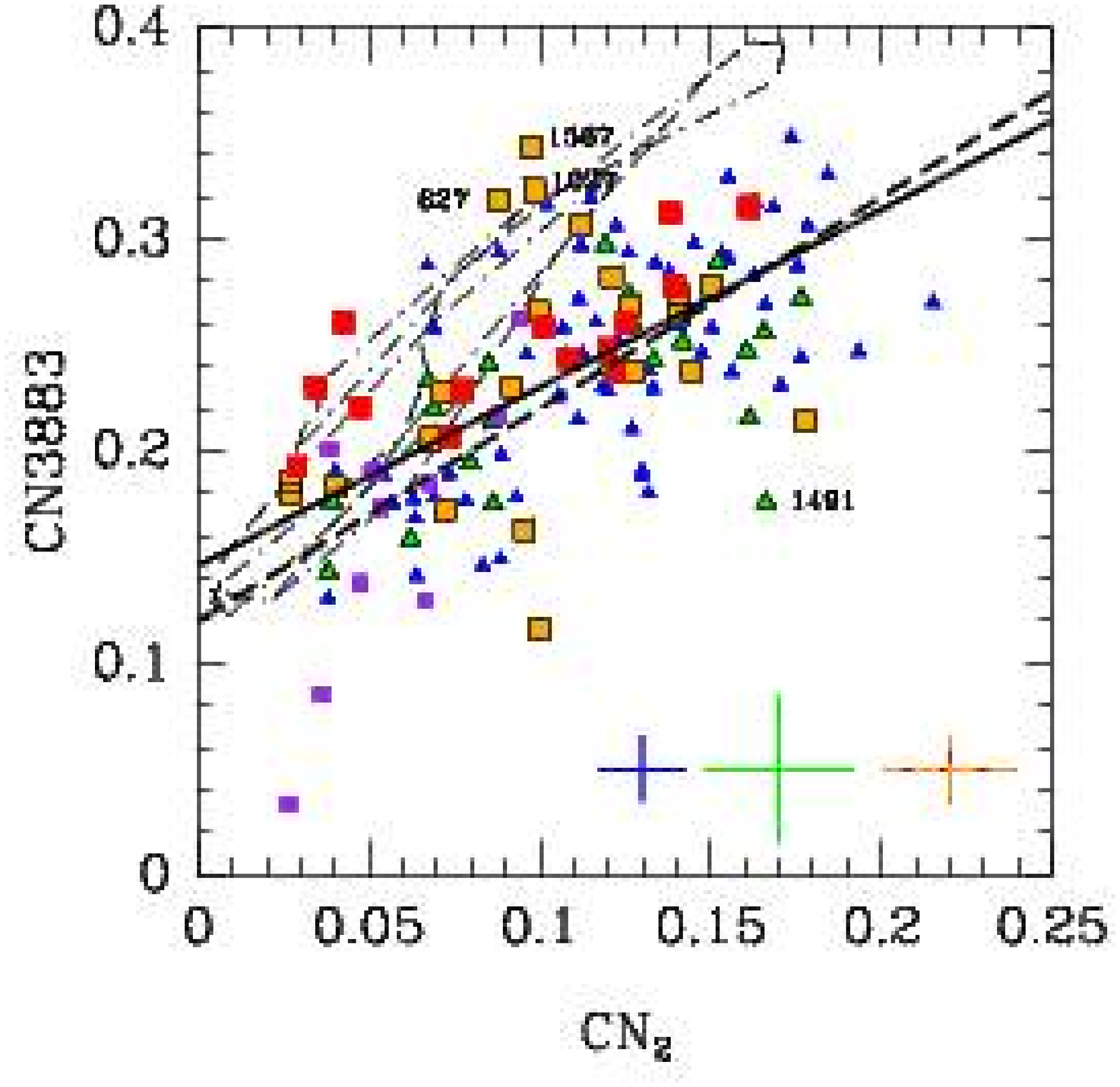}
\caption[]{CN3883 versus CN$_2$ for galaxies without significant emission.
Blue triangles -- low redshift sample; green triangles -- MS0451.6--0305;
orange boxes -- RXJ0152.7--1357; red boxes -- RXJ1226.9+3332; 
purple boxes -- non-members in the fields of MS0451.6--0305, 
RXJ0152.7--1357 and RXJ1226.9+3332. 
Typical uncertainties are shown, see Table \ref{tab-compline} in the appendix.
We adopt the same uncertainties for RXJ0152.7--1357 as for RXJ1226.9+3332.
Black solid line -- best fit relation, Eq.\ 4. 
Black dashed line -- best fit if excluding RXJ0152.7--1357 and RXJ1226.9+3332 member
from the fit, see text.
Dot-dashed lines -- model predictions based on model spectra from Maraston \& St\"{o}mb\"{a}ck (2011),
see text for discussion.
The four labeled galaxies were omitted from the fit, as the indices may be affected by
systematic residuals from the sky subtraction. The relation shown as the solid line is used to transform model
predictions for CN$_2$ to CN3883.
\label{fig-CN} }
\end{figure}

\begin{figure*}
\epsfxsize 16.5cm
\epsfbox{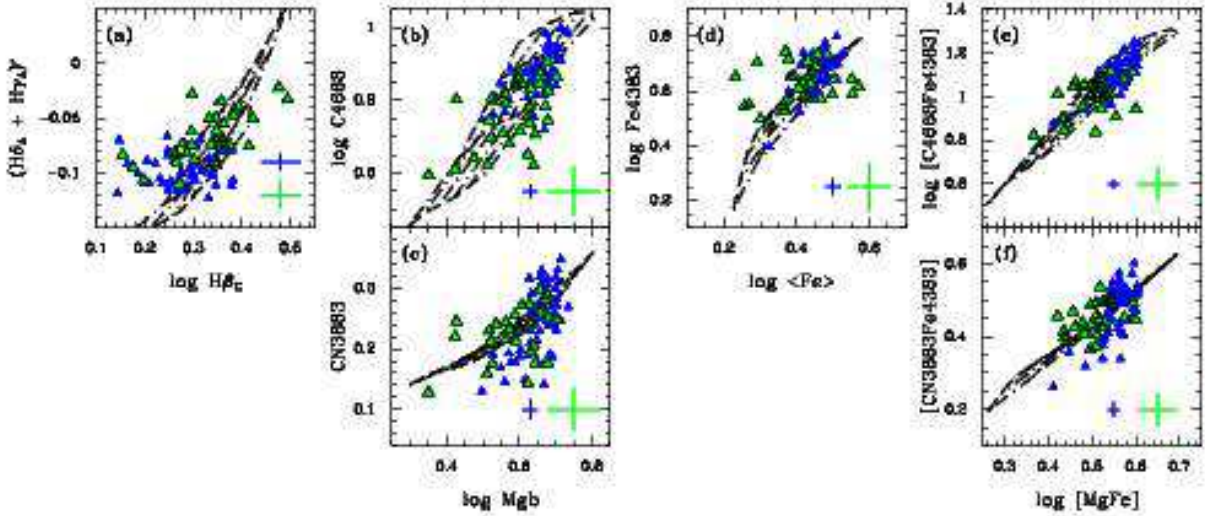}
\caption[]{Line indices in the visual versus the possible equivalents in the blue 
wavelength region.
Blue triangles -- low redshift sample; green triangles -- MS0451.6--0305.
Stellar population models from Thomas et al.\ (2011) for $[\alpha /\rm{Fe}] = 0.3$ 
are overlayed.
Typical uncertainties are shown on each panel.
As not all galaxies have all indices measured, it may be noticed that some galaxies are seemingly missing
from some panels.
\label{fig-visblue} }
\end{figure*}

\section{Models \label{SEC-MODELS}}

\subsection{Single stellar population models}

In the analysis of the line index strengths we use 
the single stellar population (SSP) models from Thomas et al.\ (2011)
for a Salpeter (1955) initial mass function (IMF). 
These authors model the Lick/IDS indices on a flux calibrated
system, which makes them ideal for comparison with our data. They also provide models
for different abundance ratios. We use their base models for different values of 
$[\alpha /\rm{Fe}]$, where the $\alpha$-elements should be understood as including
the carbon, nitrogen and sodium.
Further, we use the carbon and nitrogen enhanced models. We downloaded the model tables
from the web site provided by D.\ Thomas (http://www.icg.port.ac.uk/\~{ }thomasd).

The models do not provide predictions for the M/L ratios and CN3883.
Maraston (2005) modeled the M/L ratios for solar abundance ratios and otherwise
used the same model ingredients as Thomas et al.
The dependency of the M/L ratios on varying abundance ratios (at fixed metallicity) 
are in general poorly understood
and often assumed to be insignificant as done by Maraston.
In our discussion we also investigate the effect of differences in the IMF slope.
To do so, we use the models from Vazdekis et al.\ (2010), but note that these models cover
solar abundance ratios, only.

The strong index CN3883 is not yet modeled by any of the commonly used stellar population
models, which also vary the abundance ratios. 
Rather than use the weaker indices CN$_1$ and CN$_2$
we instead estimate how CN3883 depends on age, metallicity and abundance ratios using
two approaches. First we use the model spectra used by Maraston \& Str\"{o}mb\"{a}ck (2011) to
derive CN3883 and CN$_2$. 
The models are based on the same ingredients as the models from Thomas et al.\ (2011),
but cover only solar $[\alpha /\rm{Fe}]$.
Second we determine an empirical relation between CN3883 and CN$_2$ and then transform
the stellar population model values for CN$_2$ to the equivalent CN3883.
CN$_1$ and CN$_2$ probe the same features in the spectra. 
However, CN$_2$ excludes H$\delta$ in its blue continuum
passband and is therefore preferred over CN$_1$ (Worthey et al.\ 1994).

Figure \ref{fig-CN} shows CN3883 versus CN$_2$ and the best fit relation
\begin{equation}
{\rm CN3883} = (0.84 \pm 0.13)\, {\rm CN_2} +0.146
\label{eq-CN}
\end{equation}
with an rms scatter of 0.041. The residuals were minimized in CN3883 as the goal is
to predict CN3883 based on the CN$_2$ values.
The relation has an intrinsic scatter of about 0.03 in CN3883.
This relation has not previously been established in the literature.
As the passbands for CN3883 contain some higher order Balmer lines, one might expect 
that (part of) the intrinsic scatter of the relation is due to differences in the strengths
of the Balmer lines. However, we found no dependence of the residuals on the Balmer line
index $\rm (H\delta _A + H\gamma _A)' $.
Similarly including $\rm (H\delta _A + H\gamma _A)' $ in as a parameter in the fit does not 
lower the scatter of the established relation.
For the redshifts of RXJ0152.7--1357 and RXJ1226.9+3332 the passband for CN$_2$ 
coincides with the strong telluric absorption around 760-770 nm. However, CN$_2$ is 
only used to establish Eq.\ \ref{eq-CN} and not in the analysis itself. 
Excluding RXJ0152.7--1357 and RXJ1226.9+3332 from the fit results in slightly
steeper relation with a slope of $1.00 \pm 0.13$. This relation is also shown on 
Figure \ref{fig-CN}. None of the results in our analysis change significantly if 
we were to adopt this steeper slope. Thus, we adopt Eq.\ \ref{eq-CN} as the best fit.
Figure \ref{fig-CN} also shows the model values based on Maraston \& Str\"{o}mb\"{a}ck (2011).
These implicate stronger CN3883 at CN$_2 > 0.08$ than found from the data.
It is beyond the scope of this paper to investigate the reason for this. However, 
we note that CN3883 derived from the Maraston \& Str\"{o}mb\"{a}ck model spectra
would indicate roughly a factor two stronger dependency on age and metallicity than if 
instead we use the transformation in Eq.\ \ref{eq-CN}. 
However, because the values from the Maraston \& Str\"{o}mb\"{a}ck model spectra 
do not reproduce our data, we choose to use Eq.\ \ref{eq-CN}
to transform the model values for CN$_2$ (from Thomas et al.\ 2011) to CN3883, and
through that transformation achieve estimates of how CN3883 depends on age, metallicity
and $[\alpha /\rm{Fe}]$.

To enable studying the metallicity independent of the  $\rm [\alpha /Fe]$ abundance ratios, we 
use the combined magnesium-iron index $\rm [MgFe] \equiv (Mg{\it b} \cdot \langle Fe \rangle)^{1/2}$
first used by Gonz\'{a}lez (1993).
The models predict this index to be independent of the $\rm [\alpha /Fe]$ abundance ratios. 
Similarly we experimented to find a combination of C4668 and Fe4383, which the models also predict to be independent
of $\rm [\alpha /Fe]$. The combined index
\begin{equation}
\rm {[C4668\,Fe4383] \equiv C4668 \cdot (Fe4383)^{1/3} }
\end{equation}
has this property, as fitting this index as a function of age, metallicity and $\rm [\alpha /Fe]$
shows in an insignificant dependence on $\rm [\alpha /Fe]$.
This combined index makes it possible for us to carry out similarly $\rm [\alpha /Fe]$ independent
analysis of the higher redshift clusters as is possible at lower redshift using $\rm [MgFe]$.
We have also tested a similarly defined index using CN3883 in place of C4668, 
\begin{equation}
\rm {[CN3883\,Fe4383] \equiv CN3883 + 0.33 \log Fe4383 }
\end{equation}
Under the assumption that the transformation from CN$_2$ to CN3883 is valid (Eq.\ \ref{eq-CN}),
then fitting this index as a function of age, metallicity and $\rm [\alpha /Fe]$
also results in an insignificant dependence on $\rm [\alpha /Fe]$.
While this index is independent of $\rm [\alpha /Fe]$, its metallicity dependency relative to its
age dependency is only about a factor 1.5, while the for [C4668\,Fe4383] the
same ratio is about 2.8. Thus, the models are better separated in age and metallicity in
the [C4668\,Fe4383] versus $(\rm H\delta _A + H\gamma _A)'$ diagram than when using
[CN3883\,Fe4383] versus $(\rm H\delta _A + H\gamma _A)'$.

Because our data for the clusters RXJ0152.7--0152 and RXJ1226.9+3332 do not 
include line indices in the visible (H$\beta _{\rm G}$, Mg{\it b}, and $\rm \langle Fe \rangle$), we have
evaluated to what extent line indices in the blue can be used to trace the same properties
of the stellar populations as traced by the indices in the visible wavelength region.
Figure \ref{fig-visblue} shows the best choices in the blue versus the line indices in the visible.
The higher order Balmer lines and H$\beta _{\rm G}$ are expected to be closely correlated 
(Figure \ref{fig-visblue}a). However, the data show galaxies with
weaker H$\beta _{\rm G}$ than expected from their $(\rm H\delta _A + H\gamma _A)'$ indices. 
While this may be due to H$\beta _{\rm G}$ being partly filled in by (very) weak emission, we choose
to still include these galaxies in the analysis as the strength of their  [\ion{O}{2}] emission 
meets the sample selection criteria, see Sect.\ \ref{SEC-METHODSAMPLE}.

C4668 and CN3883 versus Mg{\it b} both approximately follow the locations predicted by the stellar populations models
for $\rm [\alpha /Fe]=0.3$ (Figure \ref{fig-visblue}b and c) and we therefore consider 
them both as viable candidates for use in the blue in place of Mg{\it b}.
However, it is important to note that the models assume that the carbon and nitrogen abundance ratios, 
[C/Fe] and [N/Fe], follow the $\rm [\alpha /Fe]$ abundance ratio. As C4668 and CN3883 both 
depend on the carbon and nitrogen abundances, the
two indices trace [C/Fe] and [N/Fe], which are then assumed to be the same as $\rm [\alpha /Fe]$.
We return to the question of additional enhancement of carbon or nitrogen in Sect.\ \ref{SEC-INDICES}.
The two iron indices $\rm \langle Fe \rangle$ and Fe4383 are expected to be closely correlated, see Figure \ref{fig-visblue}d.
The low redshift data are in agreement with this prediction, while the MS0451.6--0305 sample
show a larger scatter relative to the predicted relation between
$\rm \langle Fe \rangle$ and Fe4383.
This can be explained by the larger uncertainties on the iron indices for MS0451.6--0305.
The indices [C4668\,Fe4383] and [CN3883\,Fe4383] trace metallicity as well as [MgFe], see Figure \ref{fig-visblue}e and f.

In our analysis, we will use the models to identify 
general trends of the indices with varying ages, metallicities and 
abundance ratios. To make this more straightforward we fit the model index values as 
a function of age, total metallicity [M/H] and abundance ratio $[\alpha /\rm{Fe}]$.
All fits to the models are established for ages between 2 and 15 Gyr, [M/H] from $-0.33$ to $0.67$,
$[\alpha /\rm{Fe}]$ from $-0.3$ to 0.5. 
The fits were established as least squares fits with the residuals minimized in the line indices.
The model fits are listed in Table \ref{tab-models}. 
In addition we use the carbon and nitrogen enhanced models to establish the dependence of the indices
C4668 and CN3883 on $[{\rm C}/\alpha ]$ and  $[{\rm N}/\alpha ]$. 
The dependence of carbon and nitrogen enhancements are decoupled
from the dependence on age, metallicity and $[\alpha /\rm{Fe}]$ as inclusion of 
$[{\rm C}/\alpha ]$ and  $[{\rm N}/\alpha ]$ in the fits changes the other coefficients with 
negligible amounts.
The other indices used in our analysis depend only very weakly on $[{\rm C}/\alpha ]$ and  $[{\rm N}/\alpha ]$
(Thomas et al.\ 2011).

In the analysis we also use the models to estimate ages, metallicities [M/H] and abundance ratios $\rm [\alpha /Fe]$.
We use H$\beta_{\rm G}$ versus [MgFe] and $(\rm H\delta _A + H\gamma _A)'$ versus [C4668\,Fe4383] to estimate ages and [M/H].
The abundance ratios $\rm [\alpha /Fe]$ are derived from Mg${\it b}$ versus $\rm \langle Fe \rangle$ and from
CN3883 versus Fe4383.
As the indices in the visible and the blue do not depend on age, [M/H] and $\rm [\alpha /Fe]$ in identical ways,
and the models do not perfectly model the galaxies, we expect that there will be differences between the 
parameters derived from the visible indices and those based on the blue indices. 
For example, since CN3883 in some of the galaxies appears to be weaker than expected from the Mg${\it b}$ measurements,
given the model predictions (see Fig.\ \ref{fig-visblue}c), estimating $\rm [\alpha /Fe]$ from CN3883 versus Fe4383 may
result in lower values than when using Mg${\it b}$ versus $\rm \langle Fe \rangle$.
Further, the metal dependence relative to the age dependence is stronger for [C4668\,Fe4383] than for 
[MgFe] (see Table \ref{tab-models}), which in turn may lead to differences in both age and [M/H] estimates.
However, we do not mix the parameters derived from the visible indices with those derived from the blue
indices, and to a large extent we limit the analysis to using the differences between the clusters for 
the various parameters.
Thus, these issues do not significantly affect our conclusions.

While other stellar population models are available in the literature (e.g. Vazdekis et al.\ 2010, Schiavon 2007), 
the models from Thomas et al.\ are widely used in the field and are readily available
for fixed total metallicities, different $\rm [\alpha / Fe ]$ abundance ratios, and for carbon or nitrogen enhanced
above the $\rm [\alpha / Fe ]$ abundance ratio.
We tested if it would make a difference for our results if we had used the models from Schiavon (2007).
We used models that treat the individual elements the same as done in the base models by Thomas et al., specifically
models for which carbon, nitrogen and sodium abundances follow $[\alpha /\rm{Fe}]$. 
For each model point, total metallicities [M/H] were derived
as ${\rm [M/H] = [Fe/H]} + X$ where $X=-0.18, 0.0, 0.28$ and 0.47 for  
the four model sets of $\rm [\alpha / Fe ] = -0.2, 0.0, 0.3$ and 0.5, respectively
(R.\ Schiavon, private communication).
We then fit the model index values as function of age, [M/H], and 
$[\alpha /\rm{Fe}]$ omitting the very low [M/H] and very
young age models, as done for the models from Thomas et al.\ (2011). 
In general, the fits are very similar to those derived for the models from Thomas et al.\ 
with coefficients being within 10 per cent of those listed in Table \ref{tab-models}.
Relative to the models from Thomas et al., the exceptions are as follows. 
(1) Mg{\it b} and $\rm \langle Fe \rangle$ have a weaker dependence on $[\alpha /\rm{Fe}]$
in the models from Schiavon, while Fe4383 has a stronger $[\alpha /\rm{Fe}]$ dependence. 
This affects the exact location of the models in the Mg{\it b}-$\rm \langle Fe \rangle$ and
CN3883-Fe4383 diagrams. 
However, it does not affect our results regarding $[\alpha /\rm{Fe}]$ variations between
clusters or as a function of redshift.
(2) CN3883 and C4668 have a weaker dependence on [M/H] in the models from Schiavon.
This would result in any measured differences in [M/H] being derived larger if we used the models
from Schiavon, but it does not affect our results regarding variations between clusters or
as a function of redshift.
It should also be noted that the models from Schiavon reach [M/H] $>0.2$ only for $[\alpha /\rm{Fe}]$ 
above solar and that the models do not cover the very strong C4668 indices seen for the most massive
galaxies in our samples.
Thus, the differences in the resulting fits may, at least in part, be caused by differences in the parameter 
space covered by the two sets of models.

\subsection{Passive evolution model}

Models for passive evolution assume that the galaxies after an initial period of star formation
usually at high redshift, evolve passively without any additional star formation. 
The models are usually parameterized by a formation
redshift $z_{\rm form}$, which corresponds to the approximate epoch of the last 
major star formation episode.
As the stellar populations in the observed galaxies are assumed to age passively
and no other changes take place, 
the difference between the luminosity weighted mean ages of the stellar populations in 
the galaxies in each cluster and similar galaxies at the present is expected to be equal to the 
lookback time for that redshift. For reference, the lookback time for $z=0.89$ is 
7.3 Gyr with our adopted cosmology.

Thomas et al.\ (2005) used the properties of nearby early-type galaxies to estimate
the formation redshift $z_{\rm form}$ as a function of galaxy velocity dispersion, and 
through an empirical relation between stellar mass and velocity dispersion as a function
of stellar mass. 
We adopt their model for the high density cluster environment as our base model for passive 
evolution and show the model predictions
throughout the paper together with our data. As our relation between dynamical mass and 
velocity dispersion is slightly different than the stellar mass--velocity dispersion
relation used by Thomas et al.\ we adopt our transformation in order to show the model
predictions as a function of dynamical mass.
For the typical low and high velocity dispersion galaxies ($\log \sigma = 2.1$ and 2.35) 
in our sample the model from Thomas et al.\ implies $z_{\rm form} \approx 1.4$ and 2.2, respectively.

A more recent analysis by Thomas et al.\ (2010) used data from the SDSS
to establish the formation redshift as a function of galaxy velocity dispersion and mass,
covering primarily lower density environments.
Using this model in place of that from Thomas et al.\ (2005) does not affect our results 
significantly. Thus, we choose to use the original high density environment model from
Thomas et al.\ (2005).

In our analysis, we implicitly assume that the galaxies we observe
at high redshift are the progenitors to the galaxies in the low redshift 
comparison sample. 
This may not be the case, as discussed in detail by van Dokkum \& Franx (2001).
We return to this issue in the discussion in Section \ref{SEC-PASSIVE}.

\begin{figure}
\epsfxsize 8.5cm
\epsfbox{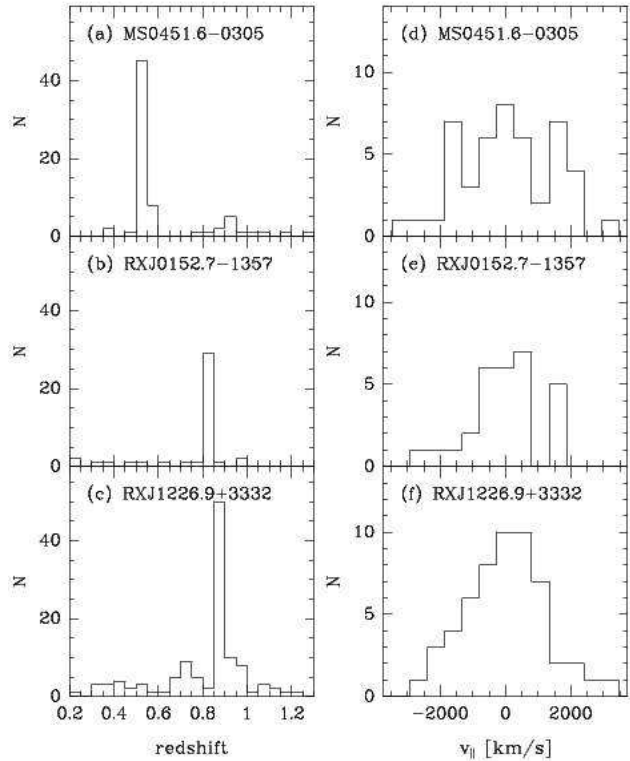}
\caption[]{(a)--(c) Redshift distribution of the spectroscopic samples in the three fields.
(d)--(f) Distribution of the radial velocities (in the rest frame of the cluster) 
relative to the cluster redshifts for cluster members, $v_{\|} = c (z - z_{\rm cluster}) / (1+z_{\rm cluster})$.
None of the radial velocity distributions for the members are significantly different
from Gaussian distributions.
\label{fig-zhist} }
\end{figure}

\begin{figure}
\epsfxsize 8.5cm
\epsfbox{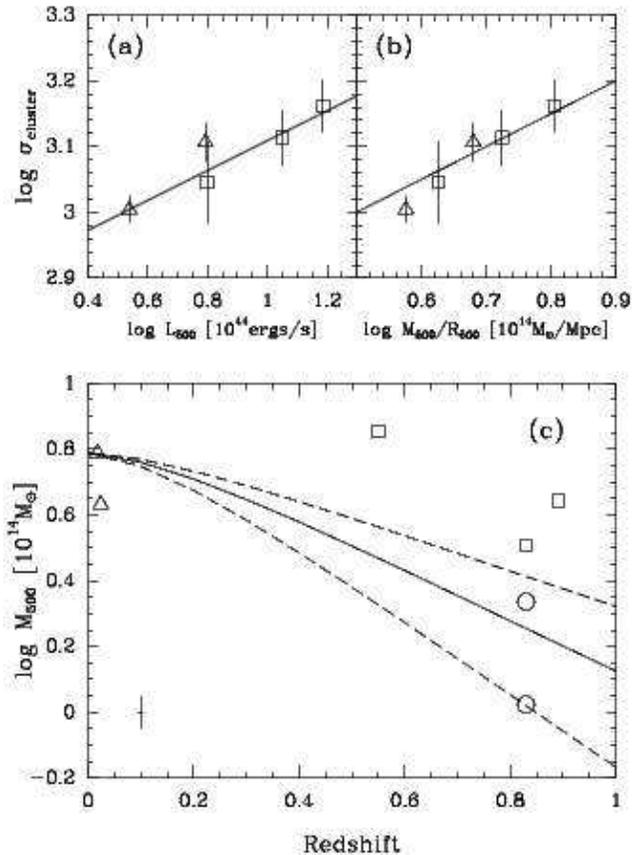}
\caption[]{Cluster velocity dispersion versus X-ray properties of the clusters. 
Triangles -- Coma and Perseus; squares -- intermediate redshift clusters.
(a) The relation shows $L_{500} \propto \sigma_{\rm cluster}^{4.4}$ at the median zero point of the clusters.
The slope is from Mahdavi \& Geller (2001).
(b) The relation shows the expected slope $M_{500}\,R_{500}^{-1} \propto \sigma_{\rm cluster}^{2}$ at the median
zero point of the clusters.
(c) The cluster mass $M_{500}$ versus the redshift. Circles -- RXJ0152.7--1357 treated as two clusters.
Solid line -- example model of cluster mass evolution based on numerical simulations from van den Bosch (2002).
Dashed lines -- expected scatter in the cluster mass evolution based on Wechsler et al.\ (2002).
\label{fig-xray} }
\end{figure}

\section{Cluster redshifts, velocity dispersions and substructure \label{SEC-CLUSTERZ}}

Cluster mass, local density and substructure are key elements of the
cluster environment experienced by the galaxies. To the extent that galaxy formation 
and evolution depend on the cluster environment, we seek to minimize differences
caused by the cluster environment in our samples, such that we can compare samples in different clusters
under the assumption that they have experienced similar cluster environments.
The clusters were therefore selected based on X-ray luminosity. 
In this section we determine the cluster redshifts and establish cluster membership. 
We determine the cluster velocity dispersions, $\sigma _{\rm cluster}$, test 
for the presence of substructure in the kinematics of the clusters, and evaluate whether the clusters follow
common relations between $\sigma _{\rm cluster}$ and the X-ray properties.

The cluster redshifts and velocity dispersions were determined using the bi-weight method
described by Beers et al.\ (1990). 
Figure \ref{fig-zhist} shows the distributions of the radial velocities of the three spectroscopic
samples. Figure \ref{fig-zhist}b and e are identical to the information shown in 
J\o rgensen et al.\ (2005) and are included here for completeness.
Table \ref{tab-redshifts} summarizes the redshifts and cluster velocity dispersions.
We note that the cluster velocity dispersion of $1450_{-159}^{+105}\, {\rm km~s^{-1}}$ for MS0451.6--0305 is higher than found by
the Borgani et al.\ (1999) for the CNOC data, who found $1330_{-94}^{+111} {\rm km~s^{-1}}$. 
The difference is due to the three galaxies
with largest $| v_{\|} |$ that we include as cluster members as we iteratively include 
galaxies as cluster members if $| v_{\|} | <3 \sigma_{\rm cluster}$. 
The three galaxies are ID=1931, 2127 and 2945.
Their inclusion as cluster members do not significantly affect the results found
for the scaling relations and stellar populations described in the following sections.
If we exclude these three galaxies, then we find $\sigma _{\rm cluster}=1262_{-103}^{+81} {\rm km~s^{-1}}$.
Moran et al.\ (2007b) obtained spectra for a large number of galaxies in MS0451.6--0305. 
Using data for the galaxies they consider members and the method from Beers et al.\ we find
$\sigma _{\rm cluster}=1363 \pm 51 {\rm km~s^{-1}}$, confirming the high velocity dispersion of this cluster.

As first found by Maughan et al.\ (2003) based on the X-ray data RXJ0152.7--1357 consists of two sub-clusters,
see J\o rgensen et al.\ (2005) for the X-ray data overlayed on our GMOS-N imaging data. 
The substructure is not reflected in the distribution of the radial velocities, which is 
consistent with being Gaussian (J\o rgensen et al.\ 2005). This may be due to the relative
orientation of the two sub-clusters.
We have tested whether the velocity distributions for the two other clusters deviate
from Gaussian distributions. We used a Kolmogorov-Smirnov test for this and find
probabilities that the samples are drawn from Gaussian distributions of 88 per cent and 98 per cent
for MS0451.6--0305 and RXJ1226.9+3332, respectively. Thus, no substructure is detectable
in these distributions.
MS0451.6--0305 and RXJ1226.9+3332 also do not show any substructure in the X-ray data, 
see Figures \ref{fig-MS0451grey} and \ref{fig-RXJ1226grey}, respectively. 
The structure seen in the outskirts of RXJ1226.9+3332 can all be associated with AGNs.
However, as noted in Sect.\ \ref{SEC-CLUSTERS} there is other evidence of more subtle substructure in 
these two clusters.

\begin{figure}
\epsfxsize 8.5cm
\epsfbox{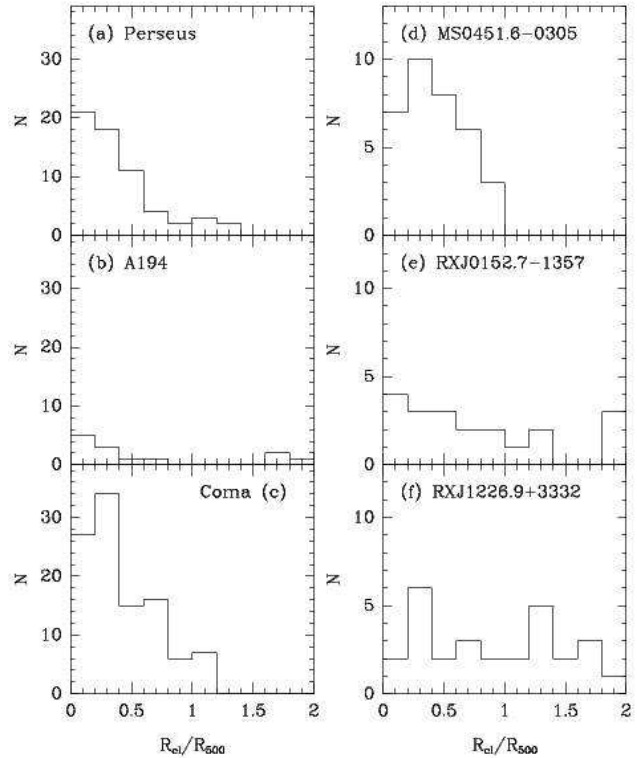}
\caption[]{Distribution of cluster center distances $R_{\rm cl}$ for the galaxies included in the analysis.
The distances have been normalized with the cluster size, $R_{500}$.
RXJ0152.7--1357 is treated as two clusters and $R_{\rm cl}$ determined relative to the nearest
of the centers of the two sub-clusters.
\label{fig-Rclhist} }
\end{figure}

On Figure \ref{fig-xray}a and b we show the cluster velocity dispersions versus the X-ray properties 
(see Table \ref{tab-redshifts} for the data). 
The clusters follow the relation between X-ray luminosity and $\sigma _{\rm cluster}$ established
by Mahdavi \& Geller (2001). They also follow the expected relation between X-ray masses, radii
and $\sigma _{\rm cluster}$. Thus, the clusters' dynamical properties match their X-ray properties.

The three intermediate redshift clusters, Coma and Perseus are all very massive. However, as cluster
masses are expected to grow over time (e.g., van den Bosch 2002), one may question if the intermediate
redshift clusters are viable progenitors for Perseus and Coma.  Figure \ref{fig-xray}c shows
the cluster mass $M_{500}$ versus redshift for all the clusters. The solid line shows
a typical model of cluster mass evolution based on numerical simulations by van den Bosch (2002) with
the dashed lines showing the expected scatter in cluster evolution (e.g.\ Wechsler et al.\ 2002).
A cluster like the Perseus cluster may originate from a cluster at $z=0.8$ that has $M_{500}$
between $10^{14} M_{\sun}$ and $10^{14.4} M_{\sun}$.
Treating RXJ0152.7--1357 as two clusters (circles on Figure \ref{fig-xray}c) brings the mass of
each of the components into this range, though of course the interaction between the two sub-clusters
is likely to affect the galaxy evolution (e.g., J\o rgensen et al.\ 2005).
RXJ1226.9+3332 and MS0451.6--0305 are both too massive to be viable progenitors for Coma and Perseus,
while RXJ1226.9+3332 may be a viable progenitor for MS0451.6--0305.
However, it remains to be seen if this also means that their galaxy populations cannot be progenitors
of those in Perseus and Coma.
Case in point, studies of nearby clusters by Smith et al.\ (2006) indicate that the local environment,
rather than the total mass of clusters, may be the determining factor for the stellar 
populations in early-type galaxies. As Smith et al.\ find that the properties of stellar populations
depend on the cluster center distance of the galaxy, $R_{\rm cl}$, normalized to the size of the cluster,
we next investigate if our samples span similar ranges in $R_{\rm cl}\, R_{500}^{-1}$. 
Figure \ref{fig-Rclhist} shows the distributions for the galaxies included in the analysis, see Section \ref{SEC-METHODSAMPLE}.
From this figure we conclude that the samples in Coma, Perseus and MS0451.6--0305 have similar
distributions in $R_{\rm cl}\, R_{500}^{-1}$. The Abell 194 sample has a few galaxies at 
$R_{\rm cl}\, R_{500}^{-1}>1$, the inclusion of which do not significantly change the relations
and distributions established for the low redshift sample in the following analysis.
The samples in  RXJ1226.9+3332 and RXJ0152.7--1357 contain a significant number of galaxies
at $R_{\rm cl}\, R_{500}^{-1}>1$. In the analysis we comment on the properties of these
galaxies relative to those closer to the cluster centers.


\begin{deluxetable*}{lrrr rrr rrr rrr}
\tablecaption{Scaling Relations \label{tab-relations} }
\tablewidth{0pc}
\tabletypesize{\scriptsize}
\tablehead{
\colhead{Relation} & \multicolumn{3}{c}{Low redshift sample} & 
  \multicolumn{3}{c}{MS0451.6--0305} &\multicolumn{3}{c}{RXJ0152.7--1357} & \multicolumn{3}{c}{RXJ1226.9+3332} \\
 & \colhead{$\gamma$} & \colhead{$N_{\rm gal}$} & \colhead{rms} 
 & \colhead{$\gamma$} & \colhead{$N_{\rm gal}$} & \colhead{rms} &  \colhead{$\gamma$} & \colhead{$N_{\rm gal}$} & \colhead{rms} 
 & \colhead{$\gamma$} & \colhead{$N_{\rm gal}$} & \colhead{rms} \\
\colhead{(1)} & \colhead{(2)} & \colhead{(3)} & \colhead{(4)} 
& \colhead{(5)} & \colhead{(6)} & \colhead{(7)} & \colhead{(8)} & \colhead{(9)} & \colhead{(10)} 
& \colhead{(11)} & \colhead{(12)} & \colhead{(13)} 
}
\startdata
$\log r_e      = (0.57 \pm 0.06) \log {\rm Mass} + \gamma$  & -5.734 & 105 & 0.16 & -5.701 & 34 & 0.17 & -5.682 & 21 & 0.11 & -5.724 & 28 & 0.19 \\   
$\log \sigma   = (0.26 \pm 0.03) \log {\rm Mass} + \gamma$  & -0.667 & 105 & 0.08 & -0.701 & 34 & 0.08 & -0.716 & 21 & 0.05 & -0.679 & 28 & 0.09 \\   
$\log \rm{M/L} = (0.24 \pm 0.03) \log {\rm Mass} + \gamma$  & -1.754 & 105 & 0.09 & -2.144 & 34 & 0.15 & -2.289 & 21 & 0.20 & -2.410 & 28 & 0.19 \\   
$\log \rm{M/L} = (1.07 \pm 0.12) \log \sigma     + \gamma$  & -1.560 & 105 & 0.11 & -1.974 & 34 & 0.13 & -2.067 & 21 & 0.18 & -2.197 & 28 & 0.17 \\   
$(\rm{H\delta _A + H\gamma _A})' = (-0.085\pm 0.015) \log \sigma + \gamma$ 
                                                          &  0.098 & 65 & 0.02 & 0.120 & 32 & 0.02 & 0.141 & 21 & 0.03 & 0.153 & 17 & 0.02 \\ 
$\log \rm{H\beta _G} = (-0.24\pm 0.04) \log \sigma + \gamma$ & 0.836 & 157 & 0.08 & 0.858 & 34 & 0.07 & \nodata & \nodata & \nodata & \nodata & \nodata & \nodata \\ 
${\rm CN3883}  = (0.29\pm 0.06) \log \sigma + \gamma$     & -0.410 & 65 & 0.05 &-0.410 & 31 & 0.03 &-0.396 & 21 & 0.05 &-0.400 & 23 & 0.04 \\ 
$\log {\rm Fe4383} = (0.19\pm 0.07) \log \sigma + \gamma$ &  0.263 & 65 & 0.06 & 0.204 & 34 & 0.07 & 0.062 & 17 & 0.30 & 0.194 & 17 & 0.09 \\ 
$\log {\rm C4668}  = (0.33\pm 0.08) \log \sigma + \gamma$ &  0.107 & 65 & 0.06 & 0.029 & 34 & 0.07 & 0.094 & 20 & 0.14 & 0.046 & 16 & 0.15 \\ 
$\log \rm{Mg{\it b}}      = (0.294\pm 0.016) \log \sigma + \gamma$ &-0.011 & 177 & 0.04 &-0.080 & 34 & 0.07 & \nodata & \nodata & \nodata & \nodata & \nodata & \nodata \\ 
$\log \rm{\langle Fe \rangle}     = (0.118\pm 0.012) \log \sigma + \gamma$ & 0.191 & 140 & 0.04 & 0.160 & 29 & 0.09 & \nodata & \nodata & \nodata & \nodata & \nodata & \nodata \\ 
\enddata
\tablecomments{(1) Scaling relation. (2) Zero point for the low redshift sample. (3) Number of galaxies
included from the low redshift sample. (4) rms in the Y-direction of the scaling relation for the low redshift sample.
(5), (6), and (7) Zero point, number of galaxies, rms in the Y-direction for the MS0451.6--0305 sample.
(8), (9), and (10) Zero point, number of galaxies, rms in the Y-direction for the RXJ0152.7--1357 sample.
(11), (12), and (13) Zero point, number of galaxies, rms in the Y-direction for the RXJ1226.9+3332 sample.}
\end{deluxetable*}

\begin{deluxetable*}{llrr}
\tablecaption{Fundamental Plane and relations for the M/L ratios \label{tab-FPfit} }
\tablewidth{0pc}
\tabletypesize{\scriptsize}
\tablehead{
\colhead{Cluster} & \colhead{Relation} & \colhead{$N_{\rm gal}$} & \colhead{rms}
}
\startdata
Coma            & $\log \rm{r_e} = (1.30 \pm 0.08) \log \sigma  - (0.82 \pm  0.03) \log \langle I \rangle _e -0.443 $ & 105 & 0.08 \\  
MS0451.6--0305  & $\log \rm{r_e} = (0.78 \pm 0.18) \log \sigma  - (0.79 \pm  0.11) \log \langle I \rangle _e +0.983 $ &  34 & 0.10 \\  
RXJ0152.7--1357 & $\log \rm{r_e} = (0.68 \pm 0.35) \log \sigma  - (0.56 \pm  0.13) \log \langle I \rangle _e +0.688 $ &  21 & 0.09 \\  
RXJ1226.9+3332  & $\log \rm{r_e} = (0.72 \pm 0.19) \log \sigma  - (0.70 \pm  0.07) \log \langle I \rangle _e +1.007 $ &  28 & 0.11 \\  
RXJ0152.7--1357,RXJ1226.9+3332\tablenotemark{a} & $\log \rm{r_e} = (0.65 \pm 0.14) \log \sigma  - (0.67 \pm  0.04) \log \langle I \rangle _e +1.070 $ & 49 & 0.09 \\ 
Coma            & $\log \rm{M/L} = (0.24 \pm 0.03) \log {\rm Mass} -1.754$ & 105 & 0.09 \\  
MS0451.6--0305  & $\log \rm{M/L} = (0.44 \pm 0.09) \log {\rm Mass} -4.499$ &  34 & 0.14 \\  
RXJ0152.7--1357 & $\log \rm{M/L} = (0.62 \pm 0.04) \log {\rm Mass} -6.593$ &  21 & 0.13 \\  
RXJ1226.9+3332  & $\log \rm{M/L} = (0.50 \pm 0.19) \log {\rm Mass} -5.272$ &  28 & 0.16 \\  
RXJ0152.7--1357,RXJ1226.9+3332\tablenotemark{a}  & $\log \rm{M/L} = (0.55 \pm 0.08) \log {\rm Mass} -5.845$ &  49 & 0.14 \\  
Coma            & $\log \rm{M/L} = (1.07 \pm 0.12) \log \sigma -1.560$ & 105 & 0.11 \\  
MS0451.6--0305  & $\log \rm{M/L} = (1.47 \pm 0.29) \log \sigma -2.894$ &  34 & 0.13 \\  
RXJ0152.7--1357 & $\log \rm{M/L} = (2.45 \pm 0.32) \log \sigma -5.203$ &  21 & 0.14 \\  
RXJ1226.9+3332  & $\log \rm{M/L} = (1.85 \pm 0.31) \log \sigma -3.900$ &  28 & 0.16 \\  
RXJ0152.7--1357,RXJ1226.9+3332\tablenotemark{a}  & $\log \rm{M/L} = (2.26 \pm 0.32) \log \sigma -4.782$ &  49 & 0.17 \\ 
\enddata
\tablenotetext{a}{RXJ0152.7--1357 and RXJ1226.9+3332 treated as one sample.}
\end{deluxetable*}

\section{The methods and the final galaxy sample \label{SEC-METHODSAMPLE} }

We characterize the stellar populations in the three clusters galaxies by 
(1) establishing the FP and other scaling relations, and 
(2) comparing the line index measurements with SSP models and deriving
the distributions of the ages, metallicities [M/H], and abundance ratios $\rm [\alpha /Fe]$
determined from the line indices using the SSP models.
We use the effective radii and surface brightnesses derived from the fits with $r^{1/4}$ profiles
since the low redshift comparison sample was fit with $r^{1/4}$ profiles. 
However, none of the results depend significantly on this choice.

The fitting technique to establish the FP and the other scaling relations
minimize the sum of the absolute residuals and the zero points are derived as the median.
This is the same fitting technique as we used in J\o rgensen et al.\ (2005).
The technique is very robust to the effect of outliers.
As also done in J\o rgensen et al.\ (2005), we derive the uncertainties of the 
slopes using a boot-strap method.
Except for relations involving $\rm (H\delta _A + H\gamma _A)'$, the relations
were fit by minimizing the residuals perpendicular to the relation.
For $(\rm H\delta _A + H\gamma _A)'$ we determine the fit by minimizing the 
residuals in $(\rm H\delta _A + H\gamma _A)'$. 

The random uncertainties on the zero point differences, $\Delta \gamma$, 
between the intermediate redshift and low redshift samples are derived as
\begin{equation}
\sigma _{\Delta \gamma} = \left ( {\rm rms}_{{\rm low-}z}^2/N_{{\rm low-}z}
  +  {\rm rms}_{{\rm int-}z}^2/N_{{\rm int-}z} \right )^{0.5}
\end{equation}
where subscripts ``low-$z$'' and ``int-$z$'' refer to the low redshift sample
and one of the intermediate redshift clusters, respectively.
In the discussion of the zero point differences (Section \ref{SEC-EVOLUTION}) we show
only the random uncertainties on the figures.
The systematic uncertainties on the zero point differences are expected to
be dominated by the possible inconsistency in the calibration of the 
velocity dispersions, 0.026 in $\log \sigma$ (cf.\ J\o rgensen et al.\ 2005), and may 
be estimated as 0.026 times the coefficient for $\log \sigma$, see Table \ref{tab-relations}.

The galaxies included in the analysis are required to meet the following selection criteria: 
(1) $\log {\rm Mass} \ge 10.3$, (2) $n_{\rm ser} \ge 1.5$, (3)
spectroscopy with S/N $\ge$ 20 per \AA\ in the restframe, and (4)
equivalent with of [\ion{O}{2}], EW[\ion{O}{2}] $\le$ 5 \AA . 
Further, MS0451.6--0305 ID=600, 1156, 1753, 1002 and 1331 are excluded.
ID=600, 1156 and 1753 have very close neighbors readily visible in the {\it HST}/ACS imaging (Figure \ref{fig-stampsMS0451}),
but too close to separate in the ground-based spectroscopy. Thus, the spectra are contaminated
by the neighboring galaxy, in general leading to systematically too weak line indices.
ID=1002 and 1331 show spiral arms in the {\it HST}/ACS imaging (Figure \ref{fig-stampsMS0451}).
The wavelength region of the spectra of these two galaxies does not include [\ion{O}{2}].
All other galaxies with spiral arms visible in the  {\it HST}/ACS imaging are already excluded
from the analysis based on the presence of [\ion{O}{2}] emission. 

Figures \ref{fig-stampsMS0451} and \ref{fig-stampsRXJ1226} in the appendix are labeled
to show which galaxies are included in the analysis.
The final samples for MS0451.6--0307 and RXJ1226.9+3332 are also marked with Figures \ref{fig-CMMS0451}a
and \ref{fig-CMRXJ1226}.

The number of points from each cluster varies slightly from plot to plot as not all 
galaxies have determinations of all line indices, see Tables \ref{tab-MS0451line} and 
\ref{tab-RXJ1226line}, as well as J\o rgensen et al.\ (2005).

\section{The scaling relations \label{SEC-SCALINGREL}}

\begin{figure}
\epsfxsize 8.5cm
\epsfbox{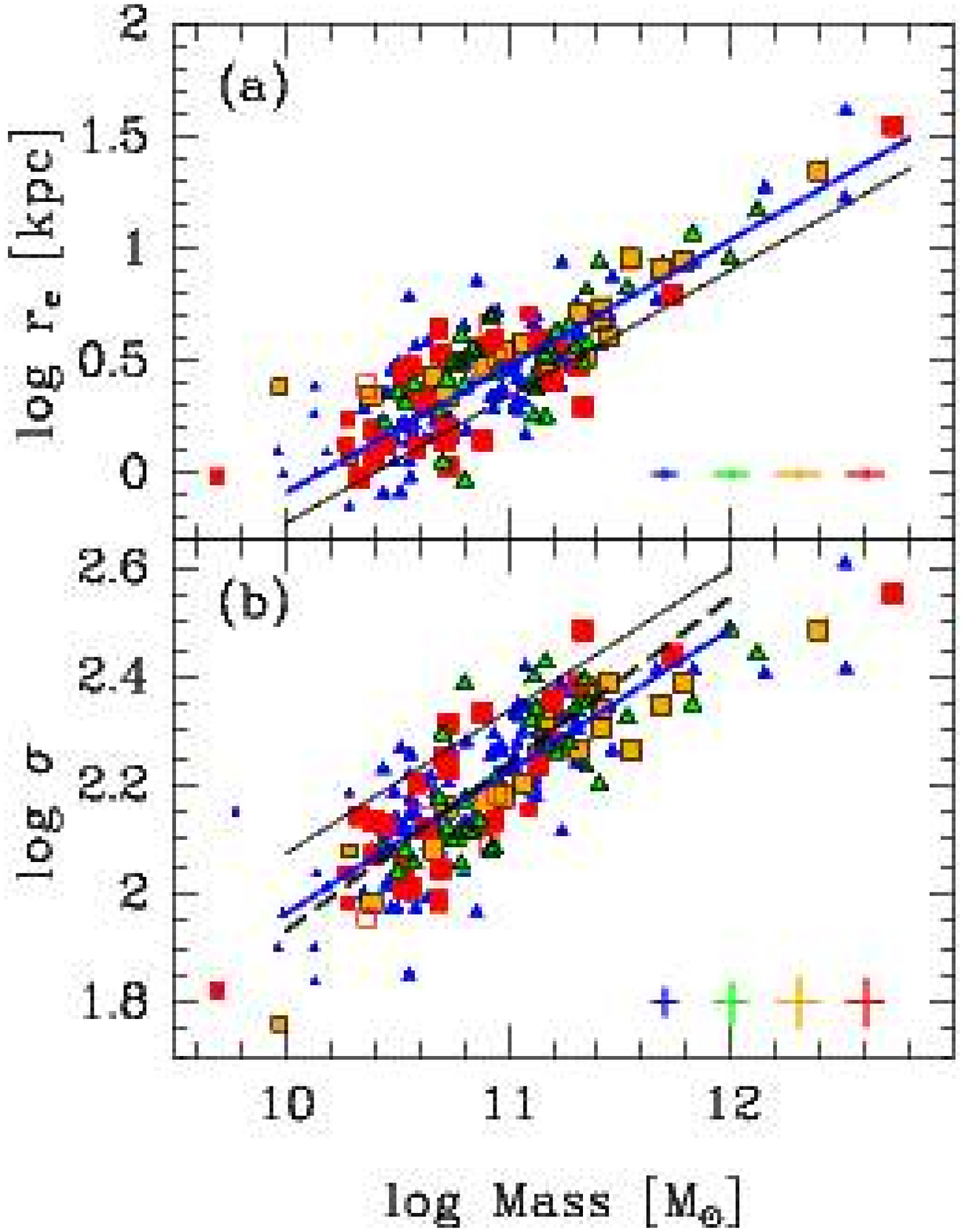}
\caption[]{(a) Effective radius versus dynamical galaxy mass. (b) Velocity dispersion
versus dynamical galaxy mass. 
The galaxy masses are derived using the approximation ${\rm Mass} = 5 r_{\rm e} \sigma ^2 {\rm G^{-1}}$
(Bender et al.\ 1992).
Blue triangles -- Coma cluster galaxies; green triangles -- MS0451.6-0305 galaxies;
orange squares -- RXJ0152.7--1357 galaxies; red squares -- RXJ1226.9+3332 galaxies.
Open points -- galaxies with significant [\ion{O}{2}] emission but EW[\ion{O}{2}] $\le$ 5 \AA.
Smaller points - galaxies excluded from analysis (log Mass $< 10.3$ and/or $n_{\rm sersic}<1.5$).
Blue solid lines -- best fit relations for the Coma cluster. Galaxies with masses above $10^{12}\, M_{\sun}$ 
have been omitted from the fit on panel (b) as the relation is non-linear at high masses.
The thick dashed black line on panel (b) shows the best fit to all clusters, $\log \sigma = (0.31 \pm 0.03) \log {\rm Mass} - 1.15$. 
Thin black lines -- the predicted location of the relations for $z=0.86$ if the size and velocity dispersion 
evolution found by Saglia et al.\ (2010) were adopted.
The intermediate redshift cluster galaxies follow the same relations as the Coma cluster galaxies 
and no significant evolution of size or velocity dispersion at a given dynamical mass is
found for this sample.
\label{fig-size} }
\end{figure}

\begin{figure}
\epsfxsize 8.5cm
\epsfbox{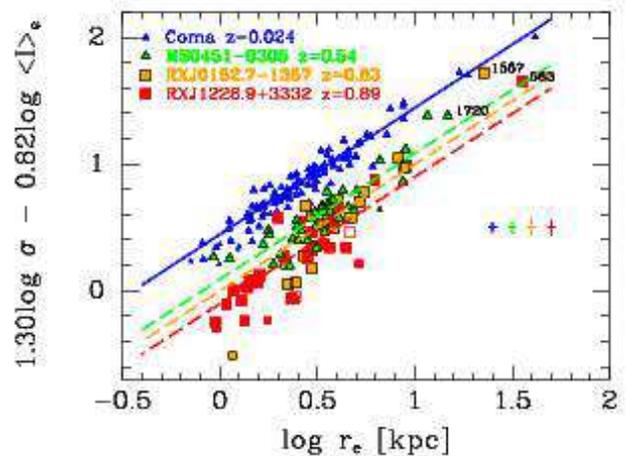}
\caption[]{The Fundamental Plane for the three clusters and the Coma cluster. 
Symbols as on Figure \ref{fig-size}.
The BCG in the three high redshift clusters are labeled.
Blue solid line -- best fit relation for the Coma cluster; 
green line -- best fit relation for the Coma cluster offset to the median zero point of MS0451.6-0305; 
orange line -- best fit relation for the Coma cluster offset to the median zero point of RXJ0152.7--1357;
red line -- best fit relation for the Coma cluster offset to the median zero point of RXJ1226.9+3332. 
\label{fig-FPonly} }
\end{figure}

\begin{figure*}
\epsfxsize 17cm
\epsfbox{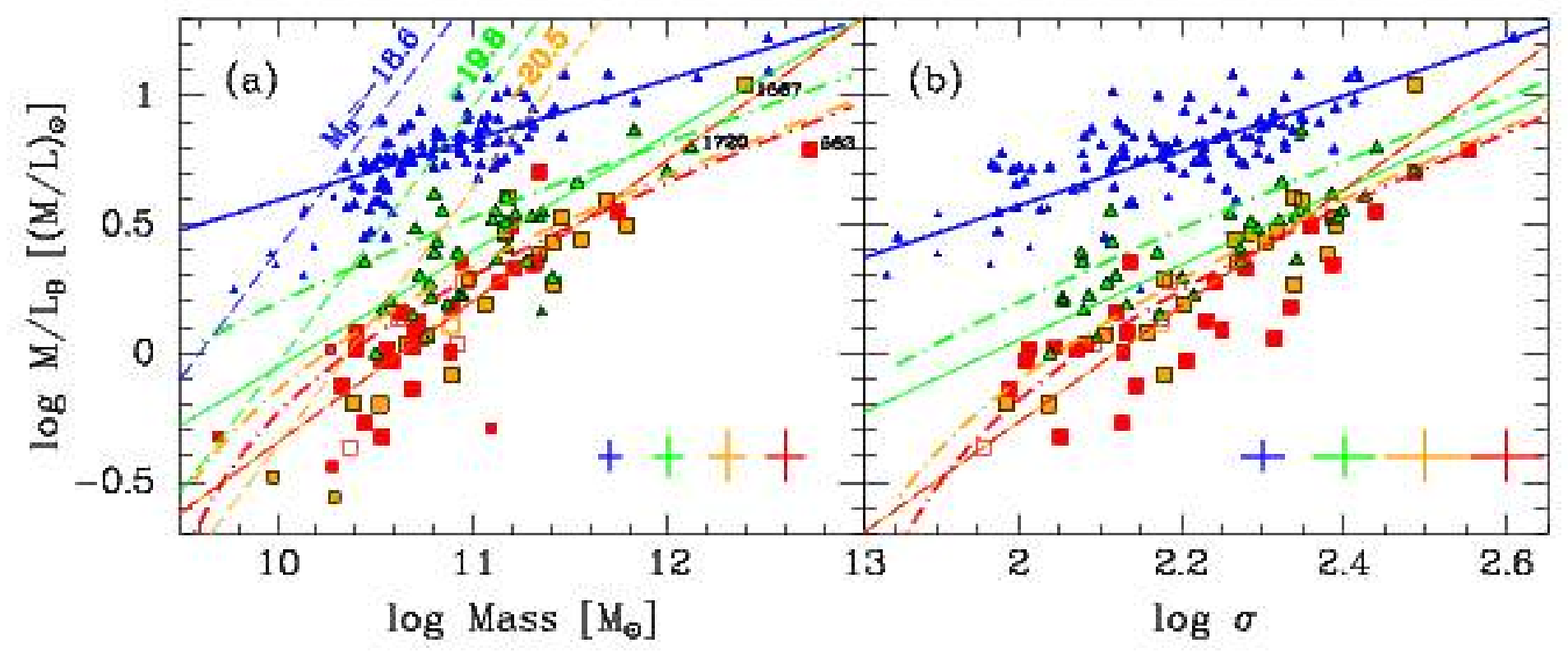}
\caption[]{The M/L ratios versus (a) masses and (b) velocity dispersions.
Symbols as on Figure \ref{fig-size}.
Blue solid lines -- best fit relations for the Coma cluster; 
green solid lines --  best fit relations for the MS0451.6--0305 galaxies;
red/orange solid lines -- best fit relations for the RXJ0152.7--1357 and RXJ1226.9+3332 galaxies. 
Thin colored dashed lines on panel (a) show the approximate magnitude limits for the Coma cluster
sample (blue), the MS0451.6--0305 sample (green) and the combined RXJ0152.7--1357 and RXJ1226.9+3332 (red) samples.
The thick dot-dashed lines show the predicted location of the relations based on the models
by Thomas et al.\ (2005) for each of cluster redshifts. 
Green -- MS0451.6--0305; orange -- RXJ0152.7--1357; red -- RXJ1226.9+3332.
\label{fig-MLonly} }
\end{figure*}

Using the fitting method and the samples described in Section \ref{SEC-METHODSAMPLE} 
we establish (1) the scaling relations for the radii and the velocity dispersions 
as a function of galaxy masses; (2) the Fundamental Plane (FP), the M/L-mass and 
M/L-velocity dispersion relations;
and (3) the scaling relations between absorption line indices and the velocity dispersions.
Tables \ref{tab-relations} and \ref{tab-FPfit} summarize the derived scaling relations.
Figures \ref{fig-size} -- \ref{fig-line_sigma} show the data and the best fit relations. 
The galaxy masses are derived using the approximation ${\rm Mass} = 5 r_{\rm e} \sigma ^2 {\rm G^{-1}}$ 
(Bender et al.\ 1992; see also Sect.\ \ref{SEC-NONHOM}).
The main results are described in the following sections.

\subsection{Radii and velocity dispersions as a function of masses \label{SEC-SIZEMASS} }

We have investigated the possible evolution of sizes and velocity dispersions
as a function of redshift by establishing relations with the dynamical galaxy masses. 
The relations and the zero points for the clusters are summarized in Table \ref{tab-relations}.
The galaxies in the samples for the higher redshift clusters follow the
same relations as found for the Coma cluster sample, with no significant differences
in the zero points, see Figure \ref{fig-size}. 
Thus, for these samples no significant evolution of size or velocity dispersion 
at a given dynamical mass with redshift is found.
For the two highest redshift clusters treated as one sample,
3-sigma zero point differences would be $\Delta \log r_e = 0.083$ and $\Delta \log \sigma = 0.041$.

Before further addressing the issue of possible evolution of the relations with redshift, we 
compared our relations to relations derived based on the SDSS data as made available
by Thomas et al.\ (2010). Using their data, taking into account the difference in adopted
cosmology and limiting the sample to early-type galaxies with Mass $\ge 10^{10.3} {\rm M_{\sun}}$, 
we find no significant difference in the slope of the radius-mass relation and a
zero point difference of only 0.03 in $\log r_{\rm e}$. Using the SDSS data we 
find a slightly steeper relation between the velocity dispersions and the masses than found
from our Coma sample (a slope of 0.35 compared to 0.26 for the Coma sample). 
However, adopting our slope the zero point difference in
$\log \sigma$ at 0.006 is well below the systematic errors on the velocity dispersions.
Thus, we conclude that our low redshift results are in agreement with relations
based on the SDSS data. 

Many authors have studied galaxy sizes as a function of redshift and found significant
evolution in the sense that galaxies at a given mass are smaller at higher redshift 
(e.g.\ Trujillo et al.\ 2007; van Dokkum et al.\ 2010; Saglia et al.\ 2010; Newman et al.\ 2012).
Of these studies the work by Saglia et al.\ most closely resembles our work, as these authors also
determine dynamical masses and study galaxy clusters up to $z=1$. They then derive
the evolution of both sizes and velocity dispersions as a function of redshift.
On Figure \ref{fig-size} the thin solid lines show the predicted locations of the relations for $z=0.86$ 
(the average of the two highest redshift clusters in our sample) 
based on the results from  Saglia et al.\ of $r_{\rm e} \propto (1+z)^{-0.5}$ and $\sigma \propto (1+z)^{0.41}$.
The predicted $\log r_{\rm e}$ offset is equivalent to a 5-sigma effect in our data,
while the predicted $\log \sigma$ offset if present would be detected in our data as an 8-sigma detection.
Other authors find larger size evolution at a given mass as a function of redshift, though
these authors use masses determined from photometry using stellar population models.
For example Newman et al., who finds the same slope of the size-mass relation as we do,
determine an evolution of the sizes with redshift that would lead to an offset 
$\Delta \log r_{\rm e} =-0.224$ at $z=0.86$. An offset this size is equivalent to an 8-sigma detection 
in our data.

We note that our result and difference in conclusion with Saglia et al.\ (and other results showing
even stronger size evolution) do not depend on our sample limit in the effective radii. 
Specifically, the sample used by Saglia et al.\ has the same limit in $r_{\rm e}$ as ours,
$r_{\rm e} \ge 1$ kpc. We therefore conclude that the disagreement between our results and 
that by Saglia et al.\ cannot be due to a selection effect.
Previous results showing evolution of the galaxy sizes and velocity dispersions as a function
of redshift were focused on field galaxies and less rich cluster environments than those of
the clusters in the present paper. Further, except for two, the clusters included in Saglia et al.\
have cluster velocity dispersions of 300-700 $\rm km\,s^{-1}$ (S\'{a}nchez-Bl\'{a}zquez et al.\ 2009). 
Thus, the difference in results may be related to the difference in cluster environment where
possibly the galaxies in dense cluster environments reach the present day sizes at earlier epochs
than do galaxies in less dense environments.

In summary, we do not detect in our data any significant evolution of sizes and velocity dispersions
as a function of redshift at a given galaxy mass. 
Therefore, we do not to make any correction for this type of evolution in the following analysis.

\subsection{The Fundamental Plane and relations for the M/L ratios \label{SEC-FP} }

Table \ref{tab-FPfit} lists the fits of the FP for each of the clusters as well as the 
fits of the M/L ratio as a function of galaxy mass and as a function of velocity dispersion,
while the zero points relative to the low redshift best fit relations are 
included in Table \ref{tab-relations}.
Figures \ref{fig-FPonly} and \ref{fig-MLonly} show the data and the relations.

The samples in two highest redshift clusters RXJ0152.7--0305 and RXJ1226.9+3332 were fit
both separately and together. By fitting two parallel relations to
the two samples we tested if there were significant zero point offsets between them and found none.
Thus, for the purpose of discussing the change in slope of the relations with redshift, we 
will consider the fits established for the joint sample of the two clusters.
The slightly larger sample of RXJ0152.7--1357 and RXJ1226.9+3332 confirms our results
from J\o rgensen et al.\ (2006, 2007) of a ``steeper'' FP at $z=0.86$ compared to that
of the Coma cluster. The MS0451.6--0305 also shows a steeper FP, but to a lesser extent.

On Figure \ref{fig-MLonly} we show the model prediction from Thomas et al.\ (2005) for each of 
the cluster redshifts for models in which high mass galaxies form the majority of 
their stars at high redshifts while lower mass galaxies experience later star formation. 
The models reproduce the general trends of the  M/L--mass and 
M/L--velocity dispersion relations for the intermediate redshift clusters,
though the data show a larger evolution in M/L ratios that predicted by the models,
especially for low mass (low velocity dispersions) galaxies.
Thus, in the case of passive evolution
this would mean a lower formation redshift than $z_{\rm form}=1.4$, 
as assumed by the models for $\log \sigma =2.1$ galaxies.
Using the best fit relations for $\log \sigma$ versus $\log {\rm M/L}$ to determine the 
differences in $\log {\rm M/L}$ and therefore the formation redshifts we find for $\log \sigma =2.1$
(Mass=$10^{10.55} {\rm M_{\sun}}$) a formation redshift of $z_{\rm form} \approx 1.2$.
At $\log \sigma =2.35$ (Mass=$10^{11.36} {\rm M_{\sun}}$) the difference in $\log {\rm M/L}$ corresponds
to $z_{\rm form} \approx 2.45$. Both of these estimates are based on the fits to the low redshift sample and the
joint sample of the two highest redshift clusters.
The fit to the MS0451.6--0305 sample implies lower formation redshifts, $\approx 0.95$ and 1.2 for low and high
velocity dispersion galaxies, respectively.

\begin{figure*}
\epsfxsize 16.0cm
\epsfbox{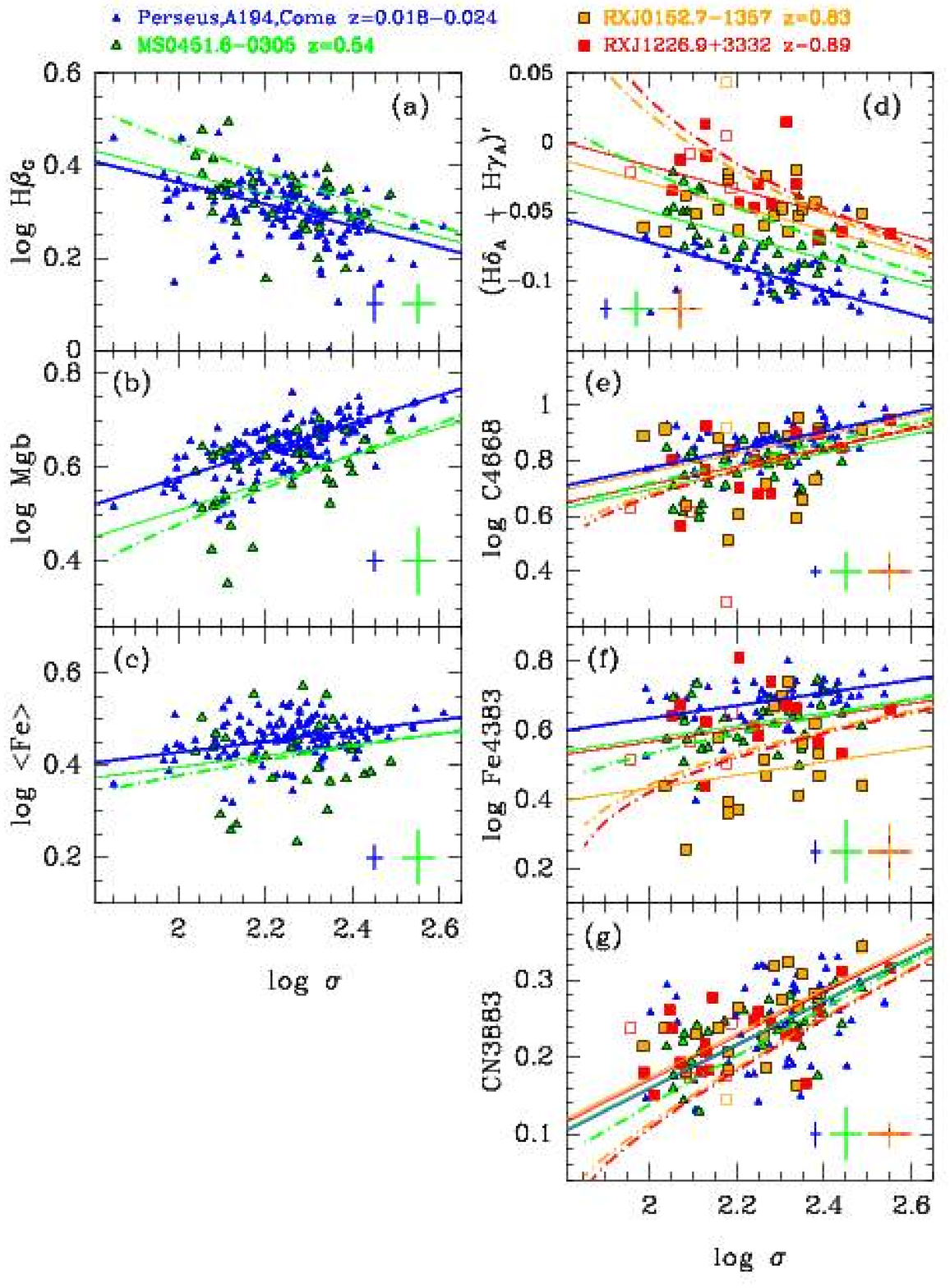}
\caption[]{Line indices versus the velocity dispersions.
Blue triangles -- The low redshift comparison sample (Perseus, Abell 194 and Coma); 
green triangles -- MS0451.6-0305 galaxies;
orange squares -- RXJ0152.7--1357 galaxies; red squares -- RXJ1226.9+3332 galaxies.
Open points -- galaxies with significant [\ion{O}{2}] emission but EW[\ion{O}{2}] $\le$ 5 \AA.
Typical uncertainties are shown, see Table \ref{tab-compline} in the appendix.
We adopt the same uncertainties for RXJ0152.7--1357 as for RXJ1226.9+3332.
Solid blue lines -- best fits to the low redshift sample; 
solid green lines -- relations for MS0451.6-0305;
solid orange lines --  relations for RXJ0152.7--1357;
solid red lines -- relations for RXJ1226.9+3332.
The slopes of the relations determined from the low redshift
sample were adopted. The lines for the intermediate redshift clusters show the 
low redshift relations offset to the median zero points of the intermediate redshift clusters.
The thick dot-dashed lines show the model predictions based on models
by Thomas et al.\ (2005) for cluster redshifts, green: MS0451.6--0305, orange: RXJ0152.7--1357, red: RXJ1226.9+3332.
\label{fig-line_sigma} }
\end{figure*}

\subsection{Line indices versus velocity dispersion \label{SEC-LINE}}

Figure \ref{fig-line_sigma} shows the absorption line indices versus the velocity
dispersions for the low redshift sample together with all the intermediate redshift samples.
The relations and zero points are summarized in Table \ref{tab-relations}.

Our data for MS0451.6--0305 allow us to compare the galaxies in this cluster directly 
with the low redshift sample using the well-studied indices $\rm H\beta _{\rm G}$, $\rm Mg$b, 
$\rm \langle Fe \rangle$.
There are no significant differences in the slopes of the relations for the low redshift
sample and those for the  MS0451.6--0305 sample.
Using the median zero points for the cluster, we find that  H$\beta _{\rm G}$ is only 
marginally stronger in MS0451.6--0305 than in the low redshift sample, while both
Mg{\it b} and $\rm \langle Fe \rangle$ are weaker at the 3-5 sigma level, see Table \ref{tab-relations}.
The scatter of the relations for Mg{\it b} and  $\rm \langle Fe \rangle$ are
two to three times higher for MS0451.6--0305 than found for the low redshift sample.
This difference cannot be explained by higher measurement uncertainties and 
may indicate that the galaxies in the MS0451.6--0305 sample are more diverse than
those in the low redshift sample.

\begin{figure*}
\epsfxsize 16.5cm
\epsfbox{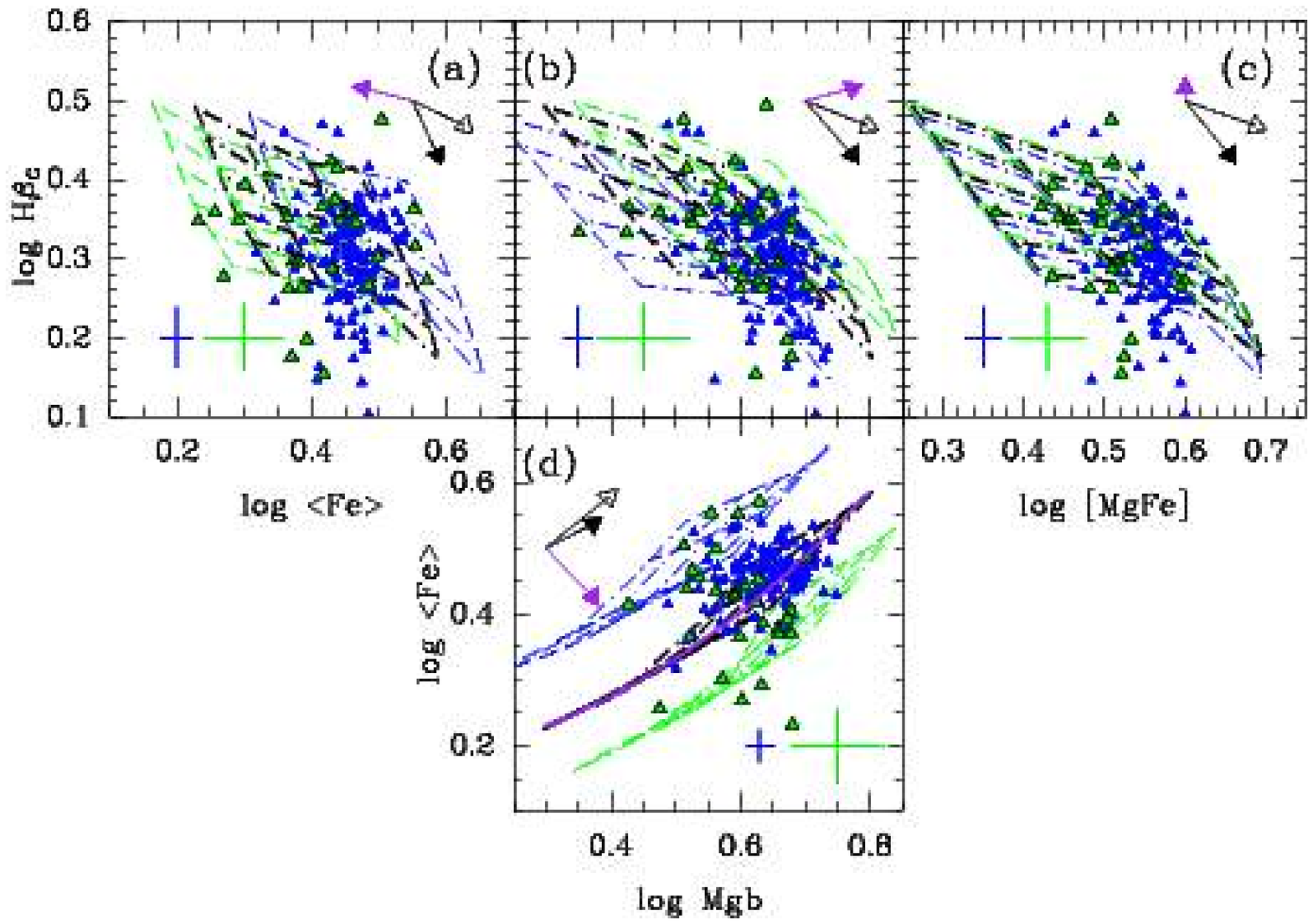}
\caption[]{Line indices in the visible versus each other for the low redshift sample and 
MS0451.6--0305.  
Blue triangles -- The low redshift comparison sample (Perseus, Abell 194 and Coma); 
green triangles -- MS0451.6-0305 galaxies.
Typical uncertainties are shown, see Table \ref{tab-compline} in the appendix.
Model grids from Thomas et al.\ (2011) are overlayed. Blue -- $\rm [\alpha /Fe] = 0$;
black -- $\rm [\alpha /Fe] = 0.3$; green --  $\rm [\alpha /Fe] = 0.5$. Models are shown for ages 
from 2 Gyr to 15 Gyr and [M/H] from $-0.33$ to 0.67.
The arrows show the model predictions for $\Delta \log {\rm age} = 0.3$ (solid); 
$\Delta {\rm [M/H]} = 0.3$ (open); $\Delta {\rm [\alpha /Fe]} = 0.3$ (purple). 
Solid purple line on panel (d) -- second order approximation to the model for $\rm [\alpha /Fe] = 0.3$ 
used for the determination of $\rm [\alpha /Fe]$, see Sect.\ \ref{SEC-INDICES}.
\label{fig-visline} }
\end{figure*}

For RXJ0152.7--1357 and RXJ1226.9+3332 only indices blue-wards of 4700 \AA\ are available.
These indices are also available for MS0451.6--0305 and for the low redshift samples in Perseus and Abell 194.
As for the line indices in the visible, we cannot detect any significant differences
of the slopes of the relations for the different clusters. We therefore use the slopes 
established for the low redshift sample and determine zero point offsets relative to these
and scatter relative to the offset relations.
Only the relation for the higher order Balmer lines shows a 
clear redshift dependence of the zero point with the intermediate redshift clusters having stronger 
Balmer lines. In J\o rgensen et al.\ (2005) we found that the RXJ0152.7--0305 galaxies have
significantly weaker Fe4383 than the low redshift sample. The other two clusters 
do not show significantly weaker Fe4383. All other relations for the metal lines show that the
$z=0.5$ to 0.9 clusters follow the same relations as the low redshift sample with the zero points
being within two sigma of each other. 
The relations for Fe4383 and C4668 show higher scatter in the intermediate redshift clusters
than found for the low redshift sample, again indicating that the samples in these
clusters may be more diverse than our low redshift comparison sample.
The samples in RXJ0152.7--1357 and RXJ1226.9+3332 contain galaxies at larger normalized cluster center
distances $R_{\rm cl}\, R_{500}^{\rm-1}$ than the other clusters.
We have tested if we in these samples can detect any correlation between $R_{\rm cl}\, R_{500}^{\rm-1}$
and the residuals for the relations between the line indices and the velocity dispersions.
We find no significant relations. There are also no significant zero point differences between
galaxies at $R_{\rm cl}\, R_{500}^{\rm-1} > 1$ and those closer to the cluster centers.
It may require larger samples in intermediate redshift clusters, going to larger cluster center
distances, to detect any correlations if these are present as they are in low redshift clusters
(Smith et al.\ 2006).

The model predictions for the passive evolution model 
from Thomas et al.\ (2005) are overlayed on all panels in Figure \ref{fig-line_sigma}.
For the Balmer lines, the models predict stronger lines than measured in the galaxies in the 
intermediate redshift clusters. Under the assumption of passive evolution, the formation redshifts
for these galaxies would have to be higher than assumed by the models. 
This in turn is inconsistent with the results based on the FP.
For the Mg{\it b} and $\rm \langle Fe \rangle$ indices the passive evolution models reproduce
the data well, while for the blue metal indices Fe4383 and CN3883 the models predict larger
zero point offsets than measured from the data. For CN3883 in particular the data 
show no significant zero point offsets.
We return to the zero point differences as a function of redshift in the discussion of 
the passive evolution model in Sect.\ \ref{SEC-PASSIVE}.

\section{Line indices and stellar population models \label{SEC-INDICES} }

\begin{figure*}
\epsfxsize 16.5cm
\epsfbox{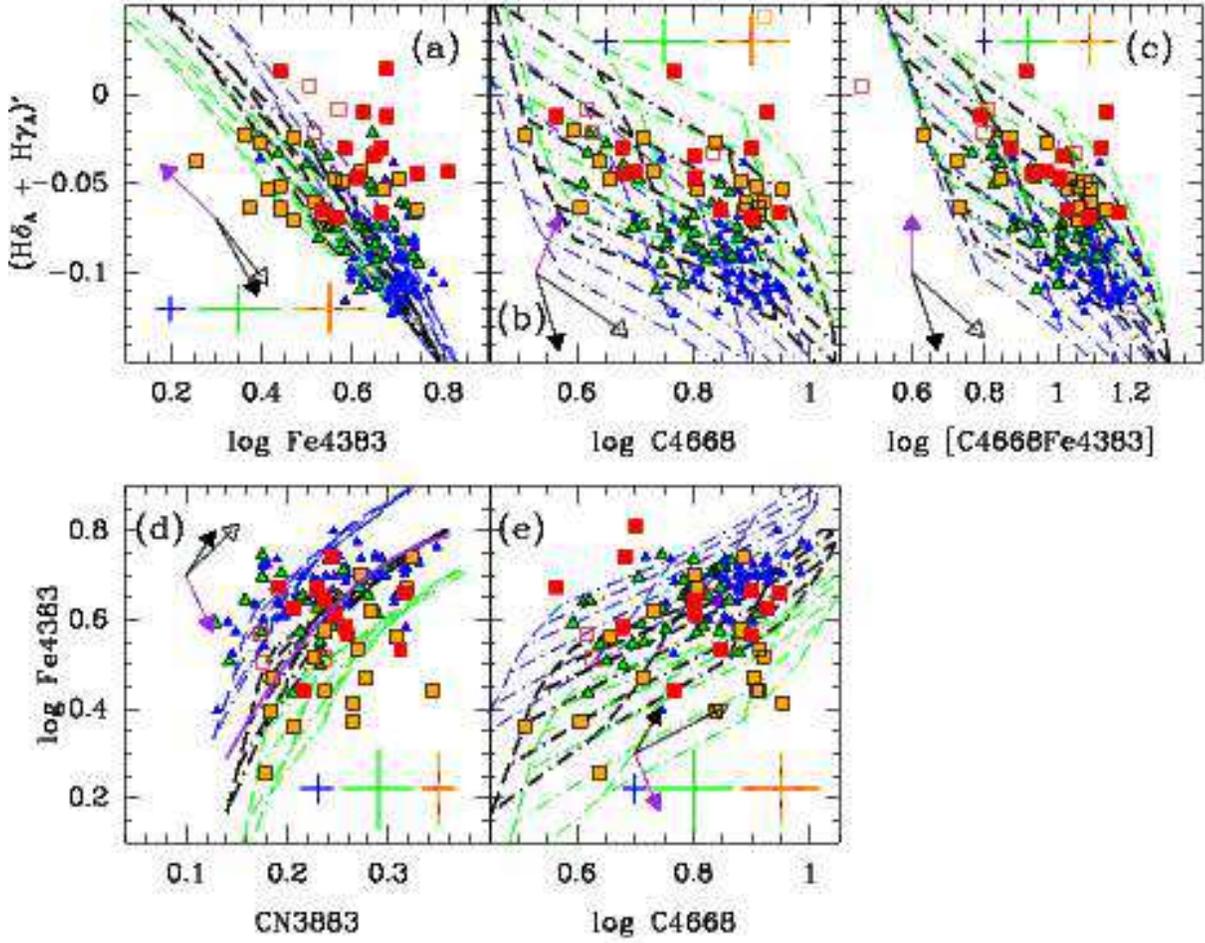}
\caption[]{Line indices in the blue versus each other.
Symbols as in Figure \ref{fig-line_sigma}.
Typical uncertainties are shown, see Table \ref{tab-compline} in the appendix.
We adopt the same uncertainties for RXJ0152.7--1357 as for RXJ1226.9+3332.
Model grids from Thomas et al.\ (2011) are overlayed. Blue -- $\rm [\alpha /Fe] = 0$;
black -- $\rm [\alpha /Fe] = 0.3$; green --  $\rm [\alpha /Fe] = 0.5$. Models are shown for ages 
from 2 Gyr to 15 Gyr and [M/H] from $-0.33$ to 0.67.
The arrows show the model predictions for $\Delta \log {\rm age} = 0.3$ (solid); 
$\Delta {\rm [M/H]} = 0.3$ (open); $\Delta {\rm [\alpha /Fe]} = 0.3$ (purple). 
Solid purple line on panel (d) -- second order approximation to the model for $\rm [\alpha /Fe] = 0.3$ 
used for the determination of $\rm [\alpha /Fe]$, see Sect.\ \ref{SEC-INDICES}.
\label{fig-blueline} }
\end{figure*}

In this section we use the SSP model predictions and the measured line indices to
derive the distributions of ages, metallicities [M/H] and abundance ratios $\rm [\alpha /Fe]$
for each of the clusters. 
Figures \ref{fig-visline} and \ref{fig-blueline} show the line indices versus each other 
with SSP models from Thomas et al.\ (2011) overlayed.

The stellar population models in general span the range of the visible indices 
(H$\beta _{\rm G}$, Mg{\it b}, $\rm \langle Fe \rangle$) fairly well,
though both the low redshift sample and the MS0451.6--0305 sample contain galaxies 
with weaker H$\beta _{\rm G}$ than predicted by the models. 
These galaxies do not have detectable emission lines, though very low level emission fill-in
of the H$\beta$ line cannot be ruled out.
The indices [MgFe] and H$\beta _{\rm G}$ make it possible to derive ages and metallicities [M/H]
without significant dependence on the assumption about $\rm [\alpha /Fe]$.
The Mg{\it b} versus $\rm \langle Fe \rangle$ diagram offers the possibility of estimating 
the abundance ratio $\rm [\alpha /Fe]$ independent of age and total metallicity. 
We return to this later in this section.

For the blue indices the models give a reasonable match to the strengths of C4668, CN3883, Fe4383 and 
$(\rm H\delta _A + H\gamma _A)'$, though a significant number of the galaxies in
the high redshift cluster RXJ1226.9+3332 have strong $(\rm H\delta _A + H\gamma _A)'$
while also having strong Fe4383, and thus being in an area of the parameter space
not covered by the models (see Figure \ref{fig-blueline}a).
From the indices [C4668\,Fe4383] and $(\rm H\delta _A + H\gamma _A)'$ we can derive
ages and metallicities while limiting dependence on the assumption regarding $\rm [\alpha /Fe]$.
CN3883 versus Fe4383 offers a similar possibility of estimating $\rm [\alpha /Fe]$
as the (Mg{\it b},$\rm \langle Fe \rangle$)-diagram. However, CN3883 is only a tracer of $\rm [\alpha /Fe]$
to the extent that carbon and nitrogen enhancements follow the $\rm [\alpha /Fe]$ abundance ratio.

\begin{figure*}
\epsfxsize 16.5cm
\epsfbox{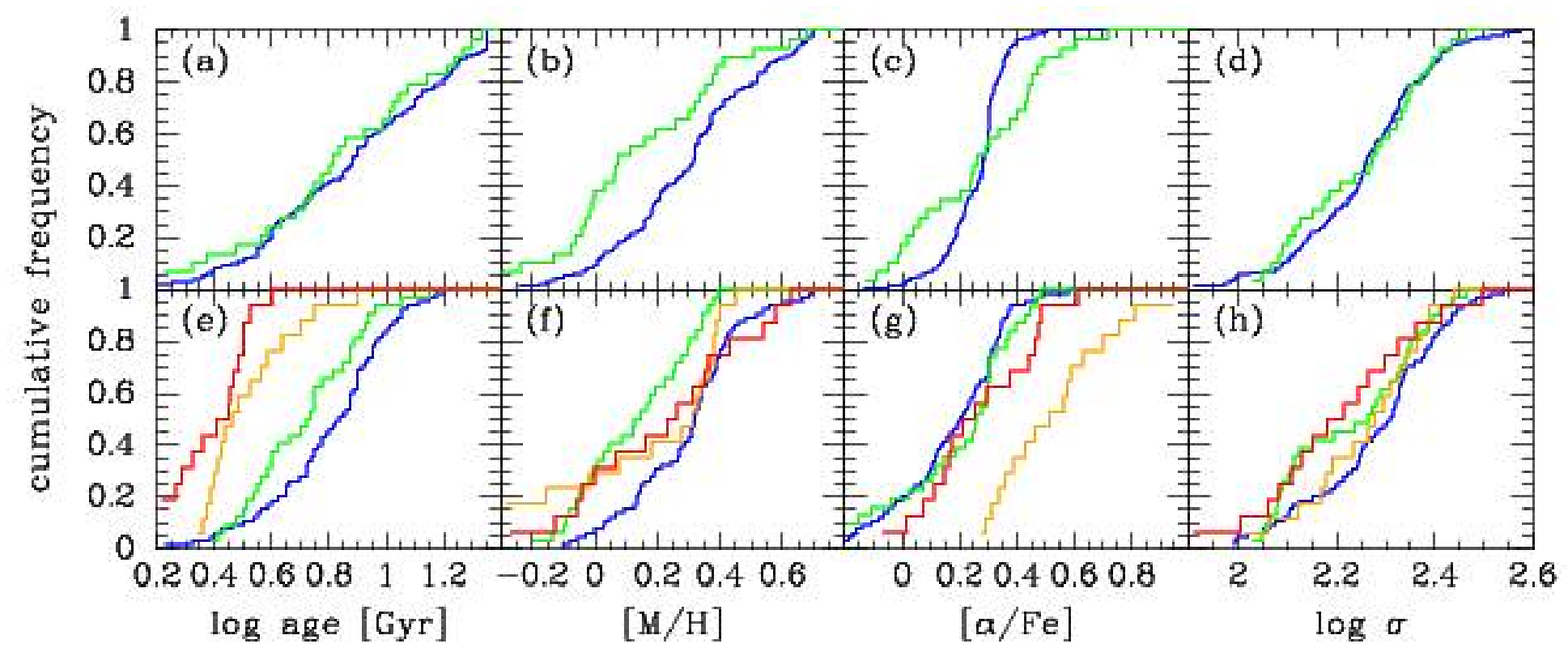}
\caption[]{Cumulative distributions of of age, [M/H], and $\rm [\alpha /Fe]$.
The ages and [M/H] are derived by inverting (H$\beta_{\rm G}$, [MgFe]) (panels a-b) or
($(\rm{H\delta _A + H\gamma _A})'$, [C4668\,Fe4383]) (panels e-f), under the assumption
that $\rm [\alpha /Fe]=0.25$. 
The abundance ratios $\rm [\alpha /Fe]$ are derived from (Fe,Mg{\it b}) or (Fe4383,CN3883), see Sect.\ \ref{SEC-INDICES}.
Panels (d) and (h) show the distributions of the velocity dispersion of the samples used for the 
determination of the ages, metallicities and abundance ratios, see text for discussion.
Blue line -- low redshift sample; green line -- MS0451.6--0305;
orange line -- RXJ0152.7--1357; red line -- RXJ1226.9+3332.
\label{fig-lage_MH_alpha_dist} }
\end{figure*}

In order to establish the distribution of mean ages, mean metallicities [M/H], and
mean abundance ratios $\rm [\alpha /Fe]$, we proceed as follows.
We use a Balmer line index and together with a combination metal index independent of $\rm [\alpha /Fe]$
to derive mean ages and mean metallicities [M/H] by interpolation in the model grid from Thomas et al.\ (2011).
We choose to use a model grid for $\rm [\alpha /Fe] = 0.25$ for this, but because of the indices
used this choice is not critical.
In the visible we use (H$\beta _{\rm G}$,[MgFe]),
while in the blue we use $(\rm H\delta _A + H\gamma _A)'$,[C4668\,Fe4383]).
The resulting cumulative distributions are shown on Figure \ref{fig-lage_MH_alpha_dist}.
To derive the abundance ratio $\rm [\alpha /Fe]$ we use Mg{\it b} versus $\rm \langle Fe \rangle$ in the visible
and CN3883 versus Fe4383 in the blue. 
For both diagrams we fit a second order polynomial to the model points for $\rm [\alpha /Fe] = 0.3$. 
Then we determine the distance from each data point
to this model approximation, measured along the vector of the $\rm [\alpha /Fe]$ dependency of the indices.
The polynomials and the vectors are shown in purple on Figures \ref{fig-visline}d and \ref{fig-blueline}d.
The vector distance is then converted to $\rm [\alpha /Fe]$ using the coefficients for the model
fits in Table \ref{tab-models}. 
The resulting distributions of $\rm [\alpha /Fe]$ are shown on Figure \ref{fig-lage_MH_alpha_dist}.

Independent of the absolute accuracy of the models, these estimates of $\rm [\alpha /Fe]$
convert the two-dimensional parameter spaces of (Mg{\it b},$\rm \langle Fe \rangle$) and 
(CN3883,Fe4383) into one-dimensional parameter space of the estimated $\rm [\alpha /Fe]$
from which we can with better accuracy test if the distributions vary from cluster to cluster.
The samples are in general too small to be suitable for a two-dimensional Kolmogorov-Smirnov test.
It should also be noted that the measurements should be considered
relative, rather than absolute measurements, and that deriving $\rm [\alpha /Fe]$ from
the blue indices assumes that the carbon and nitrogen abundances track the $\alpha$-element
abundances.

The age distributions derived from the 
visible indices show an insignificant age different between the low redshift
sample and the MS0451.6--0305 sample, as also found when using the H$\beta _{\rm G}$ 
versus the velocity dispersion relation.
The age distributions derived from the blue indices show lower median ages 
for the intermediate redshift clusters, restating the result from the $(\rm H\delta _A + H\gamma _A)'$
versus velocity dispersion relation. 

The distributions of [M/H] derived from the blue indices show that the distributions
for RXJ0152.7--1357 and RXJ1226.6+3332 are not significantly different from the distribution for the low redshift
sample. For MS0451.6--0305 the distribution of [M/H], whether derived from the visible
or from the blue indices, shows a significantly lower metallicity than the low redshift sample. 
We tested  the significance of these differences using a one-dimensional Kolmogorov-Smirnov test and
find probabilities that the [M/H] distributions for the MS0451.6--0305 and
the low redshift sample are drawn from the same parent distribution of 1 per cent and $<0.2$ per cent
when using the visible and the blue indices, respectively.
The difference in [M/H] is $\approx 0.2$ dex.

The distributions of $\rm [\alpha /Fe]$ from the blue indices show that the distributions
for MS0451.6--0305 and RXJ1226.6+3332 are not significantly different from the distribution for the low redshift sample.
For RXJ0152.7--1357 significantly higher $\rm [\alpha /Fe]$ are found, as we also noted in J\o rgensen et al.\ (2005).
The median difference between RXJ0152.7--1357 and the low redshift sample is $\approx 0.3$ dex,
A one-dimensional Kolmogorov-Smirnov test shows that the probability the two distributions 
are drawn from the same parent distribution is $\ll 0.01$ per cent.

Ages, [M/H] and $\rm [\alpha /Fe]$ have been found to correlate with the 
velocity dispersions (e.g., Thomas et al.\ 2010). Our current samples at intermediate redshifts
are not large enough to detect these correlations as they have quite high scatter.
However, to ensure that the cluster differences in distributions found for [M/H] and $\rm [\alpha /Fe]$
are not simply caused by differences in the distributions of the velocity dispersions
we also show the distributions of the velocity dispersions for the galaxies included
in the tests (see Figure \ref{fig-lage_MH_alpha_dist}d and h). 
For the distributions based on the blue indicies there is a tendency for 
the intermediate redshift samples to contain a larger fraction of low velocity dispersion
galaxies than is the case for the low redshift comparison sample. 
To correct for the possible effect introduced by these differences, we normalized
the derived ages, [M/H] and $\rm [\alpha /Fe]$ to the median velocity dispersion of the sample
($\log \sigma = 2.24$), using the relations between velocity dispersion, age, M/H] and $\rm [\alpha /Fe]$
given by the two studies by Thomas and collaborators (Thomas et al.\ 2005, 2010).
We then repeated the Kolmogorov-Smirnov tests described above. 
With these normalized values, the probabilities that the [M/H] distributions for the MS0451.6--0305 and
the low redshift sample are drawn from the same parent distribution of is 0.6-1 per cent and $2.5$ per cent
when using the visible and the blue indices, respectively.
When comparing the distributions of the normalized $\rm [\alpha /Fe]$ values for RXJ0152.7--1357 and
the low redshift sample the Kolmogorov-Smirnov test shows that the probability the two distributions
are drawn from the same parent distribution is $\ll 0.01$ per cent.
Thus, both results based on the ``raw'' measurements of [M/H] and $\rm [\alpha /Fe]$ are maintained
when correcting for the correlation of the parameters with velocity dispersion.

Finally, we used the models from Thomas et al.\ (2011) to investigate if 
the carbon or nitrogen enhanced stellar populations would be consistent with the data.
If we use the models with $\rm [C / \alpha] =0.3$ then C4668 is predicted significantly stronger than
observed. 
The dependency of C4668 on $\rm [C / \alpha]$ is so strong (Table \ref{tab-models})
that any carbon enhancement, $\rm [C / \alpha]$, of more than about 0.1 dex affecting all the galaxies
can be ruled out if the total metallicities are above $\rm [M/H] = -0.33$. 
The model with $\rm [N / \alpha] =0.3$ (and $\rm [\alpha /Fe] =0.3$) may be a possibility of 
explaining the strong CN3883 indices for some galaxies, though a general enhancement of nitrogen
$\rm [N / \alpha]$ of more than about 0.1 dex will move the location of the models
off the distribution of data in C4668 versus CN3883.
Based on these considerations, we do not consider the carbon or nitrogen enhanced models
(above the $\alpha$-enhancement) to provide a viable modeling of the line indices.
However, it should still be kept in mind that using C4668 and CN3883 in place of Mg{\it b} 
will provide information about carbon and nitrogen, rather than the $\alpha$-element magnesium,
and only to the extent that enrichment in these elements track each other 
can the indices be used interchangeable.

\begin{figure*}
\epsfxsize 16.5cm
\epsfbox{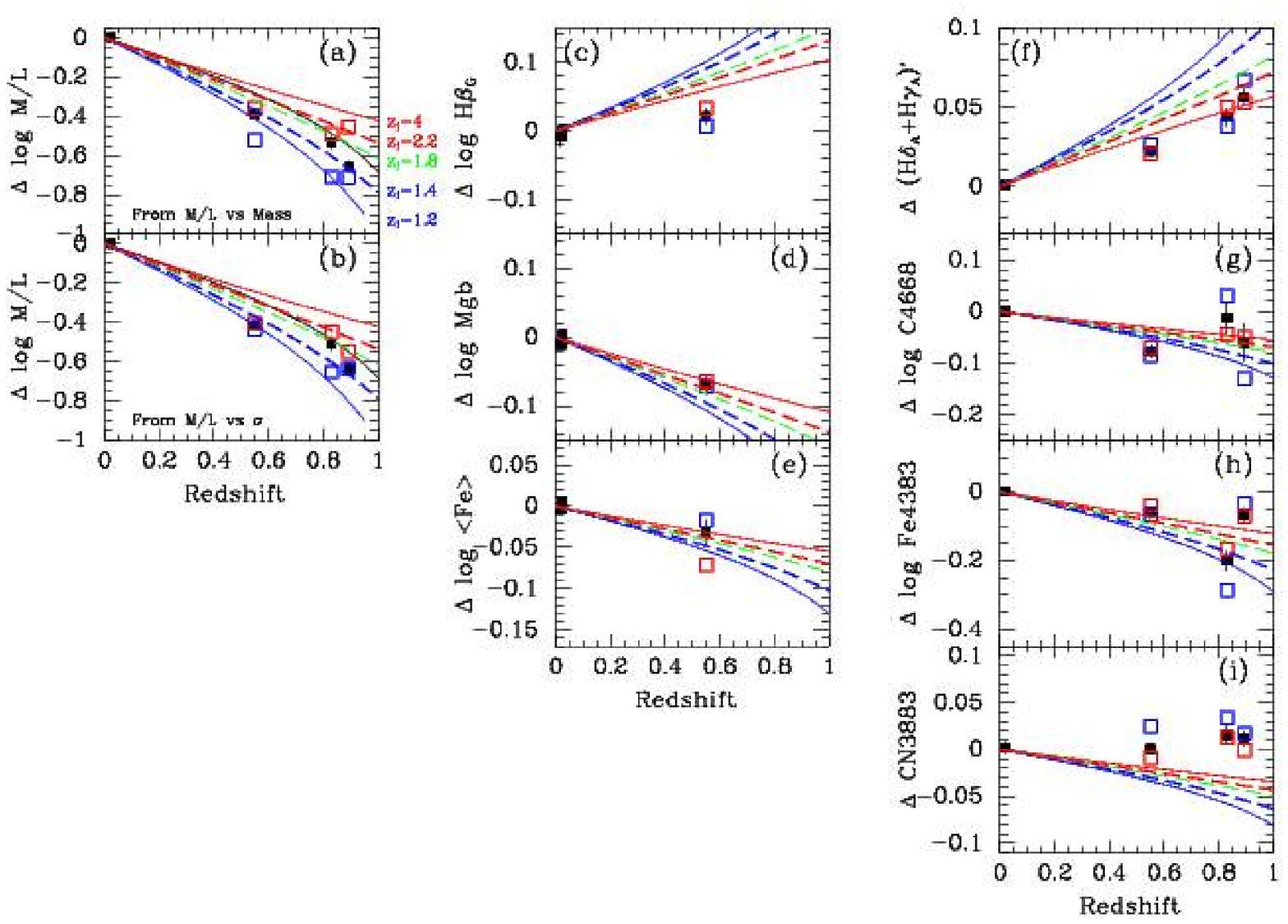}
\caption[]{Zero point differences of the scaling relations versus redshift for each of the clusters, and for
the two subsamples within each cluster.
Black -- all galaxies in a given cluster with the error bars showing the random uncertainties only, cf.\ Eq.\ 7; 
red -- galaxies with $\log \sigma \ge 2.24$ or for panel (a) $\log {\rm Mass} \ge 11$;
blue -- galaxies with $\log \sigma < 2.24$ or for panel (a) $\log {\rm Mass} < 11$.
In all cases the zero point difference is derived as $\rm \Delta zp = zp_{\rm high-z} - zp_{\rm low-z}$,
where the zero points for the full cluster samples are listed in Table \ref{tab-relations}.
Predictions for passive evolution models are overplotted. 
Solid blue line -- Salpeter IMF, $z_{\rm form}=1.2$;
dashed blue line -- Salpeter IMF, $z_{\rm form}=1.4$;
dashed green line -- Salpeter IMF, $z_{\rm form}=1.8$;
dashed red line -- Salpeter IMF, $z_{\rm form}=2.2$;
solid red line -- Salpeter IMF, $z_{\rm form}=4.0$.
Formation redshifts of $z_{\rm form}=1.4$ and 2.2 correspond to the low mass (velocity dispersion) 
and high mass (velocity dispersion) formation redshifts predicted by Thomas et al.\ (2005).
On panels (a) and (b) the solid black lines shows the prediction for a steeper IMF (x=2.3) and 
$z_{\rm form}=1.2$ using models from Vazdekis et al.\ (2010).
\label{fig-zp_redshift} }
\end{figure*}

In summary, the line indices show that (1) the median age varies with redshifts of the clusters, (2)
the MS0451.6--0305 galaxies have [M/H] $\approx$ 0.2 dex lower than the low redshift sample, 
(3) the RXJ0152.7--1357 galaxies have  $\rm [\alpha /Fe]$ $\approx$ 0.3 dex higher than 
the low redshift sample. The latter two points are in disagreement with a passive evolution model.
We will return to this in the discussion in Sect.\ \ref{SEC-DISCUSSION}.

\section{Evolution as a function of redshift \label{SEC-EVOLUTION}}

To study the evolution of the galaxies both as a function of redshift and as function of velocity
dispersion (or mass), we divide the samples in each cluster into two subsamples according to 
velocity dispersions.
We could instead have divided according to the mass. However, since mass and velocity dispersion
are tightly correlated and that relation is the same for all clusters in the sample (see Section \ref{SEC-SIZEMASS})
the two choices are equivalent.
The two subsamples are defined as follows:
\begin{eqnarray}
{\rm Low:} \log \sigma < 2.24 \nonumber \\
{\rm High:} \log \sigma \ge 2.24 
\end{eqnarray}
\setcounter{equation}{8}
The limit in $\log \sigma$ are equivalent to $\log {\rm Mass}$ = 11.0.
As such the high velocity (mass) sample is equivalent to the samples of 
high mass galaxies for which the evolution with redshift has been studied by other groups, e.g.\
van Dokkum \& van der Marel (2007).
The sample limits in $\log \sigma$ also match those used in the analysis of the EDisCS data by
S\'{a}nchez-Bl\'{a}zquez et al.\ (2009).
Galaxies with $\log {\rm Mass} < 10.3$ were excluded from the analysis, cf.\ Section \ref{SEC-METHODSAMPLE}.
Each of the subsamples contain roughly half of the full sample in each cluster.
The approximate average velocity dispersions for galaxies in the two sub-samples are $\log \sigma =2.1$ and 2.35.

We use the scaling relations for the low redshift sample as the reference, see Table \ref{tab-relations}.
For each of the two subsamples in each cluster, and for the full cluster sample, 
we determine the zero point differences to the scaling relations established for the low redshift sample.
The values for the full cluster samples are listed in Table \ref{tab-relations}.

Figure \ref{fig-zp_redshift} shows the zero point differences versus the redshifts of the clusters.
The curves on the figure are models for passive evolution with formation redshifts between
$z_{\rm form}=1.2$ and 4.0, based on the SSP models from Maraston (2005) and Thomas et al.\ (2011).
The zero point differences for the M/L ratios as a function of redshift in general indicate a
lower formation redshift than found from the line indices. 
We discuss this in detail in Section \ref{SEC-PASSIVE}.

To directly estimate the mean ages, metallicities and abundance ratios for the subsamples,
we use the scaling relation zero points for the subsamples for each cluster together with 
the slopes for the low redshift sample to determine the strength of each line at velocity 
dispersions of $\log \sigma = 2.1$ and 2.35, respectively. 
Using the zero points for the full sample in each cluster together with the slope for the 
low redshift sample, we also determine the strength of each line at velocity dispersions
of $\log \sigma = 2.24$. These three velocity dispersion values correspond to masses of
$\log {\rm Mass} = 10.55$, 11.0 and 11.36, respectively.
Thus, for each cluster we have three mean values of each line index, tracing the dependence on 
the velocity dispersion (or alternatively the mass).

We have then used the models to estimate the mean age and [M/H], and $\rm [\alpha /Fe]$ corresponding
to the mean values for the line indices, using the same technique as for the individual galaxies
(see Sect.\ \ref{SEC-INDICES}). This leads to two sets of values 
for (age, [M/H], $\rm [\alpha /Fe]$) for each sample in the low redshift
clusters and MS0451.6--0305. One set comes from using 
(H$\beta _{\rm G}$, Mg{\it b} $\langle \rm Fe \rangle$), while the other set comes from using
($(\rm{H\delta _A + H\gamma _A})'$, C4668, Fe4383). 
For the intermediate redshift clusters RXJ0152.7--1357 and RXJ1226.9+3332 only the latter set of 
values can be derived. 

\begin{figure*}
\epsfxsize 11.5cm
\epsfbox{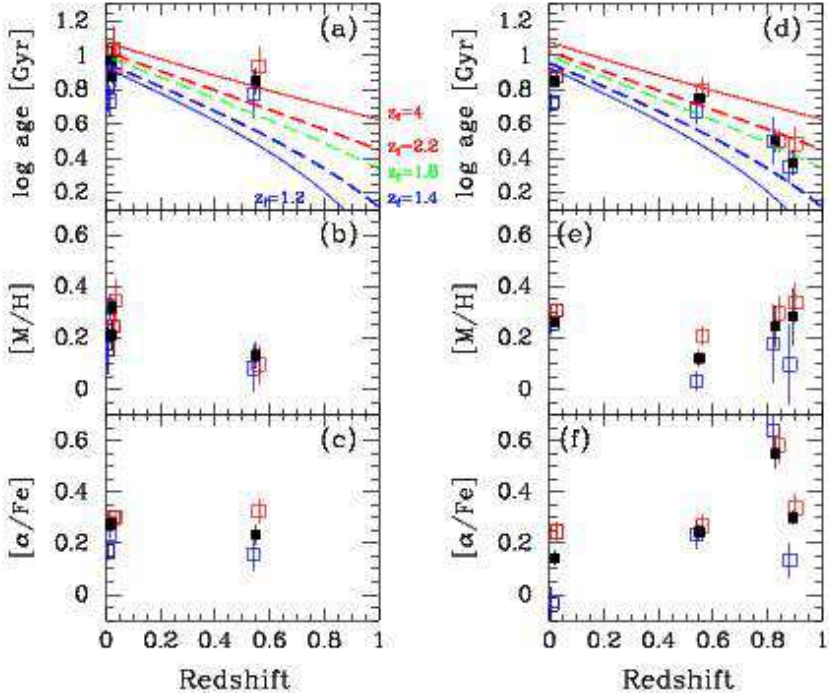}
\caption[]{Estimates of age, [M/H], and $\rm [\alpha /Fe]$ derived from the mean line index values.
The ages and [M/H] are derived by interpolation in the models for (H$\beta_{\rm G}$, [MgFe]) (panels a-b) or
($(\rm{H\delta _A + H\gamma _A})'$, [C4668\,Fe4383]) (panels d-e), under the assumption
that $\rm [\alpha /Fe]=0.25$. 
The abundance ratios $\rm [\alpha /Fe]$ are derived from (Fe,Mg{\it b}) or (Fe4383,CN3883), see Section \ref{SEC-EVOLUTION}.
The error bars correspond to the scatter within the samples of the indices used for the estimates.
Symbols as in Fig.\ \ref{fig-zp_redshift}.
Expected ages for passive evolution models are shown on panels (a) and (d):
Solid blue line -- Salpeter IMF, $z_{\rm form}=1.2$;
dashed blue line -- Salpeter IMF, $z_{\rm form}=1.4$;
dashed green line -- Salpeter IMF, $z_{\rm form}=1.8$;
dashed red line -- Salpeter IMF, $z_{\rm form}=2.2$;
solid red line -- Salpeter IMF, $z_{\rm form}=4.0$.
Formation redshifts of $z_{\rm form}=1.4$ and 2.2 correspond to the low mass (velocity dispersion) 
and high mass (velocity dispersion) formation redshifts predicted by Thomas et al.\ (2005).
\label{fig-lage_MH_alpha} }
\end{figure*}

In Figure \ref{fig-lage_MH_alpha} we show the derived parameters as a function of redshift.
This figure shows the same general trends as are evident in the detailed distributions of the parameters,
Figure \ref{fig-lage_MH_alpha_dist}. 
The ages when derived from the blue indices vary with redshift, in agreement
with a passive evolution model with a formation redshift of $z_{\rm form} \approx 1.8$ with some indication that 
the formation redshift is higher for higher mass galaxies than for the lower mass galaxies.
For MS0451.6--0305 the ages derived from the visual indices are higher than those derived from the blue indices.
However, the difference between the ages for the MS0451.6--0305 samples and the low redshift samples 
are independent of which set of indices are used.
In both cases the ages for the low redshift samples are lower than expected from the 
passive evolution models. We return to this issue in the discussion of the passive evolution model
in Section \ref{SEC-PASSIVE}. 

The metallicity [M/H] for MS0451.6--0305 whether derived from visible or blue indices, is $\approx$ 0.15 dex 
below that of the low redshift sample. Using the uncertainties derived on [M/H] and based on the scatter in
the samples, the difference is significant at the 4 sigma level.
Using the blue indices, all three intermediate redshift clusters
show that the low mass galaxies are less metal rich than the high mass galaxies. The mean difference
is $\approx 0.25$ dex. 
RXJ0152.7--1357 is found to have higher $\rm [\alpha /Fe]$
than the rest of the clusters, $\approx 0.3$ dex, in agreement with the result from the
distribution of the values for the individual galaxies.
The difference is significant at the 5 sigma level.

\section{Discussion \label{SEC-DISCUSSION} }

In this discussion, we initially assume that the galaxies in the high redshift 
clusters are the progenitors of the low redshift
sample and that at each redshift they represent a stage in the evolution that early-type galaxies 
have to pass through.
Specifically, we assume that the galaxies in these clusters have the similar evolutionary
histories, as a function of velocity dispersion (or mass), and that they
travel along similar evolutionary paths to reach the same properties at the present.

The issues we will consider are as follows: 
(1) The effect of selection effects;
(2) possible non-homology among the E/S0 galaxies;
(3) age effects, specifically testing the model of passive evolution;
(4) effects of the assumed IMF;
and (5) metallicity and abundance ratio differences with redshift or among clusters.
Each of these are discussed in the following subsections.

\subsection{Selection effects}

Before we discuss the physical effects, it is important to understand to what extent
the results may be affected by selection effects.
The spectroscopic sample selection was described in Section \ref{SEC-SPECSEL}. Most important
for the current discussion is that each cluster sample has a luminosity limit
but that high redshift samples are not complete to that limit. In addition only
galaxies with early-type morphologies and EW[\ion{O}{2}] $\le 5$ {\AA}ngstrom are included in the sample. 

We focus on how the selection effects may affect the results for the FP
offset and slope. In J\o rgensen et al.\ (2006, 2007) we presented results of
a bootstrap simulation, drawing samples of the same size as the high redshift
sample from both the Coma cluster sample and the high redshift sample and fitting
these. The results showed a significance of the slope difference at the 96-98 per cent level.
However, one could also argue that the difference is slope may be caused by systematically 
missing fainter galaxies in the high redshift sample at a given mass. 

\begin{figure}
\epsfxsize 8.5cm
\epsfbox{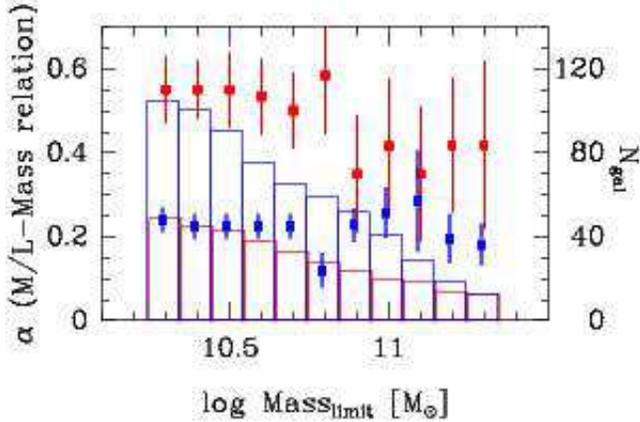}
\caption[]{The slope $\alpha$ of the relation between log M/L and log Mass as a function of the
adopted limit in log Mass. 
Red boxes -- coefficient for fit to the joint sample of RXJ0152.7--1357 and RXJ1226.9+3332.
Blue boxes -- coefficient for fit to the Coma cluster sample, for clarity slightly offset 
from the high redshift sample in log Mass.
The histograms show the number of galaxies included in the fits for each of the limits in log Mass;
red -- the sample of RXJ0152.7--1357 and RXJ1226.9+3332; blue -- the Coma cluster sample.
The difference in slope is significant for all limits of log Mass = 10.8 or smaller.
\label{fig-fpcoeff} }
\end{figure}

We tested this by determining if the difference in slope depends on the chosen
mass limit. Due to the slope of the luminosity limit lines in the M/L--mass plane (Figure \ref{fig-MLonly}a) 
the effect of the limiting magnitude will decrease with increasing mass limits.
The test was done using the joint sample
of the galaxies in RXJ0152.7--1357 and RXJ1226.9+3332, since our analysis of the FP shows
that these two samples follow the same FP. We applied lower limits in log Mass to both 
this sample and the Coma sample from 10.4 to 11.3 in steps of 0.1 and refit each
of these subsamples. Figure \ref{fig-fpcoeff} shows the coefficient of log M/L in the mass-M/L ratio as a 
function of the value of the applied mass limit. For limits of log Mass = 10.8 or smaller the 
difference in the slope for the intermediate redshift sample and the Coma sample is significant
approximately at the same level as for the full sample (also shown on the figure at the limit of
log Mass=10.3). In all cases the slope difference is about 0.32. 
If the sample is limited at log Mass = 10.9 or higher, the slope difference is about 0.16,
and the uncertainties larger due to the smaller samples. Thus, formally the slope difference
is not statistically significant for limits of log Mass $\ge$ 10.9.
However, given that the slope difference is very stable for sample limits of 
log Mass $\le$ 10.8 we conclude that the result is not an effect of the selection effects due
to the luminosity limits of the data.

\subsection{Non-homology among the E/S0 galaxies \label{SEC-NONHOM}}

The mass determination used here, ${\rm Mass} = 5 r_e \sigma^2 {\rm G^{-1}}$ (Bender et al.\ 1992), 
assumes homology among the galaxies.
The current data set does not offer the possibility of explicitly testing the 
effect of possible non-homology among the galaxies.
Van Dokkum et al.\ (2010) propose that a better mass estimate may be achieved
by using the S\'{e}rsic (1968) index, $n$, to determine the coefficient in the mass determination.
Using their equation the coefficient would be 3.24 at $n=2$ and 2.58 at $n=6$, rather
than the constant value of 5 that we have used.
If we determine the masses using the equation from van Dokkum et al.\ (2010) 
and then refit the M/L--mass relation then we find for the sample of the two highest redshift clusters
\begin{equation}
\log M/L = (0.54 \pm 0.18) \log {\rm Mass} - 3.266
\end{equation}
Comparing to the M/L--mass relation for the two highest redshift clusters
listed in Table \ref{tab-FPfit}, we conclude that 
this different mass estimate does not result in a significantly different slope for 
the relation.

Cappellari et al.\ (2006) found from the integral-field-unit (IFU) data of early-type galaxies
that any non-homology is an insignificant contribution to the difference between the FP slope 
and the slope expected from the virial relation, thus leading these authors to conclude that
the difference is due to real M/L variations. 
The same conclusion is reached by Bolton et al.\ (2006) using SDSS data to derive 
galaxy masses based on strong gravitational lensing.
Cappellari et al.\ (2006) also conclude that the approximation 
$\rm Mass = 5 r_e \sigma^2/G$ provides a reasonable mass estimate when data
are not available for more detailed modeling.
Finally, we note that there is no significant correlation between M/L--mass relation residuals and 
the S\'{e}rsic index, using the relations listed in Table \ref{tab-FPfit}.
Since the slope of the M/L--mass relation is the same using mass estimates taking the S\'{e}rsic index
into account compared to using our simpler approximation, and the low redshift IFU studies
confirm that the simpler approximation is valid, 
we will in the following discussion assume that the galaxies in our sample form a homologous class of galaxies.

\subsection{Passive evolution \label{SEC-PASSIVE}}

The simplest of models for galaxy evolution is passive evolution of the stellar populations
over the redshift interval covered by the data.
The FP (Figure \ref{fig-FPonly}), the M/L--mass relation and the M/L--velocity dispersion relation
(Figure \ref{fig-MLonly}) are consistent with passive evolution 
with the low mass (low velocity dispersion) galaxies being younger than those of higher mass (higher velocity dispersion). 
We discussed this result in detail in J\o rgensen et al.\ (2006, 2007) for the two highest 
redshift clusters RXJ0152.7--1357 and RXJ1226.9+3332. 
The data for MS0451.6--0305 are also consistent with this interpretation.
The zero point differences for the subsamples described in Sect.\ \ref{SEC-EVOLUTION} 
implies a formation redshift for the low velocity dispersion galaxies ($\log \sigma =2.1$; $\rm Mass= 10^{10.55} M_{\sun}$)
of $z_{\rm form}=1.24 \pm 0.05$. 
The uncertainty takes into account only the scatter of the data relative to the relations.
For the high velocity dispersion galaxies ($\log \sigma =2.35$; $\rm Mass= 10^{11.36} M_{\sun}$) we find
$z_{\rm form}=1.95_{-0.20}^{+0.30}$. 
The result for the high velocity dispersion galaxies can directly be compared to the result from
van Dokkum \& van der Marel (2007) who find $z_{\rm form}=2.23_{-0.18}^{+0.24}$, before correcting for
progenitor bias. Half of the difference between our result and that of van Dokkum \& van der Marel is
due to a difference in the assumed dependency of the M/L on the age of the stellar population.
As such the two results agree within the uncertainties.

The other major age indicator is the strengths of the Balmer lines. 
The Balmer lines for the intermediate redshift clusters are too weak compared 
to what is predicted by the models for passive evolution with formation redshifts as derived from the M/L ratios.
We focus here on $(\rm{H\delta _A + H\gamma _A})'$, which we have measured for all the clusters. 
As we do not detect a change in slope of the $(\rm{H\delta _A + H\gamma _A})'$--velocity dispersion relation with 
redshift,
we only give the formation redshift for the full sample. For the $z=0.8-0.9$ clusters we find
$z_{\rm form} > 2.8$. This is consistent with the result from Kelson et al.\ (2001) who finds
$z_{\rm form} > 2.4$ or $>2.9$ depending on which SSP model is used for the interpretation.
The formation redshift for the galaxies in MS0451.6--0305 would have to be even higher, see Figure \ref{fig-zp_redshift}f,
for the galaxies to be able to passively evolve into their low redshift counterparts.  

If the offset and slope differences of the M/L-mass relation are purely due to 
age differences, and therefore differences in formation redshift as a function of mass, we also
expect to see offsets and (to lesser extent) slope differences in the relations
between the metal line indices and the velocity dispersion.
As the metal lines are predicted to be weaker with younger ages, we expect
the intermediate redshift samples to show offsets to weaker metal lines.
Figure \ref{fig-zp_redshift} show the zero points together with the model predictions from 
passive evolution models.
The measured offsets do not allow a straightforward interpretation
within the passive evolution model or a conversion of the offsets to a formation redshift.
In general the offsets are smaller than predicted from the passive evolution model from Thomas et al.\ (2005).
In particular, all the intermediate redshift clusters have significantly stronger CN3883 than expected from the 
passive evolution model. CN3883 for the $z=0.8-0.9$ clusters represent 4 sigma
deviations from the model with $z_{\rm form} = 4$.
If we were to adopt the age dependence of CN3883 from the model spectra by Maraston \& Str\"{o}mb\"{a}ck (2011),
the difference would be even larger due to the stronger age dependence predicted for CN3883, see Section \ref{SEC-MODELS}.
%

Part of the difference between the formation redshifts as found from the FP and from the 
high order Balmer lines $(\rm{H\delta _A + H\gamma _A})'$ may be due to different sensitivity
in the two methods to progenitor bias. 
We found that the
mean ages for the low redshift samples are lower than expected from the passive evolution model (Fig.\ \ref{fig-lage_MH_alpha})
and that some of the galaxies have ages too low for them to be the progenitors of the intermediate redshift sample galaxies. 
For example, roughly 40 per cent of the low redshift sample
have ages of less than $\approx 7.1$ Gyr, which is the look back time corresponding to $z=0.86$. 
We have performed a simple test for the effect of the progenitor bias, by
excluding these galaxies from the estimates of the mean ages based on the blue indices. This results result in 
a mean age for the low redshift sample of $\approx 10$ Gyr, and bring the data in agreement with 
the passive evolution model (Fig.\ \ref{fig-lage_MH_alpha}). 
However, excluding these galaxies from the determination of the zero points for the scaling
relations for M/L and $(\rm{H\delta _A + H\gamma _A})'$ versus velocity dispersion lead to insignificant 
changes, only. It also does not bring the zero point differences between the low redshift sample 
and the intermediate redshift samples for these scaling relations into agreement with each other.

\begin{figure}
\epsfxsize 8.5cm
\epsfbox{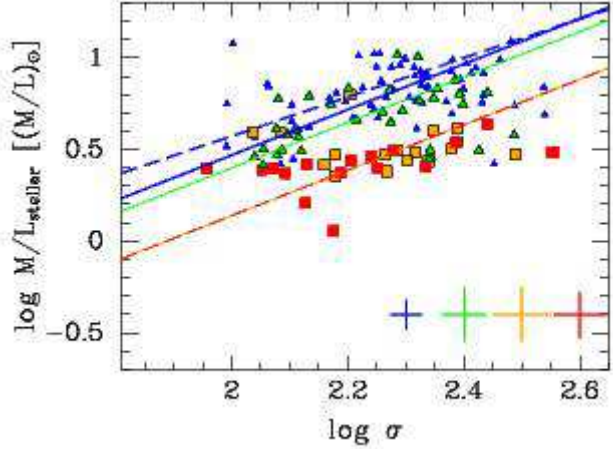}
\caption[]{The stellar M/L ratio as derived from the ages and metallicities resulting from 
$(\rm{H\delta _A + H\gamma _A})'$ and [C4668Fe4383] versus the velocity dispersions.
Symbols as in Figure \ref{fig-blueline}. The solid lines are best fit relations assuming the clusters
follow relations with identical slopes (blue -- low redshift sample; green -- MS0451.6--0307;
orange-red -- RXJ0152.7--1357 and RXJ1226.9+3332 treated as one sample).
Dashed blue line -- for reference the fit to the Coma cluster sample using dynamical M/L ratios, 
Figure \ref{fig-MLonly}b.
The zero point differences indicate formation redshifts of $z_{\rm form}>4$, see text.
\label{fig-MLstar} }
\end{figure}

As a different way of illustrating the disagreement between the FP and the strength of 
$(\rm{H\delta _A + H\gamma _A})'$ we use the SSP models (Maraston 2005) to predict the stellar M/L ratios, given
the ages and metallicities derived from the line indices. As the possible dependency of the M/L ratios
on the abundance ratios $\rm [\alpha /Fe]$ has not been modeled, we will assume that it is insignificant.
On Figure \ref{fig-MLstar} we show the resulting M/L ratio versus velocity dispersion. 
In this view there is no change in the slope with redshift. We fit the samples with a set 
of parallel lines as shown on the figure. The zero point differences agree with the high
the formation redshift found from $(\rm{H\delta _A + H\gamma _A})'$ versus velocity dispersion.
Formally we find $z_{\rm form}> 4$.
Taken at face value, Figure \ref{fig-MLstar} compared with the FP (Figure \ref{fig-MLonly}) indicate that
the stellar M/L ratios change significantly less than the dynamical M/L ratios since $z\approx 0.9$.
We note that Toft et al.\ (2012) found a similar difference in the change of M/L ratios for massive 
galaxies between $z\approx 2$ and the present. These authors speculated that the effect 
may be due to a change in the dark matter content.
In our view, it is premature to conclude that a change in dark matter content can lead to 
the effect we find for the low mass galaxies in these intermediate redshift clusters. However, we
plan to return to this issue using our full galaxy sample in the GCP.

Even if the difference in the formation redshift derived from the FP and from $(\rm{H\delta _A + H\gamma _A})'$ 
versus velocity dispersion may be consolidated by invoking a change in dark matter content,
passive evolution implies that the metallicity [M/H] and the abundance ratio
$\rm [\alpha /Fe]$ do not vary from cluster to cluster. However, both the distributions of
the estimates of [M/H] and $\rm [\alpha /Fe]$ for the individual galaxies (Figure \ref{fig-lage_MH_alpha_dist})
and the mean values derived from the mean indices (Figure \ref{fig-lage_MH_alpha}) 
show that the mean [M/H] for MS0451.6--0305 is 0.2 dex below the other clusters,
while the mean $\rm [\alpha /Fe]$ for RXJ0152.7--1357 is 0.3 dex above
the other clusters. Both of these results contradict the model of passive evolution.

In summary, it is not possible within the current SSP models using one IMF slope (Salpeter) 
for all the galaxies to find a passive evolution model that agrees with the data 
within the measurement uncertainties. While the slope changes with redshift of 
the M/L--mass and M/L--velocity dispersion relations indicate a difference in formation redshift
for low and high mass galaxies, this is not consistent with the $(\rm{H\delta _A + H\gamma _A})'$
data for the intermediate redshift clusters. 
It is also inconsistent with the strong metal lines in general 
found for all the intermediate redshift clusters. Further, the formation redshift implied
by the M/L ratios is significantly lower than found from the higher order Balmer lines.
This result does not change even if we remove all galaxies in the low redshift sample with
ages too young to have progenitors included in the intermediate redshift samples.
Future investigations may show if these two results can be consolidated by the presence of 
a change in the dark matter content, primarily for the low mass galaxies, between $z\approx 0.9$ and the present.

\subsection{Initial Mass Function}

We have investigated what effect differences in the IMF as a function of galaxy mass
would have on the observed evolution. For simplicity we assume that the galaxies evolve passively,
but that the IMF slope depends on the galaxy mass or the velocity dispersion.
This idea is based on recent results indicating that at low redshift higher mass galaxies 
may have steeper IMFs than the low mass galaxies 
(van Dokkum \& Conroy 2010, 2011; Conroy \& van Dokkum 2012ab; Cappellari et al.\ 2012). 
However, an apparently contradicting result was found by van Dokkum (2008) for the evolution
of the IMF, showing that a bottom-light IMF for high mass galaxies can explain the redshift
evolution of both the FP zero point and the (U-V) colors.

To investigate the effect of differences in the IMF between low and high mass galaxies we used
the models from Vazdekis et al.\ (2010). 
Using a Kroupa (2001) ``universal'' IMF instead of a Salpeter IMF leads to the same age 
dependence on the M/L ratio, while there naturally is a zero point difference in the resulting
M/L ratios for the two IMFs due to the presence of more low mass stars in models using a Salpeter 
IMF than those using a Kroupa IMF. However, the evolution in the M/L ratio over the length of time
probed by our data is not significantly different for the two choices of IMF.

We then tested the difference between using a Salpeter IMF with $x=1.3$ as is used in
the default models and a steeper and therefore dwarf rich IMF with a slope of $x=2.3$.
The result is shown on Figure \ref{fig-zp_redshift} as the expected evolution of the M/L ratio.
With the dwarf rich IMF ($x=2.3$) the M/L ratio evolves much slower for a given formation redshift. 
If we hypothesize that the IMF slope is $x=1.3$ for the low mass galaxies and increases slowly to 
$x=2.3$ for the high mass galaxies, then it would reproduce the change in slope for the 
M/L--mass relation (and the FP) with redshift and lead to a formation redshift 
of $z \approx 1.2$ for all the galaxies.
The high order Balmer lines are not significantly affected by the slope of the IMF (Vazdekis et al.\ 2010).
While we in principle can find a solution with a mass dependent slope of the IMF that will mostly
remove the age differences between low and high mass galaxies, the result is that their luminosity
weighted mean ages are all fairly young, about 2 Gyr for the $z=0.8-0.9$ galaxies and about 4 Gyr for the $z=0.54$ galaxies.
The model predictions for strengths of the higher order Balmer line for such young ages 
are in disagreement with our data, and would also predict a much larger change in the strength
of these lines with redshift than seen from the data, see Figure \ref{fig-zp_redshift}f.
Thus, a simple model where the IMF slope depends on the galaxy masses (or velocity dispersions) 
does not seem to be viable solution to a consistent model for the observed evolution of the zero points 
of the M/L--mass relation and the $(\rm{H\delta _A + H\gamma _A})'$--velocity dispersion relation, under the 
assumption of otherwise passive evolution. This does not rule out that there could be IMF variations 
as a function of galaxy masses in addition to more complex models for the galaxy evolution.

\subsection{Metallicities and abundance ratios \label{SEC-METAL}}

We now turn to the possible differences in metallicities and abundance ratios.
The results in Sections \ref{SEC-INDICES} and \ref{SEC-EVOLUTION} 
indicate that the cluster MS0451.6--0305 has [M/H] approximately 0.2 dex lower than the other clusters,
while RXJ0152.7--1357 has $\rm [\alpha /Fe]$ approximately 0.3 dex
higher than the other clusters.
The results indicate that passive evolution is too simple a model to explain the data.

It may be possible to envision that the galaxies in these clusters can evolve into
their low redshift counterparts in the time available. For MS0451.6--0305 this could happen
through additional star formation to increase [M/H], as long as such new star formation
does not lead to ages younger than expected at redshift zero.
For RXJ0152.7--1357 a significant change in $\rm [\alpha /Fe]$ may only
be possible through merging of the early-type galaxies with lower $\rm [\alpha /Fe]$ galaxies
that are not part of the present sample. Thus, these galaxies have to be either spiral 
galaxies, galaxies with significant emission, or galaxies fainter than the sample limit.
In the first two cases, the galaxies are expected to contain younger stellar populations than the 
early-type galaxies in our samples, and the merger products will therefore have 
younger ages than if the early-type galaxies evolved passively.
Any merger scenario also has to maintain the relation between sizes and masses, as 
we detect no evolution in this relation, cf.\ Sect.\ \ref{SEC-SIZEMASS}.

Additional star formation and/or mergers in some clusters and not in others then lead to 
different paths of evolution for the early-type galaxies in intermediate redshift 
clusters in order to reach the properties of 
the low redshift counterparts and one has to ask what the underlying reason for this may be.

Of the clusters in the present sample RXJ0152.7--1357 has the clearest evidence for substructure,
see Sections \ref{SEC-CLUSTERS} and \ref{SEC-CLUSTERZ}. 
The difference in evolutionary path may be related to the interactions of the galaxies
with the intracluster gas as the two substructures interact.
This idea is supported by the results from Demarco et al.\ (2010) who finds that the 
details of the stellar populations appear related to the local cluster density. In particular,
they find that their stacked spectrum of galaxies from low density environment has 
weaker Fe4383 than the rest of the galaxies.
The cluster MS0451.6--0305 
is significantly more massive than the other clusters. It may be that the cluster environment 
in this case led to a truncated star formation history such that the resulting early-type
galaxies have higher ages and lower [M/H] than expected if one just evolved 
RXJ1226.9+3332 passively to $z=0.54$. 
This view is supported by the results from Moran et al.\ (2007b) based on their analysis of
the D4000 break and presence (or rather absence) of emission lines in a very large sample
of galaxies in the cluster.
Moran et al.\ (2007b) also studied another rich intermediate redshift cluster
(CL0024+1654) and found a different star formation history for that cluster best explained
the data, indicating that cluster-to-cluster differences among rich clusters are present at $z\approx 0.4-0.5$.

Even with surveys like the NOAO Fundamental Plane Survey
(Smith et al.\ 2004, 2006; Nelan et al.\ 2005) our knowledge of possible
cluster-to-cluster variations in rich low redshift clusters ($z \le 0.1$) is limited.
The X-ray catalog from Piffaretti et al.\ (2011) contains 26 clusters at $z \le 0.1$ 
with X-ray luminosities at least as high as that of the Coma cluster. 
Those with $z \le 0.067$ are also included in the NOAO FPS sample. 
However, the published analysis of these data does not address possible 
cluster-to-cluster differences. 
Thus, it cannot be ruled out that there are cluster-to-cluster differences among the most massive
galaxy clusters.

\section{Conclusions \label{SEC-CONCLUSION}}

We have presented redshifts, velocity dispersions and absorption line indices
for 102 cluster galaxies at $z> 0.5$. Together with our previously published data for
an additional 29 galaxies,
this represents the largest galaxy sample to date at $z >0.5$ for which velocity dispersions
and line indices are determined for individual galaxies based on spectra with S/N $\ge$ 20 per {\AA}ngstrom in the rest frame.
We used publicly available
{\it HST}/ACS imaging to derive effective radii and surface brightnesses for the sample galaxies.
We have established the size-mass and size-velocity dispersion relations, 
the Fundamental Plane (FP), the M/L-mass and the M/L-velocity dispersion relations, 
and scaling relations between
line strengths and velocity dispersions as a function of redshift.
From the location of the galaxies in index-versus-index parameter space compared
to predictions from single stellar population models we have assessed variations 
in ages, [M/H] and $\rm [\alpha /Fe]$ among the clusters 
and as a function of redshift.
Our main conclusions are as follows:

\begin{enumerate}
\item
The sample shows
no significant evolution of galaxy size or velocity dispersion at a given mass
between $z=0.9$ and the present.

\item
We confirm the change in the slope of the FP and M/L-mass relation as a function of redshift
with the slope being steeper at higher redshift than at $z=0$. The original result was in our study
of the two high redshift clusters RXJ0152.7--1357 and RXJ1226.9+3332 (J\o rgensen et al.\ 2006, 2007). 
In this paper we have added MS0451.6--0305 at $z=0.54$.
This cluster follow relations with slopes between those of RXJ0152.7--1357 and RXJ1226.9+3332 and 
that of the low redshift reference sample. The FPs for all the clusters are consistent
with passive evolution with formation redshifts depending on the mass (or alternatively on
the velocity dispersion), such that low mass galaxies have a more 
recent formation redshift than that of higher mass galaxies. 
Under the assumption of passive evolution
we find that at a velocity dispersion of $\sigma = 125\,\rm km\,s^{-1}$ (Mass=$10^{10.55} {\rm M_{\sun}}$) the formation redshift 
is $z_{\rm form}=1.24 \pm 0.05$,
while at $\sigma = 225\,\rm km\,s^{-1}$ (Mass=$10^{11.36} {\rm M_{\sun}}$ $z_{\rm form}=1.95_{-0.2}^{+0.3}$.

\item
The scaling relations for the line indices cannot easily be reconciled with the results from 
the FP and M/L-mass relation for the clusters studied.
In particular, in the intermediate redshift clusters: 
(1) The Balmer lines are too weak and their scaling relations with the velocity dispersion 
do not undergo the slope changes with redshift as one would expect if the M/L-mass 
zero point differences and slope change are due to age differences, only.
Interpreting the difference in the higher order Balmer lines between the intermediate
redshift clusters and the low redshift sample within the model of passive evolution leads
to a formation redshift $z_{\rm form}> 2.8$.
(2) The metal lines are in general stronger than expected from the predictions for passive evolution,
and we find no significant changes with redshift in the slopes of the scaling relations between
metal line indices and the velocity dispersion.

\item
Using stellar M/L ratios based on the ages and metallicities derived from the high order 
Balmer lines and the metal index [C4668Fe4383] reproduces a high formation redshift,
formally $z_{\rm form}> 4$. It may be worth further investigation to assess if the difference
in how the dynamical and stellar M/L ratios change with redshift 
could be related to a change in the dark matter content of the galaxies.

\item
Under an assumption of passive evolution, significant changes in the IMF slope with mass 
cannot easily be reconciled with the changes with redshift in the 
M/L-velocity dispersion (or mass) relation and the relations between the Balmer lines
and the velocity dispersion. A steeper IMF for higher mass galaxies leads to lower
formation redshift for the high mass galaxies, which does not agree with the relatively
weak Balmer lines observed.
Adopting a Kroupa (2001) IMF instead of a Salpeter (1955) IMF has no significant effect
on the derived formation redshifts.

\item
The data show significant differences in [M/H] for MS0451.6--0305 and in $\rm [\alpha /Fe]$ 
for RXJ0152.7--1357 compared to the other clusters, including the low redshift sample.
MS0451.6--0305 has on average [M/H] approximately 0.2 dex below that of the other clusters,
while $\rm [\alpha /Fe]$ for RXJ0152.7--1357 is approximately 0.3 dex
above that of the other clusters.
Therefore, in order for the intermediate redshift samples to evolve into galaxies similar
to those in the low redshift sample in the time available,
they would have to experience low level star formation (MS0451.6--0305) or merging
(RXJ0152.7--1357) to change 
[M/H] and $\rm [\alpha /Fe]$ of a large fraction of the galaxies.

\end{enumerate}

Based on these results it is not clear if the galaxies in the studied intermediate redshift
clusters share a common evolutionary path 
that will allow them to evolve from what is observed at intermediate redshift to the
properties of their low redshift counterparts in the time available.
If indeed additional data confirms that cluster galaxies 
in different clusters will have to follow different paths of evolution, then these
differences may be related to the details of the cluster environments.
Thus, it remains to be seen if a common evolutionary path taking into account the properties 
of the cluster environment may be established using a larger sample of rich clusters 
both at intermediate redshifts and at $z<0.1$. \\ 

\vspace{0.5cm}

Acknowledgments:
Karl Gebhardt is thanked for making his kinematics software available.
Sean Moran is thanked for making available his unpublished data for MS0451.6--0305 
derived from the {\it HST}/ACS imaging of the cluster and for answering numerous questions.
Ricardo Schiavon is thanked for numerous comments and discussions that helped
clarify the presented results. Richard McDermid and Marcel Bergmann are thanked for
comments on the final manuscript. David Crampton is thanked for information about 
the CDE correction used in the GDDS project.
We thank the anonymous referee for constructive suggestions that helped improve
this paper.
The Gemini TACs and the former Director Matt Mountain are thanked for generous time allocations
to carry out the Gemini/HST Galaxy Cluster Project.

Based on observations obtained at the Gemini Observatory, which is operated by the
Association of Universities for Research in Astronomy, Inc., under a cooperative agreement
with the NSF on behalf of the Gemini partnership: the National Science Foundation (United
States), the Science and Technology Facilities Council (United Kingdom), the
National Research Council (Canada), CONICYT (Chile), the Australian Research Council
(Australia), Minist\'{e}rio da Ci\^{e}ncia e Tecnologia (Brazil) 
and Ministerio de Ciencia, Tecnolog\'{i}a e Innovaci\'{o}n Productiva  (Argentina)

The data presented in this paper originate from the following Gemini programs:
GN-2001B-DD-3, GN-2002B-DD-4, GN-2002B-Q-29, GN-2003B-DD-4, GN-2003B-Q-21, GN-2004A-Q-45, and 
the two engineering programs GN-2002B-SV-90 and GN-2003A-SV-80.
Data for the programs GN-2003A-C-1 and GN-2004A-C-8 were acquired through the Gemini Science Archive.
Observations have been used that were obtained with {\it XMM-Newton}, an ESA science mission 
funded by ESA member states and NASA.
In part, based on observations made with the NASA/ESA Hubble Space Telescope, 
obtained from the data archive at the Space Telescope Science Institute. 

Photometry from SDSS is used for comparison purposes. Funding for the SDSS and SDSS-II 
has been provided by the Alfred P. Sloan Foundation, the Participating Institutions, 
the National Science Foundation, the U.S. Department of Energy, the National 
Aeronautics and Space Administration, the Japanese Monbukagakusho, the Max 
Planck Society, and the Higher Education Funding Council for England. 
The SDSS Web Site is http://www.sdss.org/.



\appendix

\section{A. Photometric Data \label{SEC-IMAGING}}

The ground-based photometric data used in this study are processed as described in detail in
J\o rgensen et al.\ (2005). 
Tables \ref{tab-photdataMS0451} and \ref{tab-photdataRXJ1226} list the derived 
parameters for MS0451.6--0305 and RXJ1226.9+3332, respectively.
The colors are derived inside apertures with diameters of approximately twice the FWHM.
We used 1.45 arcsec and 1.6 arcsec for MS0451.6--0305 and RXJ1226.9+3332, respectively.
The data for RXJ0152.7--1357 used in the present paper are from J\o rgensen et al.\ (2005).

The {\it HST}/ACS imaging data and derived parameters for galaxies in
RXJ0152.7--1357 and RXJ1226.9+3332 are published in Chiboucas et al.\ (2009). 
The {\it HST}/ACS imaging data for MS0451.1--0306 were processed using the same techniques. 
Table \ref{tab-photMS0451HST} lists the derived parameters.

In the following we describe the details of the calibration of the photometric data
as well as comparisons with external data.

\subsection{A.1. Photometry from GMOS-N: Calibration and external comparisons}

The GMOS-N photometry was calibrated to the standard SDSS system using the 
calibrations detailed in J\o rgensen (2009). The absolute uncertainty on this
calibration is 0.05 mag.  
The total magnitudes ({\it mag\_Best} from SExtractor) and aperture colors were 
then compared to a selection of available photometry of the various fields. 

For MS0451.6--0305 we have compared our photometry with (1)
Gunn $g$ and $r$ photometry (in the following denoted ${g_G}$ and $r_G$) from the CNOC 
survey (Ellingson et al.\ 1998), and (2) VRI photometry from observations with
the Subaru telescope of the field by Moran and collaborators made available through 
their web site at http://www.caltech.edu/\~{ }smm/clusters/.
For these comparisons transformations between the different photometric 
systems are necessary. We use the transformations from Jordi et al.\ (2006)
between the SDSS system and VRI (Table 3 in their paper). To determine the expected
offsets between the CNOC photometry and ours, we use the transformations between
Gunn $g$ and $r$ and Johnson V and R from J\o rgensen (1994) together with the relevant 
transformations from Jordi et al. 

For RXJ1226.9+3332 photometry is available from the Sloan Digital Sky Survey, Data Release 7
(Abazajian et al.\ 2009, SDSS in the following).
For galaxies ({\it class\_star}<0.8) we adopt the SDSS magnitudes {\it modelMag}, 
while for stars we adopt the {\it psfMag} as the best measures of the total magnitudes.

\begin{figure*}
\epsfxsize 17.5cm
\epsfbox{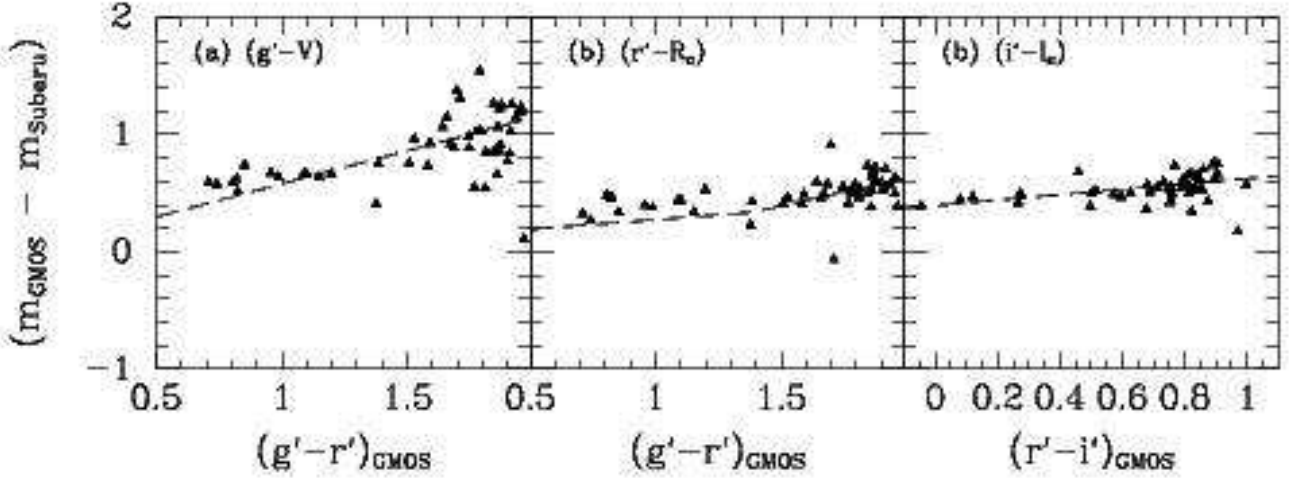}
\caption[]{
Photometry from GMOS-N ({\it mag\_Best} from SExtractor) versus photometry from Moran for galaxies in the
MS0451.6--0305 field. Dashed lines -- expected relations based on transformations from
Jordi et al.\ (2006): $(g'-V)=0.565(g'-r')+0.016$; 
$(r'-R_c)=0.162(g'-r')+0.110$ and $(r'-R_c)=0.468(g'-r')-0.305$ 
for $(g'-r')\le 1.39$ and $(g'-r')>1.39$, respectively; 
$(i'-I_c)=0.232(r'-i')+0.393$.
\label{fig-photsubMS0451} }
\end{figure*}

\begin{figure*}
\epsfxsize 17.5cm
\epsfbox{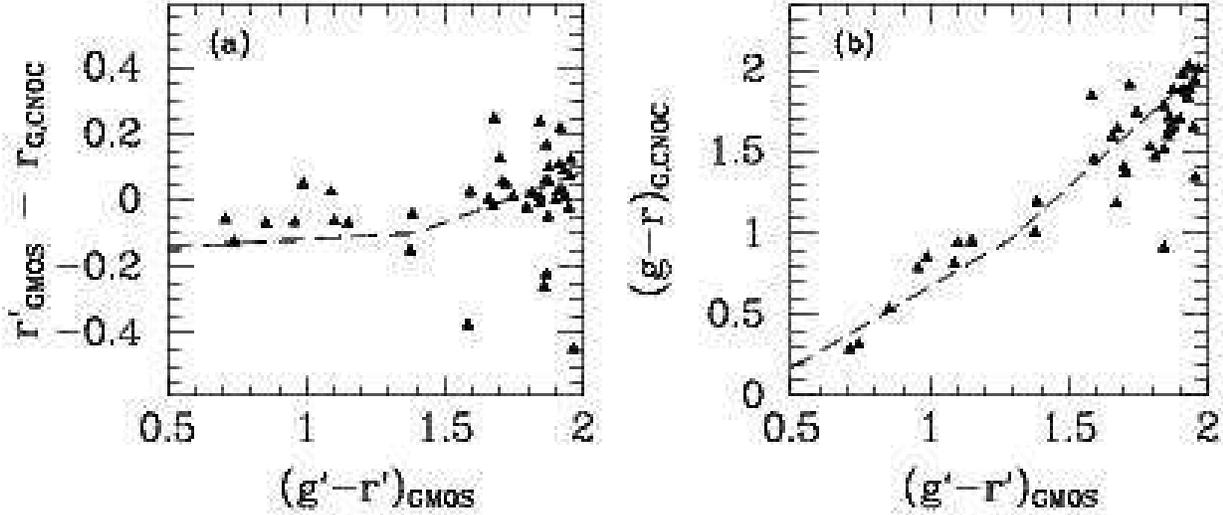}
\caption[]{
Photometry from GMOS-N versus photometry from Ellingson et al.\ (1998) (CNOC) for the
MS0451.6--0305 field. Dashed lines -- expected relations based on transformations
from J\o rgensen (1994) and Jordi et al.\ (2006): 
$(r'-r_G)=0.050(g'-r')-0.170$ and
$(r'-r_G)=0.299(g'-r')-0.512$ 
for $(g'-r')\le 1.39$ and $(g'-r')>1.39$, respectively; 
$(g_G-r_G)=1.005(r'-g')-0.330$ and
$(g_G-r_G)=1.520(r'-g')-0.995$ 
for $(g'-r')\le 1.39$ and $(g'-r')>1.39$, respectively. 
\label{fig-photcnocMS0451} }
\end{figure*}

\begin{figure*}
\epsfxsize 17.5cm
\epsfbox{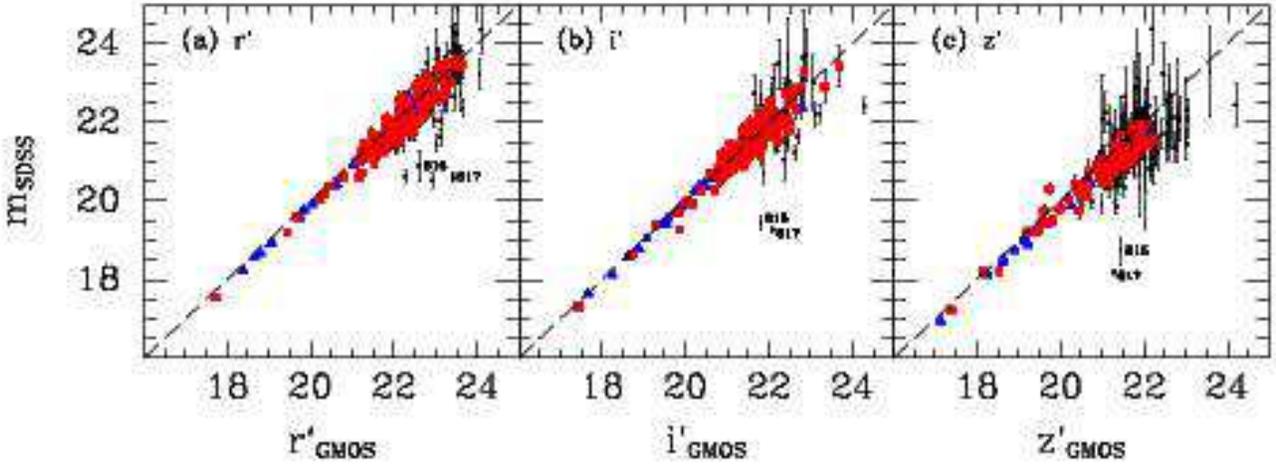}
\caption[]{
Photometry from GMOS-N ({\it mag\_Best} from SExtractor) versus photometry from SDSS/DR7 for the
RXJ1226.9+3332 field.
Blue -- stars included in the determination of the offsets; 
red -- galaxies included in the offsets. 
Small black point -- objects excluded due to large uncertainties or differences between GMOS and SDSS photometry.
Dashed lines -- one-to-one relations.
The two galaxies ID=617 and 815 deviate in all three comparisons. This is most likely due to bright close neighbors, 
which is affecting the SDSS photometry, while being taken better into account in the SExtractor processing
of the GMOS-N photometry.
\label{fig-photsdss} }
\end{figure*}

Figures \ref{fig-photsubMS0451}--\ref{fig-photsdss} show the comparisons. 
The dashed lines on the figures show the expected relations. 
Tables \ref{tab-photcomp} summarizes the offsets and scatter relative to the expected relations.
Moran and collaborators determined total magnitudes using SExtractor in a very similar
fashion as done for our GMOS-N data. The median offsets are 0.04 mag or smaller, indicating
a very good consistency of the two data sets.
The total magnitudes from the CNOC survey are measured from growth curves of aperture
magnitudes with typical aperture corrections of 0.05 mag (Yee et al.\ 1996). 
Further, Ellingson et al.\ (1998) note that 
the systematic uncertainty on $(g_G-r_G)$ is of the order 0.07 mag. Thus, our data
and the CNOC data are consistent within the expected systematic error of the CNOC data.
The SDSS magnitudes are expected to be total magnitudes, while
the SExtractor {\it mag\_Best} is expected on average to be 0.06 mag fainter than
the true total magnitudes (Bertin \& Arnouts 1996). Thus, the expected offset when comparing to the 
SDSS photometry is 0.06 mag. The average of the offsets relative to the SDSS photometry 
is 0.11, indicating that the two datasets are consistent within
0.05 mag, which is in agreement with the expected accuracy of the GMOS-N photometric
calibration (J\o rgensen 2009).

In summary, we conclude that our GMOS-N photometry is calibrated to an 
absolute accuracy of 0.05 mag. 
Tables \ref{tab-photdataRXJ1226} and \ref{tab-photdataMS0451} 
list standard calibrated the magnitudes and colors for the
spectroscopic samples of MS0451.6--0305 and RXJ1226.6+3332, respectively.
The photometry in these tables is not corrected for Galactic extinction.

\begin{deluxetable*}{lrrrrl}
\tablecaption{Comparison of GMOS-N and Other Ground-based Photometry\label{tab-photcomp} }
\tablewidth{0pc}
\tablehead{
\colhead{Cluster} & \colhead{Comparison} & \colhead{Median $\Delta$\tablenotemark{a}} & \colhead{rms} & \colhead{$N$} }
\startdata
MS0451.6--0305 & $(g'-V)$   & $-0.02$\tablenotemark{a} & 0.25 & 50 & Moran, Subaru data\\
MS0451.6--0305 & $(r'-R_c)$ &   0.04\tablenotemark{a} & 0.14 & 50 & Moran, Subaru data\\
MS0451.6--0305 & $(i'-I_c)$ & $-0.01$\tablenotemark{a} & 0.11 & 50 & Moran, Subaru data\\
MS0451.6--0305 & $(r'-r_G)$ & 0.017 \tablenotemark{b} & 0.14 & 42 & CNOC\\
MS0451.6--0305 & $(g'-r')$\tablenotemark{c}  & -0.05\tablenotemark{b} & 0.23 & 42 & CNOC\\
RXJ1226.9+3332 & $(r'-r'_{SDSS})$ & 0.12\tablenotemark{d} & 0.25 & 149 & SDSS, DR7 \\
RXJ1226.9+3332 & $(i'-i'_{SDSS})$ & 0.06\tablenotemark{d} & 0.26 & 155 & SDSS, DR7 \\
RXJ1226.9+3332 & $(z'-z'_{SDSS})$ & 0.18\tablenotemark{d} & 0.24 &  99 & SDSS, DR7 \\
\enddata
\tablenotetext{a}{Comparison relative to expected relations shown on Figure \ref{fig-photsubMS0451}.}
\tablenotetext{b}{Comparison relative to expected relations shown on Figure \ref{fig-photcnocMS0451}.}
\tablenotetext{c}{Offset and scatter of $(g_G-r_G)$ relative to the expected relation 
shown on Figure \ref{fig-photcnocMS0451}b.}
\tablenotetext{d}{Objects with uncertainty larger than 0.5 mag on  $m_{\rm SDSS}$
and objects with $|\Delta | > 0.7$ have been excluded.}
\end{deluxetable*}

\begin{deluxetable*}{rrrrrrrrrr}
\tablecaption{MS0451.6--0305: GMOS-N Photometric Data for the Spectroscopic Sample \label{tab-photdataMS0451} }
\tablewidth{0pc}
\tablehead{
\colhead{ID} & \colhead{RA (J2000)} & \colhead{DEC (J2000)\tablenotemark{a}} &
\colhead{$g'_{\rm total}$} & \colhead{$r'_{\rm total}$} & 
\colhead{$i'_{\rm total}$} & \colhead{$z'_{\rm total}$} & 
\colhead{$(g'-r')$} & 
\colhead{$(r'-i')$} & 
\colhead{$(i'-z')$} }
\startdata
   12&   4 54 12.63&   -3 03 20.2&  22.40&  21.14&  20.35&  19.77&  1.867&  0.893&  0.607\\
  153&   4 54 08.37&   -3 03 16.1&  21.67&  20.89&  20.13&  19.75&  1.377&  0.821&  0.451\\
  220&   4 54 13.35&   -3 03 14.2&  22.60&  21.97&  21.04&  20.56&  0.805&  0.839&  0.469\\
  234&   4 54 07.54&   -3 03 10.3&  21.98&  21.30&  21.04&  20.73&  0.852&  0.080&  0.322\\
  258&   4 54 01.09&   -3 03 08.3&  23.52&  22.13&  21.63&  21.14&  1.528&  0.458&  0.417\\
  299&   4 54 11.91&   -3 02 40.3&  21.64&  20.00&  19.00&  18.52&  1.917&  0.904&  0.497\\
  309&   4 54 07.88&   -3 03 00.6&  23.92&  22.48&  21.60&  21.23&  1.641&  0.711&  0.399\\
  323&   4 54 07.33&   -3 02 55.8&  22.65&  21.31&  20.13&  19.66&  1.719&  0.984&  0.470\\
  386&   4 54 06.49&   -3 02 55.9&  24.39&  23.99&  22.89&  22.17&  1.658&  1.003&  0.804\\
  468&   4 54 07.33&   -3 02 47.9&  24.67&  23.63&  22.67&  21.63&  1.182&  0.781&  1.027\\
  554&   4 54 18.72&   -3 02 36.7&  23.61&  22.20&  21.32&  20.86&  1.685&  0.732&  0.454\\
  600&   4 54 15.14&   -3 02 33.2&  22.90&  21.62&  20.74&  20.34&  1.706&  0.654&  0.445\\
  606&   4 54 05.40&   -3 02 29.5&  23.29&  21.81&  20.86&  20.44&  1.746&  0.790&  0.438\\
  684&   4 54 12.44&   -3 02 16.9&  23.38&  21.78&  21.03&  20.59&  1.775&  0.680&  0.436\\
  716&   4 54 13.68&   -3 02 21.8&  23.50&  21.86&  21.03&  20.57&  1.789&  0.756&  0.454\\
  722&   4 54 12.33&   -3 02 24.4&  23.81&  22.68&  22.43&  22.02&  1.232&  0.300&  0.326\\
  833&   4 54 21.27&   -3 02 17.2&  25.25&  24.43&  23.03&  22.31&  1.609&  1.051&  0.878\\
  836&   4 54 02.97&   -3 02 10.3&  21.76&  21.25&  21.14&  20.84&  0.741& -0.045&  0.253\\
  897&   4 54 20.36&   -3 02 03.8&  23.22&  21.76&  20.81&  20.40&  1.845&  0.811&  0.422\\
  901&   4 54 08.24&   -3 02 04.9&  22.61&  21.99&  21.13&  20.34&  0.822&  0.767&  0.788\\
  921&   4 54 15.29&   -3 01 57.0&  22.14&  20.82&  19.84&  19.37&  1.864&  0.897&  0.443\\
  971&   4 54 17.86&   -3 01 46.6&  22.15&  20.55&  19.60&  19.14&  1.875&  0.859&  0.460\\
 1002&   4 54 17.22&   -3 01 56.6&  23.08&  21.58&  20.81&  20.43&  1.794&  0.794&  0.392\\
 1082&   4 54 11.32&   -3 01 40.8&  22.84&  21.13&  20.20&  19.72&  1.878&  0.849&  0.473\\
 1156&   4 54 04.37&   -3 01 41.6&  22.24&  21.18&  20.51&  20.04&  1.384&  0.688&  0.487\\
 1204&   4 54 10.86&   -3 01 36.2&  23.13&  21.37&  20.44&  20.00&  1.933&  0.807&  0.467\\
 1331&   4 54 16.78&   -3 01 30.7&  22.76&  21.37&  20.77&  20.43&  1.672&  0.502&  0.351\\
 1491&   4 54 08.53&   -3 01 18.7&  22.39&  21.83&  21.27&  20.97&  1.407&  0.671&  0.427\\
 1497&   4 54 12.47&   -3 01 07.7&  22.16&  20.86&  20.07&  19.75&  1.859&  0.752&  0.441\\
 1500&   4 54 12.18&   -3 01 00.9&  21.72&  20.75&  19.87&  19.64&  1.965&  0.874&  0.476\\
 1507&   4 54 12.44&   -3 01 16.6&  23.11&  21.61&  20.64&  20.20&  1.927&  0.840&  0.467\\
 1584&   4 54 15.48&   -3 01 12.9&  26.35&  23.81&  22.83&  21.95&  1.947&  0.968&  0.869\\
 1594&   4 54 14.80&   -3 01 05.6&  22.62&  21.14&  20.15&  19.69&  1.948&  0.839&  0.483\\
 1638&   4 54 09.60&   -3 01 00.5&  23.02&  21.43&  20.44&  20.00&  1.867&  0.834&  0.473\\
 1720&   4 54 10.84&   -3 00 51.6&  21.26&  20.00&  19.00&  18.81&  1.698&  0.890&  0.436\\
 1723&   4 54 18.74&   -3 00 55.0&  21.49&  20.13&  19.45&  19.10&  1.591&  0.626&  0.399\\
 1753&   4 54 10.40&   -3 00 46.7&  21.84&  20.63&  19.77&  19.58&  1.582&  0.677&  0.450\\
 1823&   4 54 09.22&   -3 00 50.3&  23.08&  21.33&  20.32&  19.91&  1.955&  0.831&  0.458\\
 1904&   4 54 05.71&   -3 00 40.4&  23.04&  21.59&  20.71&  20.18&  1.872&  0.822&  0.495\\
 1931&   4 54 14.26&   -3 00 40.6&  22.77&  21.41&  20.45&  20.00&  1.813&  0.841&  0.475\\
 1952&   4 54 14.47&   -3 00 32.0&  22.49&  21.15&  20.13&  19.69&  1.900&  0.855&  0.486\\
 2031&   4 54 08.70&   -3 00 35.5&  23.11&  21.56&  20.56&  20.12&  1.930&  0.813&  0.467\\
 2032&   4 54 01.85&   -3 00 29.4&  21.76&  21.24&  20.94&  20.95&  0.709&  0.121& -0.012\\
 2127&   4 54 15.08&   -3 00 22.6&  21.16&  20.34&  19.93&  19.73&  0.957&  0.265&  0.259\\
 2166&   4 54 04.49&   -3 00 13.9&  21.39&  20.54&  20.05&  19.70&  1.152&  0.497&  0.424\\
 2223&   4 54 09.72&   -3 00 15.1&  23.41&  21.96&  21.04&  20.60&  1.748&  0.807&  0.460\\
 2230&   4 54 14.35&   -3 00 17.9&  25.12&  23.07&  21.99&  21.25&  2.219&  1.112&  0.868\\
 2240&   4 54 10.57&   -3 00 21.4&  24.83&  23.02&  22.07&  21.69&  1.886&  0.814&  0.436\\
 2491&   4 54 14.70&   -2 59 07.2&  23.83&  22.92&  21.94&  21.17&  1.198&  0.905&  0.750\\
 2561&   4 54 12.26&   -2 59 11.8&  23.67&  23.09&  22.69&  21.73&  0.714&  0.459&  0.906\\
 2563&   4 54 14.13&   -2 59 29.4&  22.34&  20.73&  19.85&  19.38&  1.844&  0.813&  0.442\\
 2645&   4 54 05.55&   -2 59 18.5&  20.87&  20.10&  19.66&  19.37&  1.088&  0.517&  0.389\\
 2657&   4 54 03.88&   -2 59 18.4&  21.50&  20.61&  20.04&  19.73&  1.101&  0.571&  0.359\\
 2689&   4 54 01.52&   -2 59 24.0&  22.85&  21.60&  20.72&  20.32&  1.677&  0.777&  0.401\\
 2788&   4 54 10.52&   -2 59 32.2&  23.36&  21.72&  20.84&  20.28&  1.912&  0.798&  0.516\\
 2913&   4 54 04.89&   -3 00 01.3&  22.50&  21.04&  20.08&  19.60&  1.909&  0.832&  0.481\\
 2945&   4 54 07.01&   -2 59 45.0&  23.70&  21.80&  20.91&  20.44&  1.953&  0.746&  0.472\\
 3005&   4 54 16.65&   -2 59 46.0&  25.34&  23.16&  21.95&  21.03&  2.282&  1.074&  0.934\\
 3124&   4 54 12.07&   -3 00 01.1&  22.87&  21.50&  20.64&  20.14&  1.844&  0.806&  0.468\\
 3260&   4 54 07.57&   -3 00 08.2&  23.96&  22.52&  21.57&  21.13&  1.827&  0.784&  0.473\\
 3521&   4 54 20.99&   -2 58 30.7&  22.60&  20.92&  20.18&  19.77&  1.657&  0.597&  0.436\\
 3610&   4 54 15.02&   -2 58 17.0&  22.50&  21.61&  21.21&  20.92&  0.988&  0.272&  0.292\\
 3625&   4 54 16.85&   -2 58 18.4&  25.63&  23.04&  22.05&  21.02&  2.403&  1.023&  1.035\\
 3635&   4 54 13.07&   -2 58 24.3&  25.30&  23.81&  22.65&  22.01&  1.930&  1.164&  0.712\\
 3697&   4 54 07.07&   -2 58 37.5&  22.93&  22.42&  22.00&  21.84&  0.966&  0.311&  0.206\\
 3724&   4 54 06.85&   -2 58 59.3&  23.14&  21.98&  21.09&  20.68&  1.811&  0.758&  0.459\\
 3749&   4 54 06.86&   -2 59 03.8&  22.99&  21.91&  21.06&  20.69&  1.768&  0.755&  0.444\\
 3792&   4 54 02.74&   -2 58 40.4&  23.84&  23.26&  22.30&  21.89&  0.825&  0.755&  0.472\\
 3857&   4 54 09.85&   -2 58 44.7&  23.74&  22.04&  21.22&  20.70&  1.860&  0.732&  0.465\\
 3906&   4 54 10.38&   -2 58 51.4&  24.66&  23.62&  22.93&  22.01&  1.193&  0.539&  0.919\\
\enddata
\tablenotetext{a}{Positions are consistent with USNO (Monet et al.\ 1998), with an rms scatter of $\approx 0.7$ arcsec.}
\tablecomments{Units of right ascension are hours, minutes, and seconds, and
units of declination are degrees, arcminutes, and arcseconds.}
\end{deluxetable*}

\begin{deluxetable*}{rrrrrrrr}
\tablecaption{RXJ1226.9+3332: GMOS-N Photometric Data for the Spectroscopic Sample \label{tab-photdataRXJ1226} }
\tablewidth{0pc}
\tablehead{
\colhead{ID} & \colhead{RA (J2000)} & \colhead{DEC (J2000)\tablenotemark{a}} &
\colhead{$r'_{\rm total}$} & \colhead{$i'_{\rm total}$} &
\colhead{$z'_{\rm total}$} & \colhead{$(r'-i')$} & \colhead{$(i'-z')$} }
\startdata
   18&  12 27 06.81&   33 35 29.7&  23.08&  21.86&  21.35&  1.160&  0.676\\
   38&  12 27 05.65&   33 35 28.6&  23.61&  22.43&  21.77&  1.191&  0.729\\
   53&  12 27 06.23&   33 35 22.4&  22.48&  22.47&  22.20& -0.107&  0.357\\
   55&  12 26 53.16&   33 35 13.7&  23.09&  21.80&  21.16&  1.271&  0.707\\
   56&  12 27 08.86&   33 35 19.9&  23.07&  21.79&  21.66&  1.166&  0.691\\
   91&  12 27 06.64&   33 35 08.2&  24.37&  23.49&  23.18&  0.696&  0.287\\
  104&  12 26 59.78&   33 35 02.1&  23.68&  22.50&  21.79&  1.207&  0.725\\
  122&  12 27 01.32&   33 34 56.3&  23.96&  22.44&  22.10&  1.255&  0.664\\
  132&  12 26 47.77&   33 34 48.3&  22.16&  21.23&  20.95&  1.203&  0.458\\
  138&  12 27 00.63&   33 34 46.9&  22.31&  21.03&  20.68&  1.219&  0.466\\
  153&  12 26 52.68&   33 34 46.3&  22.37&  22.11&  21.98&  0.242&  0.148\\
  154&  12 26 47.80&   33 34 44.6&  23.45&  22.37&  21.79&  1.099&  0.403\\
  178&  12 26 48.35&   33 34 40.2&  22.23&  21.95&  21.96&  0.271&  0.043\\
  185&  12 26 53.32&   33 34 36.2&  22.80&  21.51&  21.09&  1.329&  0.559\\
  203&  12 26 59.35&   33 34 19.9&  21.77&  21.06&  21.02&  0.783&  0.246\\
  220&  12 26 55.16&   33 34 25.5&  23.22&  22.02&  21.66&  1.218&  0.499\\
  229&  12 27 07.35&   33 34 24.3&  23.57&  22.28&  21.87&  1.229&  0.704\\
  245&  12 27 07.42&   33 34 19.7&  23.45&  22.31&  21.75&  1.181&  0.751\\
  247&  12 26 51.34&   33 34 14.9&  20.23&  19.73&  19.47&  0.480&  0.381\\
  249&  12 26 52.08&   33 34 12.5&  20.81&  20.37&  20.22&  0.417&  0.237\\
  293&  12 27 05.85&   33 34 05.2&  23.00&  21.77&  21.27&  1.266&  0.725\\
  295&  12 27 02.18&   33 34 06.4&  22.93&  21.67&  21.40&  1.153&  0.651\\
  297&  12 26 49.78&   33 34 10.4&  24.22&  23.63&  23.10&  0.444&  0.695\\
  309&  12 27 04.45&   33 34 06.8&  24.44&  23.23&  22.46&  1.166&  0.743\\
  310&  12 26 53.25&   33 34 05.7&  24.11&  22.84&  22.21&  1.237&  0.716\\
  316&  12 26 59.58&   33 34 04.3&  23.51&  22.96&  22.59&  0.616&  0.485\\
  329&  12 26 54.54&   33 33 56.6&  21.51&  21.04&  21.11&  0.507&  0.181\\
  333&  12 27 00.44&   33 33 59.2&  23.92&  23.12&  23.19&  0.729&  0.140\\
  347&  12 27 04.39&   33 33 52.6&  21.30&  20.89&  20.54&  0.334&  0.422\\
  349&  12 27 01.33&   33 33 53.4&  22.18&  21.27&  20.89&  0.840&  0.478\\
  359&  12 26 54.38&   33 33 51.0&  22.46&  21.67&  21.54&  0.777&  0.297\\
  374&  12 27 02.07&   33 33 43.6&  21.99&  21.56&  21.55&  0.402&  0.308\\
  386&  12 26 54.48&   33 33 40.7&  23.05&  22.18&  21.91&  0.815&  0.420\\
  408&  12 26 54.25&   33 33 34.8&  23.54&  22.37&  21.88&  1.132&  0.693\\
  423&  12 26 53.15&   33 33 31.4&  23.53&  22.23&  21.64&  1.245&  0.727\\
  441&  12 26 53.86&   33 33 28.9&  24.37&  23.08&  22.77&  1.253&  0.718\\
  446&  12 27 06.76&   33 33 27.1&  23.45&  22.14&  21.66&  1.209&  0.662\\
  452&  12 26 53.99&   33 33 23.1&  23.42&  22.31&  21.98&  1.194&  0.614\\
  462&  12 26 52.47&   33 33 24.2&  23.75&  22.67&  21.92&  1.182&  0.772\\
  470&  12 26 54.90&   33 33 24.5&  24.97&  23.67&  23.10&  1.231&  0.714\\
  491&  12 26 57.55&   33 33 14.0&  23.35&  21.95&  21.46&  1.326&  0.698\\
  499&  12 27 07.72&   33 33 14.3&  24.12&  23.02&  22.39&  1.147&  0.598\\
  500&  12 26 53.55&   33 33 10.1&  22.09&  21.37&  21.10&  0.735&  0.445\\
  512&  12 26 55.67&   33 33 13.1&  24.91&  23.55&  22.91&  1.303&  0.744\\
  523&  12 26 57.18&   33 33 05.1&  23.26&  21.94&  21.64&  1.253&  0.738\\
  527&  12 27 07.77&   33 33 11.1&  24.67&  24.40&  23.56&  0.252&  0.697\\
  528&  12 27 04.49&   33 33 07.8&  22.96&  21.90&  21.24&  1.092&  0.690\\
  529&  12 27 01.38&   33 33 04.8&  22.92&  21.55&  21.03&  1.288&  0.734\\
  534&  12 26 56.83&   33 33 06.3&  23.97&  22.47&  21.90&  1.293&  0.728\\
  547&  12 26 54.23&   33 32 53.6&  23.26&  22.20&  21.82&  1.085&  0.577\\
  557&  12 27 07.72&   33 32 56.1&  23.50&  22.32&  21.84&  1.222&  0.740\\
  563&  12 26 58.29&   33 32 49.0&  21.17&  19.86&  18.15&  1.352&  0.719\\
  572&  12 27 02.65&   33 32 57.5&  22.88&  22.11&  21.98&  0.704&  0.341\\
  592&  12 26 50.22&   33 32 53.3&  22.31&  22.56&  22.05& -0.282&  0.513\\
  593&  12 26 57.08&   33 32 53.5&  24.38&  22.80&  22.40&  1.359&  0.712\\
  602&  12 26 53.01&   33 32 49.9&  22.96&  21.80&  21.23&  1.194&  0.610\\
  608&  12 26 50.63&   33 32 46.5&  22.95&  21.87&  21.39&  1.189&  0.711\\
  630&  12 26 57.35&   33 32 46.9&  24.82&  23.34&  23.08&  1.318&  0.725\\
  641&  12 26 59.28&   33 32 40.9&  23.37&  21.95&  21.97&  1.298&  0.683\\
  647&  12 27 00.00&   33 32 40.9&  23.35&  22.15&  22.46&  1.069&  0.655\\
  648&  12 26 50.41&   33 32 41.7&  23.41&  22.11&  21.56&  1.296&  0.694\\
  649&  12 26 56.33&   33 32 41.2&  24.38&  23.06&  22.48&  1.232&  0.728\\
  650&  12 27 10.57&   33 32 41.9&  23.99&  22.87&  22.32&  1.133&  0.709\\
  656&  12 26 56.72&   33 32 40.7&  23.47&  22.20&  21.73&  1.243&  0.727\\
  675&  12 26 54.68&   33 32 35.6&  23.63&  22.52&  22.13&  1.259&  0.709\\
  685&  12 26 54.80&   33 32 34.4&  24.05&  23.21&  22.71&  0.814&  0.644\\
  689&  12 26 55.30&   33 32 32.8&  23.77&  22.54&  22.07&  1.248&  0.722\\
  703&  12 26 56.12&   33 32 23.4&  22.22&  20.83&  20.44&  1.265&  0.734\\
  709&  12 26 57.13&   33 32 27.8&  23.52&  22.22&  21.81&  1.254&  0.693\\
  711&  12 27 00.52&   33 32 30.1&  24.94&  23.75&  23.29&  1.000&  0.500\\
\enddata
\tablenotetext{a}{Positions are consistent with USNO (Monet et al.\ 1998), with an rms scatter of $\approx 0.7$ arcsec.}
\tablecomments{Units of right ascension are hours, minutes, and seconds, and
units of declination are degrees, arcminutes, and arcseconds.}
\end{deluxetable*}
\begin{deluxetable*}{rrrrrrrr}
\tablecaption{RXJ1226.9+3332: \em -- Continued}
\tablewidth{0pc}
\tablenum{14}
\tablehead{
\colhead{ID} & \colhead{RA (J2000)} & \colhead{DEC (J2000)\tablenotemark{a}} &
\colhead{$r'_{\rm total}$} & \colhead{$i'_{\rm total}$} &
\colhead{$z'_{\rm total}$} & \colhead{$(r'-i')$} & \colhead{$(i'-z')$} }
\startdata

  739&  12 27 04.09&   33 30 56.1&  23.89&  22.66&  21.80&  1.251&  0.799\\
  754&  12 26 58.54&   33 31 03.1&  21.92&  21.83&  21.67& -0.016&  0.231\\
  757&  12 27 04.65&   33 31 01.2&  24.11&  23.03&  22.35&  1.101&  0.755\\
  760&  12 27 03.83&   33 31 01.5&  23.52&  22.16&  21.53&  1.294&  0.701\\
  781&  12 26 46.73&   33 31 03.8&  23.60&  22.35&  21.82&  1.249&  0.505\\
  798\tablenotemark{b}&  12 26 56.11&   33 31 09.5&  25.80&  24.95&  24.54&  0.919&  0.725\\
  801&  12 27 07.23&   33 31 15.9&  23.04&  21.95&  21.44&  1.140&  0.718\\
  805&  12 27 04.35&   33 31 21.6&  22.85&  22.02&  21.79&  0.798&  0.483\\
  824&  12 27 01.58&   33 31 18.4&  21.52&  20.83&  20.76&  0.632&  0.360\\
  841&  12 27 04.49&   33 31 32.9&  21.75&  21.45&  21.26&  0.296&  0.348\\
  861&  12 27 04.90&   33 31 32.5&  21.99&  21.36&  21.13&  0.564&  0.424\\
  863&  12 27 00.87&   33 31 27.1&  22.99&  21.79&  21.55&  1.146&  0.505\\
  872&  12 26 58.08&   33 31 34.6&  23.12&  22.48&  22.20&  0.340&  0.033\\
  883&  12 26 56.32&   33 31 37.2&  24.59&  23.34&  22.78&  1.210&  0.718\\
  899&  12 26 45.63&   33 31 40.7&  23.54&  22.41&  20.83&  1.177&  0.719\\
  907&  12 27 05.19&   33 31 42.8&  23.01&  22.15&  21.93&  1.002&  0.484\\
  928&  12 27 09.15&   33 31 47.1&  20.76&  20.19&  19.82&  0.578&  0.544\\
  933&  12 26 52.98&   33 31 46.9&  24.16&  23.05&  22.59&  1.167&  0.673\\
  934&  12 26 53.53&   33 31 50.2&  21.29&  20.88&  21.03&  0.428&  0.250\\
  960&  12 27 04.98&   33 31 56.2&  22.37&  21.42&  21.30&  0.891&  0.363\\
  968&  12 26 55.83&   33 31 53.4&  24.06&  22.71&  22.07&  1.240&  0.737\\
  982&  12 27 03.02&   33 31 57.3&  19.72&  19.39&  19.38&  0.154&  0.050\\
  995&  12 26 46.13&   33 32 01.5&  23.03&  21.89&  21.28&  1.268&  0.664\\
  996&  12 26 55.64&   33 32 13.1&  21.92&  20.71&  19.61&  1.260&  0.687\\
  999&  12 26 51.27&   33 32 05.2&  24.27&  22.97&  22.75&  1.320&  0.603\\
 1001&  12 26 59.24&   33 32 12.6&  21.81&  21.54&  21.47&  0.292&  0.154\\
 1005&  12 26 54.82&   33 32 21.8&  23.12&  22.22&  22.14&  0.801&  0.311\\
 1022&  12 27 08.20&   33 32 07.5&  24.27&  23.96&  24.45&  0.524& -0.094\\
 1025&  12 26 56.57&   33 32 19.7&  23.66&  22.67&  22.31&  1.192&  0.666\\
 1027&  12 27 03.68&   33 32 10.5&  23.03&  22.34&  21.77&  0.707&  0.525\\
 1029&  12 27 06.01&   33 32 12.0&  23.16&  22.64&  22.46&  0.505&  0.431\\
 1042&  12 26 53.54&   33 32 19.3&  24.23&  22.98&  22.33&  1.223&  0.615\\
 1047&  12 26 56.08&   33 32 15.7&  23.25&  21.92&  21.47&  1.252&  0.676\\
 1057&  12 27 05.52&   33 30 43.9&  23.98&  23.51&  23.18&  0.444&  0.651\\
 1080&  12 27 01.74&   33 30 25.3&  22.83&  22.22&  22.72&  0.434&  0.058\\
 1083&  12 26 49.64&   33 30 19.4&  22.18&  21.80&  21.81&  0.384&  0.127\\
 1091&  12 27 08.23&   33 32 13.1&  23.64&  22.66&  22.10&  0.934&  0.717\\
 1103&  12 27 02.61&   33 30 13.7&  23.89&  23.12&  22.90&  0.738&  0.328\\
 1157&  12 27 03.56&   33 30 14.2&  24.08&  23.33&  22.75&  0.798&  0.543\\
 1164&  12 26 55.69&   33 30 37.9&  23.98&  22.82&  22.22&  1.172&  0.700\\
 1170&  12 26 49.39&   33 30 31.4&  23.26&  21.99&  21.33&  1.258&  0.703\\
 1175&  12 26 57.81&   33 30 37.8&  23.32&  22.12&  21.58&  1.207&  0.692\\
 1182&  12 26 53.88&   33 30 44.9&  24.59&  23.87&  23.30&  0.709&  0.474\\
 1196&  12 27 01.79&   33 30 22.2&  24.53&  23.12&  23.01&  1.444&  0.559\\
 1199&  12 27 08.57&   33 30 48.2&  23.86&  22.75&  22.12&  1.075&  0.773\\
 1251&  12 26 56.33&   33 32 41.8&  24.38&  23.06&  22.48&  1.232&  0.728\\
 1252&  12 27 03.15&   33 31 55.4&  23.17&  21.77&  21.26&  1.281&  0.552\\
 1253&  12 27 07.72&   33 32 54.1&  23.72&  22.81&  22.42&  0.955&  0.648\\
 1254\tablenotemark{b}&  12 26 56.11&   33 31 09.5&  23.41&  22.56&  22.16&  0.919&  0.725\\
\enddata
\tablenotetext{a}{Positions are consistent with USNO (Monet et al.\ 1998), with an rms scatter of $\approx 0.7$ arcsec.}
\tablenotetext{b}{ID=798 and 1254 are listed at the same position as the galaxies are superimposed,
as seen by two overlapping spectra in the slit.}
\tablecomments{Units of right ascension are hours, minutes, and seconds, and
units of declination are degrees, arcminutes, and arcseconds.}
\end{deluxetable*}

\begin{deluxetable*}{rrrrrrrrrr}
\tablecaption{MS0451.6--0305: Photometric Parameters from {\it HST}/ACS data in F814W \label{tab-photMS0451HST} }
\tablewidth{0pc}
\tabletypesize{\scriptsize}
\tablehead{
\colhead{ID} & \colhead{$m_{\rm tot,dev}$} & \colhead{$\log r_{\rm e,dev}$} & \colhead{$<\mu >_{\rm e,dev}$}
& \colhead{$m_{\rm tot,ser}$} & \colhead{$\log r_{\rm e,ser}$} & \colhead{$<\mu >_{\rm e,ser}$}
& \colhead{$n_{\rm ser}$} & \colhead{PA} & \colhead{$\epsilon$} \\
\colhead{(1)} & \colhead{(2)} & \colhead{(3)} & \colhead{(4)} & \colhead{(5)} & \colhead{(6)} & \colhead{(7)} & \colhead{(8)} & \colhead{(9)} & \colhead{(10)}
}
\startdata
   12&  19.53&  0.299&  23.02&  19.70&  0.135&  22.38& 2.32&  -53.1&   0.72\\
  153&  20.19& -0.275&  20.81&  20.19& -0.275&  20.81& 4.00&    9.1&   0.41\\
  220&  20.20&  0.188&  23.14&  20.64& -0.148&  21.90& 1.41&   40.9&   0.12\\
  234&  20.57& -0.190&  21.61&  20.94& -0.470&  20.58& 1.40&   71.3&   0.34\\
  258&  20.82& -0.185&  21.89&  21.20& -0.465&  20.87& 1.23&    7.8&   0.49\\
  299&  18.54&  0.151&  21.30&  18.56&  0.136&  21.24& 3.88&   52.9&   0.12\\
  309&  21.17& -0.254&  21.90&  21.21& -0.286&  21.78& 3.70&  -14.5&   0.19\\
  323&  19.94& -0.370&  20.09&  19.77& -0.230&  20.62& 5.61&   74.3&   0.26\\
  386&  22.27& -0.508&  21.73&  22.33& -0.557&  21.55& 3.50&   27.1&   0.14\\
  468&  21.82& -0.220&  22.72&  22.03& -0.384&  22.11& 2.38&   42.0&   0.23\\
  554&  20.81& -0.383&  20.89&  20.96& -0.497&  20.47& 2.64&  -10.2&   0.10\\
  600&  21.13& -0.446&  20.89&  21.04& -0.376&  21.15& 4.59&   53.8&   0.26\\
  606&  20.59& -0.664&  19.26&  20.53& -0.616&  19.44& 4.86&  -36.6&   0.59\\
  684&  20.49& -0.191&  21.53&  20.65& -0.311&  21.09& 2.80&  -26.0&   0.34\\
  716&  20.55& -0.102&  22.03&  20.71& -0.227&  21.57& 2.87&  -84.7&   0.24\\
  722&  21.41&  0.086&  23.83&  22.08& -0.407&  22.04& 0.60&  -24.4&   0.40\\
  833&  22.74& -0.608&  21.70&  22.74& -0.602&  21.72& 4.14&   -1.9&   0.15\\
  836&  20.45&  0.149&  23.19&  21.13& -0.324&  21.50& 0.82&   43.0&   0.53\\
  897&  20.48& -0.526&  19.84&  20.52& -0.559&  19.72& 3.49&  -17.2&   0.72\\
  901&  20.94& -0.767&  19.10&  20.60& -0.458&  20.31& 8.85&   63.6&   0.16\\
  921&  19.48& -0.143&  20.77&  19.53& -0.177&  20.64& 3.68&   69.1&   0.20\\
  971&  19.01&  0.137&  21.69&  19.12&  0.052&  21.37& 3.42&   12.3&   0.21\\
 1002&  20.10&  0.146&  22.82&  20.31& -0.005&  22.28& 2.89&  -54.1&   0.28\\
 1082&  19.79& -0.224&  20.67&  19.76& -0.200&  20.76& 4.24&    0.2&   0.20\\
 1156&  20.28& -0.173&  21.41&  20.46& -0.305&  20.92& 2.73&   -4.8&   0.61\\
 1204&  20.08& -0.308&  20.53&  20.06& -0.289&  20.60& 4.20&   13.6&   0.34\\
 1331&  20.38& -0.198&  21.39&  20.52& -0.317&  20.93& 2.54&   53.5&   0.65\\
 1491&  21.46& -0.836&  19.27&  21.54& -0.893&  19.07& 2.84&   82.7&   0.54\\
 1497&  19.95& -0.429&  19.80&  19.87& -0.373&  20.00& 4.69&   -8.2&   0.45\\
 1500&  19.60&  0.018&  21.69&  19.88& -0.190&  20.93& 2.50&  -75.5&   0.09\\
 1507&  20.39& -0.551&  19.63&  20.40& -0.556&  19.61& 3.93&   13.1&   0.35\\
 1584&  22.48& -0.582&  21.56&  22.47& -0.575&  21.59& 4.08&   19.3&   0.42\\
 1594&  19.87& -0.237&  20.68&  19.75& -0.141&  21.05& 4.92&  -34.9&   0.19\\
 1638&  20.12& -0.172&  21.26&  19.95& -0.022&  21.83& 5.40&   40.8&   0.07\\
 1720&  18.47&  0.367&  22.29&  18.12&  0.588&  23.06& 5.28&  -70.1&   0.30\\
 1723&  18.98&  0.017&  21.06&  19.27& -0.194&  20.29& 1.51&   76.1&   0.80\\
 1753&  20.16& -0.298&  20.66&  19.83& -0.020&  21.73& 6.37&  -61.7&   0.30\\
 1823&  19.98& -0.353&  20.21&  19.89& -0.278&  20.50& 4.84&  -70.2&   0.29\\
 1904&  20.35& -0.408&  20.30&  20.32& -0.389&  20.37& 4.22&  -38.8&   0.29\\
 1931&  20.11& -0.266&  20.78&  20.20& -0.333&  20.53& 3.31&  -57.1&   0.31\\
 1952&  19.88& -0.247&  20.64&  19.89& -0.255&  20.61& 3.82&  -41.6&   0.16\\
 2031&  20.41& -0.553&  19.64&  20.38& -0.527&  19.74& 4.34&  -70.5&   0.51\\
 2032&  21.01& -0.675&  19.63&  21.03& -0.778&  19.13& 1.10&   60.3&   0.16\\
 2127&  19.36&  0.244&  22.57&  19.91& -0.162&  21.09& 0.89&  -71.2&   0.51\\
 2166&  20.07& -0.015&  21.99&  20.04& -0.022&  21.92& 1.11&  -67.1&   0.41\\
 2223&  20.89& -0.401&  20.88&  20.71& -0.387&  20.77& 2.77&   10.3&   0.34\\
 2230&  21.60& -0.344&  21.88&  21.54& -0.289&  22.09& 4.51&   -6.2&   0.12\\
 2240&  21.60& -0.581&  20.70&  21.75& -0.680&  20.34& 2.40&   85.9&   0.62\\
 2491&  21.43& -0.316&  21.85&  21.43& -0.354&  21.66& 2.54&  -58.2&   0.15\\
 2561&  22.17& -0.379&  22.27&  22.06& -0.282&  22.65& 4.96&  -85.5&   0.46\\
 2563&  19.40&  0.262&  22.70&  19.37&  0.267&  22.70& 3.86&  -28.9&   0.12\\
 2645&  19.01&  0.335&  22.68&  19.43&  0.020&  21.52& 1.45&   77.5&   0.47\\
 2657&  19.25&  0.419&  23.34&  19.91& -0.058&  21.62& 0.78&  -48.7&   0.13\\
 2689&  20.39& -0.598&  19.39&  20.41& -0.616&  19.33& 3.71&  -47.4&   0.68\\
 2788&  20.57& -0.309&  21.01&  20.40& -0.275&  21.02& 2.77&   74.5&   0.13\\
 2913&  19.79& -0.315&  20.21&  19.78& -0.305&  20.26& 4.16&   -2.1&   0.19\\
 2945&  20.58& -0.445&  20.36&  20.67& -0.518&  20.08& 2.73&   28.2&   0.38\\
 3005&  21.18& -0.192&  22.22&  21.38& -0.348&  21.64& 2.69&  -32.3&   0.21\\
 3124&  20.07& -0.114&  21.50&  20.23& -0.234&  21.05& 2.90&  -28.3&   0.58\\
 3260&  21.27& -0.752&  19.50&  21.17& -0.679&  19.77& 5.43&   36.4&   0.69\\
 3521&  19.75&  0.019&  21.84&  19.95& -0.132&  21.28& 2.68&   38.8&   0.25\\
 3610&  20.54&  0.159&  23.33&  21.22& -0.362&  21.41& 0.39&  -57.7&   0.80\\
 3625&  21.54& -0.388&  21.60&  21.36& -0.238&  22.17& 5.36&  -62.9&   0.13\\
 3635&  22.34& -0.430&  22.18&  22.13& -0.434&  21.95& 3.75&   69.4&   0.73\\
 3697&  22.16& -0.632&  20.99&  22.47& -0.846&  20.24& 1.32&   48.3&   0.15\\
 3724&  20.93& -0.492&  20.47&  20.87& -0.442&  20.66& 4.59&  -50.8&   0.15\\
 3749&  20.90& -0.283&  21.48&  20.77& -0.180&  21.87& 5.06&   82.4&   0.18\\
 3792&  21.39&  0.161&  24.18&  22.05& -0.344&  22.32& 0.50&   18.5&   0.48\\
 3857&  20.84& -0.408&  20.79&  20.87& -0.436&  20.69& 3.68&  -34.4&   0.46\\
 3906&  22.51& -0.375&  22.63&  22.57& -0.420&  22.46& 3.61&   81.1&   0.14\\
\enddata
\tablecomments{Col.\ (1) Galaxy ID; col.\ (2) total magnitude from fit with $r^{1/4}$ profile;
col.\ (3) logarithm of the effective radius in arcsec from fit with $r^{1/4}$ profile; 
col.\ (4) mean surface brightness in $\rm mag\,arcsec^{-2}$ within the effective radius, from fit with $r^{1/4}$ profile;
col.\ (5), (6), (7) total magnitude, logarithm of the effective radious and the mean surface brightness, from fit with a S\'{e}rsic profile;
col.\ (8) S\'{e}rsic index; col.\ (9) position angle of major axis measured from North through East; col.\ (9) ellipticity.  
}
\end{deluxetable*}


\subsection{A.2. Photometry from {\it HST}/ACS: Calibration and external comparison}

Total magnitudes, effective radii and mean surface brightnesses for the spectroscopic
sample in MS0451.6--0305 were derived using the fitting program GALFIT (Peng et al.\ 2002)
and the techniques established for our project and described in detail in Chiboucas et al.\ (2009).
Table \ref{tab-photMS0451HST} lists the resulting parameters for the fit with the $r^{1/4}$ profile and with
a S\'{e}rsic (1968) profile.
The magnitudes are on the AB system, using a zero point for F814W of 25.937 (Sirianni et al.\ 2005).
The photometry has not been corrected for Galactic extinction.

\begin{figure*}
\epsfxsize 17.5cm
\epsfbox{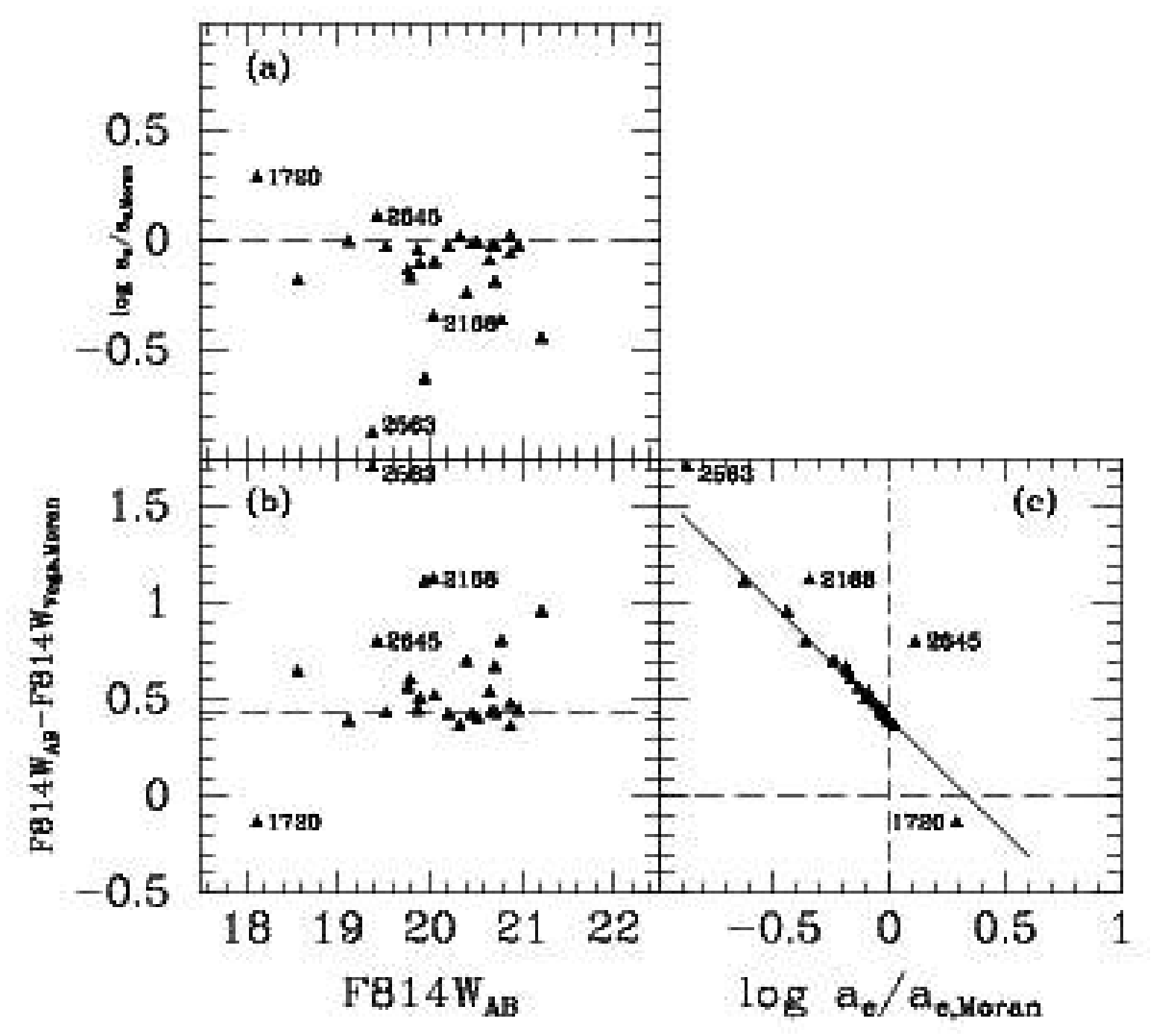}
\caption[]{
Photometry from {\it HST}/ACS F814W versus similar photometry from Moran (private communication)
for the MS0451.6--0306 field.
Dashed line on (b) marks the expected offset between our AB magnitudes and the Vega 
magnitudes from Moran.
(c) Difference in magnitudes versus difference in the logarithm to the semi major axes.
The errors on these parameters are correlated. From the linear fit (solid line) we determine
a magnitude offset between the two data sets of 0.401 mag, see text for details.
\label{fig-phothstMS0451} }
\end{figure*}

We have compared our GALFIT results with data from S.\ Moran (private communication)
who also used GALFIT to process the {\it HST}/ACS data, but with a different PSF model.
Figure \ref{fig-phothstMS0451} and Table \ref{tab-phothstMS0451} summarize the comparison.
As the errors in the total magnitude and the effective radii are correlated (e.g., J\o rgensen et al.\ 1995a)
we also show the comparison as the difference in magnitude versus the difference in effective
radius. 
The four galaxies deviating from the correlation are labeled. 
Two of these (ID=2166 and 2645) are
spirals or irregulars possibly causing the large difference between the two sets of data.
The other two (ID=1720 and 2563) have close neighbors. The detailed handling of these
when running GALFIT is likely to affect the resulting parameters and could be the 
cause of the large differences between the two sets of data. We also note that for ID=2563
the GALFIT magnitude from Moran deviates with more than 1 magnitude from the Subaru 
ground based data made available at Moran's web site http:www.astro.caltech.edu/\~{ }smm/clusters/.
The photometry from Moran is on the Vega system with an F814W zero point of 25.501.
Thus, the expected offset is 0.436 mag. Excluding the four outliers marked on the figure,
we determine the offset between the two data sets by fitting a line to
the data shown on Figure \ref{fig-phothstMS0451}c. 
The intersect between the fit and the y-axis gives an offset of 0.401 mag.
This method of determining the offset
between the magnitudes is more robust than a direct comparison as it eliminates
the effect of the correlated errors. Thus, our magnitudes are 0.035 mag brighter 
than those determined by Moran. This difference is most likely due to difference in the PSFs
used in our work and that by Moran, see Chiboucas et al.\ (2009).

\begin{deluxetable*}{lrrrr}
\tablecaption{Comparison of {\it HST}/ACS Photometry for MS0451.6--0305\label{tab-phothstMS0451} }
\tablewidth{0pc}
\tablehead{
\colhead{Comparison} & \colhead{Median $\Delta$} & \colhead{rms} & \colhead{$N$} }
\startdata
$(\log a_e - \log a_{e,Moran})$  & -0.06 & 0.16 & 23 \\
$(F814W_{AB}-F814W_{Vega,Moran})$ & 0.48 & 0.19 & 23 \\
$(F814W_{AB}-F814W_{Vega,Moran})$\tablenotemark{a} & 0.40 & 0.02 & 23 \\
\enddata
\tablenotetext{a}{Magnitude difference determined from the intersect of fit shown on 
Figure \ref{fig-phothstMS0451}.}
\end{deluxetable*}

For the purpose of calibrating the {\it HST}/ACS photometry to the rest frame
we have chosen first to calibrate the F775W (RXJ0152.7--1357) 
and F814W (MS0451.6--0305 and RXJ1226.9+3332) AB magnitudes to $i'$ magnitudes. 
We use the synthetic transformations from Sirianni et al.\ (2005) 
listed in their Table 3 together with transformations from Jordi et al.\ (2006)
in order to transform from $I_c$ to $i'$ ($I_c = i'-0.245(r'-i')-0.387$). 
While one could argue
that a more direct calibration would be preferable, we judge that our method do not
add significantly to the systematic effects on the data as these are dominated
by the underlying assumptions of the SEDs.
When applying the transformation, we use the $(r'-i')$ colors from our GMOS photometry.

\begin{figure*}
\epsfxsize 17.5cm
\epsfbox{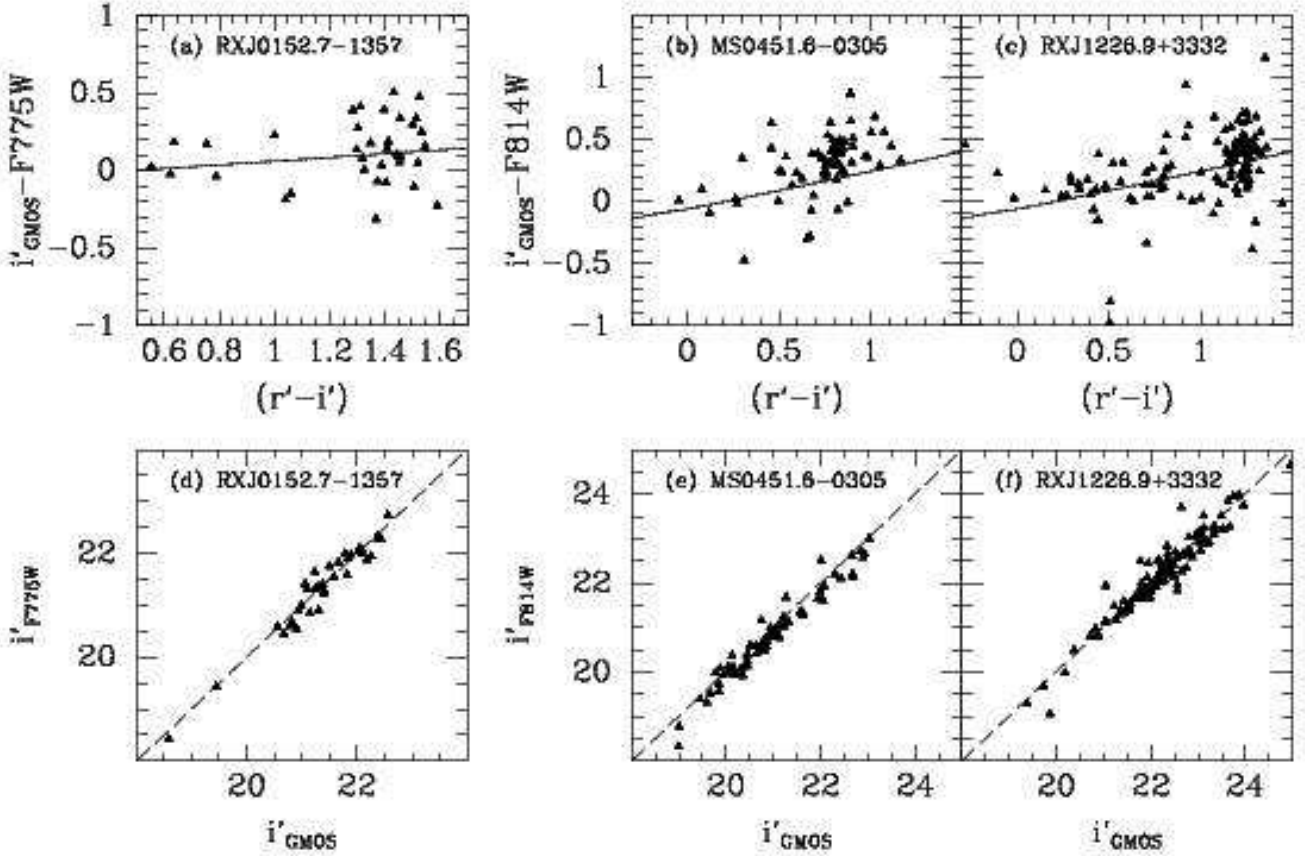}
\caption[]{
Top panels -- differences between total magnitudes from the S\'{e}rsic fits to the {\it HST}/ACS images and the total magnitudes
{\it mag\_Best} from the GMOS-N $i'$-band images. Solid lines show the expected relations, see text.
Bottom panels -- {\it HST}/ACS total magnitudes calibrated to the $i'$-band versus the total magnitudes
{\it mag\_Best} from the GMOS-N $i'$-band images. Dashed lines show the one-to-one relations.
\label{fig-photgmoshst} }
\end{figure*}

Figure \ref{fig-photgmoshst} and Table \ref{tab-photgmoshst} summarize the comparison
of our GMOS-N photometry (derived with SExtractor) and the {\it HST}/ACS photometry 
(derived with GALFIT) calibrated to the $i'$-band.
The expected offset between the magnitudes is 0.06 mag, due to the systematically missed
flux in the best magnitudes from SExtractor (Bertin \& Arnouts 1996). 
For RXJ0152.7--1357 and RXJ1226.9+3332 the magnitudes agree within
the uncertainties, while the median offset for MS0451.6--0305 deviates with 0.08 mag from
the expectations, roughly a 3-sigma difference given the scatter in the comparison.
We take this as the upper limit on systematic errors affecting the photometry.

\begin{deluxetable*}{lrrrr}
\tablecaption{Comparison of GMOS-N and {\it HST}/ACS Photometry\label{tab-photgmoshst} }
\tablewidth{0pc}
\tablehead{
\colhead{Cluster} & \colhead{Comparison} & \colhead{Median $\Delta$} & \colhead{rms} & \colhead{$N$} }
\startdata
MS0451.6--0305 & $(i'-i'_{F814W})$  & 0.14 & 0.20 & 70 \\
RXJ0152.7--1357 & $(i'-i'_{F775W})$ & 0.02 & 0.20 & 36 \\
RXJ1226.9+3332 & $(i'-i'_{F814W})$  & 0.05 & 0.26 & 118 \\
\enddata
\end{deluxetable*}

\subsection{A.3. Calibration of the photometry to the rest frame}

We have derived total magnitudes in the rest frame B-band from the observed $i'$ magnitudes
and colors, using calibrations established based on
Bruzual \& Charlot (2003) stellar population models spanning the observed color range,
see J\o rgensen et al.\ (2005) for details. 
For MS0451.6--0305 we use $(r'-i')$ in the calibration, while for the two higher redshift
clusters $(i'-z')$ was used.  Table \ref{tab-restB} lists the calibration to the rest frame B magnitudes, $B_{\rm rest}$,
for each of the three clusters at the cluster redshifts, as well as
the distance moduli, $DM(z)$, for our adopted cosmology. 
The absolute B-band magnitude, $M_{\rm B}$, is then derived as
\begin{equation}
M_{\rm B} = B_{\rm rest} - DM(z).
\end{equation}
Techniques for how to calibrate to a ``fixed-frame'' photometric system are described in
detail by Blanton et al.\ (2003).

\begin{deluxetable*}{lrrr}
\tablecaption{Calibration to rest frame B magnitudes\label{tab-restB} }
\tablewidth{0pc}
\tablehead{
\colhead{Cluster} & \colhead{Redshift} & \colhead{$DM(z)$} & \colhead{$B_{\rm rest}$} }
\startdata
MS0451.6--0305  & 0.540 & 42.46 & $i'+0.5672+0.5912(r'-i')-0.1180(r'-i')^2$ \\
RXJ0152.7--1357 & 0.835 & 43.62 & $i'+0.8026-0.4268(i'-z')-0.0941(i'-z')^2$ \\
RXJ1226.9+3332  & 0.891 & 43.79 & $i'+0.7966-0.5258(i'-z')-0.0988(i'-z')^2$ \\
\enddata
\end{deluxetable*}

\begin{figure*}
\epsfxsize 17.5cm
\epsfbox{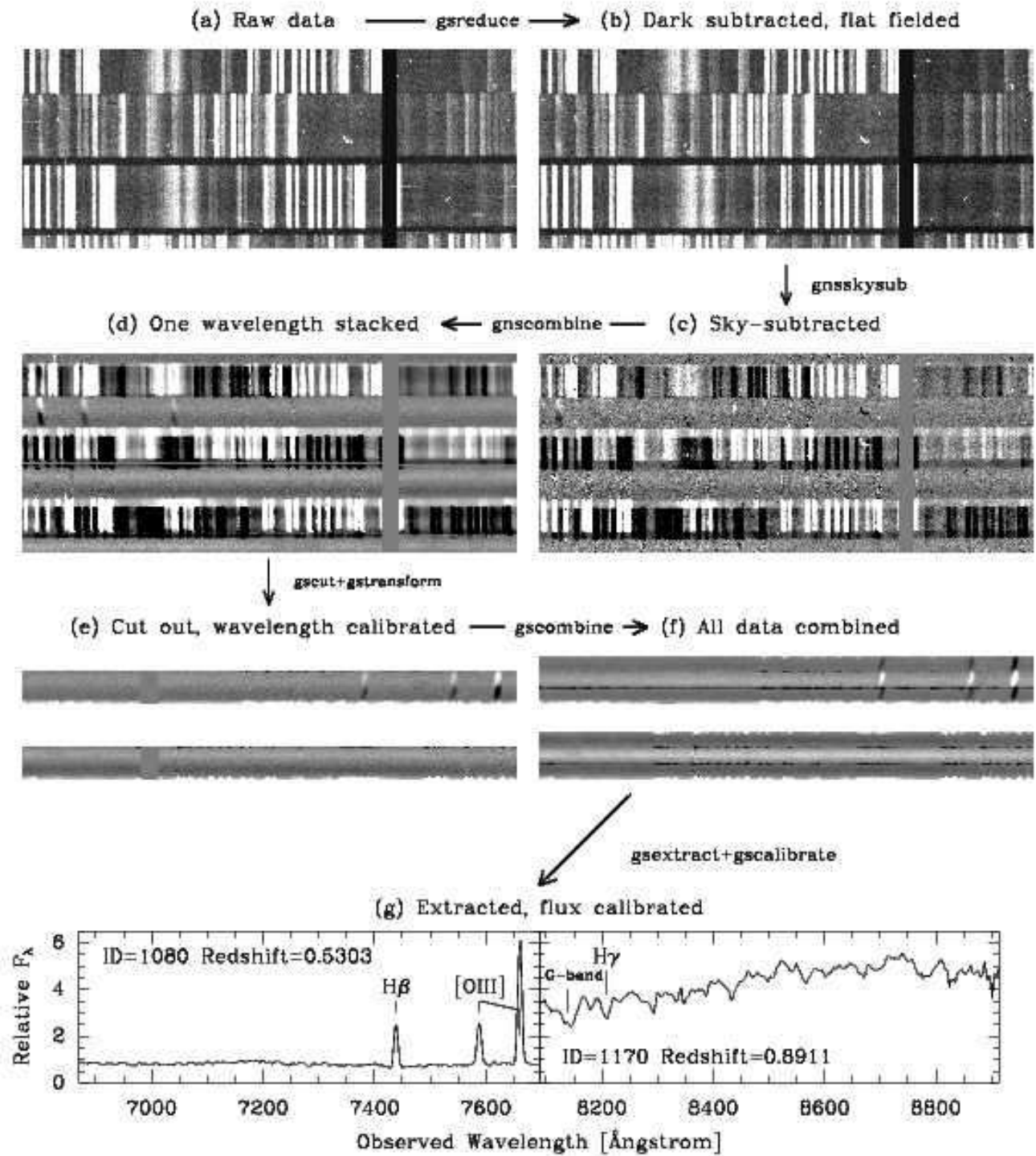}
\caption[]{
Steps in the reductions of GMOS-N nod-and-shuffle data. A small sub-frame of 
$600 \times 240$ pixels of a full frame of the RXJ1226.9+3332 field is shown through the 
various reduction steps. 
The arrows indicate the process direction and the task names from the Gemini IRAF package are listed at the arrows. 
See text for details.
\label{fig-specred} }
\end{figure*}

\section{B. Spectroscopic Data \label{SEC-SPECTROSCOPY}}

\subsection{B.1. Observations and raw data from the GCP}

The GCP spectroscopic data for MS0451.6--0305 were obtained in a similar fashion
as the RXJ0152.7--1357 and were reduced using the same techniques, 
see J\o rgensen et al.\ (2005).
The observations were obtained in individual exposures with exposure time of 2700 seconds.
The slit lengths were between 4.1 and 13.5 arcsec. Longer slits were used for larger
galaxies to facilitate better sky subtraction. A few of the slits also contained two
objects. The two masks used for the field cover a total of 70 galaxies and two blue stars.
Three M-stars were included in the mask by mistake due to limitations in the star-galaxy
separation. The blue stars are used for telluric absorption line correction, as described
in detail in J\o rgensen et al.\ (2005). The faintest objects were included in both masks.
About half the data were taken with a central wavelength set at 7330 {\AA}, the other half with
7410 {\AA}. This facilitates full wavelength coverage across the gaps between the GMOS-N
CCDs.
Between each observation the telescope was nodded along the slit. Three different positions were used: 
zero and $\pm 1.25$ arcsec. 
This was done to improve the sky subtraction, though in retrospect it did not do so, and
in some cases it complicated the sky subtraction as the spectra were very close
to the ends of the slits. 

The GCP spectroscopic data for  RXJ1226.9+3332 were obtained using the GMOS-N in nod-and-shuffle mode.
Glazebrook \& Bland-Hawthorn (2001) describe the principles of the nod-and-shuffle
technique, while Abraham et al.\ (2004) show illustrations and give a detailed
description of GMOS-N nod-and-shuffle data obtained in a fashion very similar
to what was used for the GCP observations.
In the following sections we describe the reduction of the nod-and-shuffle data 
obtained for RXJ1226.9+3332, while we refer to J\o rgensen et al.\ (2005) for those 
steps in the reduction not specific to the nod-and-shuffle mode.
Figure \ref{fig-specred} shows a subframe of observations through one of the masks at different steps
in the data reduction process. We refer to each of these steps in the following 
description.

The data for RXJ1226.9+3332 were obtained in individual 
exposures with exposure times of 1800 seconds. 
We used a slit length of 2.75 arcsec and a shuffle distance of 38 pixels. 
The sub-exposure time was 60 seconds.
The telescope nod along the slit was 1.25 arcsec. This places the two images
of the object 0.625 arcsec from the center of the slit and 0.75 arcsec 
from the ends of the slit. 
Two masks were used for RXJ1226.9+3332 covering a total of 49 galaxies and two
blue stars. The blue stars are used for correction for the telluric absorption lines in the spectra.
The faintest objects were included in both masks. 
About half the data were taken with the central wavelength 
set at 8150 {\AA} and the other half at 8200 {\AA} in order to
obtain full spectral coverage across the gaps between the CCDs.

\subsection{B.2. Charge traps and nod-and-shuffle darks}

As described in Abraham et al.\ (2004), the GMOS-N CCDs have charge traps
that show up as short pairs of elevated signal in the dispersion direction,
separated by the shuffle distance. 
The areas affected by the charge traps are one unbinned pixel in the spatial direction and 6-60 
unbinned pixels in the dispersion direction. 
Figure \ref{fig-specred}a of the raw data shows some of these charge traps.  
Due to the geometry of the charge traps, it is recommended to obtain nod-and-shuffle 
data unbinned in the Y-direction.
Partial correction for the charge traps are achieved by subtraction of nod-and-shuffle 
dark exposures. These exposures were taken with the shutter closed, and the same
exposure time and shuffle configuration as used for the science data.
For RXJ1226.9+3332 we obtained 15 darks. 
The dark exposures are stacked and cleaned for cosmic-ray-events as if they were bias images.
They are used in place of the bias images in the basic reduction of the data.
The signal in the charge traps changes slowly with time, thus it is recommended
to use nod-and-shuffle darks obtained as close as possible in time to the science data, 
while still using a sufficient number to get a high S/N combined dark frame.
Our data for  RXJ1226.9+3332 had only one set of nod-and-shuffle darks, 
so some charge traps are still visible after the dark subtraction.
Figure \ref{fig-specred}b shows the data after subtraction of the nod-and-shuffle
dark (and flat fielding). 

To further eliminate the effect of the charge traps, the science data are taken
at three different detector positions in the spacial direction. This is achieved
by moving the detector translation assembly (DTA) in steps corresponding to integer
unbinned pixels. Data were obtained at DTA positions corresponding to positions
on the detector of 0 (home position), --6 and +6 pixels.
The stacking of the individual exposures is done after the flat field correction
and is described in Sect.\ \ref{SEC-STACK}.

\subsection{B.3. Flat field and charge diffusion correction}

\begin{figure*}
\epsfxsize 17.5cm
\epsfbox{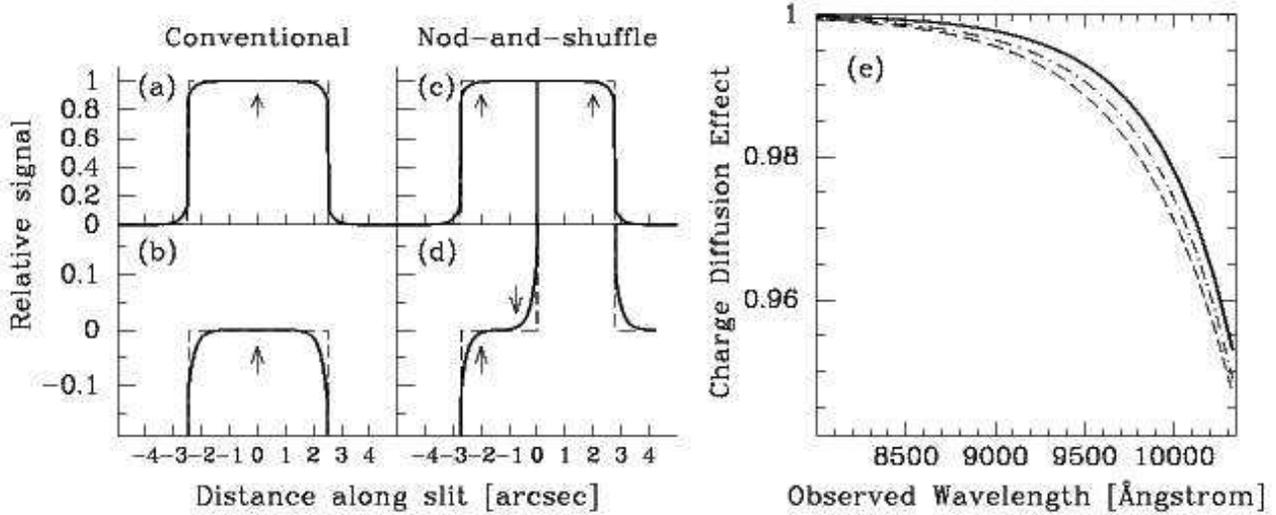}
\caption[]{
(a)-(d) Sketches illustrating the CDE as cuts through spectra in the spatial direction.
The sketches are made based on Eq.\ \ref{eq-redfix} assuming a wavelength of 9500 {\AA}.
Dashed lines -- the spectra in the absence of the CDE.
Solid lines -- the spectra affected by the CDE. The location of the object signal in
the GCP data are indicated with errors.
(a) and (b) show sketches for observations that do not use nod-and-shuffle,
(c) and (d) show sketches for observations using nod-and-shuffle.
(a) and (c) show sketches before processing, while (b) and (d) show the observations
after sky subtraction. In the nod-and-shuffle data the CDE if uncorrected
leads to an oversubtraction of the sky signal. \\
(e) The CDE shown as the correction to the normalized
flat fields, integrated over the apertures used for the targets. 
The CDE on the targets depends the position within the slitlets and the extraction
apertures. 
Solid line -- effect on the GCP data. The effect is about 0.2\% at 8930 {\AA}, increasing
to about 2.5\% at 10000 {\AA}.
Dashed line -- correction used by Abraham et al.\ (2004).
Dot-dashed line -- effect on the data from Abraham et al.\ estimated from our modeling
of the effect.
\label{fig-redfix} }
\end{figure*}

The GMOS-N CCDs show a wavelength dependent charge diffusion effect (CDE). The
effect becomes stronger at longer wavelengths and is seen as charge diffused from
each individual pixel into all neighboring pixels.
The effect is described in Abraham et al.\ (2004).
Figure \ref{fig-redfix} illustrates how the CDE affects the GCP spectroscopic
data in the spatial direction. 
The effect on the science spectra depends on the exact configuration of the 
observations. For the GCP data, and the data described in  Abraham et al.,
the CDE will lead to an oversubtraction of the sky signal at long wavelengths
if no correction is applied for the effect.
The approach used here to correct for the CDE is to model the size of the 
effect from a Quartz-Halogen flat field, and then use the model to correct otherwise normalized
flat fields such that the CDE is corrected for in the flat fielding of the data.
The model contains no correction for the CDE in the dispersion direction.
The CDE in the dispersion direction will lead to a slightly degraded spectral 
resolution at long wavelengths. However, the measurements of the spectral resolution 
as a function of wavelength show that the variation is insignificant for the 
spectral parameters measured in this work.

The details in the modeling of the CDE are as follows.
First the Quartz-Halogen flat field is bias subtracted and normalized slit-by-slit 
(the normal procedure is to normalize row by row). It is known from twilight flat fields 
that the illumination from Quartz-Halogen lamp in the Gemini calibration unit 
can be considered flat on scales of 2.75 arcsec.  Thus, any variation in signal along the 
slits in this normalized flat field is assumed to be due to the charge diffusion.
The flat field is then wavelength calibrated using a matching arc exposure.
The flat field is mosaiced and the slitlets are cut out using the standard tasks
from the Gemini IRAF {\tt gmos} package ({\tt gmosaic} and {\tt gscut}). All the slitlets are combined using
the wavelength information to obtain one flat field spectrum covering the full
wavelength range of all the slitlets. This combined spectrum shows the effect of
the CDE as a variation along the slit, the size of which increases
with increasing wavelength and decreases with distance $\Delta y$ from the slit end.
The variation is fit by a smooth function taking both of these variables into account.
The best fitting function is
%
%
\begin{equation}
\rm{CDE} = 1. - exp ( \psi + (-8.367 \times 10^{-5} - 8.692 \times 10^{-5} \psi ) \lambda)
\label{eq-redfix}
\end{equation}
where
\begin{equation}
\psi = -11.80 - 1.829 \Delta y
\end{equation}
$\lambda$ is the wavelength in {\AA}ngstrom, $\Delta y$ is the distance from the slit end
in unbinned pixels.
The CDE is a multiplicative effect. The normalized flat fields are corrected for the effect by
multiplying these with the function in Eq.\ \ref{eq-redfix}. The function does not depend
on the grating or wavelength setting of the grating. The CDE is present in all GMOS-N
data. 
The size of the CDE for a specific dataset depends on the object positions 
within the slitlets and the extraction aperture.
Figure \ref{fig-redfix} illustrates the size of the CDE for
the configuration used for the GCP data. 
The figure also compares the correction
used by Abraham et al.\ (2004) (D.\ Crampton, private communication) with the correction
resulting from Eq.\ \ref{eq-redfix} for the configuration of slit lengths,
object positions and apertures used by Abraham et al. Due to the shorter slitlets used
by Abraham et al.\ the CDE is slightly larger for those data than for our data.
For nod-and-shuffled data with the configurations used for our observations the
effect becomes larger than 0.2\% red-wards of 8930{\AA}.  
Just red-wards of this wavelength the typical signal from galaxy $i'=22.5$ mag galaxy in our
sample is only $\approx$ 2\% of the signal in the strong sky lines. Thus, uncorrected CDE would 
lead to systematic sky subtraction errors of 10\%.
Eq.\ \ref{eq-redfix} can be used to correct all GMOS-N spectroscopic data taken with 
the E2V detector array for the CDE.
Alternatively, for spectroscopic data obtained without nod-and-shuffle can be sky-subtracted
fitting a second-order polynomial to the sky signal along the spacial direction of the 
spectra, as done in J\o rgensen et al.\ (2005), in which case no correction for the
CDE is needed.

For nod-and-shuffle data, the flat fields are derived by first bias correcting and normalizing
the flat fields row-by-row. Because the flat fields are taken without nod-and-shuffle
they are then ``doubled'' by shifting the flat field by the shuffle distance,
and combining the un-shifted and shifted flat field such that the resulting flat field
contains the resulting flat field correction at both shuffle positions.
The CDE correction is only applied to the ``outer'' sections of a pair of spectra, since
as shown on Figure \ref{fig-redfix}a the effect cancels out on the ``inner'' sections.

\subsection{B.4. Sky subtraction, stacking and calibrations \label{SEC-STACK}}

After the dark subtraction and flat fielding (processing with the task {\tt gsreduce})
the frames are sky subtracted using the task {\tt gnsskysub}. This task makes 
a shifted copy of each frame and subtracts it off the original frame, such that the two
shuffled images of a given slitlet is differenced. The result is a sky subtracted image
with the object appearing positive from one nod-position and negative from the other
nod-position, see Figure \ref{fig-specred}c.
With the knowledge of the DTA positions, all exposures obtained at the same wavelength
are then stacked, shifted to the DTA zero position. In the process cosmic-ray events and
any remaining signal from the charge traps is rejected. The frames are given weight according
to their signal as determined from the two blue stars included in the mask. A custom version of the
task {\tt gnscombine} was used to accomplish all of these processing steps.
The clean stack is shown in 
Figure \ref{fig-specred}d. There are separate stacks for different wavelength settings and
different masks. 

The wavelength calibrations are established in the standard way from CuAr lamp exposures.
The stacked science frames are cut into slitlets and each slitlet is wavelength calibrated.
The cutting is done using the task {\tt gscut} which automatically finds the edges of 
the slitlets based on a matching flat field frame. The wavelength calibration is 
applied using the task {\tt gstransform}. The resulting frames have one science extension
per slitlet. Figure \ref{fig-specred}e shows the result for two objects.

The subframe shown on Figure \ref{fig-specred} contains one of the gaps between the CCDs. 
This is seen as the dark vertical area in panels (a) and (b) and the light grey vertical 
area in panels (c)--(e). 
At this point the two wavelength settings are combined and a the same time the positive and 
negative nod-position are combined. This is done using a custom script {\tt gscombine} which
handles all the offsets as well as weights the input relative to the signal in these.
The resulting frames have one science extension per slitlet (Figure \ref{fig-specred}f) 
with the full signal is in the center of the extension. The outer edges in the spatial 
direction still shows the negative object signal from the nodding. Because of exposures
being taken at two wavelength settings, the resulting spectra have no gaps in the 
wavelength coverage, though of course the S/N is slightly lower where only signal from
one wavelength setting contributes.

The spectra are then traced and extracted to one-dimensional spectra in a standard way
using the tasks {\tt gsextract}. We use an aperture size of 0.85 arcsec. 
The spectra are corrected for the telluric absorption lines using the blue stars in
the masks and a relative flux calibration is applied using the task {\tt gscalibrate}, 
see J\o rgensen et al.\ (2005) for details.
Figure \ref{fig-specred}g shows the extractions of the wavelength regions of the 
targets shown throughout the figure.

\begin{figure*}
\epsfxsize 17.5cm
\epsfbox{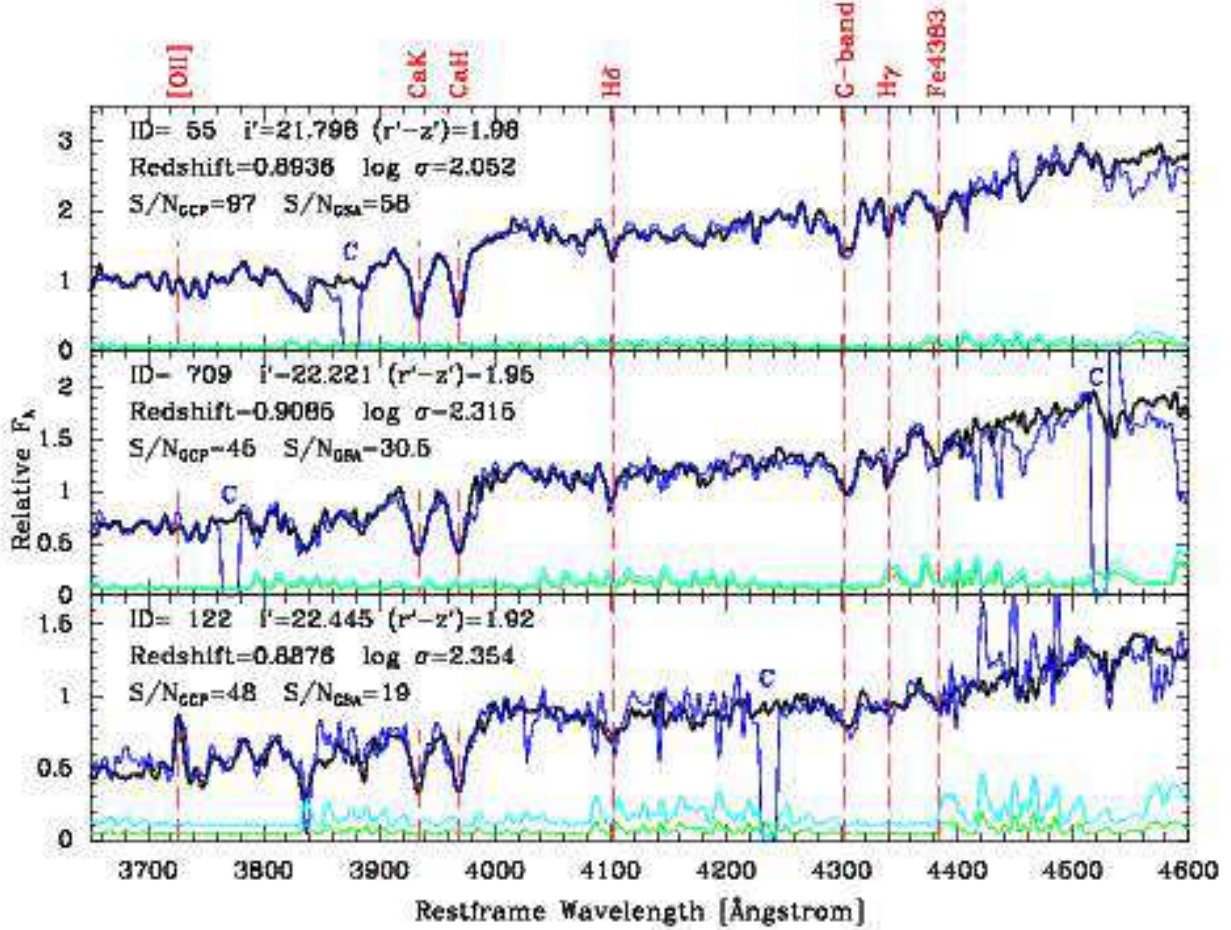}
\caption[]{Spectra from the GCP data and the GSA data of the same three galaxies.
The GSA spectra have been scaled to the level of the GCP spectra.
Black -- GCP spectra; dark blue -- GSA spectra; green -- four times the noise of the GCP spectra; 
light blue -- four times the noise of the GSA spectra. 
The chip gaps in the GSA data are marked with `C'.
\label{fig-sncutoff} }
\end{figure*}

\subsection{B.5. Considerations for the spectroscopic GSA data}

The observations for the GSA data did not make use of the GMOS nod-and-shuffle mode.
We therefore reduced the data using the methods described in J\o rgensen et al.\ (2005). 
However, many of the objects were close to the ends of the slits, presumably to 
make it possible to cover more objects with a single MOS masks.
This led to significantly larger systematic sky subtraction errors
than for the GCP data.
The systematic errors in the sky subtraction make the spectra unreliable at 
wavelengths longer than $\approx$8500 {\AA}. This corresponds to 4500 {\AA} in the 
rest frame of RXJ1226.9+3332.
In this paper, the data are not used beyond $\approx$8500 {\AA} in the observed frame.

The GSA data include no blue stars in the masks suitable for deriving the telluric correction.
We therefore used the telluric correction spectrum from the GCP data, convolved to the 
resolution of the GSA data, and scaled to optimize the correction.

\subsection{B.6. Instrumental resolution}

The instrumental resolution for each spectrum was derived from the sky lines of spectra
stacked in the same way as the science spectra, see J\o rgensen et al.\ (2005) for details.
The median resolution for each data set is listed in Table \ref{tab-spdata}. The GSA data have lower 
spectral resolution ($\approx$ 3.9{\AA}) than the GCP data ($\approx$ 3.1{\AA}). 
This is due to the wider slits used for the GSA data. 
There are small differences
in the spectral resolution for the various datasets taken through the same nominal slit widths.
In particular, the average difference between the RXJ1226.9+3332 data and the MS0451.6--0305 data
is about 3 percent, which is equivalent to 20 micron difference in slit width at the focal plane of GMOS-N.
This difference may be caused by a small difference in the focus setting for the laser cutter.
Further, as was seen for the RXJ0152.7--1357 data (J\o rgensen et al.\ 2005), the resolution
was found to vary with X-position on the array, while there is no dependency on the Y-position on
the array. The range of the variations is approximately $\pm $8 percent.
As noted in J\o rgensen et al.\ (2005), it is not known why this dependency is present,
though it is possible that it is due to a small tilt of the mask in the laser cutter during
the cutting, such that effectively the laser is slightly out of focus on one side of the mask. 
%
%
Independent of the cause, it makes it necessary to ensure that the individual values for the 
instrumental resolution are used when deriving the velocity dispersions of the galaxies.



\begin{figure*}
\epsfxsize 16.5cm
\epsfbox{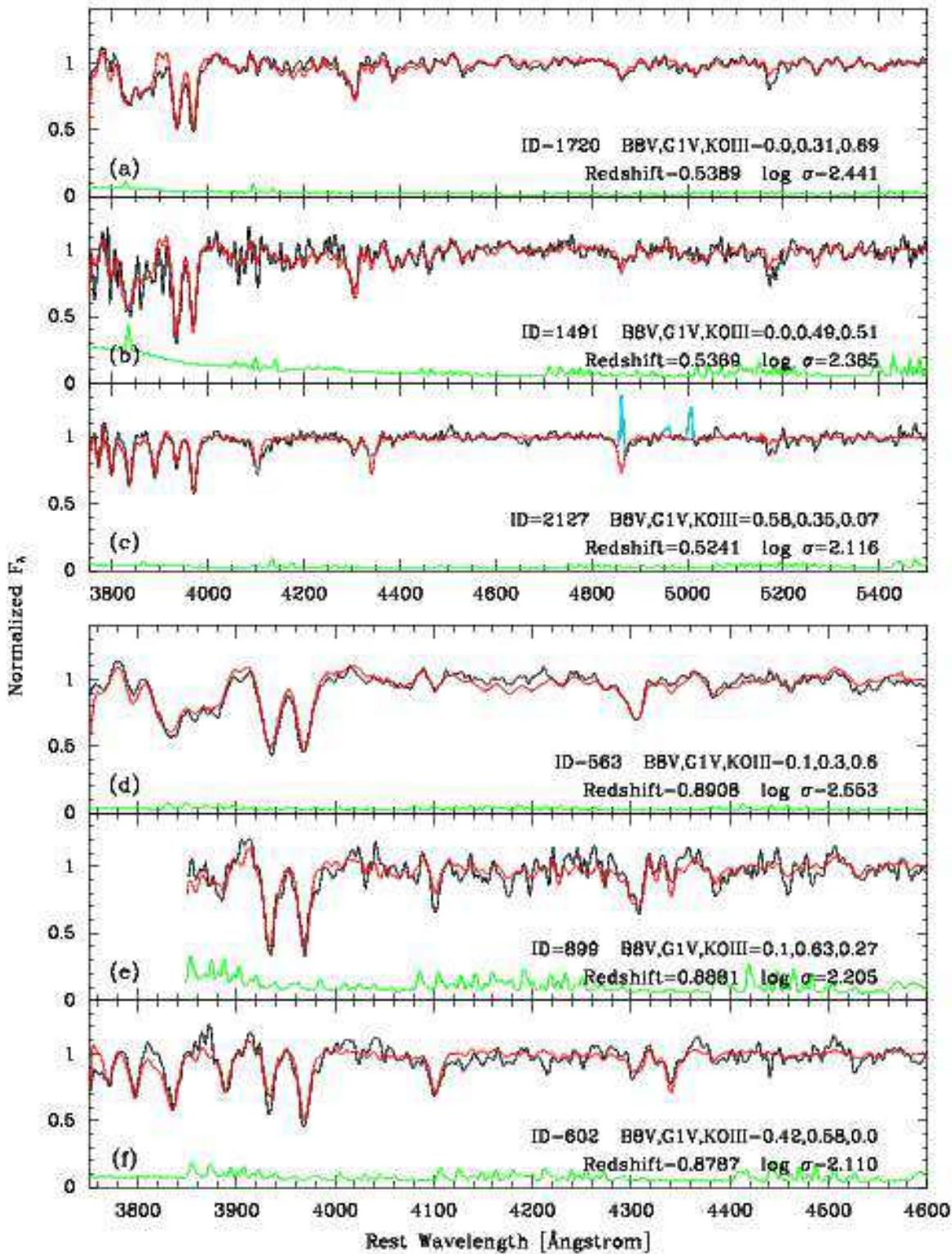}
\caption[]{Sample spectra from MS0451.6--0305 (panels (a)-(c)) and RXJ1226.9+3332 (panels (d)-(f))
with the best fit from the kinematics fitting over-plotted.
Black lines -- observed spectra, with cyan pieces marking emission lines excluded from the fits;
red lines -- best fit;
green lines -- four times the random noise in the spectra.
The figures demonstrated that spectra from those with dominating metal lines and weak Balmer lines
(panels (a) and (c)) to those with very strong Balmer lines (panels (c) and (f)) are well-modeled
with the three template stars used for the fitting.
\label{fig-kinfit_output} }
\end{figure*}

\begin{figure*}
\epsfxsize 17.5cm
\epsfbox{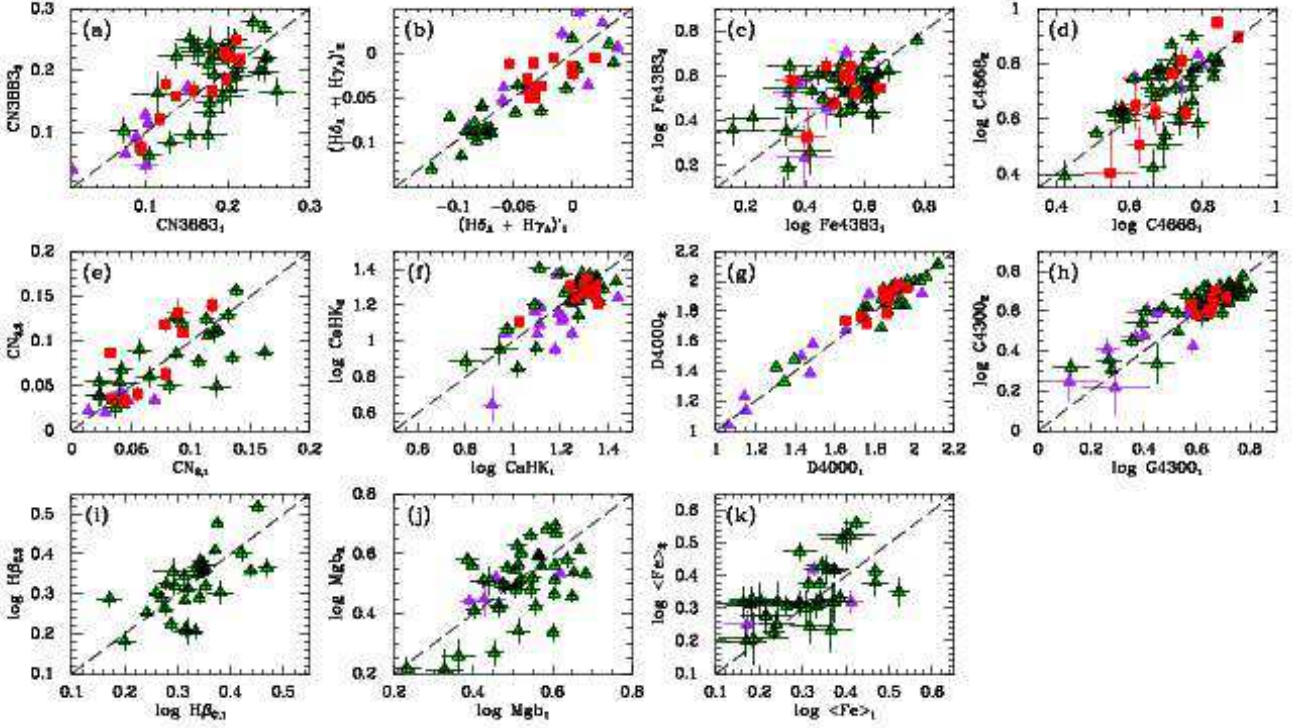}
\caption[]{Internal comparison of line indices derived from sub-stacks of the available data.
Green triangles -- members of MS0451.6--0305; red boxes -- members of  RXJ1226.9+3332;
purple triangles -- non-members in both fields.
Dashed lines show the one-to-one relations. The scatter around these is summarized in Table \ref{tab-compline}.
\label{fig-compline} }
\end{figure*}

\begin{deluxetable*}{llrrrr}
\tablecaption{Internal comparison of line indices \label{tab-compline} }
\tablewidth{0pc}
\tablehead{
\colhead{Cluster} & \colhead{Index} & \colhead{N} & \colhead{rms} & \colhead{$\sigma _{\rm adopt}$} & \colhead{Ratio} \\
\colhead{(1)} & \colhead{(2)} & \colhead{(3)} & \colhead{(4)} & \colhead{(5)} & \colhead{(6)} }
\startdata
MS0451.6--0305  & CN3883 & 33 & 0.050 & 0.035 & 1.9 \\
 &  log CaHK       &     43 & 0.085 & 0.060 &  2.7 \\
 &  D4000          &     22 & 0.072 & 0.051 &  4.9 \\
 &  $(\rm{H\delta _A + H\gamma _A})'$       &     27 & 0.018 & 0.013 &  2.8 \\
 &  CN$_2$         &     25 & 0.032 & 0.022 &  2.6 \\
 &  log G4300      &     44 & 0.079 & 0.056 &  2.1 \\
 &  log Fe4383     &     41 & 0.127 & 0.090 &  2.1 \\
 &  log C4668      &     40 & 0.097 & 0.069 &  2.2 \\
 &  log H$_{\beta}$   &     35 & 0.059 & 0.042 &  2.4 \\
 &  log Mg{\it b}  &     42 & 0.102 & 0.072 &  3.3 \\
 &  log $\rm \langle Fe \rangle$       &     37 & 0.084 & 0.059 &  1.4 \\
RXJ1226.9+3332  & CN3883 & 12 & 0.022 & 0.016 & 1.9 \\
 &  log CaHK       &     13 & 0.063 & 0.044 &  4.1 \\
 &  D4000          &     12 & 0.047 & 0.033 &  5.9 \\
 &  $(\rm{H\delta _A + H\gamma _A})'$          &     13 & 0.020 & 0.014 &  3.4 \\
 &  CN$_2$         &      9 & 0.027 & 0.019 &  2.7 \\
 &  log G4300      &     10 & 0.039 & 0.028 &  1.7 \\
 &  log Fe4383     &      9 & 0.112 & 0.079 &  1.8 \\
 &  log C4668      &      9 & 0.094 & 0.066 &  1.4 \\
\enddata
\tablecomments{Col.\ (1) Galaxy cluster; col.\ (2) line index; col.\ (3) number of galaxies in comparison;
col.\ (4) rms scatter of comparison;
col.\ (5) adopted uncertainty on the line index; 
col.\ (6) ratio between the adopted uncertainty and the median internal uncertainties on the line indices.}
\end{deluxetable*}

\subsection{B.7. Spectroscopic parameters \label{SEC-SPECPARAM} }

The redshifts and the velocity dispersions were determined by fitting
a mix of three template stars to the spectra. 
As in J\o rgensen et al.\ (2005), we used software made available by Karl Gebhardt
(Gebhardt et al.\ 2000, 2003) and three template stars of spectral types K0III, G1V and B8V.
The velocity dispersions have been corrected for the aperture size using the technique
from J\o rgensen et al.\ (1995b).
Tables \ref{tab-MS0451kin} and \ref{tab-RXJ1226kin} in 
Appendix \ref{SEC-SPECTROSCOPY} summarize results from the template fitting.
Measured velocity dispersions as well as aperture corrected velocity dispersions
are listed for the cluster members. 
The velocity dispersions for galaxies in MS0451.6--0305 were also corrected for the 
systematic errors due to the lower resolution of these spectra. The correction is
described in detail in the next section.
For galaxies that are not members of the clusters, we give the redshift. The detailed data for these galaxies
will be discussed in a future paper.
The wavelength intervals used for the fits are 3750-5600 {\AA} and 3750-4650 {\AA} for MS0451.6--0305
and RXJ1226.9+3332, respectively. In the tables we include a measure of the goodness of the fit
derived as
\begin{equation}
\chi ^2 = N^{-1} \, \sum  [(S_{\rm obs,i} - S_{\rm fit,i}) / N_{\rm obs,i}]^2
\end{equation}
where $S_{\rm obs,i}$ is the observed spectrum, $S_{\rm fit,i}$ is the best fit, $N_{\rm obs,i}$ is the noise,
the sum is over the spectral pixels inside the fitting wavelength range, and $N$ is the number
of spectra pixels.
The mean $\chi ^2$ for the galaxies included in the analysis is 4.2 and 4.4 for MS0451.6--0305
and RXJ1226.9+3332, respectively. The $\chi ^2$ values are generally large than one due to a 
combination of slight template mismatch and underestimated uncertainties due to sky subtraction
noise at the strong sky lines.
On Figure \ref{fig-kinfit_output} we show three sample spectra from each cluster with the best fit
over-plotted. The main purpose of using a set of template stars is to span the range of stellar
populations represented in the spectra. The figure demonstrates that three template stars
successfully accomplish this, except for a slight template mismatch at the strong magnesium
lines at 5170 {\AA}. Such a mismatch is also seen if we use the SSP model spectra from
Maraston \& Str\"{o}mb\"{a}ck (2011). Further, using these model spectra result is similar quality
of fits as measured by $\chi ^2$ as using the three template stars. Thus, for consistency with
previously published results, we choose to continue to use the template stars for the kinematics fits.

The large wavelength range of the MS0451.6--0307 spectra allow us to test if excluding the 
calcium H and K lines from the fit leads to significantly different velocity dispersions.
We find that not to be the case, the mean difference in $\log \sigma$ is $0.020 \pm 0.010$ with
and rms scatter of 0.066. The difference is within the expected systematic differences between data sets
of 0.026 and also only significant at the 2-sigma level. The wavelength coverage for RXJ1226.9+3332
does not allow a similar test. For consistency we include the calcium H and K lines in 
all the kinematics fits.

The Lick/IDS absorption line indices CN$_1$, CN$_2$, G4300, Fe4383, C4668 (Worthey et al.\ 1994),
as well as the higher order Balmer line indices $\rm H\delta _A$ and $\rm H\gamma _A$
(Worthey \& Ottaviani 1997) were derived for both clusters.
We have also determined the D4000 index (Bruzual 1983; Gorgas et al.\ 1999),
and the blue indices CN3883 and CaHK (Davidge \& Clark 1994).
For MS0451.6--0305 the wavelength coverage allowed determination of the 
Lick/IDS absorption line indices $\rm H\beta$, Mg{\it b}, Fe5270 and Fe5335, as well as 
$\rm H\beta _G$ (see Gonz\'{a}lez 1993; J\o rgensen 1997).
Because of the lower relative uncertainties on $\rm H\beta _G$ compared to $\rm H\beta$,
$\rm H\beta _G$ is used in the analysis. A transformation between the two 
indices can be found in J\o rgensen (1997). 
The indices have been corrected for the aperture size and for the effect of the
velocity dispersions as described in J\o rgensen et al.\ (2005), see also J\o rgensen et al.\ (1995b).

We refer the reader to Table 13 in J\o rgensen et al.\ (2005) for the coefficients adopted for the
aperture corrections for the velocity dispersions and the line indices.
The aperture corrections are the largest for the highest redshift clusters. 
For RXJ1226.9+3332 the typical correction for the velocity dispersion and the metal lines measured 
as an equivalent width is 0.025 on the logarithm of the measurement. The correction on CN3883 is 
of a similar size.  The Balmer lines have insignificant aperture corrections.
The coefficients given in J\o rgensen et al.\ (2005) are based on radial profiles of nearby galaxies.
While one could argue that we do not know that radial profiles of intermediate redshift galaxies
are similar, no better information exists at this point. We also note that the luminosity 
profiles for the intermediate redshift galaxies are similar to those found for low redshift
samples (well-modeled by S\'{e}rsic profiles with an index of roughly 2 to 6), lending
support to the assumption that line index profiles are also similar.
In the extreme one could use the size of the aperture corrections as a limit on the consistency
between the data sets. 

We have estimated the uncertainties on the line indices using two methods: (1)
uncertainties derived from the local S/N in the passbands following
the method described in Cardiel et al.\ (1998), and (2) from internal comparisons of 
indices derived from stacking subsets of the frames available for each cluster.

We first note that method (1) leads to slightly larger uncertainties
than the method use in J\o rgensen et al.\ (2005). We have therefore rederived the uncertainties
for the RXJ0152.7--1357 galaxies. The effect on the uncertainties on H$\delta _A$ and H$\gamma _A$ 
is negligible. The uncertainties on Fe4383, C4668 and CN3883 are $\approx 40$ per cent larger than
those listed in J\o rgensen et al.\ (2005).

The internal comparisons of line indices are based on galaxies included in both masks for each
of the clusters and for RXJ1226.9+3332 also on line indices derived from each of the two
stacks of data for the two wavelength settings.
Figure \ref{fig-compline} and Table \ref{tab-compline} summarize the comparisons. Only galaxies for which the S/N
per {\AA}strom in the rest frame is above 20 in the sub-stacks are included. 
We use the scatter in the comparisons to estimate the uncertainties on the indices. Column (5) in
Table \ref{tab-compline} lists the adopted uncertainties. For the indices used in the analysis 
these are a factor 1.5 to 3 larger than the uncertainties based on the local S/N of the spectra. 
The larger uncertainties resulting from this method is most likely due to increased noise in the
sky subtraction around the strong sky lines. However, there is no indication of general over-
or under-subtraction of the sky signal blue-wards of 9290 {\AA}.
In the analysis we adopt the uncertainties listed in Table \ref{tab-compline} for the line indices and show these as
typical error bars on all relevant figures.

At wavelengths longer than 9290 {\AA} systematic errors in the sky subtraction of the GCP data
become evident. This is the wavelength at which very strong OH bands start 
affecting the data. Even with nod-and-shuffle and the correction for the CDE it is
clear from our data that systematic errors dominate red-wards of this wavelength.
Therefore only line indices that fall at shorter wavelengths than 9290 {\AA} in the observed frame
are considered reliable and included in the tables and the analysis.
For RXJ1226.9+3332 this wavelength corresponds to 4910 {\AA} the rest frame,
while for MS0451.6--0305 it is 6032 {\AA}.
Figure \ref{fig-specRXJ1226} shows the RXJ1226.9+3332 spectra to 5200 {\AA} 
in the rest frame to make it possible to assess the presence of emission 
in H$\beta$ as well as the [\ion{O}{3}] lines at 4949 {\AA} and 5007 {\AA}.
Line index measurements considered unreliable due to systematic errors in the sky subtraction
are not listed in the tables.


The systematic sky subtraction errors for the GSA data are larger than those of the GCP data
due to the poorer sky subtraction.
Figure \ref{fig-sncutoff} shows representative spectra from the GSA data with $S/N \approx 20$,
30, and $S/N>50$, together with the higher S/N spectra of the same 
galaxies from the GCP data. The plot shows the consistency of the wavelength calibrations,
also confirmed from redshift comparisons and the consistency of the relative flux
calibration which is important for consistent determination of the line indices.
Based on inspection of the spectra
and the corresponding sky spectra, we conclude in the observed wavelength intervals
7700--8035 {\AA} and 8268--8475 {\AA}, which contain very strong sky lines, 
the weak or narrow metal line indices
(CN$_1$, CN$_2$, Fe4383) and the Balmer line indices can be significantly affected
by the systematic errors.
We use the S/N per {\AA}ngstrom in the rest frame for the full spectrum to 
determine if the indices are reliable. 
For indices to be reliable in the observed wavelength interval 7700--8035 {\AA} we 
require $S/N \ge 30$, while the requirement for indices in the interval 8268--8475 {\AA} is $S/N \ge 50$.
For spectra with $S/N<10$, only the redshifts have listed in Table \ref{tab-RXJ1226kin} 
as both the velocity dispersions and the indices have very large uncertainties.

\begin{figure*}
\epsfxsize 17.5cm
\epsfbox{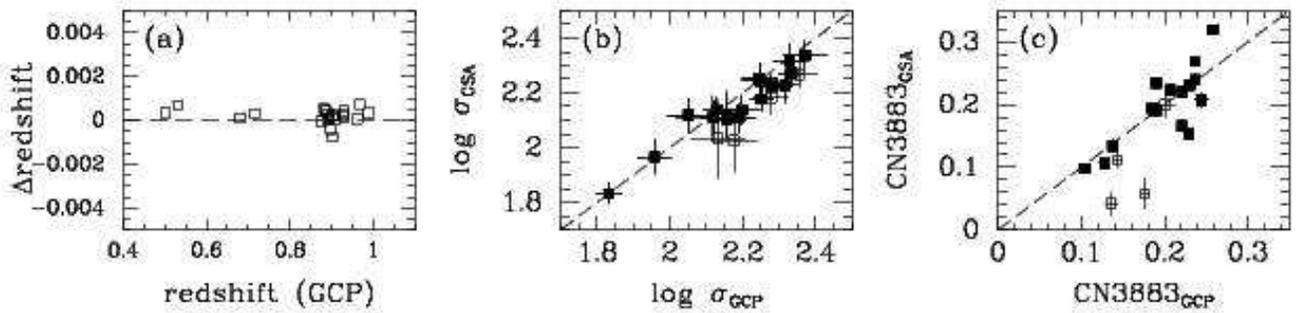}
\caption[]{Comparison of the GCP data and the GSA data for the spectral parameters used in the analysis
and with the most galaxies in common. 
On panels (b) and (c): filled boxes -- $\rm S/N_{\rm GSA} \ge 20$ per {\AA}ngstrom in the rest frame,
open boxes -- $\rm S/N_{\rm GSA} < 20$.
\label{fig-speccomp} }
\end{figure*}

Finally, we compare the GCP and the GSA data to ensure consistency and in order to provide
better estimates of the uncertainties on the line indices from GSA data than those resulting
from estimates based on the local S/N of the spectra.
There are 23 galaxies in common between the GCP data and the GSA data for RXJ1226.9+3332.
Our approach is to calibrate the GSA data to consistency
with the GCP data, since the latter has higher S/N and better resolution.
Figure \ref{fig-speccomp} shows the comparison of the redshifts, velocity dispersions and CN3883
for these two data sets.  Table \ref{tab-speccomp} summarizes the comparisons. 
The velocity dispersions derived from the GSA data are systematically 0.022 smaller in $\log \sigma$
than found from the GCP data. Even though this is below the 0.026 possible systematic 
errors in the low redshift comparison samples (cf.\ J\o rgensen et al.\ 2005), we have chosen to correct the GSA data for the offset.
The scatter for the comparison of the velocity dispersions is within the expected scatter based
on the uncertainties as derived from the kinematics fits.
None of the offsets for the line indices are significant. As for the GCP data, we use the scatter
to estimate the uncertainties. We use the adopted uncertainties for the GCP data (Table \ref{tab-compline}).
The resulting uncertainties listed in Table \ref{fig-speccomp} are a factor two to three larger
than the uncertainties on the GCP data. The higher order Balmer lines are 
not included in the table as none of the GSA spectra result in measurements if these indices
used in the analysis. 

\begin{deluxetable}{lrrrr}
\tablecaption{Spectroscopic comparisons \label{tab-speccomp} }
\tablewidth{0pt}
\tablehead{Parameter & N & GCP$-$GSA\tablenotemark{a} & rms & $\sigma _{\rm adopt}$\tablenotemark{b} }
\startdata
Redshift        & 23 & 0.0003 & 0.0003 & \\
$\log \sigma$   & 23 & 0.035 &  0.054 & \\
$\log \sigma$\tablenotemark{c}   & 18 & 0.022 & 0.046 & \\
CN3883          & 20 & 0.002 & 0.045 & 0.042 \\
log CaHK        & 19 & -0.005 & 0.071 & 0.056 \\
D4000           & 21 & -0.032 & 0.084 & 0.077 \\
log G4300       & 14 & -0.034 & 0.10 & 0.096 \\
log Fe4383      &  4 & -0.10 & 0.20 & 0.18 \\
\enddata
\tablenotetext{a}{Median difference of data from the GCP data and the GSA data.}
\tablenotetext{b}{Adopted uncertainties on line indices from the GSA data.}
\tablenotetext{c}{$\rm S/N_{GSA} \ge 20$.}
\end{deluxetable}

For galaxies with detectable emission from  [\ion{O}{2}] we determined the equivalent width 
of the [\ion{O}{2}]$\lambda\lambda$3726,3729 doublet.
With an instrumental resolution of $\sigma \approx 3$\,{\AA} (FWHM $\approx 7$\,{\AA}), 
the doublet is not resolved in our spectra and we refer to it simply as the
``[\ion{O}{2}] line''.

\begin{figure*}
\epsfxsize 17.5cm
\epsfbox{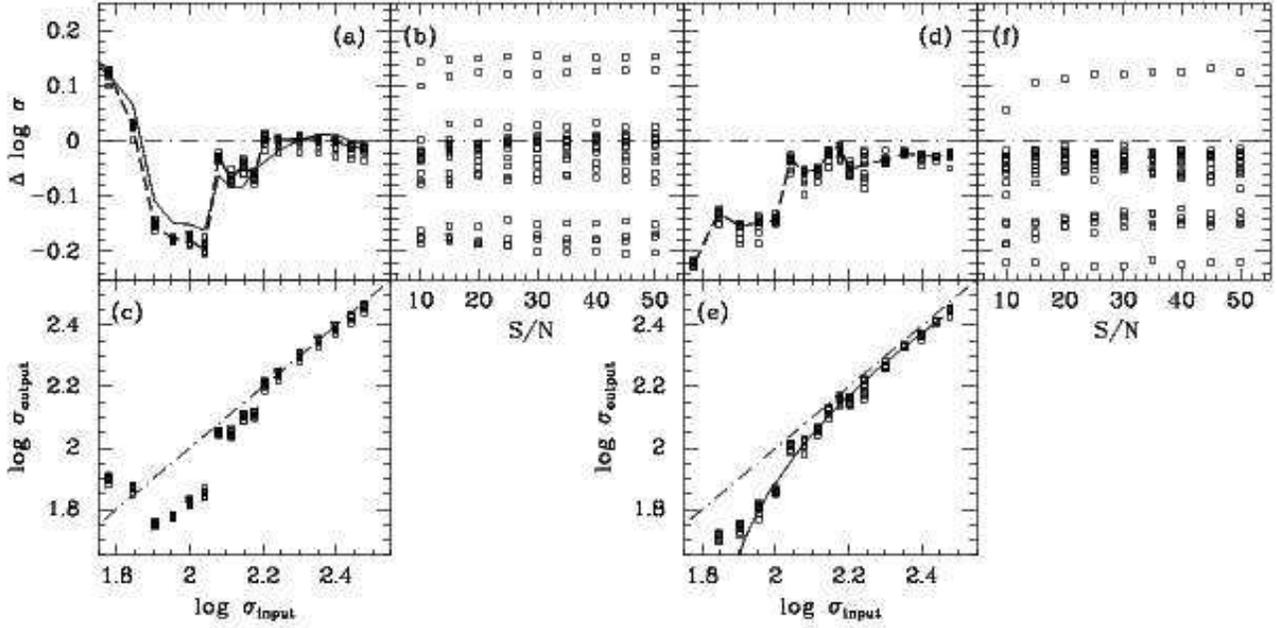}
\caption[]{Results from simulations. The systematic error on $\log \sigma$, 
$\Delta \log \sigma = \log \sigma _{\rm output} - \log \sigma _{\rm input}$ as a function of the 
input values $\log \sigma _{\rm input}$ (a) and (d) and as a function of the S/N \AA$^{-1}$ (b) and (f). 
Panels (c) and (e) shows input versus output values. The simulations in
panels (a)--(c) apply to the RXJ1226.9+3332 GSA data, while panels (d)--(e) apply to the MS0451.6--0305 data.
The dashed lines on (a) and (d) mark the median values of $\Delta \log \sigma$. The solid line on (a) shows the 
simulations for the RXJ0152.7--1357 data from J\o rgensen et al.\ (2005). 
The solid line on (e) shows the adopted correction for the systematic errors for the MS0451.6--0305 data, 
see Eq.\ \ref{eq-sigmasys}.
\label{fig-velsim} }
\end{figure*}

\subsection{B.8. Systematic effects in derived velocity dispersions \label{SEC-SIM}}

In J\o rgensen et al.\ (2005) we used simulated spectra to determine how the input S/N
and the galaxy velocity dispersion relative to the instrumental resolution affected the determined
velocity dispersions. Those simulations are also applicable to the RXJ1226.9+3332 data
obtained for the GCP, as these have similar instrumental resolution and S/N as the RXJ0152.7--1357
data in J\o rgensen et al.
The instrumental resolution of the GCP data for RXJ1226.9+3332 and RXJ0152.7--1357 is about 120 $\rm km~s^{-1}$.
However, both the RXJ1226.9+3332 GSA data and the MS0451.6--0305 GCP data have worse instrumental
resolutions, both equivalent to about 145 $\rm km~s^{-1}$ at 4300 \AA\ in the rest frame of the galaxies
in the clusters. Further, the GSA data in general have lower S/N than the GCP data
and are subject to larger effects from systematic sky subtraction errors.
To estimate the effect of the worse instrumental resolution we ran simulations in a similar
fashion as described in J\o rgensen et al., using the observed noise spectrum and the 
instrumental resolution relevant for the data set. We then simulated spectra with S/N between
10 and 50 \AA$^{-1}$ in the rest frame, and input velocity dispersions between 50 and 300 $\rm km~s^{-1}$.
For these simulations we used a mix of the G1V and K0III template stars each contributing half
the flux to the simulated spectrum.
Each combination of S/N and input velocity dispersion was realized 100 times. 
We use the median of the output velocity dispersion in the following assessment.

Figure \ref{fig-velsim} shows the result of the two sets of simulations.
The main difference between the simulations for the two high redshift clusters
and for MS0451.6--0305 is difference in the length of the wavelength interval used for the kinematic fitting.
The simulations matching the RXJ1226.9+3332 GSA dataset (Fig.\ \ref{fig-velsim}a--c) show a similar dependence
on the input velocity dispersions as found for our previous simulations for RXJ0152.7--1257.
The thin solid line on Figure \ref{fig-velsim}a shows the simulations for RXJ0152.7--1257 
from J\o rgensen et al.
Since the dependence on the input velocity dispersion is approximately is the same 
for the two sets of simulations, any difference between velocity dispersions derived
from the GCP data and the GSA data is expected to be small, as was found in the previous section. 

The simulations matching the MS0451.6--0305 dataset (Fig.\ \ref{fig-velsim}d--f) show that velocity dispersions are
systematically derived too small, and that the systematic error increases with decreasing input
velocity dispersion. For input velocity dispersion $\log \sigma >1.9$ we establish a 
correction function:
\begin{equation}
\label{eq-sigmasys}
\log \sigma _{\rm corrected} =\log \sigma _{\rm output} +0.0424 - 0.2138(\log \sigma _{\rm output}-2.1) + 0.5426(\log \sigma _{\rm output}-2.1)^2
\end{equation}
The function is shown as the solid line on Figure \ref{fig-velsim}f.
Since the size of this correction is similar to or larger than typical inconsistencies 
between different datasets, we have chosen to apply the correction to
the measured velocity dispersions for MS0451.6--0305 galaxies. 
The correction is applied to the value in column $\log \sigma _{\rm cor}$ (together
with the aperture correction) in Table \ref{tab-MS0451kin},
while the column $\log \sigma$ lists the raw measure values.

\subsection{B.9. Final calibrated spectroscopic parameters} 

Tables \ref{tab-MS0451kin}--\ref{tab-RXJ1226line} list the results from the template fitting 
and the derived line indices. For RXJ1226.9+3332 galaxies included in both the GCP and the GSA data, only the 
values from the GCP data are listed. The GSA velocity dispersions are calibrated to consistency
with the GCP data as explained in Sec.\ \ref{SEC-SPECPARAM}.

\section{C. Presentation of the imaging and spectra for MS0451.6--0305 and RXJ1226.9+3332 \label{SEC-SPECIM}}

\begin{figure*}
\epsfxsize 17.5cm
\epsfbox{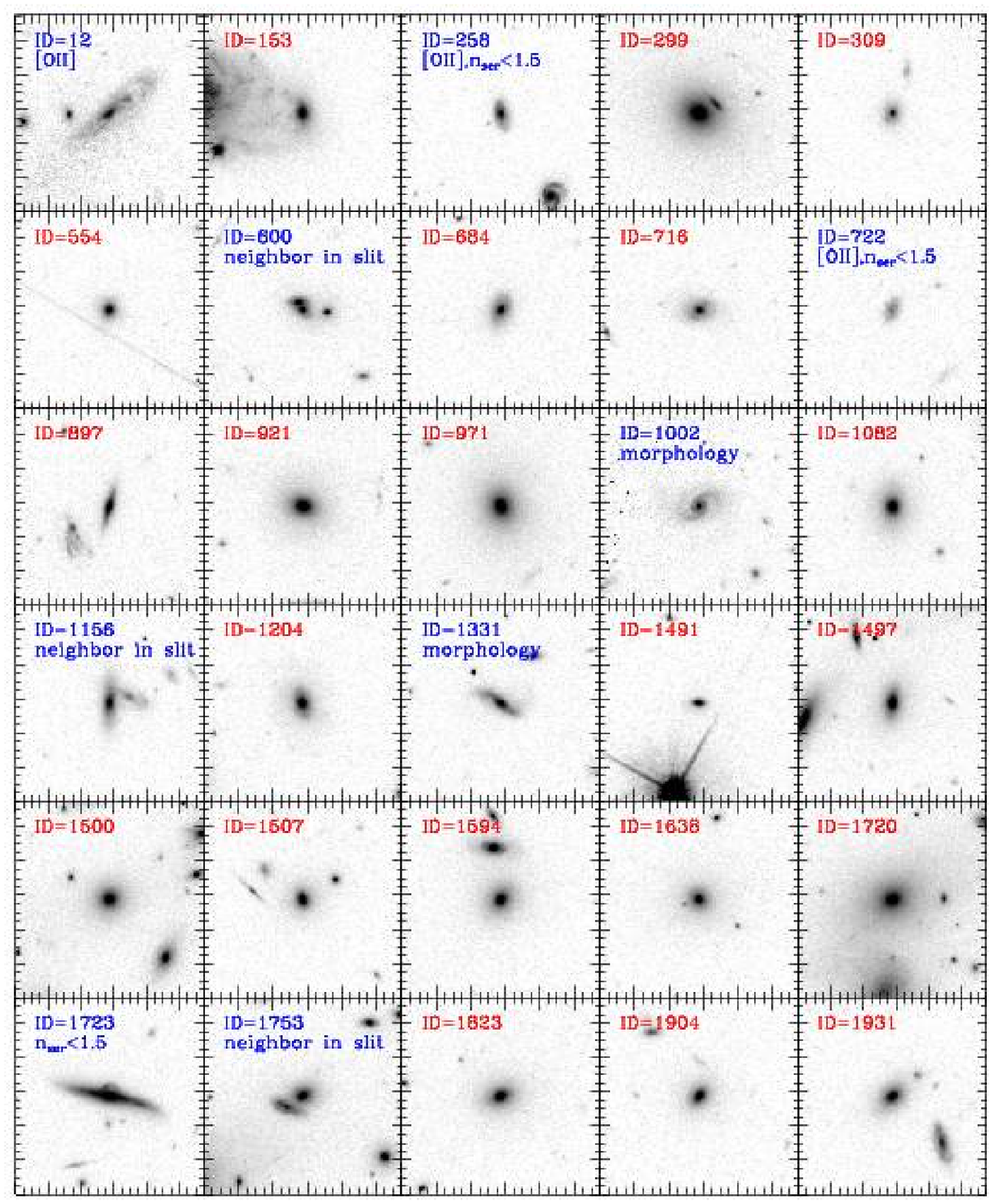}
\caption[]{
{\it HST}/ACS images of the MS0451.6--0305 sample in the F814W filter. Each image is 11.5 $\times $ 11.5 arcsec.
At the distance of  MS0451.6--0305 this corresponds to 75 kpc $\times$ 75 kpc for our adopted cosmology. North is up, East to the left.
In each panel, the galaxy with the ID listed above the image is located in the center of that image.
Red ID -- galaxy is included in the analysis; 
Blue ID -- galaxy is excluded from the analysis, the reason is listed, see Section \ref{SEC-METHODSAMPLE} for details.
\label{fig-stampsMS0451}
}
\end{figure*}

\begin{figure*}
\epsfxsize 17.5cm
\epsfbox{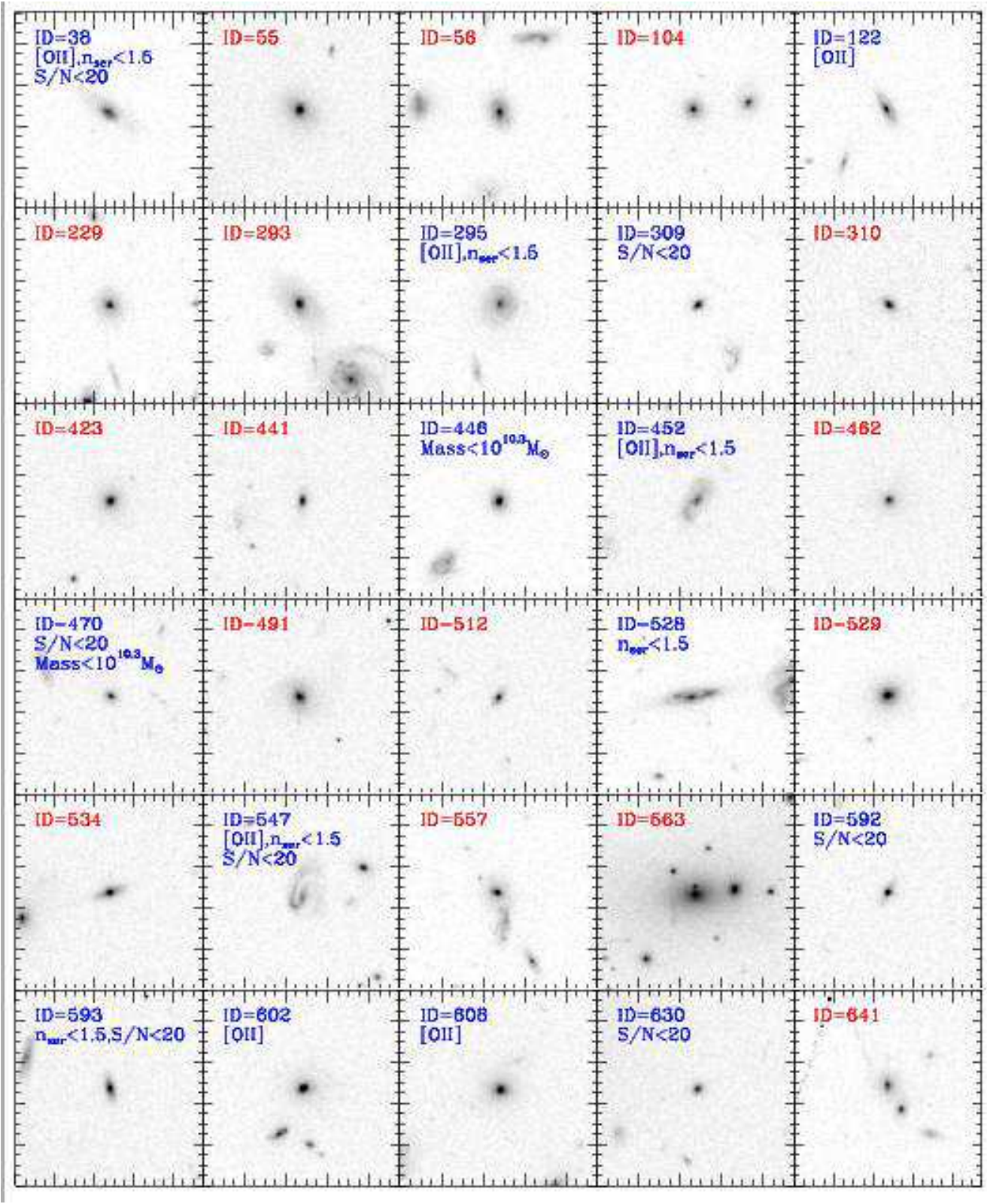}
\caption[]{
{\it HST}/ACS images of the RXJ1226.9+3332 sample in the F814W filter. Each image is 9.5 $\times $ 9.5 arcsec.
At the distance of  RXJ1226.9+3332 this corresponds to 75 kpc $\times$ 75 kpc for our adopted cosmology. North is up, East to the left.
In each panel, the galaxy with the ID listed above the image is located in the center of that image.
Red ID -- galaxy is included in the analysis; 
Blue ID -- galaxy is excluded from the analysis, the reason is listed, see Section \ref{SEC-METHODSAMPLE} for details.
\label{fig-stampsRXJ1226}
}
\end{figure*}

\begin{figure*}
\epsfxsize 16.5cm
\epsfbox{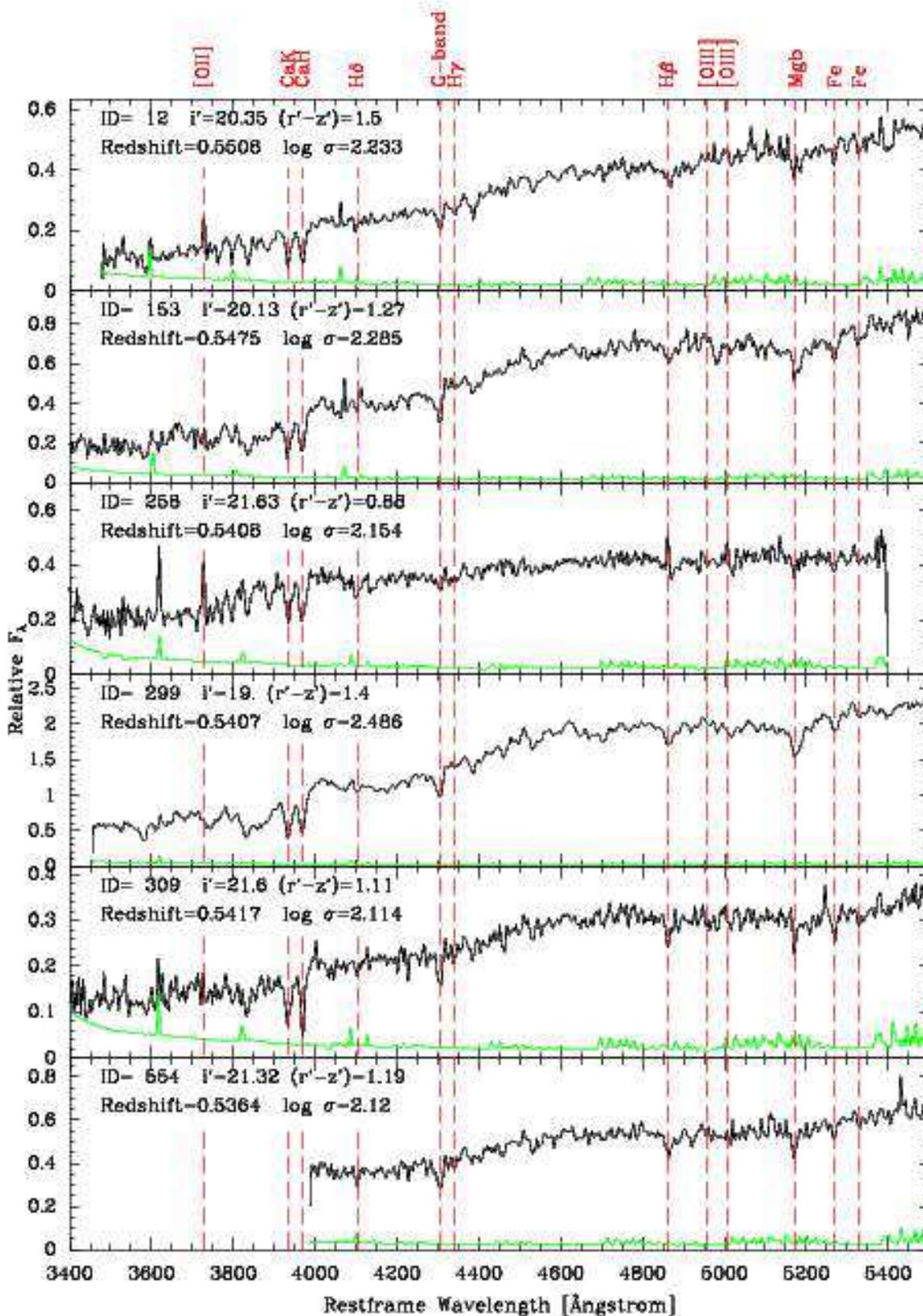}
\vspace{-1cm}
\caption[]{Spectra of the galaxies that are considered members of MS0451.6--0305.
Black lines -- the galaxy spectra; green lines -- the random noise multiplied with four.
At the strong sky lines, the random noise underestimates the real noise due to systematic
errors in the sky subtraction. 
Some of the absorption lines are marked. The location of the emission lines [\ion{O}{2}],
[\ion{O}{3}]$\lambda$4959, and [\ion{O}{3}]$\lambda$5007 are also marked, though these lines are only present in some of
the galaxies. The spectra shown in this figure have been processed as described
in the text, including resampling to just better than critical sampling. 
Only the first page of this figure appears in print, the remainder are available in the online version of the Journal.
\label{fig-specMS0451} }
\end{figure*}

\begin{figure*}
\epsfxsize 16.5cm
\epsfbox{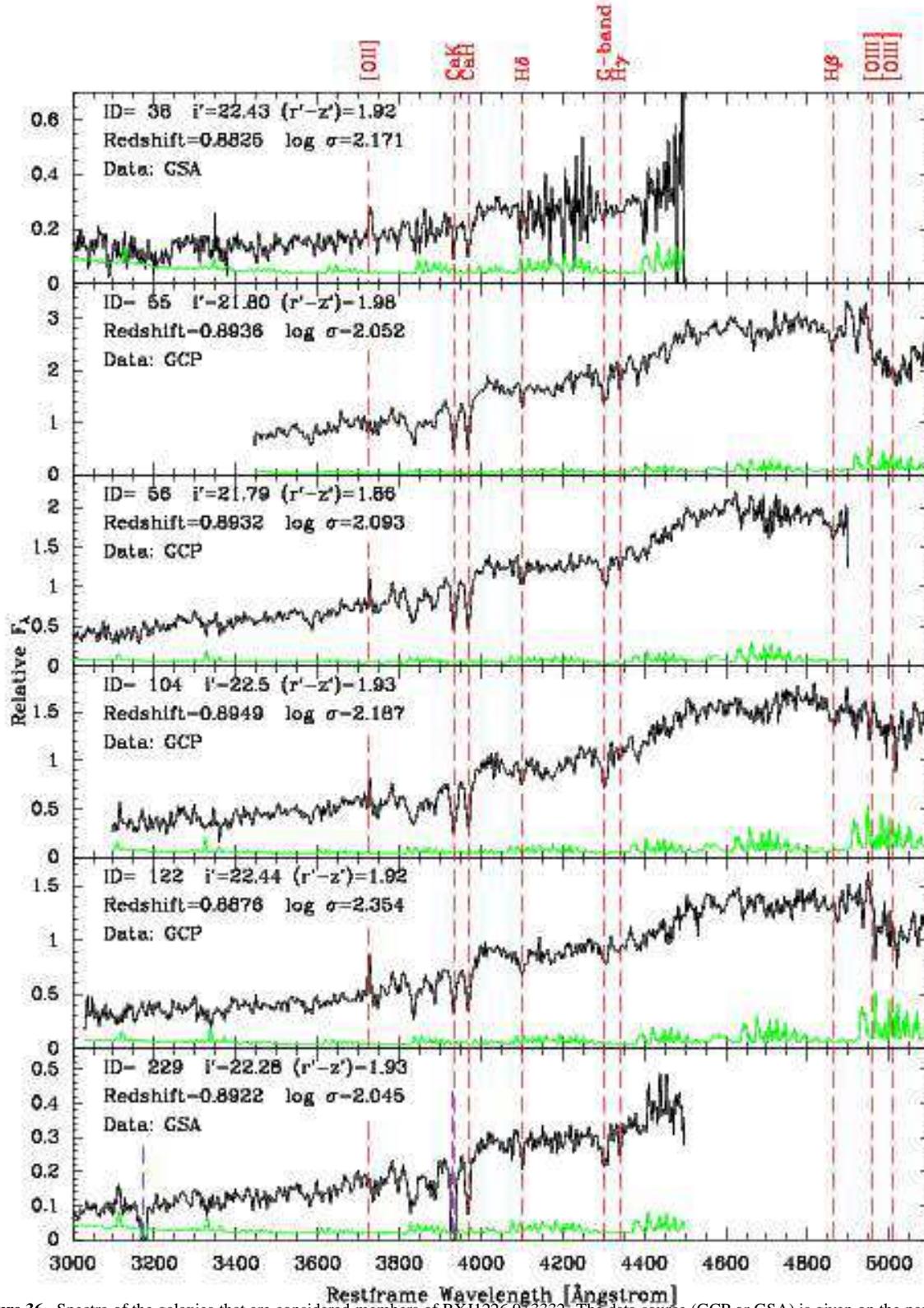}
\vspace{-1cm}
\caption[]{Spectra of the galaxies that are considered members of RXJ1226.9+3332.
The data source (GCP or GSA) is given on the panels.
Lines and labeling as on Figure \ref{fig-specMS0451}.
The GSA data are not plotted for rest frame wavelength larger
than 4500 \AA as the spectra are affected significantly by systematic sky subtraction errors, see text for
details. 
For GSA data taken with only one wavelength setting there are gaps in the spectra due to the gaps
between the GMOS-N CCDs. These gaps are marked with blue dashed lines.
Only the first page of this figure appears in print, the remainder are available in the online version of the Journal.
\label{fig-specRXJ1226} }
\end{figure*}

Stamp-sized images of the galaxies from the {\it HST}/ACS imaging are shown on 
Figures \ref{fig-stampsMS0451} and \ref{fig-stampsRXJ1226}.
The stamps cover the equivalent of 75 kpc $\times$ 75 kpc at the distance of the clusters.
The panels are labelled to show which galaxies are included in the analysis. The selection criteria
are detailed in Section \ref{SEC-METHODSAMPLE}.
The spectra of the cluster members of MS0451.6--0306 and RXJ1226.9+3332 are shown in 
Figures \ref{fig-specMS0451} and \ref{fig-specRXJ1226}. 
Full figures are available in the online version of the Journal.

\begin{deluxetable*}{rrrrrrrrrrr}
\tablecaption{MS0451.6--0305: Results from Template Fitting \label{tab-MS0451kin} }
\tablewidth{400pt}
\tabletypesize{\scriptsize}
\tablehead{
\colhead{ID} & \colhead{Redshift} & \colhead{Member\tablenotemark{a}} &  
\colhead{$\log \sigma$} & \colhead{$\log \sigma _{\rm cor}$\tablenotemark{b}} & 
\colhead{$\sigma _{\log \sigma}$} & \multicolumn{3}{c}{Template fractions} & \colhead{$\chi^2$} & \colhead{S/N\tablenotemark{c}} \\
\colhead{}&\colhead{} &\colhead{} &\colhead{} &\colhead{} &\colhead{} & \colhead{B8V} & \colhead{G1V} & \colhead{K0III} & }
\startdata
   12& 0.5508&  1&  2.178&  2.233&  0.044&   0.33&   0.26&   0.40&    2.6&   44.7\\
  153& 0.5475&  1&  2.236&  2.285&  0.029&   0.00&   0.59&   0.41&    6.7&   77.2\\
  220& 0.8259&  0&  \nodata&  \nodata&  \nodata&  \nodata&  \nodata&  \nodata&  \nodata&   45.4\\
  234& 0.3554&  0&  \nodata&  \nodata&  \nodata&  \nodata&  \nodata&  \nodata&  \nodata&   75.4\\
  258& 0.5408&  1&  2.081&  2.154&  0.056&   0.35&   0.58&   0.07&    2.1&   41.4\\
  299& 0.5407&  1&  2.429&  2.486&  0.022&   0.12&   0.31&   0.58&   12.6&  197.5\\
  309& 0.5417&  1&  2.026&  2.114&  0.035&   0.00&   0.74&   0.26&    1.5&   34.2\\
  323& 0.5891&  0&  \nodata&  \nodata&  \nodata&  \nodata&  \nodata&  \nodata&  \nodata&  114.3\\
  386& 0.8849&  0&  \nodata&  \nodata&  \nodata&  \nodata&  \nodata&  \nodata&  \nodata&   14.7\\
  468& 1.1792&  0&  \nodata&  \nodata&  \nodata&  \nodata&  \nodata&  \nodata&  \nodata&   25.5\\
  554& 0.5364&  1&  2.035&  2.120&  0.035&   0.03&   0.70&   0.27&    2.5&   50.5\\
  600& 0.5371&  1&  1.998&  2.094&  0.076&   0.00&   0.68&   0.32&    4.3&   92.2\\
  606& 0.5684&  0&  \nodata&  \nodata&  \nodata&  \nodata&  \nodata&  \nodata&  \nodata&   71.1\\
  684& 0.5423&  1&  1.936&  2.054&  0.064&   0.18&   0.30&   0.52&    3.4&   52.8\\
  716& 0.5474&  1&  1.972&  2.077&  0.060&   0.00&   0.57&   0.43&    2.3&   46.4\\
  722& 0.5473&  1&  2.092&  2.162&  0.088&   0.47&   0.53&   0.00&    2.3&   29.2\\
  833& 0.9496&  0&  \nodata&  \nodata&  \nodata&  \nodata&  \nodata&  \nodata&  \nodata&   16.1\\
  836& 0.3676&  0&  \nodata&  \nodata&  \nodata&  \nodata&  \nodata&  \nodata&  \nodata&   44.0\\
  897& 0.5457&  1&  2.106&  2.174&  0.040&   0.00&   0.49&   0.51&    4.0&   65.5\\
  901& 1.0656&  0&  \nodata&  \nodata&  \nodata&  \nodata&  \nodata&  \nodata&  \nodata&   95.0\\
  921& 0.5494&  1&  2.223&  2.274&  0.036&   0.00&   0.73&   0.27&   11.9&  139.0\\
  971& 0.5437&  1&  2.140&  2.201&  0.039&   0.00&   0.66&   0.34&   10.4&  103.4\\
 1002& 0.5477&  1&  1.974&  2.078&  0.034&   0.00&   0.81&   0.19&    2.1&   44.9\\
 1082& 0.5501&  1&  2.307&  2.355&  0.029&   0.00&   0.44&   0.56&    7.7&  119.1\\
 1156& 0.5397&  1&  2.020&  2.109&  0.073&   0.09&   0.48&   0.43&    2.1&   43.4\\
 1204& 0.5342&  1&  2.002&  2.096&  0.045&   0.00&   0.62&   0.38&    4.4&   77.4\\
 1331& 0.5362&  1&  2.059&  2.137&  0.041&   0.39&   0.51&   0.09&    2.4&   74.6\\
 1491& 0.5369&  1&  2.337&  2.385&  0.035&   0.00&   0.49&   0.51&    1.9&   37.3\\
 1497& 0.5318&  1&  2.294&  2.342&  0.042&   0.00&   0.52&   0.48&    7.9&  137.3\\
 1500& 0.5391&  1&  2.276&  2.324&  0.035&   0.10&   0.22&   0.67&    5.1&   77.5\\
 1507& 0.5392&  1&  2.350&  2.399&  0.037&   0.14&   0.04&   0.82&    8.7&  105.0\\
 1584& 0.9546&  0&  \nodata&  \nodata&  \nodata&  \nodata&  \nodata&  \nodata&  \nodata&   14.5\\
 1594& 0.5373&  1&  2.221&  2.272&  0.048&   0.00&   0.61&   0.39&    5.3&   84.7\\
 1638& 0.5473&  1&  2.208&  2.260&  0.037&   0.00&   0.36&   0.64&    5.2&   85.7\\
 1720& 0.5389&  1&  2.389&  2.441&  0.022&   0.00&   0.31&   0.69&    8.9&  148.4\\
 1723& 0.5372&  1&  2.177&  2.232&  0.024&   0.21&   0.59&   0.19&    7.6&  131.0\\
 1753& 0.5305&  1&  2.221&  2.272&  0.043&   0.00&   0.40&   0.60&    3.7&   77.1\\
 1823& 0.5306&  1&  2.157&  2.215&  0.048&   0.00&   0.43&   0.57&    5.7&   90.3\\
 1904& 0.5391&  1&  2.273&  2.321&  0.021&   0.00&   0.45&   0.55&    4.6&   89.0\\
 1931& 0.5550&  1&  2.054&  2.133&  0.033&   0.00&   0.52&   0.48&    3.8&   73.9\\
 1952& 0.5494&  1&  2.292&  2.340&  0.030&   0.18&   0.20&   0.61&    8.1&  117.6\\
 2031& 0.5315&  1&  2.375&  2.426&  0.032&   0.00&   0.42&   0.58&    4.4&   84.3\\
 2032& 0.5114&  0&  \nodata&  \nodata&  \nodata&  \nodata&  \nodata&  \nodata&  \nodata&   74.3\\
 2127& 0.5241&  1&  2.031&  2.116&  0.041&   0.58&   0.35&   0.07&    4.9&  102.6\\
 2166& 0.5319&  1&  2.044&  2.126&  0.058&   0.48&   0.28&   0.24&    3.3&   62.3\\
 2223& 0.5450&  1&  1.937&  2.055&  0.037&   0.00&   0.57&   0.43&    2.2&   44.2\\
 2230& 0.9199&  0&  \nodata&  \nodata&  \nodata&  \nodata&  \nodata&  \nodata&  \nodata&   25.4\\
 2240& 0.5469&  1&  1.974&  2.078&  0.082&   0.17&   0.48&   0.36&    2.2&   30.6\\
 2491& 0.9192&  0&  \nodata&  \nodata&  \nodata&  \nodata&  \nodata&  \nodata&  \nodata&   39.2\\
 2561& 1.0493&  0&  \nodata&  \nodata&  \nodata&  \nodata&  \nodata&  \nodata&  \nodata&   16.5\\
 2563& 0.5315&  1&  2.300&  2.348&  0.049&   0.00&   0.45&   0.55&    5.1&   74.1\\
 2645& 0.5300&  1&  1.964&  2.071&  0.048&   0.37&   0.42&   0.21&    5.8&  108.3\\
 2657& 0.5787&  0&  \nodata&  \nodata&  \nodata&  \nodata&  \nodata&  \nodata&  \nodata&   47.3\\
 2689& 0.5308&  1&  2.101&  2.169&  0.037&   0.00&   0.59&   0.41&    4.8&   81.1\\
 2788& 0.5405&  1&  2.020&  2.109&  0.039&   0.00&   0.42&   0.58&    3.4&   59.7\\
 2913& 0.5420&  1&  2.344&  2.392&  0.020&   0.00&   0.57&   0.43&   10.1&  134.0\\
 2945& 0.5255&  1&  1.911&  2.039&  0.061&   0.00&   0.60&   0.40&    3.8&   60.9\\
 3005& 0.9193&  0&  \nodata&  \nodata&  \nodata&  \nodata&  \nodata&  \nodata&  \nodata&   35.5\\
 3124& 0.5469&  1&  1.987&  2.087&  0.045&   0.13&   0.49&   0.37&    2.2&   57.8\\
 3260& 0.5336&  1&  2.245&  2.294&  0.039&   0.16&   0.20&   0.64&    2.5&   42.5\\
 3521& 0.4912&  0&  \nodata&  \nodata&  \nodata&  \nodata&  \nodata&  \nodata&  \nodata&   64.2\\
 3610& 0.5663&  0&  \nodata&  \nodata&  \nodata&  \nodata&  \nodata&  \nodata&  \nodata&   32.5\\
 3625& 0.9489&  0&  \nodata&  \nodata&  \nodata&  \nodata&  \nodata&  \nodata&  \nodata&   28.9\\
 3635& 0.7668&  0&  \nodata&  \nodata&  \nodata&  \nodata&  \nodata&  \nodata&  \nodata&   13.7\\
 3697& 0.5674&  0&  \nodata&  \nodata&  \nodata&  \nodata&  \nodata&  \nodata&  \nodata&   23.3\\
 3724& 0.5417&  1&  1.977&  2.080&  0.036&   0.00&   0.58&   0.42&    2.5&   59.2\\
 3749& 0.5422&  1&  2.028&  2.115&  0.047&   0.19&   0.38&   0.44&    2.5&   47.1\\
 3792& 0.8996&  0&  \nodata&  \nodata&  \nodata&  \nodata&  \nodata&  \nodata&  \nodata&   19.4\\
 3857& 0.5348&  1&  2.103&  2.171&  0.057&   0.00&   0.47&   0.53&    3.5&   60.7\\
 3906& 1.2846&  0&  \nodata&  \nodata&  \nodata&  \nodata&  \nodata&  \nodata&  \nodata&   21.4\\
\enddata
\tablenotetext{a}{Adopted membership: 1 -- galaxy is a member of MS0451.6--0305; 
0 -- galaxy is not a member of MS0451.6--0305}
\tablenotetext{b}{Velocity dispersions corrected for systematic effects, see Sect \ref{SEC-SIM}, and corrected
to a standard size aperture equivalent to a
circular aperture with diameter of 3.4 arcsec at the distance of the Coma cluster.}
\tablenotetext{c}{S/N per {\AA}ngstrom in the rest frame of the galaxy. The wavelength 
interval was chosen based on the redshift of the galaxy as follows: 
Redshift less than 0.4 -- 4500-5500 {\AA};
redshift 0.4-0.6 -- 4100-5500 {\AA}; 
redshift 0.75-1.1 -- 3800-4300 {\AA};
redshift larger than 1.1 -- 3800-4000 {\AA}.}
\end{deluxetable*}

\begin{deluxetable*}{rrrrrrrrrrrrrrrrrr}
\tablecaption{MS0451.6--0305: Line Indices and EW[\ion{O}{2}] for Cluster Members\label{tab-MS0451line} }
\tabletypesize{\scriptsize}
\tablewidth{0pc}
\tablehead{
\colhead{ID} & \colhead{CN3883}& \colhead{CaHK}& \colhead{D4000}
& \colhead{H$\delta_A$}& \colhead{CN$_1$}& \colhead{CN$_2$} 
& \colhead{G4300}& \colhead{H$\gamma_A$}& \colhead{Fe4383}& \colhead{C4668}
& \colhead{H$\beta$} & \colhead{H$\beta _G$} & \colhead{Mg{\it b}} & \colhead{Fe5270} & \colhead{Fe5335} 
& \colhead{EW [\ion{O}{2}]}
} 
\startdata
 12& 0.129& 18.45& 1.955& 1.41& 0.037& 0.045& 3.36& -2.03& 4.11& 5.50& 1.55& 1.70
& 3.50& 1.38& \nodata& 5.6\\
 12& 0.017& 0.75& 0.010& 0.31& 0.008& 0.009& 0.22& 0.22& 0.28& 0.35& 0.14& 0.09
& 0.16& 0.14& \nodata& 0.7\\
 153& 0.242& 21.56& 2.134& -2.90& \nodata& \nodata& 5.93& -5.86& 4.14& 5.46& 1.54& 1.87
& 4.25& 3.45& 4.01& \nodata\\
 153& 0.010& 0.41& 0.007& 0.18& \nodata& \nodata& 0.12& 0.14& 0.17& 0.19& 0.08& 0.05
& 0.10& 0.09& 0.11& \nodata\\
 258& 0.039& 14.60& 1.487& 2.33& \nodata& \nodata& 1.09& -0.38& 3.08& 1.81& -0.63& -0.36
& 2.26& 1.68& 0.25& 7.4\\
 258& 0.011& 0.55& 0.005& 0.25& \nodata& \nodata& 0.22& 0.19& 0.27& 0.37& 0.16& 0.11
& 0.21& 0.19& 0.21& 0.6\\
 299& 0.289& 22.49& 2.116& -0.56& 0.123& 0.152& 5.43& -5.24& 4.09& 8.10& 1.97& 2.17
& 4.76& 3.02& 2.08& \nodata\\
 299& 0.004& 0.16& 0.002& 0.07& 0.002& 0.002& 0.05& 0.05& 0.06& 0.08& 0.03& 0.02
& 0.04& 0.04& 0.04& \nodata\\
 309& 0.129& 20.22& 1.868& \nodata& \nodata& \nodata& 3.54& -2.79& 3.92& 3.93& 2.01& 2.15
& 2.24& 2.76& \nodata& \nodata\\
 309& 0.018& 0.82& 0.010& \nodata& \nodata& \nodata& 0.27& 0.27& 0.36& 0.43& 0.18& 0.12
& 0.26& 0.21& \nodata& \nodata\\
 554& \nodata& \nodata& \nodata& 0.68& 0.040& 0.055& 4.25& -3.35& 3.51& 4.37& 2.10& 2.28
& 2.99& 2.32& 1.30& \nodata\\
 554& \nodata& \nodata& \nodata& 0.26& 0.007& 0.008& 0.19& 0.19& 0.25& 0.28& 0.11& 0.08
& 0.18& 0.15& 0.16& \nodata\\
 600& 0.090& 10.41& 1.500& -1.22& 0.033& 0.030& 1.63& -2.05& 2.44& -0.27& 1.05& 1.55
& 1.86& 1.78& 1.04& \nodata\\
 600& 0.006& 0.31& 0.003& 0.15& 0.004& 0.004& 0.10& 0.10& 0.14& 0.16& 0.07& 0.05
& 0.10& 0.09& 0.09& \nodata\\
 684& 0.159& 21.70& 1.874& 0.84& 0.029& 0.062& 3.77& -2.51& 4.40& 4.19& 2.80& 2.99
& 3.26& 3.31& 3.07& \nodata\\
 684& 0.012& 0.52& 0.007& 0.23& 0.006& 0.007& 0.17& 0.17& 0.21& 0.29& 0.12& 0.07
& 0.17& 0.14& 0.14& \nodata\\
 716& 0.220& 20.60& 1.865& -0.24& 0.043& 0.069& 4.30& -3.66& 4.36& 4.09& 1.83& 2.15
& 2.66& 2.97& \nodata& \nodata\\
 716& 0.014& 0.61& 0.008& 0.27& 0.007& 0.007& 0.19& 0.21& 0.27& 0.33& 0.14& 0.09
& 0.19& 0.16& \nodata& \nodata\\
 722& 0.096& 9.76& 1.398& 4.82& \nodata& \nodata& 0.67& 1.25& 0.14& 2.01& -3.18& -2.88
& 0.20& 0.29& \nodata& 16.8\\
 722& 0.014& 0.76& 0.006& 0.30& \nodata& \nodata& 0.27& 0.24& 0.36& 0.54& 0.28& 0.19
& 0.34& 0.28& \nodata& 0.8\\
 897& \nodata& \nodata& \nodata& \nodata& \nodata& \nodata& 5.30& -5.04& 3.88& 5.13& 1.86& 2.07
& 3.59& 3.33& 3.83& \nodata\\
 897& \nodata& \nodata& \nodata& \nodata& \nodata& \nodata& 0.14& 0.15& 0.19& 0.23& 0.10& 0.06
& 0.13& 0.11& 0.13& \nodata\\
 921& 0.274& 22.87& 2.064& -0.64& 0.105& 0.126& 4.83& -5.01& 4.90& 6.08& 2.06& 2.28
& 4.25& 3.02& 2.67& \nodata\\
 921& 0.005& 0.21& 0.003& 0.10& 0.002& 0.003& 0.07& 0.07& 0.09& 0.11& 0.04& 0.03
& 0.06& 0.05& 0.07& \nodata\\
 971& \nodata& \nodata& \nodata& -1.20& 0.115& 0.143& 5.41& -5.40& 3.79& 5.67& 1.09& 1.43
& 4.20& 2.93& 2.31& \nodata\\
 971& \nodata& \nodata& \nodata& 0.13& 0.003& 0.004& 0.09& 0.10& 0.12& 0.14& 0.06& 0.04
& 0.08& 0.07& 0.08& \nodata\\
 1002& \nodata& 20.18& \nodata& 0.96& \nodata& 0.051& 4.66& -3.05& 3.42& 5.43& 2.32& 2.57
& 3.59& 2.55& \nodata& \nodata\\
 1002& \nodata& 0.62& \nodata& 0.27& \nodata& 0.008& 0.19& 0.21& 0.28& 0.33& 0.14& 0.09
& 0.19& 0.16& \nodata& \nodata\\
 1082& 0.244& 21.64& 2.041& -0.94& 0.085& 0.108& 5.97& -6.05& 3.86& 6.55& 1.85& 2.00
& 3.65& 3.18& 3.12& \nodata\\
 1082& 0.006& 0.25& 0.004& 0.11& 0.003& 0.003& 0.07& 0.09& 0.11& 0.13& 0.05& 0.03
& 0.07& 0.06& 0.08& \nodata\\
 1156& \nodata& 8.59& 1.367& -1.31& 0.091& 0.118& 2.74& -5.68& 4.79& 6.39& 1.29& 1.36
& 2.52& 2.23& 1.28& \nodata\\
 1156& \nodata& 0.57& 0.004& 0.28& 0.007& 0.008& 0.21& 0.22& 0.28& 0.33& 0.14& 0.10
& 0.21& 0.17& 0.19& \nodata\\
 1204& 0.194& 22.33& 2.065& 0.35& 0.066& 0.080& 5.72& -5.13& 5.06& 6.79& 1.77& 2.23
& 4.29& 1.92& 2.00& \nodata\\
 1204& 0.010& 0.41& 0.006& 0.18& 0.005& 0.005& 0.12& 0.13& 0.17& 0.19& 0.08& 0.05
& 0.12& 0.10& 0.11& \nodata\\
 1331& \nodata& 18.12& \nodata& 5.10& \nodata& \nodata& 2.08& 2.06& 2.21& -0.27& 3.79& 3.96
& 2.78& 1.86& 1.99& \nodata\\
 1331& \nodata& 0.30& \nodata& 0.13& \nodata& \nodata& 0.12& 0.11& 0.16& 0.19& 0.08& 0.05
& 0.13& 0.12& 0.12& \nodata\\
 1491& 0.176& 21.04& 2.019& -2.63& 0.141& 0.166& 6.23& -4.91& 5.57& 5.57& 1.08& 1.50
& 4.77& 1.69& 3.00& \nodata\\
 1491& 0.021& 0.85& 0.013& 0.37& 0.009& 0.010& 0.25& 0.28& 0.34& 0.37& 0.16& 0.11
& 0.25& 0.20& 0.20& \nodata\\
 1497& 0.233& 23.08& 1.941& 0.43& 0.041& 0.067& 4.73& -3.47& 3.15& 4.79& 2.33& 2.47
& 3.74& 2.39& 1.61& \nodata\\
 1497& 0.005& 0.21& 0.003& 0.10& 0.003& 0.003& 0.07& 0.07& 0.09& 0.10& 0.05& 0.03
& 0.06& 0.06& 0.06& \nodata\\
 1500& 0.247& 23.35& 2.136& -1.16& 0.126& 0.160& 5.60& -6.00& 5.41& 7.34& 1.44& 1.84
& 5.04& 2.98& 3.13& \nodata\\
 1500& 0.010& 0.40& 0.007& 0.19& 0.005& 0.005& 0.12& 0.13& 0.15& 0.18& 0.08& 0.05
& 0.11& 0.09& 0.10& \nodata\\
 1507& 0.256& 22.50& 2.024& -0.93& 0.126& 0.166& 4.95& -5.57& 5.14& 6.87& 1.71& 1.93
& 3.88& 2.61& 2.78& \nodata\\
 1507& 0.007& 0.29& 0.004& 0.14& 0.004& 0.004& 0.09& 0.10& 0.11& 0.13& 0.06& 0.04
& 0.08& 0.07& 0.07& \nodata\\
 1594& 0.251& 22.21& 2.076& -1.60& 0.109& 0.142& 5.57& -5.77& 4.50& 6.68& 1.96& 2.23
& 4.79& 2.14& 1.27& \nodata\\
 1594& 0.009& 0.36& 0.006& 0.17& 0.004& 0.005& 0.11& 0.12& 0.15& 0.16& 0.07& 0.05
& 0.11& 0.09& 0.10& \nodata\\
 1638& 0.215& 20.45& 1.986& -0.92& 0.138& 0.161& 5.21& -4.93& 4.51& 6.15& 2.30& 2.37
& 3.65& 3.00& 2.45& \nodata\\
 1638& 0.008& 0.35& 0.005& 0.15& 0.004& 0.004& 0.11& 0.12& 0.15& 0.17& 0.07& 0.04
& 0.10& 0.08& 0.11& \nodata\\
 1720& 0.243& 19.15& 1.926& -1.14& 0.111& 0.133& 5.06& -5.52& 4.47& 6.22& 1.52& 1.83
& 4.31& 2.27& 2.59& \nodata\\
 1720& 0.005& 0.20& 0.003& 0.10& 0.002& 0.003& 0.06& 0.07& 0.09& 0.10& 0.04& 0.03
& 0.06& 0.05& 0.06& \nodata\\
 1723& \nodata& \nodata& \nodata& 2.79& \nodata& \nodata& 3.38& -0.93& 3.42& 3.97& 2.66& 2.90
& 3.79& 2.13& 1.83& \nodata\\
 1723& \nodata& \nodata& \nodata& 0.09& \nodata& \nodata& 0.07& 0.07& 0.10& 0.11& 0.05& 0.03
& 0.07& 0.06& 0.07& \nodata\\
 1753& 0.124& 11.80& 1.404& -0.26& 0.049& 0.040& 1.95& -3.17& 2.38& 4.14& 1.49& 1.70
& 3.24& 1.89& 1.42& \nodata\\
 1753& 0.007& 0.38& 0.003& 0.18& 0.004& 0.005& 0.12& 0.13& 0.17& 0.19& 0.09& 0.06
& 0.11& 0.11& 0.12& \nodata\\
 1823& 0.260& 22.18& 2.151& -1.33& 0.103& 0.127& 5.09& -4.55& 3.66& 7.50& 1.90& 1.98
& 4.56& 2.44& 2.27& \nodata\\
 1823& 0.009& 0.35& 0.006& 0.16& 0.004& 0.005& 0.12& 0.12& 0.14& 0.15& 0.07& 0.05
& 0.09& 0.09& 0.09& \nodata\\
 1904& 0.280& 22.52& 2.054& -2.77& \nodata& \nodata& 6.00& -5.73& 4.53& 6.07& 1.28& 1.58
& 4.70& 2.64& 2.31& \nodata\\
 1904& 0.008& 0.35& 0.005& 0.18& \nodata& \nodata& 0.11& 0.12& 0.14& 0.16& 0.07& 0.04
& 0.10& 0.08& 0.09& \nodata\\
 1931& 0.233& 21.22& 1.969& -0.72& \nodata& \nodata& 5.15& -5.06& 3.55& 5.20& 1.58& 1.89
& 4.00& 2.36& 1.36& \nodata\\
 1931& 0.009& 0.38& 0.005& 0.17& \nodata& \nodata& 0.12& 0.14& 0.18& 0.21& 0.08& 0.05
& 0.10& 0.10& 0.14& \nodata\\
\enddata
\tablecomments{The indices have been corrected for galaxy velocity dispersion and aperture corrected. The second line for each galaxy lists the uncertainties determined from the local S/N.}
\end{deluxetable*}

\begin{deluxetable*}{rrrrrrrrrrrrrrrrrr}
\tablecaption{MS0451.6--0305: \em -- Continued}
\tabletypesize{\scriptsize}
\tablenum{22}
\tablewidth{0pc}
\tablehead{
\colhead{ID} & \colhead{CN3883}& \colhead{CaHK}& \colhead{D4000}
& \colhead{H$\delta_A$}& \colhead{CN$_1$}& \colhead{CN$_2$} 
& \colhead{G4300}& \colhead{H$\gamma_A$}& \colhead{Fe4383}& \colhead{C4668}
& \colhead{H$\beta$} & \colhead{H$\beta _G$} & \colhead{Mg{\it b}} & \colhead{Fe5270} & \colhead{Fe5335} 
& \colhead{EW [\ion{O}{2}]}
} 
\startdata
 1952& 0.238& 20.77& 2.001& -0.04& 0.090& 0.119& 4.55& -4.27& 4.41& 7.29& 2.11& 2.31
& 3.95& 3.48& 3.64& \nodata\\
 1952& 0.006& 0.25& 0.004& 0.11& 0.003& 0.003& 0.08& 0.08& 0.11& 0.13& 0.05& 0.03
& 0.07& 0.06& 0.08& \nodata\\
 2031& 0.297& 22.35& 2.130& 0.03& 0.085& 0.119& 5.22& -4.90& 5.22& 7.01& 1.55& 1.98
& 4.55& 2.71& 2.08& \nodata\\
 2031& 0.010& 0.38& 0.006& 0.16& 0.004& 0.005& 0.12& 0.13& 0.15& 0.17& 0.08& 0.05
& 0.10& 0.09& 0.10& \nodata\\
 2127& 0.048& 10.43& 1.196& 4.63& \nodata& \nodata& 0.65& 1.85& 0.73& 2.41& -1.35& -1.25
& 2.77& 1.63& 1.56& \nodata\\
 2127& 0.003& 0.18& 0.001& 0.08& \nodata& \nodata& 0.08& 0.07& 0.11& 0.13& 0.07& 0.05
& 0.11& 0.10& 0.10& \nodata\\
 2166& 0.066& 10.36& 1.389& 1.39& \nodata& \nodata& 1.19& -0.04& 2.12& 3.14& -4.14& -3.99
& 3.30& 2.95& 1.84& 11.8\\
 2166& 0.007& 0.38& 0.003& 0.17& \nodata& \nodata& 0.15& 0.14& 0.20& 0.24& 0.12& 0.09
& 0.14& 0.14& 0.15& 1.5\\
 2223& 0.175& 19.80& 1.890& 1.14& \nodata& 0.039& 4.42& -4.80& 4.97& 5.09& 2.00& 2.41
& 4.40& 2.76& \nodata& \nodata\\
 2223& 0.014& 0.62& 0.008& 0.26& \nodata& 0.008& 0.21& 0.23& 0.29& 0.34& 0.14& 0.09
& 0.17& 0.16& \nodata& \nodata\\
 2240& 0.143& 21.85& 1.908& 0.38& 0.042& 0.038& 3.57& -2.53& 3.21& 4.48& 2.00& 1.98
& 4.21& 2.90& \nodata& \nodata\\
 2240& 0.022& 0.94& 0.013& 0.41& 0.010& 0.011& 0.27& 0.30& 0.41& 0.50& 0.21& 0.13
& 0.25& 0.23& \nodata& \nodata\\
 2563& 0.213& 19.82& 1.936& -1.02& \nodata& \nodata& 5.13& -4.85& 3.51& 4.97& 1.99& 2.25
& 3.33& 2.68& 1.94& \nodata\\
 2563& 0.010& 0.43& 0.006& 0.20& \nodata& \nodata& 0.14& 0.14& 0.17& 0.19& 0.09& 0.06
& 0.12& 0.11& 0.11& \nodata\\
 2645& 0.094& 13.26& 1.488& 2.02& \nodata& \nodata& 2.36& -1.43& 2.59& 2.75& -1.98& -1.76
& 3.17& 1.85& 1.82& 12.0\\
 2645& 0.004& 0.24& 0.002& 0.11& \nodata& \nodata& 0.09& 0.08& 0.11& 0.13& 0.07& 0.05
& 0.09& 0.08& 0.08& 0.7\\
 2689& 0.300& 26.61& 2.216& -2.82& \nodata& \nodata& 6.09& -6.95& 6.37& 5.65& 1.53& 1.90
& 4.89& 2.18& 2.66& 13.8\\
 2689& 0.012& 0.42& 0.008& 0.21& \nodata& \nodata& 0.13& 0.15& 0.16& 0.17& 0.08& 0.05
& 0.10& 0.10& 0.10& 1.2\\
 2788& 0.241& 20.96& 2.079& -0.24& 0.066& 0.085& 5.15& -5.10& 5.44& 6.32& 2.16& 2.29
& 3.44& 2.88& 2.85& \nodata\\
 2788& 0.013& 0.53& 0.008& 0.23& 0.006& 0.007& 0.15& 0.17& 0.21& 0.24& 0.10& 0.07
& 0.13& 0.12& 0.13& \nodata\\
 2913& 0.271& 22.46& 2.123& -1.69& 0.139& 0.176& 5.62& -5.42& 4.71& 7.02& 1.94& 2.23
& 4.05& 2.85& 2.69& \nodata\\
 2913& 0.005& 0.24& 0.004& 0.11& 0.003& 0.003& 0.07& 0.08& 0.10& 0.11& 0.05& 0.03
& 0.06& 0.05& 0.06& \nodata\\
 2945& 0.213& 22.87& 2.051& 0.73& 0.059& 0.086& 4.76& -4.64& 3.76& 7.01& 2.38& 2.65
& 3.90& 2.53& 2.84& \nodata\\
 2945& 0.014& 0.51& 0.008& 0.22& 0.006& 0.007& 0.18& 0.18& 0.22& 0.23& 0.10& 0.07
& 0.16& 0.14& 0.13& \nodata\\
 3124& 0.175& 21.38& 1.985& -0.10& 0.066& 0.086& 4.77& -3.76& 3.74& 6.41& 1.86& 2.22
& 3.36& 3.83& 2.04& \nodata\\
 3124& 0.012& 0.50& 0.007& 0.22& 0.005& 0.006& 0.15& 0.16& 0.20& 0.26& 0.11& 0.07
& 0.14& 0.12& 0.15& \nodata\\
 3260& 0.235& 20.75& 2.038& -1.91& \nodata& \nodata& 6.34& -4.47& 3.35& 7.56& 1.26& 1.83
& 3.97& 2.04& 2.61& \nodata\\
 3260& 0.019& 0.73& 0.011& 0.32& \nodata& \nodata& 0.23& 0.25& 0.32& 0.32& 0.15& 0.09
& 0.22& 0.18& 0.19& \nodata\\
 3724& 0.230& 23.02& 2.030& -1.99& \nodata& \nodata& 4.67& -3.95& 4.23& 5.57& 1.99& 2.60
& 3.29& 3.21& 2.27& \nodata\\
 3724& 0.013& 0.45& 0.007& 0.24& \nodata& \nodata& 0.16& 0.17& 0.21& 0.24& 0.10& 0.06
& 0.15& 0.12& 0.13& \nodata\\
 3749& 0.205& 21.83& 1.979& 0.26& 0.043& 0.070& 4.32& -2.80& 2.73& 4.17& 2.94& 3.12
& 4.37& 3.86& \nodata& \nodata\\
 3749& 0.015& 0.58& 0.009& 0.28& 0.007& 0.008& 0.20& 0.20& 0.27& 0.31& 0.12& 0.07
& 0.18& 0.14& \nodata& \nodata\\
 3857& 0.246& 24.01& 1.944& -1.82& \nodata& \nodata& 5.93& -4.29& 4.30& 6.34& 1.86& 2.32
& 2.68& 2.88& 2.32& \nodata\\
 3857& 0.012& 0.49& 0.007& 0.23& \nodata& \nodata& 0.16& 0.17& 0.21& 0.23& 0.10& 0.07
& 0.14& 0.13& 0.14& \nodata\\
\enddata
\tablecomments{The indices have been corrected for galaxy velocity dispersion and aperture corrected. The second line for each galaxy lists the uncertainties determined from the local S/N.}
\end{deluxetable*}

\clearpage

\begin{deluxetable*}{rrrrrrrrrrrr}
\tablecaption{RXJ1226.9+3332: Results from Template Fitting \label{tab-RXJ1226kin} }
\tablewidth{400pt}
\tablehead{
\colhead{ID} & \colhead{Redshift} & \colhead{Member\tablenotemark{a}} &  
\colhead{$\log \sigma$} & \colhead{$\log \sigma _{\rm cor}$\tablenotemark{b}} & 
\colhead{$\sigma _{\log \sigma}$} & \multicolumn{3}{c}{Template fractions} & \colhead{$\chi^2$} & \colhead{S/N\tablenotemark{c}} 
& \colhead{Data\tablenotemark{d}} \\
\colhead{}&\colhead{} &\colhead{} &\colhead{} &\colhead{} &\colhead{} & \colhead{B8V} & \colhead{G1V} & \colhead{K0III} & & }
\startdata
   18& 0.7557&  0&  \nodata&  \nodata&  \nodata&  \nodata&  \nodata&  \nodata&  \nodata&   34.9&              2\\
   38& 0.8825&  1&  2.122&  2.171&  0.150&   0.34&   0.53&   0.13&    3.3&   17.2&              2\\
   53& 1.2360&  0&  \nodata&  \nodata&  \nodata&  \nodata&  \nodata&  \nodata&  \nodata&   42.8&              2\\
   55& 0.8936&  1&  2.027&  2.052&  0.034&   0.13&   0.44&   0.43&    6.4&   96.9&              1\\
   56& 0.8932&  1&  2.068&  2.093&  0.063&   0.22&   0.53&   0.25&    6.9&   72.3&              1\\
   91& 0.9744&  0&  \nodata&  \nodata&  \nodata&  \nodata&  \nodata&  \nodata&  \nodata&   15.9&              1\\
  104& 0.8949&  1&  2.162&  2.187&  0.047&   0.16&   0.36&   0.48&    5.3&   54.2&              1\\
  122& 0.8876&  1&  2.329&  2.354&  0.052&   0.29&   0.36&   0.35&    7.4&   48.1&              1\\
  132& 0.6793&  0&  \nodata&  \nodata&  \nodata&  \nodata&  \nodata&  \nodata&  \nodata&   81.3&              1\\
  138& 0.7136&  0&  \nodata&  \nodata&  \nodata&  \nodata&  \nodata&  \nodata&  \nodata&   55.5&              2\\
  153& 1.9750&  0&  \nodata&  \nodata&  \nodata&  \nodata&  \nodata&  \nodata&  \nodata&   23.3&              2\\
  154& 0.6797&  0&  \nodata&  \nodata&  \nodata&  \nodata&  \nodata&  \nodata&  \nodata&   18.5&              2\\
  178& 0.0970&  0&  \nodata&  \nodata&  \nodata&  \nodata&  \nodata&  \nodata&  \nodata&   64.1&              1\\
  185& 0.6868&  0&  \nodata&  \nodata&  \nodata&  \nodata&  \nodata&  \nodata&  \nodata&   34.8&              2\\
  203& 0.8403&  0&  \nodata&  \nodata&  \nodata&  \nodata&  \nodata&  \nodata&  \nodata&   20.4&              2\\
  220& 0.6882&  0&  \nodata&  \nodata&  \nodata&  \nodata&  \nodata&  \nodata&  \nodata&   46.6&              2\\
  229& 0.8922&  1&  1.997&  2.045&  0.073&   0.22&   0.44&   0.35&    3.7&   32.0&              2\\
  245& 0.9748&  0&  \nodata&  \nodata&  \nodata&  \nodata&  \nodata&  \nodata&  \nodata&   35.3&              2\\
  247& 0.3849&  0&  \nodata&  \nodata&  \nodata&  \nodata&  \nodata&  \nodata&  \nodata&  172.0&              1\\
  249& 0.5003&  0&  \nodata&  \nodata&  \nodata&  \nodata&  \nodata&  \nodata&  \nodata&   44.2&              1\\
  293& 0.8838&  1&  2.254&  2.278&  0.045&   0.19&   0.41&   0.40&    2.9&   38.4&              1\\
  295& 0.8954&  1&  1.948&  1.973&  0.060&   0.29&   0.33&   0.38&    4.5&   47.6&              1\\
  297& 0.8450&  0&  \nodata&  \nodata&  \nodata&  \nodata&  \nodata&  \nodata&  \nodata&    7.2&              2\\
  309& 0.8859&  1&  2.294&  2.343&  0.147&   0.00&   0.20&   0.80&    3.2&   14.5&              2\\
  310& 0.8885&  1&  2.096&  2.145&  0.088&   0.18&   0.49&   0.33&    3.5&   31.8&              2\\
  316& 0.7315&  0&  \nodata&  \nodata&  \nodata&  \nodata&  \nodata&  \nodata&  \nodata&    5.1&              2\\
  329& 0.5883&  0&  \nodata&  \nodata&  \nodata&  \nodata&  \nodata&  \nodata&  \nodata&   48.8&              1\\
  333& 0.7146&  0&  \nodata&  \nodata&  \nodata&  \nodata&  \nodata&  \nodata&  \nodata&   17.4&              1\\
  347& 0.3682&  0&  \nodata&  \nodata&  \nodata&  \nodata&  \nodata&  \nodata&  \nodata&  103.5&              1\\
  349& 0.6871&  0&  \nodata&  \nodata&  \nodata&  \nodata&  \nodata&  \nodata&  \nodata&   47.3&              2\\
  359& 0.7026&  0&  \nodata&  \nodata&  \nodata&  \nodata&  \nodata&  \nodata&  \nodata&   23.8&              2\\
  374& 0.3394&  0&  \nodata&  \nodata&  \nodata&  \nodata&  \nodata&  \nodata&  \nodata&   14.6&              2\\
  386& 0.9895&  0&  \nodata&  \nodata&  \nodata&  \nodata&  \nodata&  \nodata&  \nodata&   15.2&              2\\
  408& 0.9890&  0&  \nodata&  \nodata&  \nodata&  \nodata&  \nodata&  \nodata&  \nodata&   57.3&              1\\
  423& 0.8946&  1&  2.309&  2.358&  0.040&   0.11&   0.54&   0.34&    2.0&   26.9&              2\\
  441& 0.8920&  1&  2.046&  2.071&  0.044&   0.10&   0.66&   0.24&    3.1&   35.1&              1\\
  446& 0.8836&  1&  1.936&  1.985&  0.067&   0.18&   0.75&   0.07&    3.7&   34.8&              2\\
  452& 0.8818&  1&  2.107&  2.132&  0.075&   0.55&   0.27&   0.18&    2.8&   35.0&              1\\
  462& 0.8929&  1&  2.071&  2.120&  0.072&   0.00&   0.40&   0.60&    3.9&   22.1&              2\\
  470& 0.8987&  1&  1.991&  2.040&  0.096&   0.16&   0.49&   0.34&    3.3&   19.8&              2\\
  491& 0.8791&  1&  2.216&  2.240&  0.046&   0.00&   0.50&   0.49&    6.3&   71.8&              1\\
  499& 0.7326&  0&  \nodata&  \nodata&  \nodata&  \nodata&  \nodata&  \nodata&  \nodata&   14.5&              2\\
  500& 0.4010&  0&  \nodata&  \nodata&  \nodata&  \nodata&  \nodata&  \nodata&  \nodata&   15.7&              2\\
  512& 0.8973&  1&  2.085&  2.133&  0.059&   0.00&   0.33&   0.67&    3.9&   20.7&              2\\
  523& 0.9297&  0&  \nodata&  \nodata&  \nodata&  \nodata&  \nodata&  \nodata&  \nodata&   29.5&              1\\
  527& 1.1280&  0&  \nodata&  \nodata&  \nodata&  \nodata&  \nodata&  \nodata&  \nodata&   10.1&              2\\
  528& 0.8873&  1&  2.297&  2.346&  0.058&   0.44&   0.27&   0.29&    2.0&   21.2&              2\\
  529& 0.8970&  1&  2.361&  2.386&  0.028&   0.00&   0.45&   0.55&   11.3&   95.5&              1\\
  534& 0.8880&  1&  2.437&  2.485&  0.051&   0.00&   0.79&   0.21&    1.9&   21.6&              2\\
  547& 0.8825&  1&  2.079&  2.127&  0.227&   1.00&   0.00&   0.00&    3.3&   12.1&              2\\
  557& 0.8909&  1&  2.205&  2.230&  0.086&   0.00&   0.59&   0.41&    4.4&   27.0&              1\\
  563& 0.8908&  1&  2.528&  2.553&  0.017&   0.10&   0.30&   0.60&   17.4&  116.3&              1\\
  572& 0.4993&  0&  \nodata&  \nodata&  \nodata&  \nodata&  \nodata&  \nodata&  \nodata&   31.8&              1\\
  592& 0.8760&  1&  \nodata&  \nodata&  \nodata&   \nodata&   \nodata&   \nodata&  \nodata&   16.0&              2\\
  593& 0.8812&  1&  2.127&  2.176&  0.074&   0.00&   0.89&   0.11&    1.8&   17.5&              2\\
  602& 0.8787&  1&  2.085&  2.110&  0.040&   0.42&   0.58&   0.00&    5.0&   49.7&              1\\
  608& 0.8752&  1&  2.160&  2.209&  0.038&   0.00&   0.89&   0.11&    5.9&   51.8&              2\\
  630& 0.8905&  1&  \nodata&  \nodata&  \nodata&  \nodata&  \nodata&  \nodata&  \nodata&    9.3&              2\\
  641& 0.8968&  1&  2.088&  2.136&  0.042&   0.00&   0.57&   0.43&    3.7&   24.4&              2\\
  647& 0.8920&  1&  1.963&  2.012&  0.098&   0.12&   0.62&   0.26&    2.1&   12.3&              2\\
  648& 0.8757&  1&  1.933&  1.957&  0.046&   0.00&   0.73&   0.27&    6.0&   67.5&              1\\
  649& 0.9254&  0&  \nodata&  \nodata&  \nodata&  \nodata&  \nodata&  \nodata&  \nodata&   18.7&              2\\
  650& 0.8937&  1&  1.963&  2.012&  0.076&   0.32&   0.47&   0.21&    2.9&   27.6&              2\\
  656& 0.9293&  0&  \nodata&  \nodata&  \nodata&  \nodata&  \nodata&  \nodata&  \nodata&   29.4&              2\\
  675& 0.8974&  1&  2.151&  2.176&  0.064&   0.22&   0.40&   0.38&    3.6&   48.8&              1\\
\enddata
\tablenotetext{a}{Adopted membership: 1 -- galaxy is a member of RXJ1226.9+3332; 
0 -- galaxy is not a member of RXJ1226.9+3332}
\tablenotetext{b}{Velocity dispersions corrected to a standard size aperture equivalent to a
circular aperture with diameter of 3.4 arcsec at the distance of the Coma cluster.}
\tablenotetext{c}{S/N per {\AA}ngstrom in the rest frame of the galaxy. The wavelength 
interval was 4100-4600 {\AA}, except when wavelength coverage did not allow this typically
due to the redshift of the galaxy the following intervals were used:
ID 178 -- 6500-7000 {\AA}.
ID 132, 247, 249, 982, 1029 and 1083 -- 5000-5500 {\AA};
ID 527, 1027, 1057 -- 3500-4000 {\AA};
ID 53, 153 -- 2500-3000 {\AA};
ID 754 -- 1500-2000 {\AA}.}
\tablenotetext{d}{Source of data -- (1) GCP, (2) GSA.}
\end{deluxetable*}

\begin{deluxetable*}{rrrrrrrrrrrr}
\tablecaption{RXJ1226.9+3332: \em -- Continued}
\tablewidth{0pc}
\tablenum{23}
\tablehead{
\colhead{ID} & \colhead{Redshift} & \colhead{Member\tablenotemark{a}} &  
\colhead{$\log \sigma$} & \colhead{$\log \sigma _{\rm cor}$\tablenotemark{b}} & 
\colhead{$\sigma _{\log \sigma}$} & \multicolumn{3}{c}{Template fractions} & \colhead{$\chi^2$} & \colhead{S/N\tablenotemark{c}} 
& \colhead{Data\tablenotemark{d}} \\
\colhead{}&\colhead{} &\colhead{} &\colhead{} &\colhead{} &\colhead{} & \colhead{B8V} & \colhead{G1V} & \colhead{K0III} & & }
\startdata
  685& 1.0179&  0&  \nodata&  \nodata&  \nodata&  \nodata&  \nodata&  \nodata&  \nodata&    6.5&              2\\
  689& 0.8958&  1&  2.337&  2.386&  0.068&   0.00&   0.41&   0.59&    1.8&   13.6&              2\\
  703& 0.8977&  1&  2.415&  2.440&  0.017&   0.12&   0.30&   0.58&    8.5&   85.7&              1\\
  709& 0.9085&  1&  2.290&  2.315&  0.038&   0.21&   0.46&   0.32&    3.4&   45.4&              1\\
  711& 0.8822&  1&  1.797&  1.822&  0.090&   0.62&   0.38&   0.00&    3.3&   16.4&              1\\
  739& 1.0556&  0&  \nodata&  \nodata&  \nodata&  \nodata&  \nodata&  \nodata&  \nodata&   14.2&              2\\
  754& 3.2000&  0&  \nodata&  \nodata&  \nodata&  \nodata&  \nodata&  \nodata&  \nodata&   71.6&              2\\
  757& 0.8856&  1&  2.230&  2.279&  0.116&   0.00&   0.74&   0.26&    1.3&   11.3&              2\\
  760& 0.8901&  1&  2.310&  2.335&  0.037&   0.15&   0.47&   0.38&   10.4&   80.0&              1\\
  781& 0.7668&  0&  \nodata&  \nodata&  \nodata&  \nodata&  \nodata&  \nodata&  \nodata&   40.0&              2\\
  798& 0.9636&  0&  \nodata&  \nodata&  \nodata&  \nodata&  \nodata&  \nodata&  \nodata&   32.3&              1\\
  801& 0.8923&  1&  1.939&  1.987&  0.076&   0.13&   0.65&   0.22&    7.2&   37.3&              2\\
  805& 0.6276&  0&  \nodata&  \nodata&  \nodata&  \nodata&  \nodata&  \nodata&  \nodata&   22.4&              2\\
  824& 0.4251&  0&  \nodata&  \nodata&  \nodata&  \nodata&  \nodata&  \nodata&  \nodata&   21.3&              2\\
  841& 0.3556&  0&  \nodata&  \nodata&  \nodata&  \nodata&  \nodata&  \nodata&  \nodata&   18.7&              2\\
  861& 0.4208&  0&  \nodata&  \nodata&  \nodata&  \nodata&  \nodata&  \nodata&  \nodata&   15.1&              2\\
  863& 0.7144&  0&  \nodata&  \nodata&  \nodata&  \nodata&  \nodata&  \nodata&  \nodata&   53.7&              1\\
  872& 0.4967&  0&  \nodata&  \nodata&  \nodata&  \nodata&  \nodata&  \nodata&  \nodata&   11.3&              1\\
  883& 0.8890&  1&  2.112&  2.160&  0.085&   0.21&   0.56&   0.24&    3.3&   18.6&              2\\
  899& 0.8881&  1&  2.181&  2.205&  0.036&   0.10&   0.63&   0.27&    3.9&   34.9&              1\\
  907& 0.7671&  0&  \nodata&  \nodata&  \nodata&  \nodata&  \nodata&  \nodata&  \nodata&    6.8&              2\\
  928& 0.3389&  0&  \nodata&  \nodata&  \nodata&  \nodata&  \nodata&  \nodata&  \nodata&   67.8&              1\\
  933& 0.9645&  0&  \nodata&  \nodata&  \nodata&  \nodata&  \nodata&  \nodata&  \nodata&   21.5&              1\\
  934& 0.4479&  0&  \nodata&  \nodata&  \nodata&  \nodata&  \nodata&  \nodata&  \nodata&   23.0&              2\\
  960& 0.7981&  0&  \nodata&  \nodata&  \nodata&  \nodata&  \nodata&  \nodata&  \nodata&   22.1&              2\\
  968& 0.9680&  0&  \nodata&  \nodata&  \nodata&  \nodata&  \nodata&  \nodata&  \nodata&   49.8&              1\\
  982& 0.0604&  0&  \nodata&  \nodata&  \nodata&  \nodata&  \nodata&  \nodata&  \nodata&   17.0&              2\\
  995& 0.9611&  0&  \nodata&  \nodata&  \nodata&  \nodata&  \nodata&  \nodata&  \nodata&   85.4&              1\\
  996& 0.9008&  1&  2.132&  2.157&  0.053&   0.21&   0.51&   0.29&   10.0&  120.8&              1\\
  999& 0.8930&  1&  2.004&  2.053&  0.118&   0.57&   0.43&   0.00&    3.7&   13.1&              2\\
 1001& 0.3377&  0&  \nodata&  \nodata&  \nodata&  \nodata&  \nodata&  \nodata&  \nodata&   10.2&              2\\
 1005& 0.8869&  1&  2.247&  2.272&  0.093&   0.59&   0.41&   0.00&    2.7&   23.8&              1\\
 1022& 0.9126&  1&  1.956&  1.980&  0.244&   1.00&   0.00&   0.00&    3.2&   12.3&              1\\
 1025& 0.9030&  1&  2.058&  2.107&  0.083&   0.10&   0.44&   0.46&    2.7&   17.6&              2\\
 1027& 1.0588&  0&  \nodata&  \nodata&  \nodata&  \nodata&  \nodata&  \nodata&  \nodata&   10.9&              2\\
 1029& 0.2300&  0&  \nodata&  \nodata&  \nodata&  \nodata&  \nodata&  \nodata&  \nodata&    9.5&              2\\
 1042& 0.7028&  0&  \nodata&  \nodata&  \nodata&  \nodata&  \nodata&  \nodata&  \nodata&    7.8&              2\\
 1047& 0.8980&  1&  2.102&  2.127&  0.052&   0.24&   0.41&   0.35&    3.3&   43.7&              1\\
 1057& 1.1810&  0&  \nodata&  \nodata&  \nodata&  \nodata&  \nodata&  \nodata&  \nodata&    6.9&              2\\
 1080& 0.5303&  0&  \nodata&  \nodata&  \nodata&  \nodata&  \nodata&  \nodata&  \nodata&   46.8&              1\\
 1083& 0.5341&  0&  \nodata&  \nodata&  \nodata&  \nodata&  \nodata&  \nodata&  \nodata&   37.0&              1\\
 1091& 0.8841&  1&  2.001&  2.050&  0.089&   0.46&   0.00&   0.54&    4.4&   22.0&              2\\
 1103& 0.7028&  0&  \nodata&  \nodata&  \nodata&  \nodata&  \nodata&  \nodata&  \nodata&   21.9&              1\\
 1157& 0.7669&  0&  \nodata&  \nodata&  \nodata&  \nodata&  \nodata&  \nodata&  \nodata&   11.8&              1\\
 1164& 0.8902&  1&  1.960&  2.009&  0.077&   0.21&   0.48&   0.31&    6.1&   28.5&              2\\
 1170& 0.8911&  1&  2.104&  2.129&  0.040&   0.00&   0.71&   0.29&    7.0&   79.1&              1\\
 1175& 0.9304&  0&  \nodata&  \nodata&  \nodata&  \nodata&  \nodata&  \nodata&  \nodata&   94.7&              1\\
 1182& 0.7340&  0&  \nodata&  \nodata&  \nodata&  \nodata&  \nodata&  \nodata&  \nodata&    5.4&              2\\
 1196& 0.9438&  0&  \nodata&  \nodata&  \nodata&  \nodata&  \nodata&  \nodata&  \nodata&    7.6&              2\\
 1199& 0.9029&  1&  2.226&  2.250&  0.032&   0.21&   0.26&   0.53&    5.6&   64.0&              1\\
 1251& 0.8928&  1&  1.966&  2.015&  0.081&   0.00&   0.90&   0.10&    3.9&   19.0&              2\\
 1252& 0.8997&  1&  2.207&  2.255&  0.080&   0.12&   0.57&   0.31&    1.2&   10.6&              2\\
 1253& 1.0600&  0&  \nodata&  \nodata&  \nodata&  \nodata&  \nodata&  \nodata&  \nodata&  \nodata&              1\\
 1254& 1.1236&  0&  \nodata&  \nodata&  \nodata&  \nodata&  \nodata&  \nodata&  \nodata&   31.8&              1\\

\enddata
\tablenotetext{a}{Adopted membership: 1 -- galaxy is a member of RXJ1226.9+3332; 
0 -- galaxy is not a member of RXJ1226.9+3332}
\tablenotetext{b}{Velocity dispersions corrected to a standard size aperture equivalent to a
circular aperture with diameter of 3.4 arcsec at the distance of the Coma cluster.}
\tablenotetext{c}{S/N per {\AA}ngstrom in the rest frame of the galaxy. The wavelength 
interval was 4100-4600 {\AA}, except when wavelength coverage did not allow this typically
due to the redshift of the galaxy the following intervals were used:
ID 178 -- 6500-7000 {\AA}.
ID 132, 247, 249, 982, 1029 and 1083 -- 5000-5500 {\AA};
ID 527, 1027, 1057 -- 3500-4000 {\AA};
ID 53, 153 -- 2500-3000 {\AA};
ID 754 -- 1500-2000 {\AA}.}
\tablenotetext{d}{Source of data -- (1) GCP, (2) GSA.}
\end{deluxetable*}

\begin{deluxetable*}{rrrrrrrrrrrrr}
\tablecaption{RXJ1226.9+3332: Line Indices and EW[\ion{O}{2}] for Cluster Members\label{tab-RXJ1226line} }
\tabletypesize{\scriptsize}
\tablewidth{0pc}
\tablehead{
\colhead{ID} & \colhead{CN3883}& \colhead{CaHK}& \colhead{D4000}
& \colhead{H$\delta_A$}& \colhead{CN$_1$}& \colhead{CN$_2$} 
& \colhead{G4300}& \colhead{H$\gamma_A$}& \colhead{Fe4383}& \colhead{C4668} & \colhead{EW [\ion{O}{2}]}
} 
\startdata
 38& \nodata& 15.34& 1.707& \nodata& 0.071& 0.082& \nodata& \nodata& \nodata& \nodata& 10.2\\
 38& \nodata& 0.92& 0.010& \nodata& 0.014& 0.017& \nodata& \nodata& \nodata& \nodata& 1.0\\
 55& 0.237& 21.42& 2.000& 0.36& 0.099& 0.122& 4.89& -3.04& 4.41& 6.33& \nodata\\
 55& 0.003& 0.16& 0.002& 0.09& 0.003& 0.003& 0.06& 0.08& 0.14& 0.20& \nodata\\
 56& 0.173& 18.70& 1.811& 1.24& 0.033& 0.055& 4.04& -1.87& 3.71& 4.15& 3.2\\
 56& 0.004& 0.22& 0.003& 0.12& 0.003& 0.004& 0.09& 0.10& 0.18& 0.28& 0.3\\
 104& 0.245& 22.07& 2.008& 0.26& 0.145& 0.153& 4.80& -2.76& 4.48& 6.84& 3.5\\
 104& 0.006& 0.29& 0.004& 0.17& 0.005& 0.005& 0.11& 0.14& 0.24& 0.33& 0.5\\
 122& 0.143& 18.99& 1.829& 3.73& \nodata& 0.026& 2.60& -0.38& 3.42& 5.35& 6.9\\
 122& 0.006& 0.31& 0.004& 0.15& \nodata& 0.006& 0.13& 0.14& 0.24& 0.35& 1.2\\
 229& 0.261& \nodata& 1.988& 0.41& \nodata& 0.043& 4.98& \nodata& \nodata& \nodata& \nodata\\
 229& 0.009& \nodata& 0.005& 0.24& \nodata& 0.008& 0.18& \nodata& \nodata& \nodata& \nodata\\
 293& 0.244& 20.42& 2.102& 0.50& 0.085& 0.109& 5.46& -3.94& 5.52& 4.81& \nodata\\
 293& 0.008& 0.41& 0.006& 0.21& 0.006& 0.008& 0.16& 0.20& 0.28& 0.46& \nodata\\
 295& 0.139& 16.72& 1.784& 1.48& 0.030& 0.050& 3.65& -1.78& 2.91& 6.40& 8.7\\
 295& 0.006& 0.34& 0.004& 0.18& 0.005& 0.006& 0.14& 0.17& 0.30& 0.36& 1.2\\
 309& 0.221& 23.67& 1.947& \nodata& \nodata& \nodata& 5.56& \nodata& \nodata& \nodata& \nodata\\
 309& 0.018& 0.90& 0.011& \nodata& \nodata& \nodata& 0.40& \nodata& \nodata& \nodata& \nodata\\
 310& 0.278& \nodata& 2.045& -1.17& 0.118& 0.139& 4.70& \nodata& \nodata& \nodata& \nodata\\
 310& 0.009& \nodata& 0.006& 0.26& 0.008& 0.009& 0.19& \nodata& \nodata& \nodata& \nodata\\
 423& 0.166& \nodata& 2.119& \nodata& \nodata& \nodata& 4.68& \nodata& \nodata& \nodata& \nodata\\
 423& 0.011& \nodata& 0.008& \nodata& \nodata& \nodata& 0.23& \nodata& \nodata& \nodata& \nodata\\
 441& 0.192& 21.53& 1.948& 1.48& \nodata& 0.029& 4.52& -2.40& 4.73& 3.67& \nodata\\
 441& 0.009& 0.44& 0.006& 0.23& \nodata& 0.008& 0.17& 0.22& 0.37& 0.54& \nodata\\
 446& 0.079& \nodata& 1.865& 1.41& \nodata& \nodata& 3.23& \nodata& \nodata& \nodata& \nodata\\
 446& 0.008& \nodata& 0.005& 0.23& \nodata& \nodata& 0.20& \nodata& \nodata& \nodata& \nodata\\
 452& 0.136& 12.50& 1.760& 4.72& \nodata& \nodata& 0.48& 3.23& 2.60& 3.17& 10.4\\
 452& 0.008& 0.45& 0.005& 0.20& \nodata& \nodata& 0.20& 0.19& 0.32& 0.51& 0.8\\
 462& 0.182& \nodata& 1.896& \nodata& \nodata& \nodata& 5.27& \nodata& \nodata& \nodata& \nodata\\
 462& 0.012& \nodata& 0.008& \nodata& \nodata& \nodata& 0.26& \nodata& \nodata& \nodata& \nodata\\
 470& \nodata& 25.34& 2.067& \nodata& \nodata& \nodata& 3.85& \nodata& \nodata& \nodata& \nodata\\
 470& \nodata& 0.85& 0.012& \nodata& \nodata& \nodata& 0.32& \nodata& \nodata& \nodata& \nodata\\
 491& 0.248& 22.59& 2.037& -0.26& 0.090& 0.120& 4.26& -3.36& 4.08& 6.35& \nodata\\
 491& 0.004& 0.22& 0.003& 0.11& 0.003& 0.004& 0.10& 0.10& 0.15& 0.23& \nodata\\
 512& \nodata& 21.01& 2.075& \nodata& \nodata& \nodata& 4.34& \nodata& \nodata& \nodata& \nodata\\
 512& \nodata& 0.83& 0.010& \nodata& \nodata& \nodata& 0.29& \nodata& \nodata& \nodata& \nodata\\
 528& 0.106& 14.62& 1.791& \nodata& \nodata& \nodata& \nodata& \nodata& \nodata& \nodata& 3.1\\
 528& 0.012& 0.72& 0.008& \nodata& \nodata& \nodata& \nodata& \nodata& \nodata& \nodata& 0.4\\
 529& 0.260& 23.46& 2.065& -1.24& 0.102& 0.125& 5.44& -4.15& 3.68& 7.92& \nodata\\
 529& 0.003& 0.17& 0.002& 0.10& 0.003& 0.003& 0.07& 0.09& 0.15& 0.18& \nodata\\
 534& \nodata& 20.59& 2.087& \nodata& \nodata& \nodata& 4.95& \nodata& \nodata& \nodata& \nodata\\
 534& \nodata& 0.68& 0.008& \nodata& \nodata& \nodata& 0.27& \nodata& \nodata& \nodata& \nodata\\
 547& \nodata& 7.00& 1.560& \nodata& \nodata& \nodata& \nodata& \nodata& \nodata& \nodata& 24.0\\
 547& \nodata& 1.28& 0.011& \nodata& \nodata& \nodata& \nodata& \nodata& \nodata& \nodata& 1.6\\
 563& 0.315& 23.38& 2.212& -0.66& 0.133& 0.161& 5.23& -4.53& 4.59& 8.86& \nodata\\
 563& 0.003& 0.14& 0.002& 0.08& 0.002& 0.002& 0.06& 0.07& 0.11& 0.14& \nodata\\
 592& \nodata& \nodata& \nodata& \nodata& \nodata& \nodata& \nodata& \nodata& \nodata& \nodata& 3.6\\
 592& \nodata& \nodata& \nodata& \nodata& \nodata& \nodata& \nodata& \nodata& \nodata& \nodata& 0.3\\
 593& \nodata& 20.54& 2.237& \nodata& 0.094& 0.154& 4.53& \nodata& \nodata& \nodata& \nodata\\
 593& \nodata& 0.91& 0.012& \nodata& 0.014& 0.017& 0.36& \nodata& \nodata& \nodata& \nodata\\
 602& 0.102& 16.24& 1.703& 4.43& \nodata& \nodata& 1.42& 2.16& 3.86& 3.64& 11.5\\
 602& 0.005& 0.32& 0.003& 0.14& \nodata& \nodata& 0.15& 0.14& 0.22& 0.39& 0.6\\
 608& 0.210& 22.07& 1.973& \nodata& \nodata& \nodata& 4.96& -4.12& 5.81& \nodata& 10.7\\
 608& 0.006& 0.32& 0.004& \nodata& \nodata& \nodata& 0.14& 0.15& 0.20& \nodata& 1.2\\
 641& 0.182& 22.00& 1.889& \nodata& \nodata& \nodata& 4.55& \nodata& \nodata& \nodata& \nodata\\
 641& 0.012& 0.63& 0.008& \nodata& \nodata& \nodata& 0.23& \nodata& \nodata& \nodata& \nodata\\
 647& 0.225& 24.43& 2.210& \nodata& \nodata& \nodata& 5.80& \nodata& \nodata& \nodata& \nodata\\
 647& 0.026& 1.35& 0.020& \nodata& \nodata& \nodata& 0.49& \nodata& \nodata& \nodata& \nodata\\
 648& 0.237& 19.13& 2.047& 1.10& 0.050& 0.073& 4.45& -2.77& 3.26& 4.25& 2.7\\
 648& 0.005& 0.25& 0.003& 0.12& 0.004& 0.004& 0.10& 0.11& 0.16& 0.27& 0.3\\
 650& \nodata& 17.99& 1.755& \nodata& \nodata& \nodata& 4.62& \nodata& \nodata& \nodata& \nodata\\
 650& \nodata& 0.56& 0.006& \nodata& \nodata& \nodata& 0.21& \nodata& \nodata& \nodata& \nodata\\
 675& 0.176& 21.29& 1.949& 2.53& \nodata& 0.035& 3.92& -2.14& 3.20& 1.95& 3.7\\
 675& 0.006& 0.32& 0.004& 0.18& \nodata& 0.006& 0.13& 0.17& 0.29& 0.43& 0.4\\
 689& 0.299& \nodata& 2.173& \nodata& \nodata& \nodata& 2.88& \nodata& \nodata& \nodata& \nodata\\
 689& 0.026& \nodata& 0.019& \nodata& \nodata& \nodata& 0.51& \nodata& \nodata& \nodata& \nodata\\
\enddata
\tablecomments{The indices have been corrected for galaxy velocity dispersion and aperture corrected. The second line for each galaxy lists the uncertainties determined from the local S/N. 
ID=557 and ID=630 have no reliable line index measurements.}
\end{deluxetable*}

\begin{deluxetable*}{rrrrrrrrrrrrr}
\tablecaption{RXJ1226.9+3332: \em -- Continued}
\tablenum{24}
\tabletypesize{\scriptsize}
\tablewidth{0pc}
\tablehead{
\colhead{ID} & \colhead{CN3883}& \colhead{CaHK}& \colhead{D4000}
& \colhead{H$\delta_A$}& \colhead{CN$_1$}& \colhead{CN$_2$} 
& \colhead{G4300}& \colhead{H$\gamma_A$}& \colhead{Fe4383}& \colhead{C4668} & \colhead{EW [\ion{O}{2}]}
} 
\startdata
 703& 0.312& 23.82& 2.170& -0.61& 0.108& 0.138& 5.68& -4.47& 3.40& 7.05& \nodata\\
 703& 0.004& 0.18& 0.003& 0.11& 0.003& 0.003& 0.07& 0.10& 0.16& 0.22& \nodata\\
 709& 0.230& 21.09& 2.007& 1.36& \nodata& 0.035& 3.73& -0.26& 4.72& \nodata& \nodata\\
 709& 0.007& 0.36& 0.005& 0.22& \nodata& 0.006& 0.14& 0.19& 0.30& \nodata& \nodata\\
 711& 0.067& 9.35& 1.461& 7.06& \nodata& \nodata& -0.51& 7.76& \nodata& \nodata& \nodata\\
 711& 0.012& 0.84& 0.007& 0.34& \nodata& \nodata& 0.42& 0.35& \nodata& \nodata& \nodata\\
 757& 0.306& 26.79& 2.007& \nodata& 0.070& \nodata& 3.93& \nodata& \nodata& \nodata& \nodata\\
 757& 0.027& 1.31& 0.019& \nodata& 0.023& \nodata& 0.61& \nodata& \nodata& \nodata& \nodata\\
 760& 0.229& 21.85& 2.014& 1.19& 0.056& 0.078& 4.80& -3.49& 4.62& 7.94& \nodata\\
 760& 0.004& 0.19& 0.003& 0.10& 0.003& 0.004& 0.08& 0.10& 0.15& 0.21& \nodata\\
 801& 0.180& 20.33& 1.964& \nodata& \nodata& \nodata& 4.58& \nodata& \nodata& \nodata& \nodata\\
 801& 0.008& 0.41& 0.005& \nodata& \nodata& \nodata& 0.17& \nodata& \nodata& \nodata& \nodata\\
 883& \nodata& \nodata& 1.875& \nodata& \nodata& \nodata& 4.50& \nodata& \nodata& \nodata& \nodata\\
 883& \nodata& \nodata& 0.008& \nodata& \nodata& \nodata& 0.33& \nodata& \nodata& \nodata& \nodata\\
 899& \nodata& 21.27& \nodata& 1.89& 0.092& 0.124& 5.02& -5.26& 6.45& 5.03& \nodata\\
 899& \nodata& 0.45& \nodata& 0.23& 0.007& 0.009& 0.18& 0.22& 0.33& 0.50& \nodata\\
 996& 0.221& 21.83& 1.994& 2.12& \nodata& 0.048& 4.41& -0.72& 2.18& 5.12& \nodata\\
 996& 0.002& 0.12& 0.002& 0.07& \nodata& 0.002& 0.05& 0.07& 0.12& 0.17& \nodata\\
 999& 0.120& 19.27& 1.646& \nodata& \nodata& \nodata& 2.11& \nodata& \nodata& \nodata& \nodata\\
 999& 0.019& 1.02& 0.010& \nodata& \nodata& \nodata& 0.45& \nodata& \nodata& \nodata& \nodata\\
 1005& \nodata& 8.32& 1.367& 3.93& \nodata& \nodata& 1.02& -1.44& 1.89& 1.25& 33.6\\
 1005& \nodata& 0.58& 0.005& 0.28& \nodata& \nodata& 0.27& 0.30& 0.55& 0.98& 6.7\\
 1022& \nodata& \nodata& 1.408& \nodata& \nodata& \nodata& \nodata& \nodata& \nodata& \nodata& 58.6\\
 1022& \nodata& \nodata& 0.009& \nodata& \nodata& \nodata& \nodata& \nodata& \nodata& \nodata& 4.4\\
 1025& \nodata& 19.24& \nodata& \nodata& \nodata& \nodata& 4.31& \nodata& \nodata& \nodata& \nodata\\
 1025& \nodata& 1.01& \nodata& \nodata& \nodata& \nodata& 0.34& \nodata& \nodata& \nodata& \nodata\\
 1047& 0.217& 22.59& 1.975& 1.87& \nodata& \nodata& 3.85& -0.89& 2.76& 5.83& \nodata\\
 1047& 0.007& 0.37& 0.005& 0.21& \nodata& \nodata& 0.15& 0.19& 0.34& 0.49& \nodata\\
 1091& 0.214& \nodata& 1.690& \nodata& 0.032& 0.033& 2.85& \nodata& \nodata& \nodata& 8.2\\
 1091& 0.011& \nodata& 0.007& \nodata& 0.011& 0.013& 0.29& \nodata& \nodata& \nodata& 1.0\\
 1164& 0.151& \nodata& 2.072& \nodata& \nodata& \nodata& 4.90& \nodata& \nodata& \nodata& \nodata\\
 1164& 0.010& \nodata& 0.007& \nodata& \nodata& \nodata& 0.21& \nodata& \nodata& \nodata& \nodata\\
 1170& 0.207& 20.47& 1.956& 1.28& 0.051& 0.074& 4.12& -2.03& 4.22& 8.45& \nodata\\
 1170& 0.004& 0.20& 0.003& 0.10& 0.003& 0.004& 0.08& 0.10& 0.16& 0.22& \nodata\\
 1199& 0.258& 22.55& 1.915& 0.11& 0.088& 0.101& 4.55& -2.43& 3.83& 4.77& \nodata\\
 1199& 0.005& 0.25& 0.003& 0.15& 0.004& 0.004& 0.10& 0.13& 0.21& 0.30& \nodata\\
 1251& 0.045& 12.13& 1.935& \nodata& \nodata& \nodata& 3.47& \nodata& \nodata& \nodata& \nodata\\
 1251& 0.009& 0.91& 0.008& \nodata& \nodata& \nodata& 0.32& \nodata& \nodata& \nodata& \nodata\\
\enddata
\tablecomments{The indices have been corrected for galaxy velocity dispersion and aperture corrected. The second line for each galaxy lists the uncertainties determined from the local S/N. ID=557 and ID=630 have no reliable line index measurements.}
\end{deluxetable*}

\clearpage

\begin{figure*}
\epsfxsize 17.5cm
\epsfbox{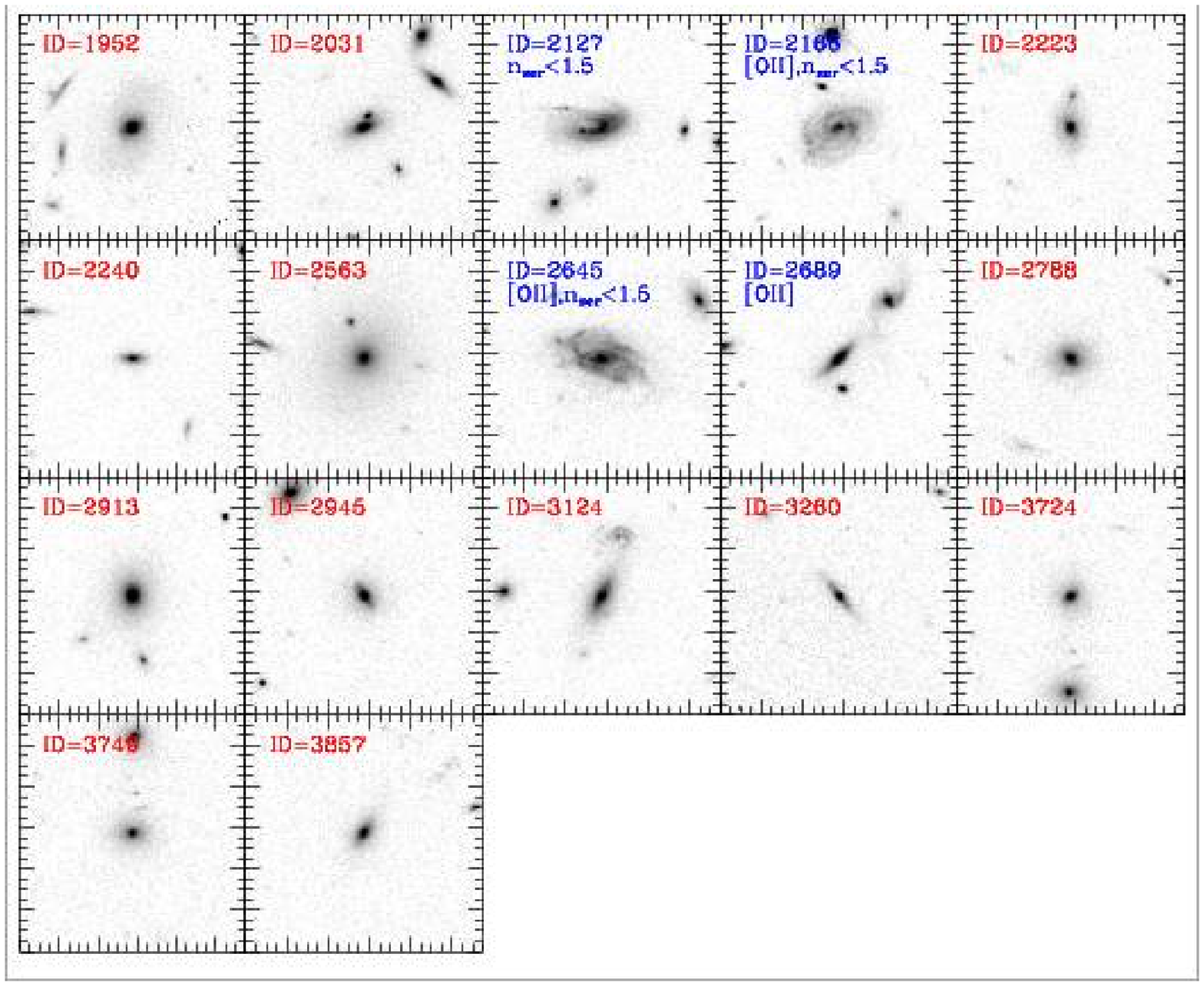}
\center{Fig.\ 33 -- {\em Continued.}}
\end{figure*}

\begin{figure*}
\epsfxsize 17.5cm
\epsfbox{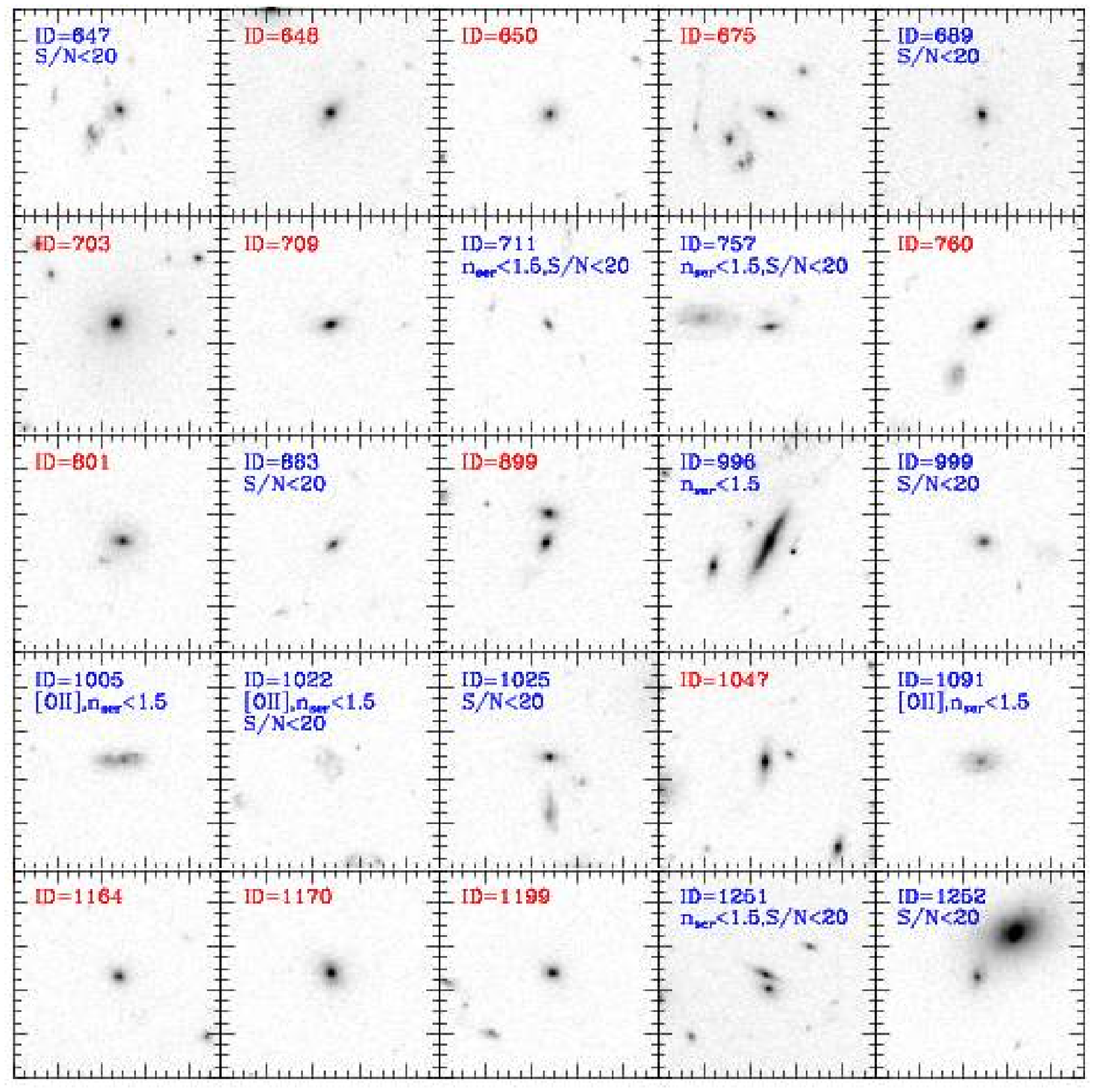}
\center{Fig.\ 34 -- {\em Continued.}}
\end{figure*}

\clearpage

\begin{figure*}
\epsfxsize 16.5cm
\epsfbox{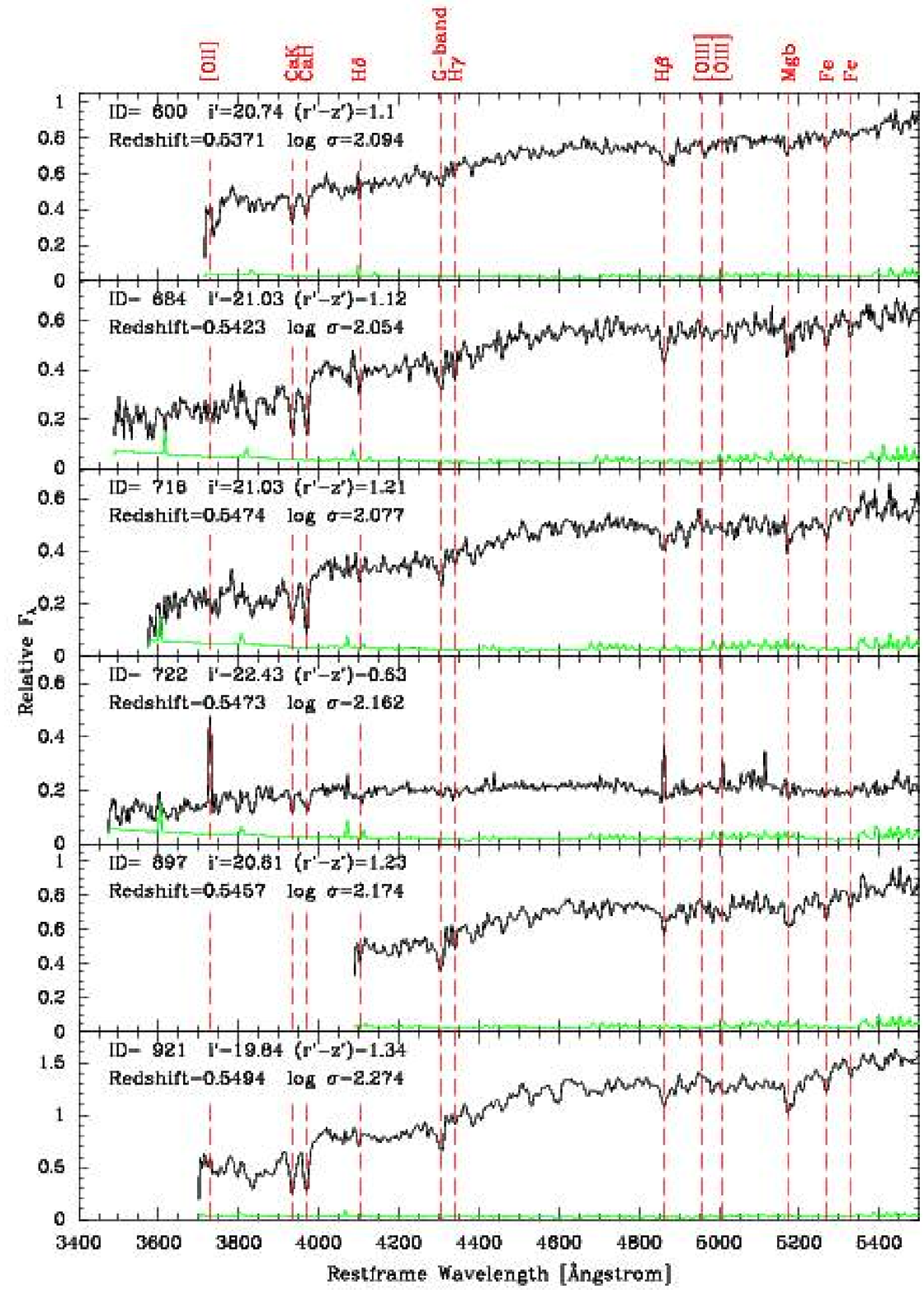}
\vspace{-1cm}
\center{Fig.\ 35 -- {\em Continued.}}
\end{figure*}

\begin{figure*}
\epsfxsize 16.5cm
\epsfbox{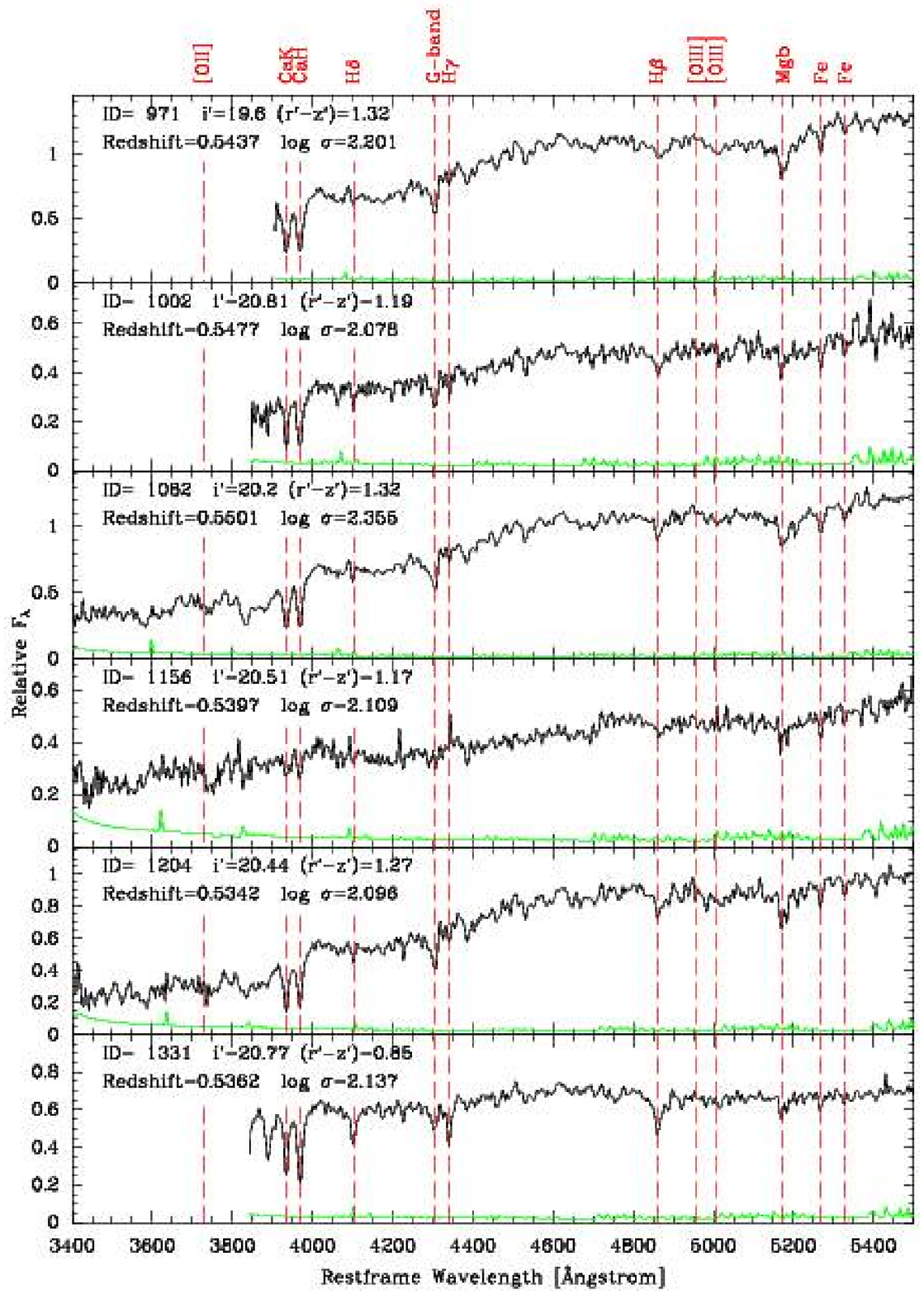}
\vspace{-1cm}
\center{Fig.\ 35 -- {\em Continued.}}
\end{figure*}

\begin{figure*}
\epsfxsize 16.5cm
\epsfbox{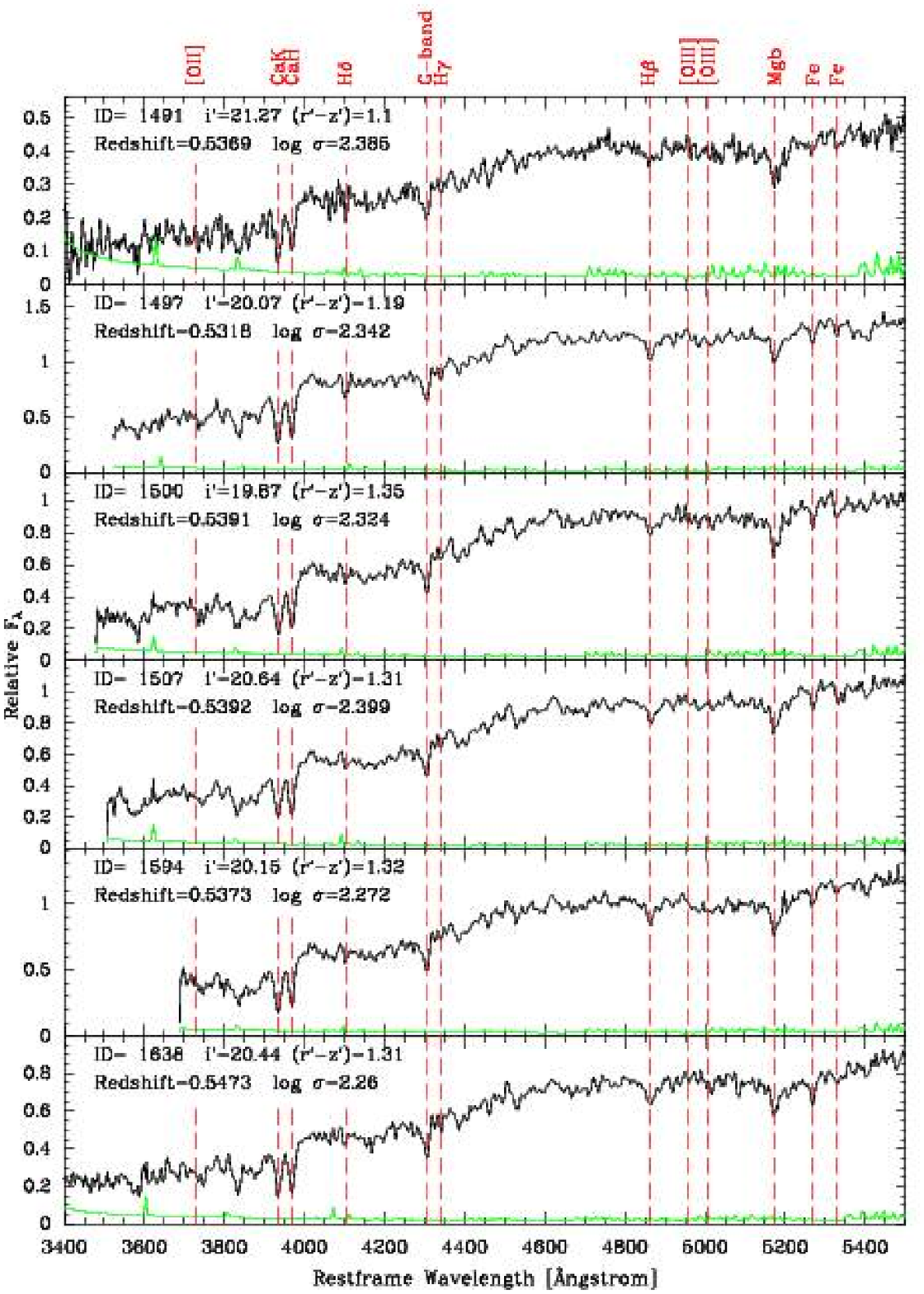}
\vspace{-1cm}
\center{Fig.\ 35 -- {\em Continued.}}
\end{figure*}

\begin{figure*}
\epsfxsize 16.5cm
\epsfbox{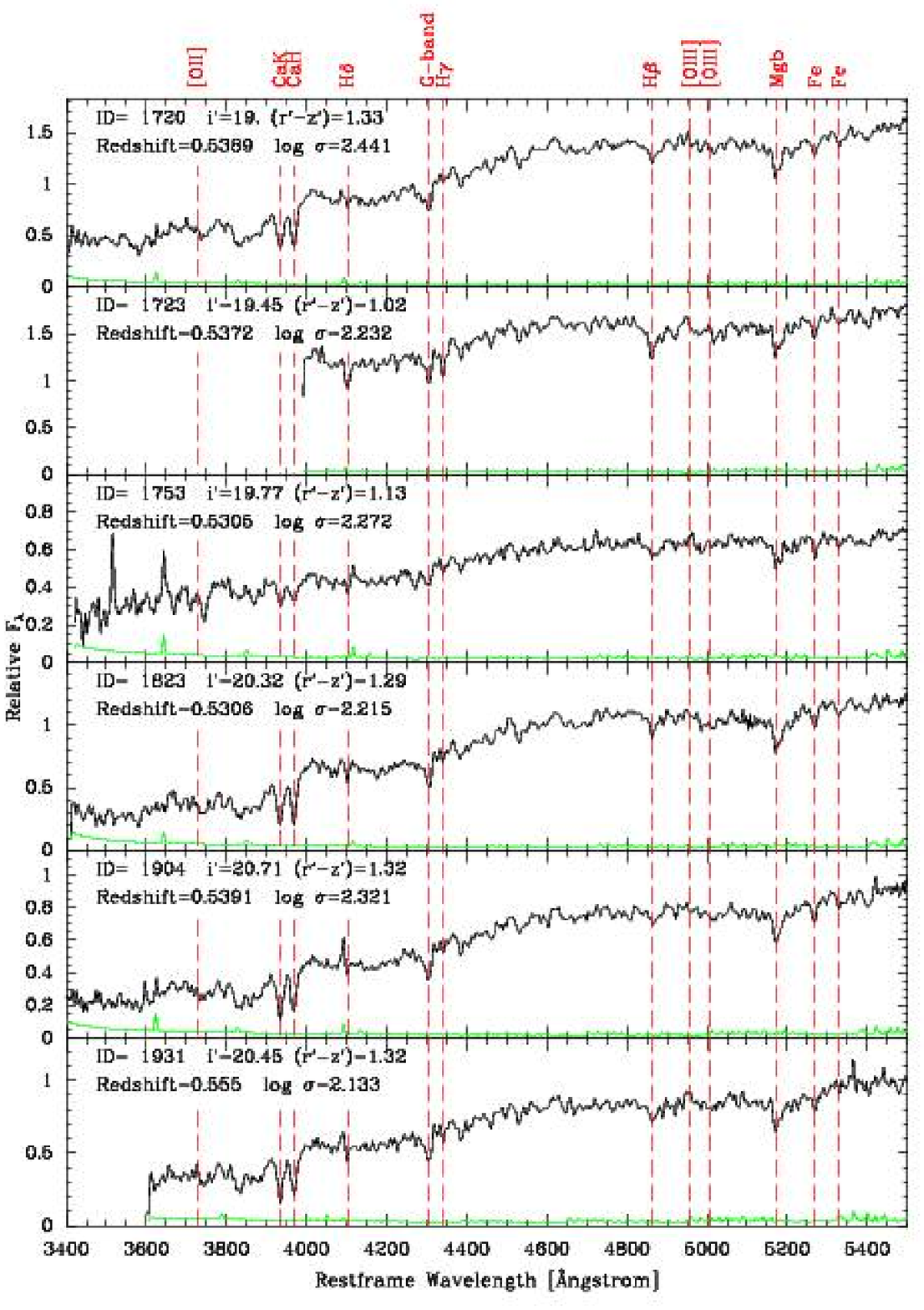}
\vspace{-1cm}
\center{Fig.\ 35 -- {\em Continued.}}
\end{figure*}

\begin{figure*}
\epsfxsize 16.5cm
\epsfbox{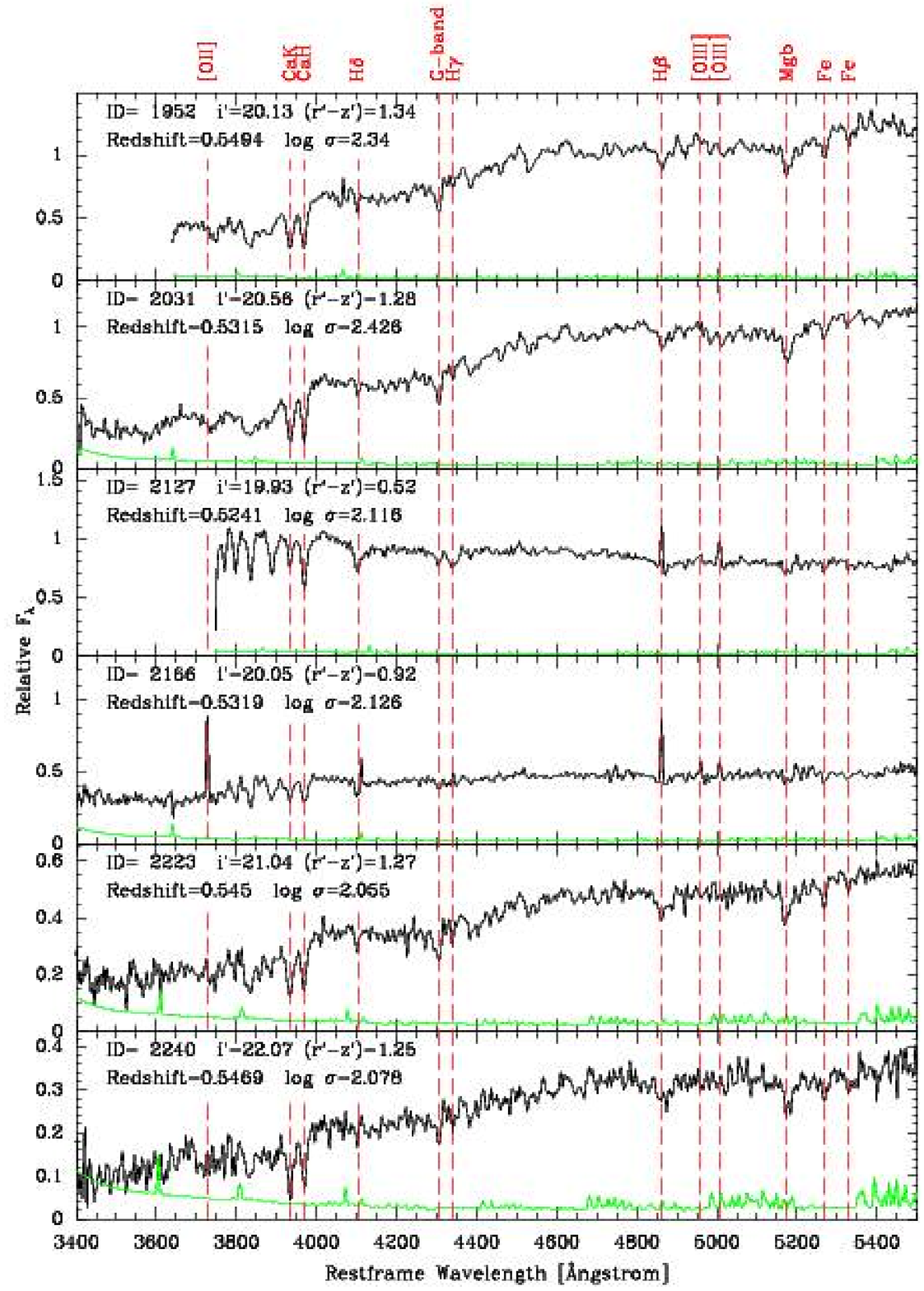}
\vspace{-1cm}
\center{Fig.\ 35 -- {\em Continued.}}
\end{figure*}

\begin{figure*}
\epsfxsize 16.5cm
\epsfbox{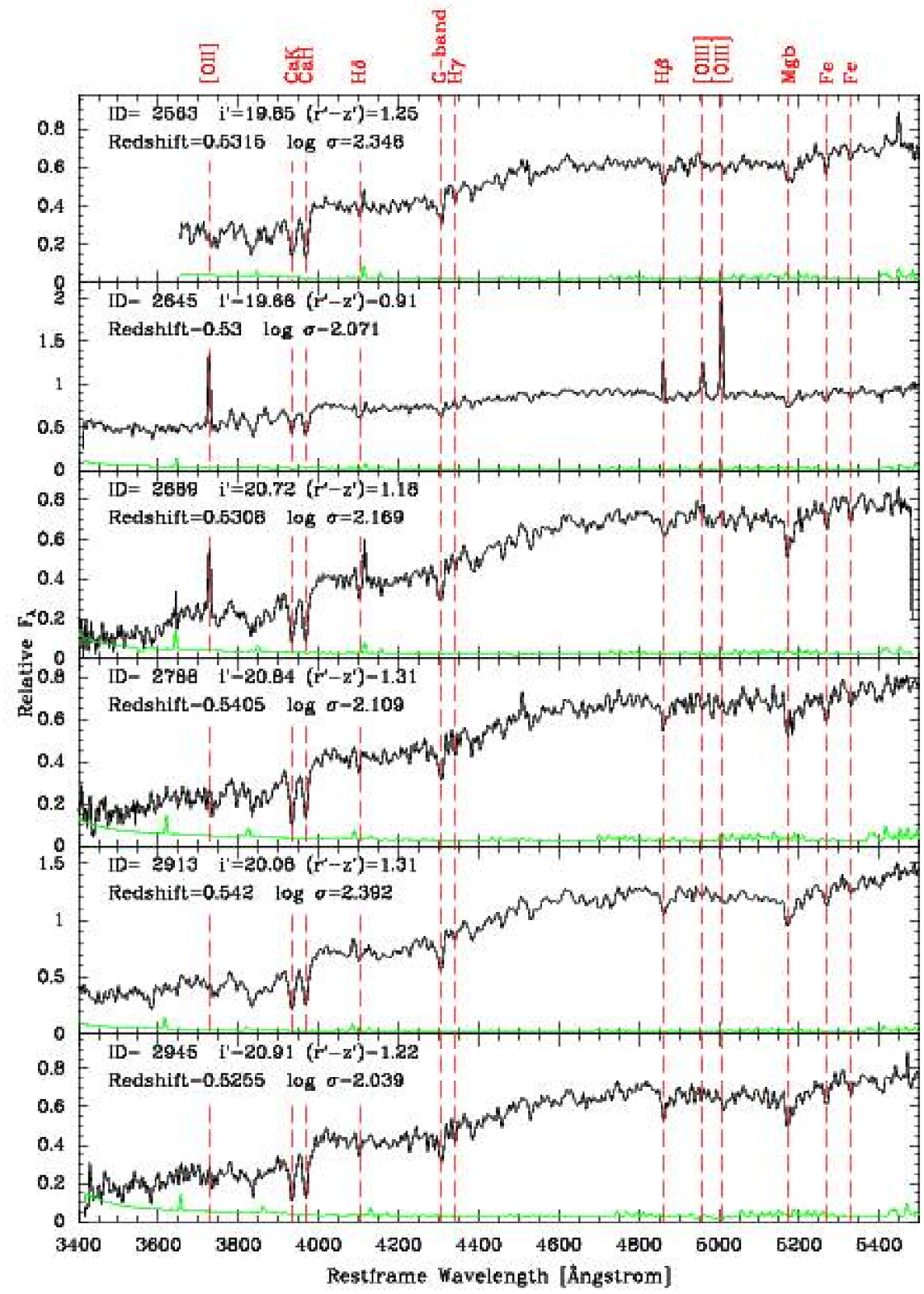}
\vspace{-1cm}
\center{Fig.\ 35 -- {\em Continued.}}
\end{figure*}

\begin{figure*}
\epsfxsize 16.5cm
\epsfbox{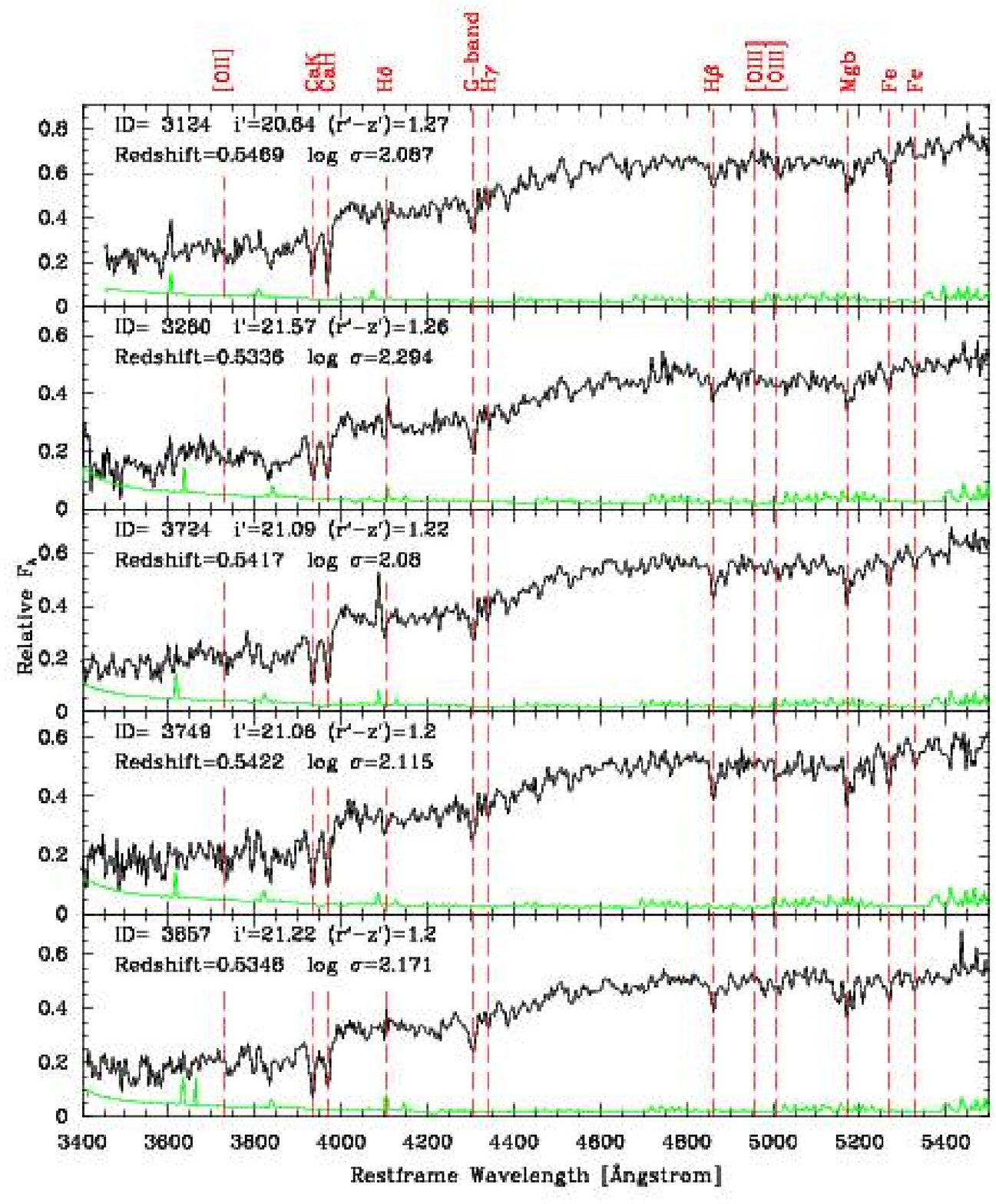}
\vspace{-1cm}
\center{Fig.\ 35 -- {\em Continued.}}
\end{figure*}

\clearpage

\begin{figure*}
\epsfxsize 16.5cm
\epsfbox{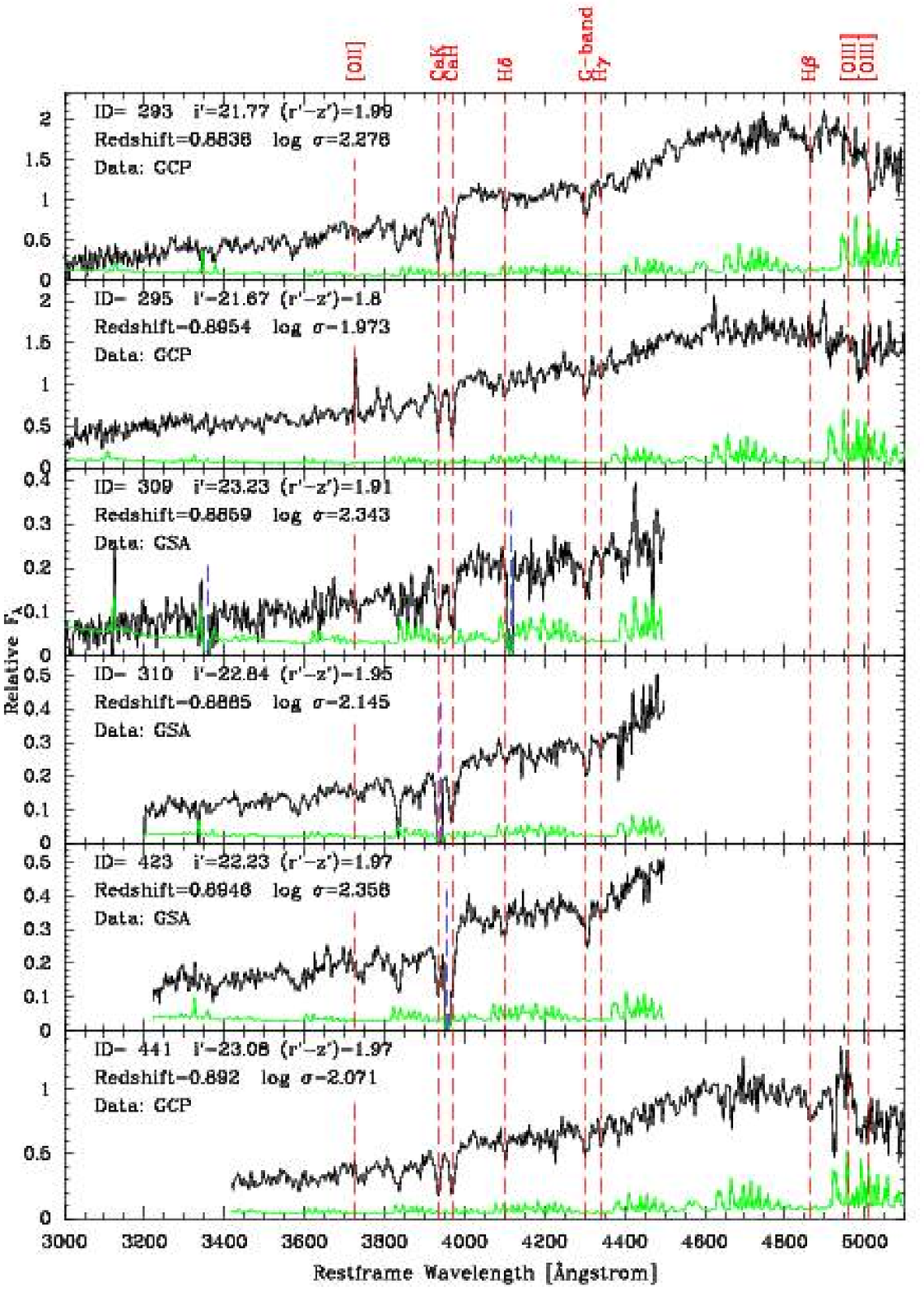}
\vspace{-1cm}
\center{Fig.\ 36 -- {\em Continued.}}
\end{figure*}

\begin{figure*}
\epsfxsize 16.5cm
\epsfbox{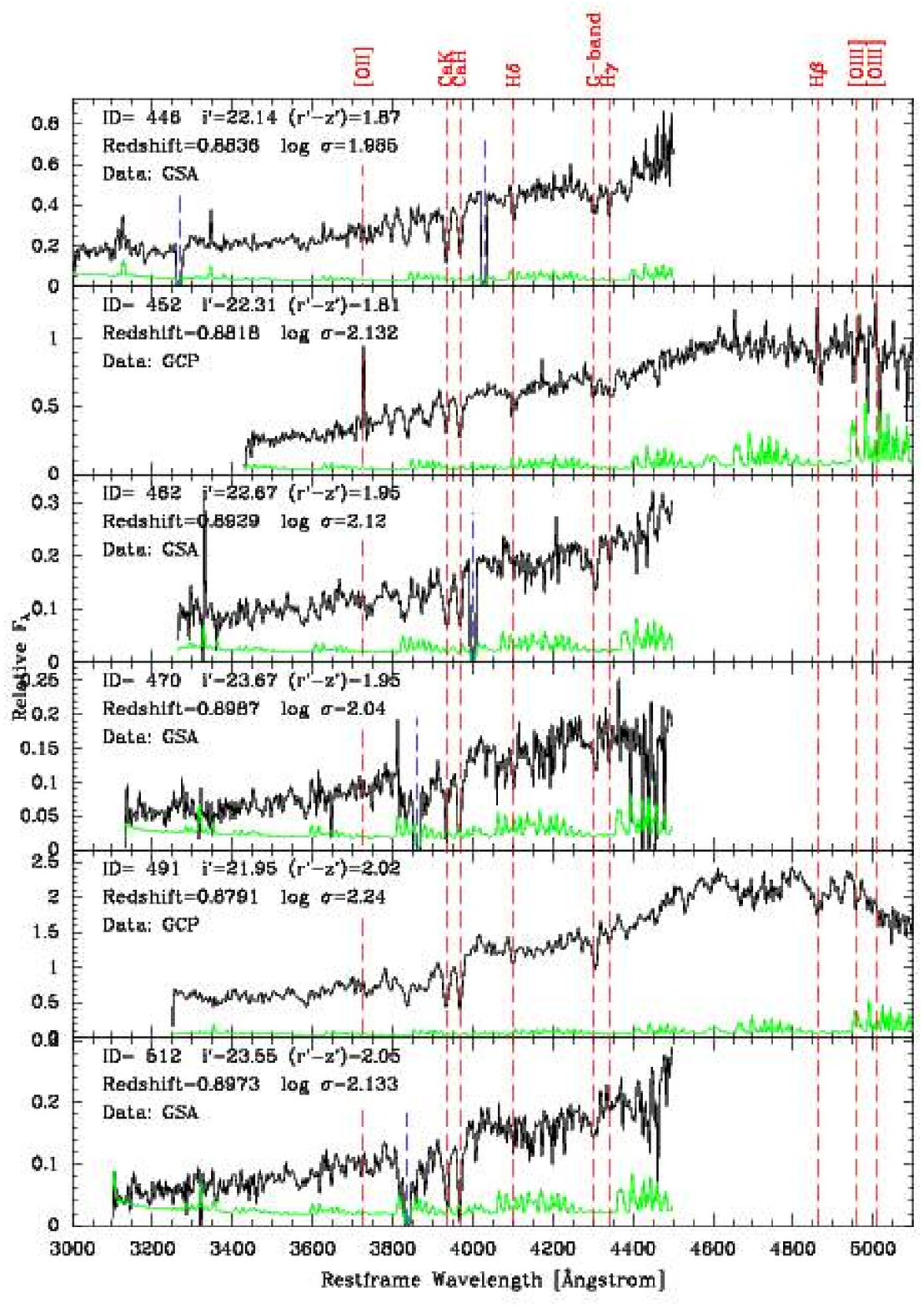}
\vspace{-1cm}
\center{Fig.\ 36 -- {\em Continued.}}
\end{figure*}

\begin{figure*}
\epsfxsize 16.5cm
\epsfbox{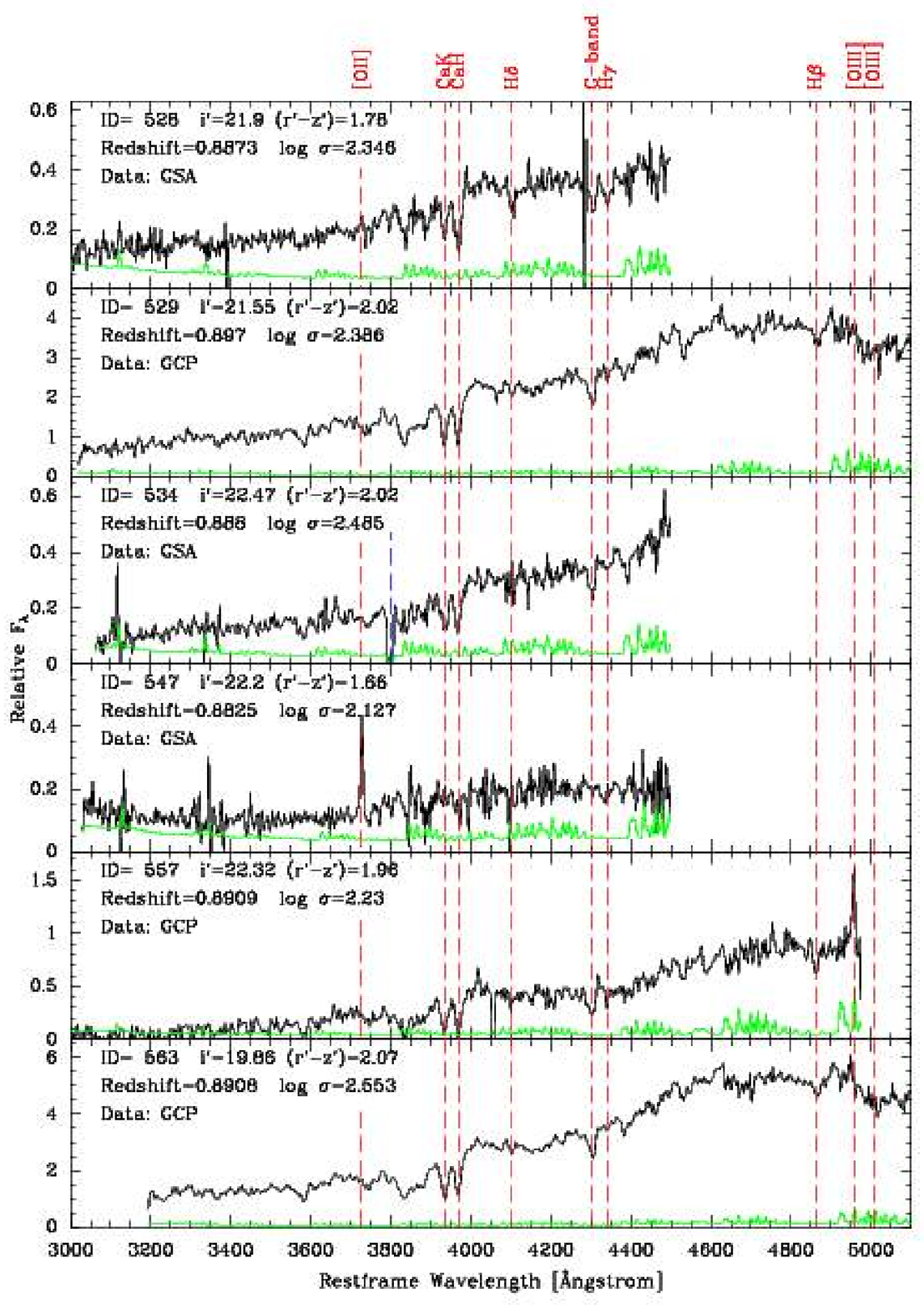}
\vspace{-1cm}
\center{Fig.\ 36 -- {\em Continued.}}
\end{figure*}

\begin{figure*}
\epsfxsize 16.5cm
\epsfbox{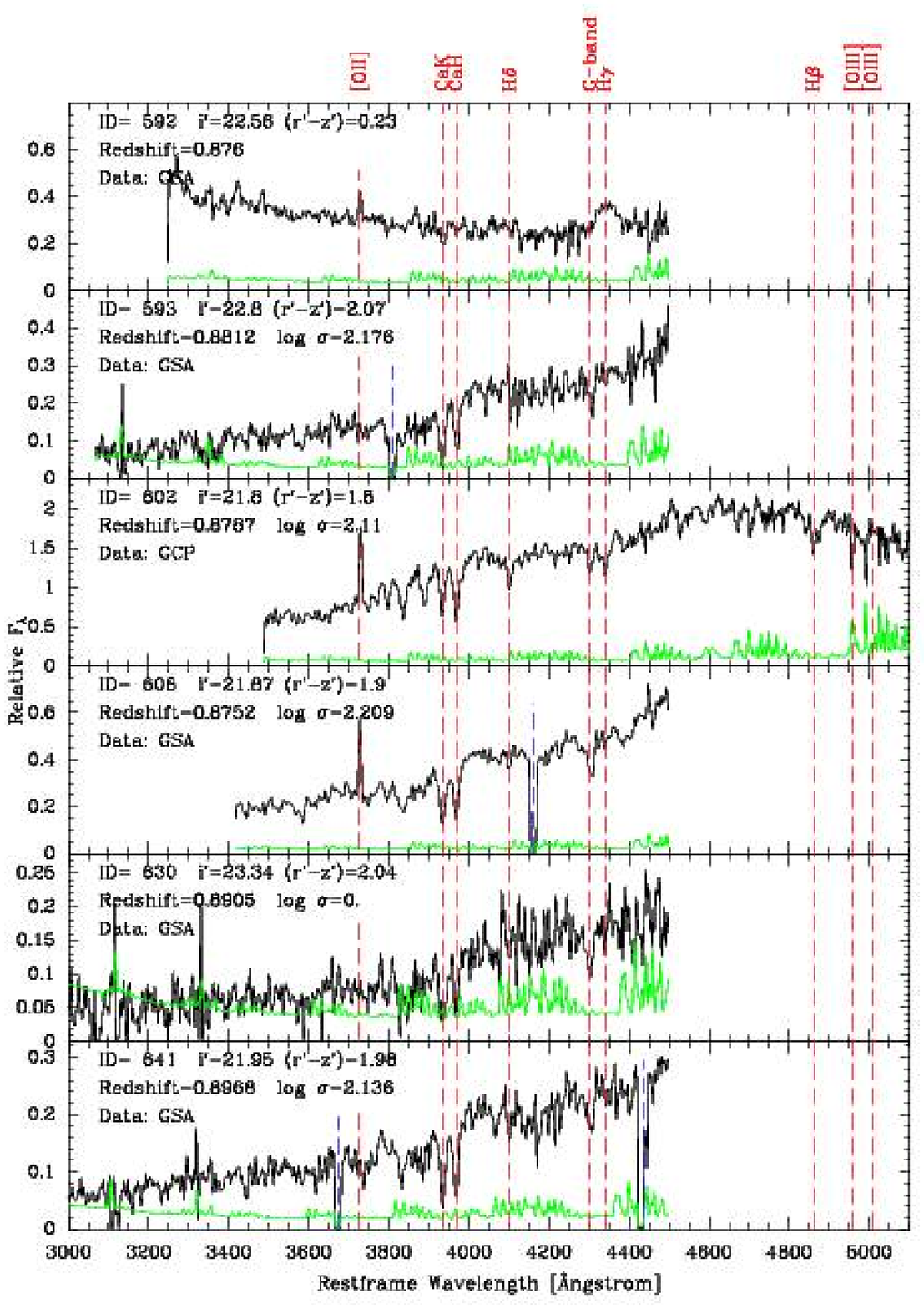}
\vspace{-1cm}
\center{Fig.\ 36 -- {\em Continued.}}
\end{figure*}

\begin{figure*}
\epsfxsize 16.5cm
\epsfbox{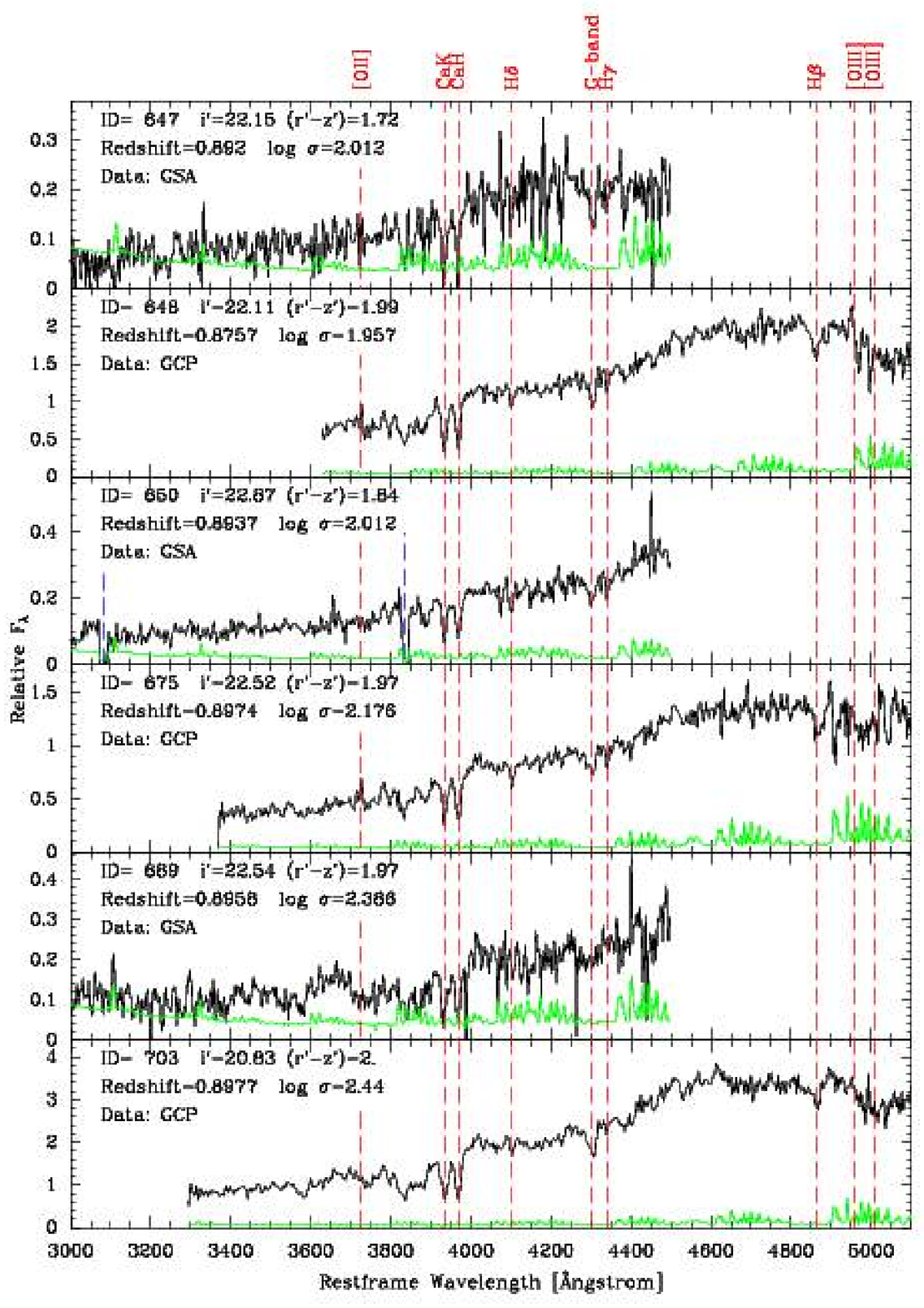}
\vspace{-1cm}
\center{Fig.\ 36 -- {\em Continued.}}
\end{figure*}

\begin{figure*}
\epsfxsize 16.5cm
\epsfbox{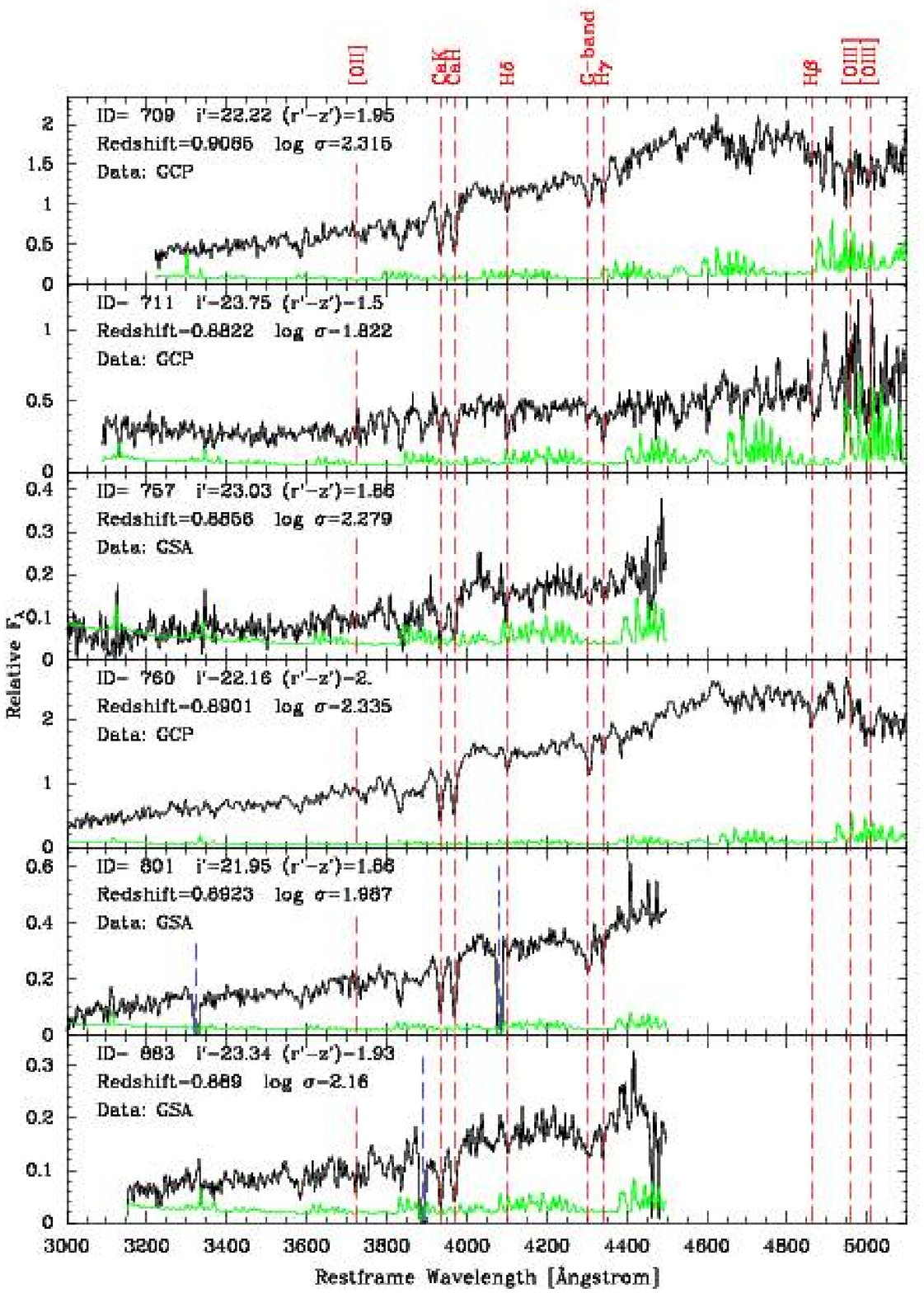}
\vspace{-1cm}
\center{Fig.\ 36 -- {\em Continued.}}
\end{figure*}

\begin{figure*}
\epsfxsize 16.5cm
\epsfbox{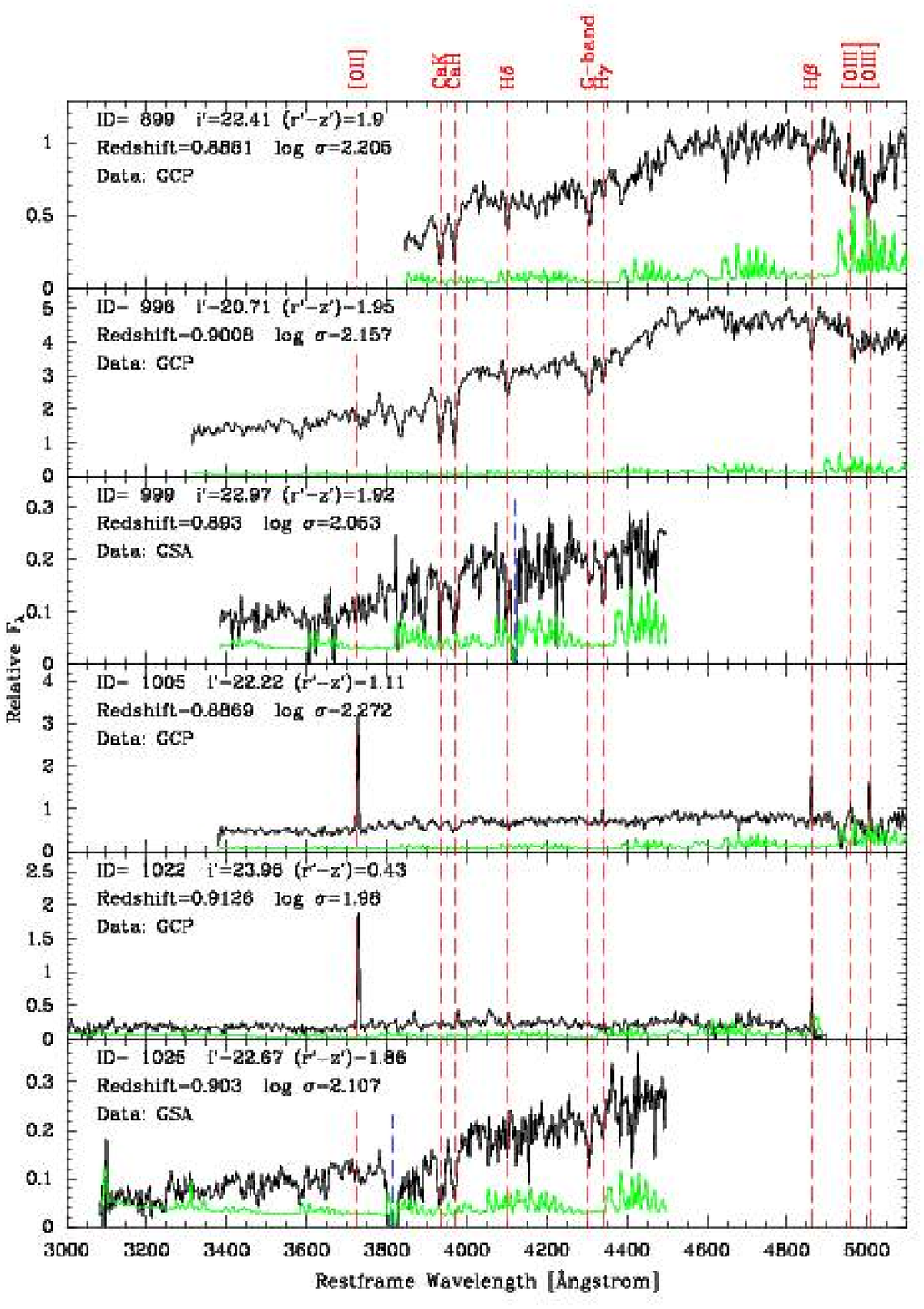}
\vspace{-1cm}
\center{Fig.\ 36 -- {\em Continued.}}
\end{figure*}

\begin{figure*}
\epsfxsize 16.5cm
\epsfbox{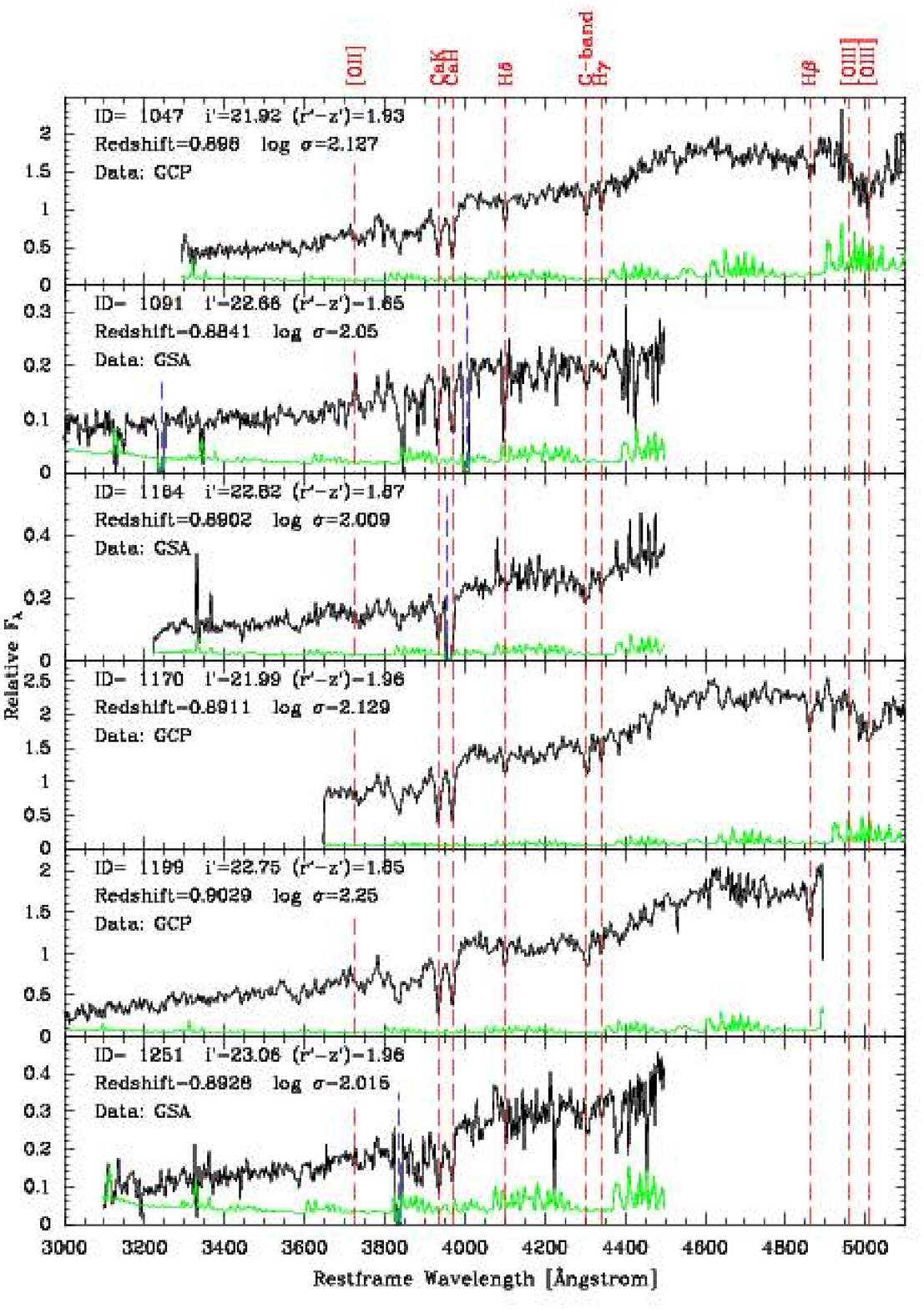}
\vspace{-1cm}
\center{Fig.\ 36 -- {\em Continued.}}
\end{figure*}

\begin{figure*}
\epsfxsize 16.5cm
\epsfbox{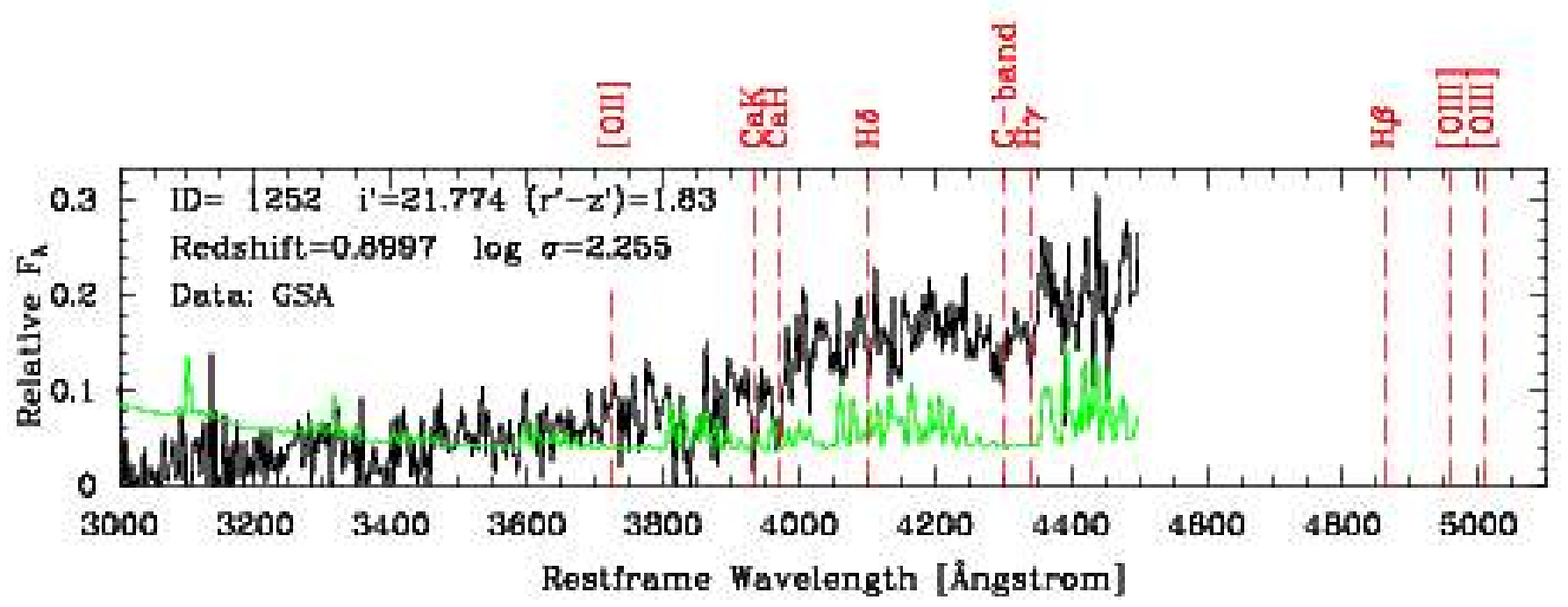}
\center{Fig.\ 36 -- {\em Continued.}}
\end{figure*}

\end{document}